\documentclass[a4paper,11pt]{article}
\pdfoutput=1 % if your are submitting a pdflatex (i.e. if you have
\usepackage{jheppub} % for details on the use of the package, please
\usepackage[T1]{fontenc} % if needed

\def\a{\alpha}
\def\b{\beta}
\def\d{\delta}
\def\D{\Delta}

\def\g{\gamma}
\def\G{\Gamma}
\def\e{\epsilon}
\def\et{\eta}

\def\l{\lambda}
\def\L{\Lambda}
\def\m{\mu}
\def\n{\nu}
\def\s{\sigma}
\def\r{\rho}
\def\ps{\psi}
\def\ph{\phi}
\def\Ph{\Phi}

\def\t{\tau}
\def\th{\theta}

\def\o{\omega}

\def\ze{\zeta}
\def\ch{\chi}
\def\abs#1{\left| #1 \right|}
\def\pd{\partial}
\def\avg#1{\langle #1 \rangle}

\def\de{\text{d}}

\def\scL{\mathcal{L}}

\def\scM{\mathcal{M}}

\def\scO{\mathcal{O}}

\def\scK{\mathcal{K}}

\def\scJ{\mathcal{J}}
\def\scI{\mathcal{I}}
\newcommand{\bigslant}[2]{{\left.\raisebox{.2em}{$#1$}\middle/\raisebox{-.2em}{$#2$}\right.}}

\def\Mgen{M_{n,\mathcal{I}}^{G}}
\def\Cbb{\mathbb{C}}

\newcommand{\be}{\begin{eqnarray}}
\newcommand{\ee}{\end{eqnarray}}

\makeatletter
\newsavebox\myboxA
\newsavebox\myboxB
\newlength\mylenA
\newcommand*\xoverline[2][0.75]{%
    \sbox{\myboxA}{$\m@th#2$}%
    \setbox\myboxB\null% Phantom box
    \ht\myboxB=\ht\myboxA%
    \dp\myboxB=\dp\myboxA%
    \wd\myboxB=#1\wd\myboxA% Scale phantom
    \sbox\myboxB{$\m@th\overline{\copy\myboxB}$}%  Overlined phantom
    \setlength\mylenA{\the\wd\myboxA}%   calc width diff
    \addtolength\mylenA{-\the\wd\myboxB}%
    \ifdim\wd\myboxB<\wd\myboxA%
       \rlap{\hskip 0.5\mylenA\usebox\myboxB}{\usebox\myboxA}%
    \else
        \hskip -0.5\mylenA\rlap{\usebox\myboxA}{\hskip 0.5\mylenA\usebox\myboxB}%
    \fi}
\makeatother

\newcommand{\drawsquare}[2]{\hbox{%
\rule{#2pt}{#1pt}\hskip-#2pt%  left vertical
\rule{#1pt}{#2pt}\hskip-#1pt%  lower horizontal
\rule[#1pt]{#1pt}{#2pt}}\rule[#1pt]{#2pt}{#2pt}\hskip-#2pt%  upper horizontal
\rule{#2pt}{#1pt}}% right vertical

\newcommand{\Yfund}{\raisebox{-.5pt}{\drawsquare{6.5}{0.4}}}%  fund
\newcommand{\Ysymm}{\raisebox{-.5pt}{\drawsquare{6.5}{0.4}}\hskip-0.4pt%
        \raisebox{-.5pt}{\drawsquare{6.5}{0.4}}}%  symmetric second rank
\newcommand{\Ythrees}{\raisebox{-.5pt}{\drawsquare{6.5}{0.4}}\hskip-0.4pt%
          \raisebox{-.5pt}{\drawsquare{6.5}{0.4}}\hskip-0.4pt% 
          \raisebox{-.5pt}{\drawsquare{6.5}{0.4}}}%  symmetric third rank
\usepackage{slashed}
\usepackage{braket}
\usepackage{color}
\usepackage{harmony}
\usepackage{amssymb}
\usepackage{mathtools}
\usepackage{tikz}
\usetikzlibrary{decorations.pathreplacing}

\usepackage{subfigure}
\usepackage{multirow}

\usepackage{array}
\usepackage{arydshln}

\usepackage{ytableau}

\title{Operator bases, $S$-matrices, and their partition functions}

\author[a]{Brian Henning,}
\author[b]{Xiaochuan Lu,}
\author[c,d,e]{Tom Melia}
\author[c,d,e]{and Hitoshi Murayama}

\affiliation[a]{Department of Physics, Yale University, New Haven, Connecticut 06511, USA}
\affiliation[b]{Department of Physics, University of California, Davis, California 95616, USA}
\affiliation[c]{Department of Physics, University of California, Berkeley, California 94720, USA}
\affiliation[d]{Theoretical Physics Group, Lawrence Berkeley National Laboratory, Berkeley, California 94720, USA}
\affiliation[e]{Kavli Institute for the Physics and Mathematics of the Universe (WPI), Todai Institutes for Advanced Study, University of Tokyo, Kashiwa 277-8583, Japan}

\emailAdd{brian.henning@yale.edu}
\emailAdd{xclu@davis.edu}
\emailAdd{tmelia@lbl.gov}
\emailAdd{hitoshi@berkeley.edu, hitoshi.murayama@ipmu.jp}

\preprinta{IPMU17-0083}

\abstract{Relativistic quantum systems that admit scattering experiments are quantitatively described by effective field theories, where $S$-matrix kinematics and symmetry considerations are encoded in the operator spectrum of the EFT. In this paper we use the $S$-matrix to derive the structure of the EFT operator basis, providing complementary descriptions in (i) position space utilizing the conformal algebra and cohomology and (ii) momentum space via an algebraic formulation in terms of a ring of momenta with kinematics implemented as an ideal. These frameworks systematically handle redundancies associated with equations of motion (on-shell) and integration by parts (momentum conservation).

We introduce a partition function, termed the Hilbert series, to enumerate the operator basis---correspondingly, the $S$-matrix---and derive a matrix integral expression to compute the Hilbert series. The expression is general, easily applied in any spacetime dimension, with arbitrary field content and (linearly realized) symmetries.

In addition to counting, we discuss construction of the basis. Simple algorithms follow from the algebraic formulation in momentum space. We explicitly compute the basis for operators involving up to $n=5$ scalar fields. This construction universally applies to fields with spin, since the operator basis for scalars encodes the momentum dependence of $n$-point amplitudes.

We discuss in detail the operator basis for non-linearly realized symmetries. In the presence of massless particles, there is freedom to impose additional structure on the $S$-matrix in the form of soft limits. The most na\"ive implementation for massless scalars leads to the operator basis for pions, which we confirm using the standard CCWZ formulation for non-linear realizations.

Although primarily discussed in the language of EFT, some of our results---conceptual and quantitative---may be of broader use in studying conformal field theories as well as the AdS/CFT correspondence.}

\begin{document}
\maketitle
\flushbottom

\section{Introduction}
The basic tenets of \(S\)-matrix theory and effective field theory are equivalent: the starting point is to take assumed particle content and parameterize all possible scattering experiments. This is dictated by kinematics and symmetry principles. In the context of effective field theory (EFT), this parameterization is embodied in the operator basis \(\mathcal{K}\) of the EFT, which is defined to be the set of all operators that lead to physically distinct phenomena.

The purpose of this work is to formalize the rules governing the operator basis and investigate the structure they induce on \(\mathcal{K}\). By considering the set \(\mathcal{K}\) in its own right, we are aiming to get as much mileage from kinematics and selection rules as possible before addressing specific dynamics. At a practical level, as well as imposing Lorentz invariance (and other internal symmetries), it amounts to dealing with redundancies associated with equations of motion (EOM) and total derivatives (also called integration by parts (IBP) redundancies). At a more fundamental level, we are accounting for Poincar\'e covariance of single particle states together with Poincar\'e invariance of the \(S\)-matrix.

We focus on  the exploration of operator bases in a wide class of relativistic EFTs in $d\ge2$ spacetime dimensions. Our main purpose is to introduce and develop a number of systematic methods to deal with EOM and IBP redundancies at all orders in the EFT expansion. This treatment of course also provides a standard procedure to determine low-order terms of an operator basis, a continuing topic of interest for various relativistic phenomenological theories as experimental energy and precision thresholds are crossed.

One basic and very useful means of studying the operator basis is to consider the partition function on \(\scK\)
\begin{equation}
H = \text{Tr}_{\scK}\widehat{w} = \sum_{\scO \in \scK} \widehat{w}\big(\scO\big),
\label{eq:H_as_trace_K}
\end{equation}
where \(\widehat{w}\) is some weighting function. We call \(H\) the \textit{Hilbert series of the operator basis}, or simply \textit{Hilbert series} for short. The Hilbert series is a counting function, just as the usual physical partition function (path integral) counts states in the Hilbert space weighted by \(e^{-\b E}\). In many ways, the Hilbert series can be thought of as a partition function of the \(S\)-matrix. Loosely speaking, it captures the rank of the \(S\)-matrix by enumerating independent observables. Hilbert series will play a central role in our study of operator bases.

Effective field theory is a description of the \(S\)-matrix, and as such can address questions concerning the use of general physical principles to derive and implement---bootstrap---consistency conditions with the goal of probing physical observables. A revival of older \(S\)-matrix theory ideas, {\it e.g.} \cite{Eden:1966dnq}, in the past 25 years has seen remarkable, useful, and beautiful results based on unitarity and causality (for recent reviews see {\it e.g.} \cite{Dixon:1613349,Elvang:2013cua}). Similar types of general considerations, also discussed last century \cite{FERRARA197277,FERRARA1973161,1974JETP...39...10P}, found concrete implementation in \(d>2\) CFTs a decade ago \cite{Rattazzi:2008pe}, spawning the modern CFT bootstrap program (see {\it e.g.} \cite{Rychkov:2016iqz} for an overview). The AdS/CFT correspondence \cite{Maldacena:1997re} has demonstrated underlying connections between these two areas \cite{Heemskerk:2009pn,Heemskerk:2010ty,Fitzpatrick:2012yx,Komargodski:2012ek,Alday:2014tsa,Li:2015rfa,Li:2015itl,Fitzpatrick:2016ive,Dyer:2016pou,Afkhami-Jeddi:2016ntf,Alday:2016htq,Aharony:2016dwx,Rastelli:2016nze,Perlmutter:2016pkf,Caron-Huot:2017vep,Chen:2017yze}.

More concretely, the AdS/CFT correspondence in the flat space limit of AdS indicates that there is a (not entirely understood) correspondence between \(d\)-dimensional scattering amplitudes and \((d-1)\)-dimensional conformal correlators~\cite{Heemskerk:2009pn,Fitzpatrick:2010zm,Penedones:2010ue,Fitzpatrick:2011ia}. In~\cite{Heemskerk:2009pn} it was shown that solutions to the crossing equations for scalar four point functions in \(\text{CFT}_{d-1}\) are in one-to-one correspondence with operators in \(d\)-dimensions consisting of four scalar fields and derivatives. In our analysis, this set is described by \(M_4\) in eq.~\eqref{eq:M_n_module}, whose solution is (see sec.~\ref{sec:rsf}) a freely generated ring in \(st + su+ tu\) and \(stu\) where \((s,t,u)\) are the usual Mandelstam variables. The initial clue in~\cite{Heemskerk:2009pn} was a counting argument, that showed the number of objects in \(M_4\) is the same as the number of solutions to the crossing equations. The Hilbert series techniques developed in this paper allow an easy and straightforward calculation of determining the numbers of operators in more general cases, and simple algorithms we present aid in their explicit construction.

In a similar vein, when an \(n\)-point conformal correlator involves spinning objects, the correlator is first decomposed into tensor structures which then multiply functions of the conformal-cross ratios~\cite{Costa:2011mg}. The number of such tensor structures is the same as the number of such structures for corresponding \(n\)-point amplitudes in one higher dimension~\cite{Hofman:2008ar,Costa:2011mg,Dymarsky:2013wla,Kravchuk:2016qvl}. A general procedure for counting the tensor structures was recently given in~\cite{Kravchuk:2016qvl}. Our Hilbert series techniques capture this information and more: they account for not only the tensor structures but also the Mandelstam invariants (which serve as generators that can be applied an infinite number of times, and appear as denominators in the Hilbert series), as well as  secondary invariants which can be used only once.\footnote{For example, something containing the epsilon tensor, like \(\e_{\m_1\dots \m_d}p_1^{\m_1}\cdots p_d^{\m_d}\), only appears once since two epsilon tensors multiplied together can be decomposed into products of the metric.}

At a superficial level, EFT may also offer some novelty in connecting modern amplitude and CFT ideas simply by the language typically used to discuss EFTs: EFTs are frequently formulated in position space and discussed in terms of operators---similar to CFT (largely on account of the operator--state correspondence)---while the actual physical object described by the EFT is the \(S\)-matrix. 

EFTs are very good at clearly isolating relevant degrees of freedom---this has given birth to an entire zoo of EFTs in high energy physics, \textit{e.g.} chiral lagrangians, HQET, SCET, SM EFT, \textit{etc}. In most discussions, however, EFTs are truncated (for very good practical reasons), which can mask analytic properties. One hope in studying the entire set of operators, embodied in \(\scK\), is that it elucidates analytic properties more clearly. On the other hand, the remarkable success of EFT is a serious hint that truncation is quantitatively sensible, potentially even for studying strongly coupled field theories~\cite{Hogervorst:2014rta,Katz:2016hxp}.

Our presentation is somewhat lengthy---as well as containing new results and techniques, it also exhibits a number of self-contained, heuristic, and physics-oriented derivations of existing results in the (mainly mathematics) literature. The following summary aims to both clarify the structure of this paper, and to make the distinction between new results and new derivations/interpretations of known results; for the reader's convenience, separate summaries are included at the end of Sections~\ref{sec:reps}--\ref{sec:non-linear} to tabulate more detailed and explicit results and formulae from each section. Section~\ref{sec:overview} aims to establish the logic of operator bases from physical principles, as well as provide the reader with an overview of the main ideas we present in the bulk of this work.

{\bf Section \ref{sec:overview}: The operator basis.}
The operator basis follows from the construction of the \(S\)-matrix. We introduce what we term `single particle modules', which consist of all possible Lorentz spin operators that interpolate an asymptotic single particle state. In essence, these modules are an abstraction of a field along with a tower of its derivatives modulo EOM {\it by definition}. From this perspective, it is clear the EOM redundancy is simply an on-shell condition; it is precisely this EOM---as used in LSZ reduction---which provides the freedom to perform field redefinitions~\cite{Haag:1958vt,Kamefuchi:1961sb,Coleman:1969sm}. We subsequently build operators via (tensor) products of these modules, thus accounting for EOM redundancy across the operator basis.

Establishing the operator basis along this line of reasoning has distinct advantages. It derives the rules governing \(\mathcal{K}\) using on-shell quantities, bypassing the introduction of a Lagrangian.\footnote{Note that we are only advocating the idea of conceptually divorcing operator content (kinematics) from Wilson coefficients (dynamics). The first can, and should, be established with only physical quantities. For the latter, the Lagrangian remains an indispensable tool.} This is not merely some pleasing alternative viewpoint---it often clarifies the validity and use of certain rules. One example is the use of the free-field EOM. Another is why covariant derivatives behave as commuting objects in our analysis. Both these rules are obvious from the perspective that \(D^k\ph\) interpolates a single particle state. Moreover, this picture allows us to give a physical derivation of the Hilbert series, which is outlined at the end of sec.~\ref{sec:overview}.

We then give an overview of the new techniques we use to further impose IBP. We find three mathematical formulations for doing so, in terms of 
\begin{enumerate}
\item[(1)] cohomology (\(\scO \in \scK\) are co-closed but not co-exact 0-forms),
\item[(2)] conformal representation theory (\(\scO \in \scK\) are scalar conformal primaries), 
\item[(3)] an algebraic description (\(\scO \in \scK\) are elements of certain kinematic polynomial rings).
\end{enumerate}
While all three formulations have been introduced at some level in our previous work \cite{Henning:2015daa,Henning:2015alf}, they are presented in full detail and in greater generality here.

Next, we discuss the treatment of linearly realized internal symmetries, covariant derivatives, and topological terms; some of this methodology was pioneered in \cite{Lehman:2015via}, and also presented in~\cite{Henning:2015alf}. A question about soft limits naturally leads to non-linearly realized internal symmetries; we give an overview of what this implies for the operator basis and take it up in full in section~\ref{sec:non-linear}.

{\bf Section \ref{sec:reps}: Conformal representations and characters.} This section reviews results in the representation theory of the conformal group in $d$ dimensions necessary for treating IBP via formulation (2) mentioned above. The use of character formulae for irreducible representations of the conformal group is of particular importance. The treatment of conformal characters in $d$ dimensions was worked out by Dolan in~\cite{Dolan:2005wy} and we have relied heavily on this reference. However,  we present a largely self-contained version for our purposes that elaborates on derivations and includes physical intuitions that supplement~\cite{Dolan:2005wy}.

As character theory plays an important role in our analysis, we note here that Appendix~\ref{app:sodspin} tabulates character formulae for the classical Lie groups. In Appendix~\ref{app:Weyl} we derive the Weyl integration formula and provide the measures for the classical Lie groups. While these results are textbook material, the treatment of the Weyl integral formula is aimed to be more physics-oriented than that which is usually encountered.

{\bf Section \ref{sec:compute}: Counting operators: Hilbert series.} 
A new and central result of this work is a matrix integral (group integral) formula for the Hilbert series of an operator basis, in \(d\ge 2\) dimensions, eq.~\eqref{eqn:Hresults}. The use of conformal representation theory---formulation (2)---to handle IBP redundancy allows us to obtain a complete derivation of the Hilbert series. This treatment is valid when the single particle modules form representations of the conformal group---this encompasses a large class of EFTs that includes gauge theories in four dimensions, and in particular the SM EFT. This was the approach used (but only outlined) in $d=4$ in our previous work~\cite{Henning:2015alf} where we gave the application to the SM EFT at mass dimension $>4$; here we show how to derive the Hilbert series in arbitrary spacetime dimension, including relevant operators with mass dimension $\le d$.

More precisely, we write \(H = H_0 + \D H\), and present a matrix integral for \(H_0\); \(\D H\) corresponds to a finite number of operators not properly accounted for in \(H_0\). It is only in formulation (2), when the single particle modules correspond to unitary representations of the conformal group, that we are able to {\it derive} an explicit formula for \(\D H\).  In the general case, where the single particle modules are not unitary representations of the conformal group, we emphasize that \(H_0\) is valid and relatively straightforward to derive (see sec.~\ref{sec:overview} for a physical derivation, and sec.~\ref{sec:non-linear} for a derivation utilizing differential forms). Moreover, our observations indicate that \(\D H\) only pertains to certain marginal or relevant operators---the cohomology picture allows us to understand its structure more clearly which we discuss in sec.~\ref{sec:non-linear}.

Hilbert series have been previously used in the particle physics literature to study flavor invariants~\cite{Jenkins:2009dy}. They were adapted in the two papers \cite{Lehman:2015via,Lehman:2015coa} to deal with the problem of finding Lorentz invariant operators involving multiple types of field, but only for the subset of the operator basis where no more than one derivative appears in an operator. Hilbert series for enumerating independent operators with any number of derivatives were studied first for scalar EFTs in $0+1$ dimensions in ref.~\cite{Henning:2015daa}, and for the Standard Model (SM) EFT in ref.~\cite{Henning:2015alf}. They have since found application in non-relativistic EFTs~\cite{Gunawardana:2017zix,Kobach:2017xkw}. Counting of conformal primaries in free CFTs in \(d=4\) has also been addressed using topological field theory~\cite{Koch:2014nka}.

We also explain in detail how to impose parity on the operator basis---moving from $SO(d)$ to $O(d)$ invariance. Previous documentation of the character theory of the parity odd and even pieces of $O(d)$ can be found, contained to a few pages, in Weyl's original work \cite{Weyl:classical}. We expand upon this, including a self-contained discussion on the `folding' of a Dynkin diagram, which explains the---at first mysterious---appearance of the $\mathfrak{sp}$ algebra in \(d=2r\) dimensions. Many of these details are contained in Appendix~\ref{app:parity}.

{\bf Section \ref{sec:rsf}: Constructing operators: kinematic polynomial rings.} This section develops an algebraic description---formulation (3)---in order to explicitly build the operator basis. The basic idea is to construct a ring in momentum whose elements are Feyman rules. EOM and IBP redundancies manifest as kinematic constraints of on-shell conditions and momentum conservation---they are implemented as an ideal in the ring of momenta. As we shall see, the Hilbert series is an indispensable tool towards explicitly constructing the operator basis: many algorithms (as well as human intuition) for finding a basis rely on input that comes from the Hilbert series.

After giving a precise definition of the rings relevant for an operator basis involving scalar fields, we use four point kinematics to showcase various aspects of the effects of dimensionality, parity, and particle indistinguishability on the construction of the operators,  and we reflect on the connection to constructing physical scattering amplitudes. We then prove some general properties of the rings (that they are Cohen-Macaulay), and discuss the universal properties and application of these scalar rings to cases involving spin. We provide a connection between the elements of the ring and conformal primaries, and finally detail a construction algorithm which we use to construct the ring with five point kinematics in the case of indistinguishable scalars.

The application of commutative algebra to the IBP problem was first explored in our previous work in $d=1$ dimension~\cite{Henning:2015daa}. In this paper we elucidate the algebraic modules (quotient rings in particle momenta) that form the operator bases of EFTs in $d\ge 2$ dimensions, given in eq.~\eqref{eq:ddimring}, thus providing a significant generalization of~\cite{Henning:2015daa}.

{\bf Section \ref{sec:case}: Applications and examples of Hilbert series.} In this section we present some novel calculations using our developed framework. Common to all of these examples is how the Hilbert series reveals structure in the operator basis, such as primary and secondary invariants (generators), information on relations between invariants (spacetime rank conditions---termed Gram conditions---are particularly interesting), \textit{etc.}. The case studies are: sec.~\ref{subsec:H_rsf_d=4}, studying Hilbert series for real scalar field in \(d=4\) dimensions, and the impact of indistinguishability on the structure of the operator basis;   sec.~\ref{subsec:H_d=2}, deriving a closed form expression for the Hilbert series of \textit{distinguishable} scalars in \(d=2,3\) dimensions, where Gram conditions are more readily understood;  sec.~\ref{subsec:hilbert_spinning}, exploring Hilbert series with spinning particles, in particular the relation to those involving only scalar particles, and how this provides information on the tensor decomposition of scattering amplitudes of spin; and sec.~\ref{subsec:quant_redund}, quantifying how much EOM, IBP, and Gram conditions ``cut down'' the operator basis (EOM and IBP give \textit{polynomial} reductions, Gram constraints give \textit{exponential} reduction).

{\bf Section \ref{sec:non-linear}: Non-linear realizations.} 
In this section we show how to include invariance of the operator basis under non-linearly realized internal symmetry groups.
We follow a linearization procedure, \`a la CCZW~\cite{Coleman:1969sm,Callan:1969sn}, and use this to give an identification of the single particle modules. Having constructed this module (and thus dealt with EOM), we appeal to formulation (1) and use Hodge theory to address the IBP redundancy. The Hilbert series takes a form very similar to the case where the single particle modules coincide with conformal irreps; here, however, we appeal to  Hodge theory to determine (what turn out to be) the relevant and marginal operators belonging to \(\Delta H\). 

We conclude with a short discussion in Section \ref{sec:discuss}.

\section{The operator basis}
\label{sec:overview}
Under some mild physical assumptions, at low enough energies a relativistic system is described either by an interacting CFT or an infrared free theory. We are concerned with the latter, and we look to build the EFT to describe the \(S\)-matrix. We proceed on this route by imposing the consequences of Poincar\'e symmetry.

\subsection{Single particle modules}
\label{sec:subsec2p1}
Assume that we are given a set of particles that we have asymptotic access to---\textit{i.e.} these are particles which can be part of in and out states in scattering experiments. The single particle states are specified by their mass \(m^2 = p^2\), spin under the relevant little group, and other possible internal quantum numbers. Let \(\ket{\mathbf{p} \,\s}\) denote such a state with spin \(\s\) (for now, we ignore other possible quantum numbers for simplicity of discussion). Following the standard pathway to field theory~\cite{Weinberg:1995mt}, local fields $\Ph_l(x)$ are constructed which interpolate this single particle state
\begin{equation}
\braket{0| \Ph_l(x)|\mathbf{p}\, \s} \sim U_l^{\s}(p)e^{-ip\cdot x},
\label{eq:interp_gen_field}
\end{equation}
where \(l\) is some unspecified Lorentz index structure. Up to some normalization (hence the \(\sim\) in the above equation), the right hand side is determined purely by properties of \(\ket{\mathbf{p} \, \s}\) under Poincar\'e transformations. In turn, this dictates the field equation(s) obeyed by \(\Ph_l(x)\). There are many---in fact, infinite---different types of interpolating fields; equivalently, there is an entire set of distinct wavefunctions \(U_l^{\s}(p)\) consistent with locality. To capture the most general set of local interactions, we must take all distinct interpolating fields.

The different types of interpolating fields are not mysterious. There is some basic field \(\ph_l\) which has a minimum number of Lorentz indices to specify the spin component \(\s\). Beyond this, the only available object on the RHS of~\eqref{eq:interp_gen_field} is the momentum \(p_{\m}\), corresponding to derivatives acting on \(\ph_l\). Importantly, these derivatives can only be added in a manner consistent with the field equations obeyed by \(\ph_l(x)\). It is worth elaborating on this point. From
\begin{equation}
\braket{0| \ph_l(x)|\mathbf{p}\, \s} \sim u_l^{\s}(p)e^{-ip\cdot x},
\label{eq:interp_phi}
\end{equation}
and assuming \(\ph_l(x)\) is linear in the creation operator \(a^{\dag}_{\mathbf{p},\s}\), the field equations can be determined via consistency with \(\ket{\mathbf{p}\, \s} \to U(\L,a) \ket{\mathbf{p}\, \s}\) with \(U(\L,a)\) a Poincar\'e transformation~\cite{Weinberg:1995mt}. Specifically, every field obeys a Klein-Gordon equation due to the on-shell condition \(p^2 = m^2\); fields with spin will have a transverse condition \(p^{\m}u_{\m\dots}^{\s} = 0\); and spinning, massless fields have further constraints in the form of Bianchi identities. Taken together, they are relativistic wave equations \cite{Bargmann01051948}. Because \(\ket{\mathbf{p} \, \s}\) is an asymptotic particle state, these look like free-field equations of motion but involve the physical mass \(m\). They are structurally equivalent to the linearized equation of motion obtained from the field theory Lagrangian.\footnote{Recall that it is the linearized EOM which determines the propagator and is used in LSZ reduction.}

We define the \textit{single particle module} \(R_{\ph_l}\) to be the set of all distinct interpolating fields for \(\ket{\mathbf{p} \, \s}\), \textit{i.e.} \(R_{\ph_l}\) consists of \(\ph_l(x)\) together with an infinite tower of derivatives on top of \(\ph_l\) modulo EOM. Equivalently, \(R_{\ph_l}\) can be represented by the set of wavefunctions \(\{U_l^{\s}(p)\}\). The correspondence between these two representations---interpolating fields \(\Ph_l(x)\) versus wavefunctions \(U_l^{\s}(p)\)---is eq.~\eqref{eq:interp_gen_field}, which is simply the starting point for the familiar correspondence between operators and Feynman rules. In practice, it can be thought of as simply a Fourier transform, although this slightly obscures the Hilbert space state \(\ket{\mathbf{p} \, \s}\) from which these objects arise. This dual momentum space picture will prove useful, especially in giving a concrete algorithm to obtain \(\mathcal{K}\). For now, however, we stick to the position space picture in terms of interpolating fields.

To clear up the abstraction, let's look at a specific example. Consider some scalar particle state, \(\s =0\). The possible interpolating fields are
\begin{equation}\def\arraystretch{1.3}
\begin{array}{rcl}
\braket{0| \ph| \mathbf{p}} & \sim &  e^{-ip\cdot x}\\
\braket{0| \pd_{\m}\ph| \mathbf{p}} & \sim &  p_{\m}e^{-ip\cdot x}\\
\braket{0| \pd_{\{\m_1}\pd_{\m_2\}}\ph| \mathbf{p}} & \sim &  p_{\{\m_1}p_{\m_2\}}e^{-ip\cdot x}\\
 & \vdots &\end{array},
\label{eq:interp_tower_scalar}
\end{equation}
where \(\{\cdots\}\) denotes the symmetric, traceless component---the trace is removed because of the on-shell condition \(p^2 = m^2\). The single particle module is
\begin{equation}\def\arraystretch{1.1}
R_{\ph} = \left(\begin{array}{c}
 \ph\\  \pd_{\m}\ph  \\ \pd_{\{\m_1}\pd_{\m_2\}}\ph  \\ \vdots 
\end{array}\right).
\label{eq:SPM_scalar}
\end{equation}
For future reference, we note that \(R_{\ph}\) coincides with the conformal representation of a free scalar field.

As a second example, consider a massless spin-one particle. The basic interpolating field is \(F_{\m\n}(x) = -F_{\n\m}(x)\) giving the wavefunction \(u_{\m\n}^{\s}(p) = p_{\m}\e_{\n}^{\s}(p) - p_{\n}\e_{\m}^{\s}(p)\) with \(\e_{\m}^{\s}(p)\) the polarization vector. By on-shell \(p^2 = 0\), transversality \(p^{\m}\e_{\m}^{\s}(p) = 0\), and (obviously) \(p_{[\m}p_{\n]} = 0\), allowed wavefunctions take the form \(p^{}_{\{\m_1}\dots p^{}_{\m_n}p^{}_{[\m\}}\e_{\n]}^{\s}\). Correspondingly, the field equations are \(\pd^{\m} F_{\m\n} = \e^{\m_1 \dots \m_d}\pd_{\m_1}F_{\m_2\m_3} = \pd^2 F_{\m\n} = 0\), leading to the single particle module
\begin{equation}\def\arraystretch{1.1}
R_{F} = \left(\begin{array}{c}
 F_{\m\n}\\  \pd_{\{\m_1}F_{\m\}\n}  \\ \pd_{\{\m_1}\pd_{\m_2}F_{\m\}\n}  \\ \vdots
\end{array}\right).
\label{eq:SPM_F}
\end{equation}
In four dimensions, \(R_F\) coincides with the conformal representation for a free vector. Outside of four dimensions \(R_F\) is \textit{not} a unitary conformal representation (one indication of this is that the coupling constant in the Maxwell action, \(\int d^dx \, \frac{1}{4g^2}F^2\), is dimensionless only in \(d = 4\)).

Let us briefly pause to emphasize a well known point concerning removing the EOM redundancy. In the full fledged field theory---which requires additional structure such as a Lagrangian and a renormalization prescription---operators proportional to the EOM generically appear in the Lagrangian as one flows between various energy scales. Such operators can always be removed by a field redefinition. (Here we emphasize that the valid operation in a Lagrangian is performing a field redefinition, not plugging in the EOM.\footnote{A generic EFT Lagrangian is a truncated expansion $\scL = \scL^{(0)} + \scL^{(1)} + \cdots + \scL^{(n)}$ with $\mathcal{L}^{(0)}$ the leading order term,  \(\scL^{(k)}\) suppressed by $\frac{1}{\L^k}$, and terms suppressed by more than \(\frac{1}{\L^n}\) dropped. An order $k>0$ redundant operator $\frac{1}{\L^k}\scO\frac{\delta\scL^{(0)}}{\delta\ph} \subset \scL^{(k)}$ (proportional to the \textit{leading order} EOM) can be eliminated in favor of higher order terms through the field redefinition $\ph\to\ph-\frac{1}{\L^k}\scO$, which changes the Lagrangian by
\begin{align*}
\Delta\scL &= \left( \frac{\d \scL^{(0)}}{\d \ph} + \frac{\d \scL^{(1)}}{\d \ph} + \cdots \right) \left(-\frac{1}{\L^k}\scO\right) + \frac{1}{2}\frac{\d^2 \scL}{\d \ph^2} \left(-\frac{1}{\L^k}\scO\right)^2 + \frac{1}{6}\frac{\d^3 \scL}{\d \ph^3} \left(-\frac{1}{\L^k}\scO\right)^3 + \cdots
\end{align*}
The very first term cancels the redundant operator, while all the other terms give its higher order compensation. This is clearly different from plugging in the full EOM, which amounts to keeping only the first variation terms in the $\Delta\scL$ above. However, in the special case of $n=k=1$, the two are equivalent.} This is well known and has been explicitly emphasized in the literature, \textit{e.g.}~\cite{Scherer:1994wi}.) The ability to perform field redefinitions is formally justified by the LSZ reduction procedure~\cite{Haag:1958vt,Kamefuchi:1961sb,Coleman:1969sm}; it tells us that in computing the $S$-matrix elements, the only thing that matters is that an interpolating field give the proper wavefunction (\(\braket{0|\ph_l(x)|\mathbf{p}\,\s}\)) for creating the single particle state. This clearly allows for redefinitions of an interpolating field.

\subsection{The operator basis via cohomology and conformal representation theory}
\label{sec:subsec2p2}

Local operators are built by taking products of the interpolating fields. Viewing composite operators as simply a product of other operators is a perturbative statement, which is justified in our framework because the asymptotic multi-particle states in scattering are built as a Fock space from the single particle states.\footnote{In other words, this is justified because we are perturbing free fields.} Let \(\scJ\) denote the set of all operators formed from products of the interpolating fields, \textit{i.e.} \(\scJ\) is the set of all local operators modulo equations of motion. Mathematically, \(\scJ\) is a differential ring formed from implementing the equations of motion as an ideal. Explicitly, for a single scalar field
\begin{align}
\scJ &= \bigslant{\mathbb{C}[\ph;\pd_{\m}]}{\avg{\pd^2\ph}} \nonumber \\
&= \mathbb{C}[\ph, \, \pd_{\m}\ph, \, \pd_{\{\m_1}\pd_{\m_2\}}\ph, \dots].
\label{eq:J_scalar}
\end{align}
Making use of the single particle modules, \(\scJ\) can equivalently be built by taking tensor products of the \(R_{\Ph_i}\). For the single scalar field we then have
\begin{equation}
\scJ = \bigoplus_{n=0}^{\infty}\text{sym}^n\big(R_{\ph}\big),
\label{eq:J_sym_prod}
\end{equation}
where we take symmetric tensor products due to Bose statistics. In the case that the single particle modules form free field representations of the conformal group, \(\scJ\) is the set of operators which participate in the operator state correspondence.\footnote{If there is a gauge symmetry, one further restricts \(\scJ\) to gauge invariant operators.}

The operator basis is clearly a subset of \(\scJ\). To select out \(\scK\subset \scJ\) we need to apply Poincar\'e invariance of the \(S\)-matrix. Use crossing symmetry to take all particles as incoming and let \(\ket{\a} = \ket{\mathbf{p}_1\s_1;\cdots; \mathbf{p}_n\s_n}\) denote an asymptotic multiparticle state. Then, in order for \(\braket{0|S|\a}\) to be non-trivial, \(\a\) must have the same quantum numbers as the vacuum. In particular, it is a Lorentz scalar and carries no momentum, \(p_1^{\m} + \cdots + p_n^{\m} = 0\). Interactions must respect this invariance as well: operators are Lorentz singlets and carry no momentum, meaning they cannot be a total derivative.

It is at least conceptually clear how to get Lorentz singlets, although determining independent scalars can be extraordinarily difficult due to \(SO(d)\) group relations like Fierz identities, Gram determinants, \textit{etc}. What is conceptually less obvious is how to determine operators up to a total derivative. This is an equivalence relation, \(\scO_l \sim \scO_l'\) if \(\nobreak{\scO_l = \scO_l' + (\pd \scO'')_l}\). Implementing this equivalence relation allows us to formally identify the operator basis as the differential ring
\begin{equation}
\scK = \left[\bigslant{\scJ}{d\scJ}\right]^{SO(d)},
\label{eq:K_differential_mod}
\end{equation}
where \(d\scJ\) denotes taking the derivative of every element of \(\scJ\), \textit{i.e.} \(\bigslant{\scJ}{d\scJ}\) is the set of equivalence classes with relation \(\sim\) defined above. The superscript \(SO(d)\) means that we apply \(SO(d)\) invariance. Since we know that we want Lorentz scalar operators---and that a scalar which is a total derivative must be the divergence of some vector---we can be even more precise in the equivalence relation: \textit{the operator basis consists of co-closed but not co-exact 0-forms}.\footnote{Some basic definitions of Hodge theory are reviewed in sec.~\ref{subsec:nonlinearHilbert}.}

The construction of the the single particle modules is a great hint at how to handle the equivalence up to total derivatives. Viewing \(R_{\ph_l}\) as some column vector with \(\ph_l\) at the top of the multiplet followed by a tower of derivatives, \textit{e.g.} eqs.~(\ref{eq:SPM_scalar},~\ref{eq:SPM_F}), we have solved the total derivative equivalence problem for operators in \(\scJ\) consisting of only one \(\ph_l\) field plus derivatives. In building \(\scJ\) via tensor products of the \(R_{\Ph_i}\) clearly we solve total derivative equivalence if we can decompose back into a sum of multiplets consisting of a non-total derivative operator \(\scO\) followed by a tower of derivatives acting on \(\scO\), schematically,
\begin{equation}
\begin{pmatrix}\ph_l \\ \pd \ph_l \\ \pd^2 \ph_l \\ \vdots \end{pmatrix}^{\bigotimes n} \sim \ \, \sum_{\scO}\  \begin{pmatrix}\scO \\ \pd \scO \\ \pd^2 \scO \\ \vdots \end{pmatrix}.
\label{eq:sketchofdecomp}
\end{equation}
These multiplets follow precisely the structure of conformal multiplets. Conformal representations are reviewed in detail in sec.~\ref{sec:reps}, but the essential feature is that the derivative is a lowering operator within the conformal algebra. What is obviously suggested is to use conformal representation theory to organize the total derivative equivalence: we take tensor products of the \(R_{\Ph_i}\) and decompose these into conformal representations. Only the primary operators are not total derivatives. Lorentz invariance means we take scalar operators. Therefore, we arrive at the conclusion that \textit{the operator basis is spanned by scalar, conformal primaries}.

These conclusions about the conformal group are established rigorously when the single particle modules are unitary representations of the conformal group. This is the case for free scalars and spinors in any dimension as well as free spin \((l,\dots,l,\pm l)\) fields in even dimensions~\cite{Siegel:1988gd} (see sec.~\ref{sec:reps} for notation). Note that this covers gauge fields in \(d = 4\) where \(F_{L,R} = F \pm i \widetilde{F}\) correspond to spin \((1,\pm1)\) fields. Generically, the \(R_{\Ph_i}\) do not coincide with unitary representations of the conformal group.\footnote{In the general case, although the single particle modules mimic the structure of conformal representations, they are decisively not unitary representations. This is due to EOM shortening conditions not present in unitary conformal representations. It is possible that these could be considered non-unitary representations, with ghosts accounting for the shortening conditions. See~\cite{ElShowk:2011gz} for a related discussion concerning Maxwell theory outside four dimensions.} 
However, our experience indicates that this largely does not matter for determining the majority of \(\scK\). We will see an example of this when we study non-linear realizations in sec.~\ref{sec:non-linear}. We will be liberal with terminology and say ``primary'' and ``descendant'', although it is to be understood that, in general, there is no special conformal generator which makes these words precise.

\subsection{The operator basis as Feynman rules}
\label{sec:subsec2p3}

The operator basis can also be described in the language of Feynman rules. The route to this description also proceeds via the single particle modules, but this time formulated in momentum space with the wavefunctions \(U_l^{\s}(p)\). Focusing on the scalar particle, the wavefunctions in eq.~\eqref{eq:interp_tower_scalar} are simply polynomials in \(p_{\m}\) with \(p^2\) component removed, implying
\begin{equation}
R_{\ph} = \bigslant{\mathbb{C}[p_{\m}]}{\avg{p^2}},
\end{equation}
from which
\begin{equation}
\text{sym}^n\big(R_{\ph}\big) = \left[\bigslant{\mathbb{C}[p_1^{\m},\cdots,p_n^{\m}]}{\avg{p_1^2,\cdots,p_n^2}}\right]^{S_n},
\label{eq:sym_Rphi_mom}
\end{equation}
where the superscript \(S_n\) means to take polynomials symmetric under permutations of the momenta. To get the independent, Lorentz invariant Feynman rules we take the \(SO(d)\) invariant polynomials and enforce equivalence under momentum conservation \(p_1^{\m} + \cdots +p_n^{\m} = 0\). This last condition is accounted for by placing \(p_1^{\m} + \cdots +p_n^{\m}\) in the ideal. Hence the independent Feynman rules in eq.~\eqref{eq:sym_Rphi_mom} are given by the ring\footnote{In Sec.~\ref{sec:rsf} we will label these rings as \(M_{n,\scK}^{SO(d)\times S_n}\); as this cumbersome notation is presently unnecessary we use the simpler \(M_n\).}
\begin{equation}
M_n \equiv \left[\bigslant{\mathbb{C}[p_1^{\m},\cdots,p_n^{\m}]}{\avg{p_1^{\m} + \cdots +p_n^{\m},p_1^2,\cdots,p_n^2}}\right]^{SO(d)\times S_n},
\label{eq:M_n_module}
\end{equation}
and the operator basis (for a single scalar field) is
\begin{equation}
\scK = \bigoplus_{n=0}^{\infty}M_n.
\end{equation}

The basic physical content of \(M_n\) is that it captures the objects an \(n\)-point amplitude can depend on. Typically we think of an amplitude as polarization tensors dotted into a Feynman amplitude of spin \(l\),
\begin{equation}
\mathcal{A}_{\s_1,\dots,\s_n}(\{\e_{\s_i}\},\{p_i^{\m}\}) = \big(\e_{\s_1}\cdots \e_{\s_n}\big)_l \scM^l(\{p_i^{\m}\}).
\label{eq:feyn_amp}
\end{equation}
The amplitude can be decomposed into a finite number of Lorentz singlet tensor structures (tensor structure refers to little group indices) multiplying amplitudes which are functions purely of the Mandelstam invariants \(s_{ij} = p_i\cdot p_j\),
\begin{equation}
\mathcal{A}_{\s_1,\dots,\s_n}(\e_i,p_i) = \sum_I g^I_{\s_1\dots \s_n}(\e_i,p_i)\mathcal{A}_I(s_{ij}).
\label{eq:amp_tens}
\end{equation}
Here, the little group tensors \(g^I_{\s_1\dots \s_n}\) are Lorentz scalar polynomials linear in each \(\e_{\s_i}\). An analogous decomposition is done for CFT correlators~\cite{Costa:2011mg}; holographic arguments suggest that the conformal decomposition in \((d-1)\) dimensions corresponds to the amplitude decomposition in \(d\) dimensions. To this point, the number of tensor structures---the number of \(g^I\) above---coincides~\cite{Hofman:2008ar,Costa:2011mg,Dymarsky:2013wla,Kravchuk:2016qvl}. A general prescription for counting the number of such structures was recently given in~\cite{Kravchuk:2016qvl}. It essentially is the naive expectation: it is the number of different helicity configurations one can take for the external particles. That is, eq.~\eqref{eq:amp_tens} is the decomposition into helicity amplitudes~\cite{Jacob:1959at}.

In terms of the operator basis, the decomposition in eq.~\eqref{eq:amp_tens} implies that the generalization of eq.~\eqref{eq:M_n_module} to include spinning particles will be some decomposition into polynomials \(f^I\) (serving as appropriate avatars for the \(g^I\)) multiplying the \(M_n\),
\begin{equation}
M_n\big(\{\Ph_{l_i}\}\big) \sim \bigoplus_{I} f^{I}\big(u_{l_i}^{\s_i},p_i\big) M_n.
\label{eq:M_n_spin}
\end{equation}
We use a tilde here because the \(M_n\) appearing above need not be exactly identical to that in eq.~\eqref{eq:M_n_module}, nor even the same for each term in the sum (it is also not clear we can get a clean direct sum decomposition). For example, \(SO(d)\) group relations could cause some changes. The essential point is that the modules \(M_n\) in eq.~\eqref{eq:M_n_module} (with possibly different permutation groups to account for identical versus distinguishable particles) are \textit{universal} to all aspects of the operator basis. This is precisely because they are analyzing possible momentum dependence, which is an ingredient in all scattering processes. We therefore aim in this work to develop a healthy understanding of the universal ingredient \(M_n\), leaving the precise description of the general case in eq.~\eqref{eq:M_n_spin} to a future work.

\subsection{Non-linear realizations}\label{sec:2.4}

Our approach thus far has been to start with an IR free theory and use the symmetries of the \(S\)-matrix---namely, Poincar\'e symmetry---to build up the operator basis. Are there any other assumptions that we can make within this framework? The hallmark of IR free theories is the absence of long range interactions. If there are massless particles, this means interactions vanish as \(p \to 0\). Suppose we have some massless scalar states \(\ket{\mathbf{p}\, \s=0\, i} \equiv \ket{\pi_{\mathbf{p}}^i}\) with \(i = 1,\dots, n\). To ensure the absence of scattering as \(p\to 0\), a sufficient condition\footnote{At present, this claim is justified \textit{a posteriori} from our field theory knowledge.} is if the interpolating fields vanish with momentum. In a suggestive notation, the first such interpolating field in eq.~\eqref{eq:interp_tower_scalar} is
\begin{equation}
\braket{0|j_{\m}^i(x)|\pi_{\mathbf{p}}^j} \sim \d^{ij}p_{\m}e^{-ip\cdot x},
\label{eq:Goldstone}
\end{equation}
which should be ringing bells about Goldstone's theorem.

At this stage the field appearing in eq.~\eqref{eq:Goldstone} is simply some low-energy vector field which (1) interpolates a scalar state and (2) is divergenceless because the scalar is massless. From these two pieces of knowledge, all we can conclude is that the field has a derivative acting in some unknown manner on a scalar function. To go further and determine this field as some specific function requires additional input. We do not pursue this avenue here; we refer to~\cite{Cheung:2014dqa,Cheung:2016drk} for the very interesting question about constructing EFTs from soft-limits.

Equation~\eqref{eq:Goldstone} tells us that any derivatives acting on the vector interpolating field must be traceless and symmetric, because \(p^2=0\). This naturally makes one guess that the appropriate single particle module is essentially the scalar module \(R_{\ph}\) in eqs.~\eqref{eq:interp_tower_scalar}-\eqref{eq:SPM_scalar}, but without the ``top'' component \(\ph \in R_{\ph}\), \textit{i.e.} delete all modes from \(R_{\ph}\) which do not vanish as \(p^{\m} \to 0\). Concretely, renaming the field in eq.~\eqref{eq:Goldstone} to \(u_{\m}^i(x)\),\footnote{The CCWZ construction does not directly work with the symmetry current \(j_{\m}^i\), but instead with the Maurer-Cartan form \(u_{\m}^i\), see sec.~\ref{sec:non-linear}. The two are related by \(j_{\m}^i = \text{Tr}\big(X^i\xi u_{\m}\xi^{-1}\big)\) with \(\xi = e^{i\pi^iX^i}\). Hence, both \(u_{\m}^i\) and \(j_{\m}^i\) interpolate the single pion state, but differ when interpolating multi-pion states.} an ansatz for the appropriate single particle module is
\begin{equation}\def\arraystretch{1.3}
\begin{array}{rcl}
\braket{0| u_{\m}^i| \pi_{\mathbf{p}}^j} & \sim &  \d^{ij}p_{\m}e^{-ip\cdot x}\\
\braket{0| D_{\{\m_1}u_{\m_2\}}^i| \pi_{\mathbf{p}}^j} & \sim &  \d^{ij}p_{\{\m_1}p_{\m_2\}}e^{-ip\cdot x}\\
 & \vdots &\end{array}
\quad \Rightarrow \quad
\def\arraystretch{1.1}
R_{u} = \left(\begin{array}{c}
  u_{\m}^i  \\ D_{\{\m_1}u_{\m_2\}}^i  \\ \vdots 
\end{array} \right).
\label{eq:SPM_non-linear}
\end{equation}
In sec.~\ref{sec:non-linear} we will show this ansatz for \(R_u\) is correct via the conventional CCWZ Lagrangian description of non-linear realizations~\cite{Coleman:1969sm,Callan:1969sn}.

\subsection{Gauge symmetries, topological terms and discrete spacetime symmetries}\label{sec:2.5}
If the EFT has an internal symmetry group \(G\), then the operators in \(\scK\) should be \(G\)-invariant. At an abstract level, we simply append an additional superscript to eq.~\eqref{eq:K_differential_mod},
\begin{equation}
\scK = \left[\bigslant{\scJ}{d\scJ}\right]^{SO(d)\times G}.
\label{eq:K_differential_mod_G}
\end{equation}
Some care should be taken to identify the appropriate degrees of freedom, such as when a symmetry is non-linearly realized, as mentioned above.

Some comments should be made about massless particles with spin, which lead to gauge symmetries. The construction outlined presently ultimately works with simply transforming objects, since it is the field strengths which interpolate single particle states. Of course, there is a gauge potential field which the field strength is some number of exterior derivatives of, schematically \(f \sim \text{d}^na\), designed to guarantee the correct little group transformation (ensuring that the non-compact directions of \(ISO(d-2) \supset SO(d-2)\) act trivially).

For concreteness, let's discuss the familiar vector case \(F = \text{d}A\). The point is we work with \(F_{\m\n}\), not \(A_{\m}\), as is apparent in eq.~\eqref{eq:SPM_F}. Of course, we also promote \(\pd_{\m} \to D_{\m} = \pd_{\m} + A_{\m}\). In the case that there are multiple vector single particle states with a symmetry imposed on them, we know this must correspond to a non-abelian gauge theory and we work with \(F_{\m\n}=F_{\m\n}^aT^a\). We first note that gauge invariance is straightforward to impose because we work with gauge covariant quantities like \(F_{\m\n}\) and \(D_{\m}\ph_l\). Second, note that for the purposes of constructing the operator basis, the covariant derivative will behave like a partial derivative
\begin{equation}
\braket{0|D_{\m}\ph_l|\mathbf{p}\,\s} = \braket{0|\pd_{\m}\ph_l|\mathbf{p}\,\s},
\end{equation}
because \(A_{\m}\) plays no role in interpolating the single particle state (for one thing, \(A_{\m}\ph_l\) has two annihilation operators). This can also be justified by noting that \([D_{\m},D_{\n}] \sim F_{\m\n}\), so we can always replace \(D_{\m}D_{\n} \to D_{(\m}D_{\n)} + F_{\m\n}\) when constructing operators.

An important question is whether we mis-identify or omit operators by working with the field strengths instead of the gauge fields. The answer is yes, and it is unsurprisingly due to topological terms. For example, in even dimensions we will pick up the theta term \(\e^{\m_1\cdots \m_d}F_{\m_1\m_2}\cdots F_{\m_{d-1}\m_d}\) even though it is a total derivative, \(F\wedge \cdots \wedge F = \text{d}\big(A\wedge \text{d}A \wedge \cdots \wedge \text{d}A\big)\). Likewise, in odd dimensions we will not pick up the Chern-Simons term \(A\wedge \text{d}A \wedge \cdots \wedge \text{d}A\) (or its non-abelian generalization), which is gauge invariant up to a total derivative. Similar comments apply to other massless spinning fields, as well as to topological terms such as theta, Hopf, or Wess-Zumino terms in non-linear realizations (where instead of working with pions, we work with the Maurer-Cartan forms).

Let us also comment on discrete spacetime symmetries \(P\), \(T\), and \(C\). For calculations in this work, we work in Euclidean space \(SO(d)\), where we can only have \(P\) and \(C\). We have made some effort to incorporate parity, primarily to allow us to probe the rings \(M_n\) in eq.~\eqref{eq:M_n_module} by exploring the difference between \(SO(d)\) and \(O(d)\). We have not, however, attempted to include charge conjugation, nor to systematically address the Wick rotation back to Minkowski space where time-reversal becomes an optional symmetry.

\subsection{A partition function for the operator basis}\label{sec:2.6}

Now that we have outlined the definition of the operator basis in sections~\ref{sec:subsec2p1}--\ref{sec:subsec2p3} above, we can sketch the form that the partition function---Hilbert series---takes on the basis. As we might expect from a partition function, we will find the Hilbert series a very interesting and useful object for studying operator bases. Deferring a more precise description to the following sections, here we outline a simple physical derivation of the Hilbert series.

Take a single particle module \(R_{\Ph}\) and label the states via a character \(\ch_{\Ph}^{}\). The character is essentially a single particle partition function with weights for the energy ($q$) and angular momentum ($x_i$) of each state in the module. The tower of states in a single particle module arise from translations, which is reflected in the character by a factor of `momentum' \(P(q;x_i)\), that is $\chi_{\Ph}^{}\propto P$.

Since asymptotic scattering states are built as a Fock space, the multi-particle partition function is constructed from the single-particle partition function using the plethystic exponential (PE), which is the familiar generating function for free systems in statistical mechanics (although the name may not be familiar).\footnote{Crossing symmetry implies we can take all particles as incoming or outgoing, which is why we only need to consider one set of multiparticle states.} That is, PE$[\chi_{\Ph}^{}]$ is the partition function on $\mathcal{J}$. 

To now get at states in the operator basis, we need to enforce momentum conservation and Lorentz invariance. Consider the multiparticle states in $\mathcal{J}$ organized as per the RHS of eq.~\eqref{eq:sketchofdecomp}. Since their characters are proportional to \(P\), it is clear that multiplying by $1/P$ will remove total derivative states from $\mathcal{J}$; in essence we are removing a factor of momentum from every multiparticle module.  Finally, Lorentz invariance is simply applied by averaging over boosts and rotations, which is implemented using an $SO(d)$ group  integral over the angular variables $x_{i}$. That is, we find the Hilbert series, $H$, to take the schematic form,
\begin{equation}
H \sim \int d\mu_{SO(d)} \frac{1}{P} \text{PE}[ \chi^{}_{\Phi}] \,.
\end{equation}

\section{Conformal representations and characters}
\label{sec:reps}
A central tool to identifying the operator spectrum is the use of characters. To each single particle module we associate a character which labels its states by their scaling dimensions and transformation properties under the Lorentz group. When the single particle module corresponds to a conformal representation, the associated character is a conformal character.

The representations of the conformal group and their characters can be used as a powerful tool to address IBP redundancy, as we will show in section~\ref{sec:compute}. For this purpose, we review and summarize the necessary results in conformal representation theory. A comprehensive and excellent treatment of conformal characters in arbitrary dimensions is given by Dolan~\cite{Dolan:2005wy}; a somewhat gentler treatment in four dimensions can be found in~\cite{Barabanschikov:2005ri}.\footnote{We adopt the conventions of~\cite{Dolan:2005wy} in our analysis, although we differ in notation on one point. For characters of long and short irreducible representations~\cite{Dolan:2005wy} uses \(\mathcal{A}_{[\D;l]}\) and \(\mathcal{D}_{[\D;l]}\), respectively, while we use \(\ch^{}_ {[\D;l]}\) and \(\widetilde{\ch}_{[\D;l]}\), respectively.}

\subsection{Unitary conformal representations}\label{subsec:unitary}

We work with the conformal group in \(d\)-dimensions \(SO(d,2) \simeq SO(d+2,\mathbb{C})\).\footnote{Our conventions on factors of \(i\) follow~\cite{Dolan:2005wy}. We work in the orthonormal basis, where the generators of the Lie algebra are split into the Cartan subalgebra together with raising and lowering operators. In this basis, weights of representations are eigenvalues of the Cartan generators. For a physics oriented introduction, see \textit{e.g.}~\cite{Georgi}. We allow spinors, so we actually work with the covering group \(Spin(d+2,\mathbb{C})\), although we will not be careful to make this distinguishment throughout the text.} The \(SO(d)\) subgroup is the usual rotation group in Euclidean space \(\mathbb{R}^d\): it has rank \(r = \lfloor\frac{d}{2} \rfloor\) which means that its representations are labeled by \(r\) spin quantum numbers \(l = (l_1,\cdots,l_r)\). The extra two directions in the conformal group increase the rank by one, thereby giving us another label \(\D\) for representations, called the scaling dimension as it is physically associated with the dilatation operator. As \(SO(d)\) is compact, the spin labels are quantized, \(l_1,\cdots,l_r \in \frac{1}{2} \mathbb{Z}\); in contrast, the scaling dimension is continuous, \(\D \in \mathbb{R}\), as it is associated with non-compact directions. In summary, irreducible representations (irreps) of the conformal group are labeled by \(r+1\) quantum numbers \((-\D,l_1,\cdots,l_r)\), where the minus sign is a convention due to our working with \(SO(d+2,\mathbb{C})\).\footnote{This minus sign (with \(\D\) positive) leads to infinite dimensional representations for \(d \ge 2\). For \(d=1\), \(\D\) can be negative (the free scalar field has scaling dimension \(-1/2\)), which leads to finite dimensional representations when \(\D \in -\mathbb{N}/2\) and explains the \(SL(2,\mathbb{C})\) structure uncovered in~\cite{Henning:2015daa}.}

Unitary representations of \(SO(d+2,\mathbb{C})\) are constructed as follows: a highest weight state is specified and the representation is filled out by applying the lowering operators. In addition to the raising and lowering operators of \(SO(d)\), in \(SO(d+2,\mathbb{C})\) the translation generators \(P_\mu\) are lowering operators while the special conformal generators \(K_\mu\) are conjugate raising operators, \textit{i.e.} \(P_\mu^{\dagger} = K_\mu\). Due to the non-compact nature, the translation operators may be applied an infinite number of times without annihilating some lowest weight state, implying unitary irreps are infinite dimensional.

Heuristically, this picture is easy to understand in terms of field theory operators. The physical intuition comes from conformal field theories, where the state-operator correspondence implies that operators are organized into irreps of the conformal group. In essence, unitary irreps consist of some operator \(\scO_l\), called \textit{primary}, of spin \(l\) and scaling dimension \(\D\) together with an infinite tower of derivatives acting on \(\scO_l\), called \textit{descendants}:
\begin{equation}
R_{[\D;l]} \sim \begin{pmatrix} \scO_l \\ \pd_{\m_1}\scO_l \\ \pd_{\m_1}\pd_{\m_2} \scO_l \\ \vdots \end{pmatrix}.
\end{equation}

Requiring a representation be unitary places conditions on \(\D\) and \(l\). The conditions on \(l\) are the familiar ones for finite dimensional irreps of \(SO(d)\): a unitary irrep is labeled by \(l = (l_1,\cdots,l_r)\), satisfies \(l_i \in \frac{1}{2}\mathbb{Z}\) and \(l_i - l_{i+1} \in \mathbb{Z}\), with \(l_1 \ge \cdots \ge l_{r-1} \ge \abs{l_r}\) for \(SO(2r)\) and \(l_1 \ge \cdots \ge l_r \ge 0\) for \(SO(2r+1)\).\footnote{These are the conditions such that the Dynkin labels $(\L_1,\cdots,\L_r)$ are non-negative integers. In even dimensions $d=2r$, the Dynkin labels are $\L_i=l_i-l_{i+1}$ for $1\le i \le r-1$ and $\L_r=l_{r-1}+l_r$. In odd dimensions $d=2r+1$, the Dynkin labels are $\L_i=l_i-l_{i+1}$ for $1\le i \le r-1$, and $\L_r=2l_r$. Under this convention, for integer values of $l_i$, $(l_1,\ldots,l_r)$ corresponds to the usual Young diagrams used to label representations with \(l_1\) boxes in first row, \(l_2\) boxes in the second row, \textit{etc.}} Then for each $l$ satisfying these conditions, $\D$ needs to satisfy a lower bound $\D \ge \D_l$ for the irrep to be unitary. This lower bound is hence called the \textit{unitarity bound}, and is given by~\cite{Ferrara:2000nu,Dolan:2005wy}
\begin{equation}\def\arraystretch{1.5}
\D \ge \D_l = \left\{\begin{array}{ll}
(d-2)/2  & \hspace{3mm} \text{for } l=(0,\cdots,0) \\
(d-1)/2  & \hspace{3mm} \text{for } l=(\frac{1}{2},\cdots, \frac{1}{2}) \\
l_1+d-p_l^{}-1 & \hspace{3mm} \text{for all other } l \\
\end{array} \right. \, , \label{eqn:unitaritybound}
\end{equation}
where $1\le p_l^{} \le r$ denotes the position (the serial number) of the last component in $l=(l_1,l_2,\cdots,l_r)$ that has the same absolute value as the first component $l_1$, namely that the components of $l$ satisfy $|l_1|=|l_2|=\cdots=|l_{p_l^{}}|>|l_{p_l^{}+1}|$.

Physically, the unitarity bound imposes representations to have scaling dimension greater than or equal to that of free fields or conserved currents. When a bound is saturated, \textit{i.e.} $\D=\D_l$, some descendant obtained by applying \(\pd_{\m}\) is annihilated: it leads to null and subsequently negative norm states~\cite{Minwalla:1997ka}. In such a case, these states are removed and the irrep is called a \textit{short} representation. Accordingly, unitary irreps that are not short are sometimes referred to as \textit{long} representations.

To gain intuition, let us specialize to \(l = (n,0,\cdots,0)\) with \(n\in\mathbb{N}\) (traceless symmetric tensors with \(n\) indices), where the bounds read:
\begin{equation}\def\arraystretch{1.2}
\D_l = \left\{\begin{array}{ll}
(d-2)/2  & \hspace{3mm} \text{for}\hspace{2mm} n=0 \\
n+d-2    & \hspace{3mm} \text{for}\hspace{2mm} n>0
\end{array} \right. \, .
\end{equation}
Saturation for \(n=0\) corresponds to the free scalar, while for \(n>0\) we have conserved currents (\(n=1\) and 2 are a conserved current \(j_{\mu}\) and stress-tensor \(T_{\mu\nu}\), respectively, while \(n\ge 3\) are generalized higher spin conserved currents). In each instance, some descendant is annihilated by the derivative action. For example, for a free scalar field \(\ph\) the EOM dictates \(\pd^2\ph = 0\); for \(T_{\m\n}\), conservation dictates \(\pd^{\m}T_{\m\n} = 0\).

Our method involves working with free fields, using them as building blocks to construct the operator basis. In odd dimensions, the only free fields are scalars and spinors; in even dimensions, we may have scalars, chiral spinors, (anti) self-dual \(\frac{d}{2}\)-form field strengths, and higher spin generalizations~\cite{Siegel:1988gd}. These representations are correspondingly labeled by  $l=(s,\cdots,s)$ with \(s = 0,\, \frac{1}{2}\) for odd $d$, and $l=(s,\cdots,s,\pm s)$ with $s \in \mathbb{N}/2$ for even $d$. Due to the free field EOM, \textit{e.g.} \(\pd^2 \ph = 0\) for scalars, \(\slashed{\pd}\ps=0\) for fermions, \textit{etc}., the free field representations are short representations.

\subsection{Character formulae}

In this section we reproduce character formulae \cite{Dolan:2005wy} for unitary irreps of the conformal group. The character for a representation \(R\) is the trace of a group element in that representation, \(\chi^{}_R(g) = \text{Tr}_R(g)\) for \(g \in G\). For a connected Lie group \(G\), any \(g \in G\) can be conjugated into the maximal torus \(T = U(1)^{\text{rank}(G)}\), \textit{i.e.} there exists \(h \in G\) such that \(h^{-1} g h \in T\). This is simply a diagonalization theorem, generalizing the familiar statement for \(G = U(N)\) that any unitary matrix can be diagonalized by another unitary matrix. Since the trace is conjugation invariant, it depends only on the \(\text{rank}(G) = \text{dim}(T)\) parameters of the torus. For \(SO(d+2,\mathbb{C})\) we parameterize the \((r+1)\)-dimensional torus (recall, \(r = \lfloor \frac{d}{2} \rfloor\)) by the following variables: $q$, associated with the scaling dimension, and $x=(x_1,\dots,x_{r})$ for the parameters of the $SO(d)$ torus, \textit{i.e.} $q=e^{i\theta_q}$ and $x_i=e^{i\theta_i}$, with $\theta_q\in\mathbb{C}$ and $\theta_i\in\mathbb{R}$.

\subsubsection*{Long representations}

Consider a unitary irrep \(R_{[\D;l]}\) of scaling dimension \(\D\) and spin \(l = (l_1,\cdots, l_r)\). If the irrep does not saturate a unitarity bound, \textit{i.e.} $\D > \D_l$, then it takes the form
\begin{equation}
R_{[\D;l]} = \begin{array}{cc|ccc} & &  & \text{scaling dim} & \text{spin} \\ \scO_l & & & \D & l \\ \pd_{\m_1}\scO_l & & &  \D + 1 & \text{sym}^1(\Yfund) \otimes l \\  \pd_{\m_1}\pd_{\m_2} \scO_l & & & \D + 2 & \text{sym}^2(\Yfund) \otimes l \\ \vdots & & &  \vdots & \vdots \end{array},
\end{equation}
where \(\text{sym}^n(\Yfund)\) is the representation formed by the \(n\)th symmetric product of the vector representation, \(\Yfund \leftrightarrow l = (1,0,\dots,0)\), of \(SO(d)\) (the derivative is a \(SO(d)\) vector and symmetrization comes because partial derivatives commute). A group element \(g \in SO(d+2,\mathbb{C})\) acting on the above representation is an infinite dimensional matrix with block diagonal components acting on the primary and descendants. Evidently, upon tracing over such a matrix the character \(\ch^{(d)}_{[\D;l]}(q;x)\) is
\begin{equation}
\ch^{(d)}_{[\D;l]}(q;x) = \sum_{n=0}^{\infty} q^{\D + n} \ch_{\text{sym}^n(\Box)}^{(d)}(x)\ch_l^{(d)}(x) = q^{\D} \ch^{(d)}_l(x) P^{(d)}(q;x) , \label{eq:chi_ddim}
\end{equation}
where $\ch^{(d)}_l(x)$ denotes the character of the spin $l$ representation of $SO(d)$ (see app.~\ref{app:sodspin} for general formulae), and we have used the fact \(\ch_{R_1 \otimes R_2}^{} = \ch_{R_1}^{} \ch_{R_2}^{}\). In the second equality above, we have defined a quantity
\begin{equation}
P^{(d)}(q;x) \equiv \sum_{n=0}^{\infty} q^n \ch_{\text{sym}^n(\Box)}^{(d)}(x) , \label{eq:symm_vec_P}
\end{equation}
which can be thought of as the momentum generating function: $q^{\D} \ch^{(d)}_l(x)$ is the contribution from the primary block, while $P^{(d)}(q;x)$ generates the contributions from all the descendants. This function can be computed as follows. In even dimensions, $d=2r$, a group element $h\in SO(2r)$ in the vector representation has eigenvalues
\be
h_{\Box}^{(2r)} \mapsto \text{diag}(x_1, x_1^{-1}, \cdots, x_r, x_r^{-1}) . \nonumber
\ee
Because the representation $\text{sym}^n(\Box)$ is formed by the $n$th fully symmetric products of the vector components, each distinct degree-$n$ monomial formed by the above eigenvalues should show up precisely once in $\ch_{\text{sym}^n(\Box)}^{(2r)}(x)$. Therefore
\be
\ch_{\text{sym}^n(\Box)}^{(2r)}(x) = \sum_{a_1+\bar{a}_1+\cdots+a_r+\bar{a}_r=n} \big(x_1\big)^{a_1} \big(x_1^{-1}\big)^{\bar{a}_1} \cdots \big(x_r\big)^{a_r} \big(x_r^{-1}\big)^{\bar{a}_r} . \nonumber
\ee
Plugging this in and performing the sum in eq.~\eqref{eq:symm_vec_P} we get
\be
P^{(2r)}(q;x) = \prod_{i=1}^{r}\frac{1}{(1-q \,x_i)(1-q / x_i)} .
\label{eq:P_even_dim}
\ee
Similarly, in odd dimensions, $d=2r+1$, we have
\begin{align}
h_{\Box}^{(2r+1)} &\mapsto \text{diag}(1, x_1, x_1^{-1}, \cdots, x_r, x_r^{-1}) , \nonumber\\
\ch_{\text{sym}^n(\Box)}^{(2r+1)}(x) &= \sum_{a_0+a_1+\bar{a}_1+\cdots+a_r+\bar{a}_r=n} \big(1\big)^{a_0} \big(x_1\big)^{a_1} \big(x_1^{-1}\big)^{\bar{a}_1} \cdots \big(x_r\big)^{a_r} \big(x_r^{-1}\big)^{\bar{a}_r} , \nonumber
\end{align}
and
\be
P^{(2r+1)}(q;x) = \frac{1}{1-q}\prod_{i=1}^{r}\frac{1}{(1-q \,x_i)(1-q / x_i)} .
\label{eq:P_odd_dim}
\ee

From the above, we note that for both even and odd dimension cases
\begin{equation}
P^{(d)}(q;x) = \sum_{n=0}^{\infty} q^n \ch_{\text{sym}^n(\Box)}^{(d)}(x) = \frac{1}{\det_{\Box}\left(1- q h(x)\right)}. \label{eqn:Pdet}
\end{equation}
This identity is central to many of the calculations in this work: a generating function for symmetric tensor products of a representation \(V\) under a group \(G\) is given by
\begin{equation}
\sum_{n=0}^{\infty} u^n \ch_{\text{sym}^n(V)}(g) = \frac{1}{\det_V(1- u g)} , \label{eqn:symGendet}
\end{equation}
for \(g \in G\). Using \(\log \det = \text{Tr} \log\), the above can be rewritten into an object called the plethystic exponential,
\begin{equation}
\frac{1}{\det_V(1- u g)} = \exp \left( \sum_{m=1}^{\infty}\frac{1}{m} u^m \text{Tr}_V(g^m) \right) \equiv \text{PE}[u \ch_V^{}(g)] . \label{eqn:symPEdef}
\end{equation}

With the plethystic exponential, one can define the $n$th symmetric product of functions, \textit{i.e.} objects like \(\text{sym}^n[f(x)]\): it is the \(u^n\) coefficient of \begin{equation}
\text{PE}[uf(x)] = \text{exp}[\sum_{m=1}^\infty\frac{u^m}{m}f(x^m)].
\end{equation}
For example, one readily finds
\begin{align}
\text{sym}^2\big[f(x)\big] &= \frac{1}{2} \big[f(x)^2 + f(x^2)\big], \nonumber \\
\text{sym}^3\big[f(x)\big] &= \frac{1}{6} \big[f(x)^3 + 3f(x)f(x^2)+2 f(x^3)\big], \nonumber
\end{align}
and so forth. With this definition, we see from eqs.~\eqref{eqn:symGendet} and~\eqref{eqn:symPEdef} that $\ch_{\text{sym}^n(V)}(x)$ is the same as $\text{sym}^n[\ch^{}_V(x)]$.\footnote{A simple exercise to familiarize these operations is to work out some examples for \(SU(2)\). The doublet character is \(\ch_\mathbf{2}^{}(\a) = \a + \a^{-1}\). Check familiar statements like \(\mathbf{2}\times \mathbf{2} = \mathbf{3} + \mathbf{1}\), \(\text{sym}^2(\mathbf{2}) = \mathbf{3}\), \textit{etc}.}

For anti-symmetric (exterior) tensor products, \(\wedge^n(V)\), a similar identity holds:
\begin{equation}
\sum_{n=0}^{\infty} u^n \ch_{\wedge^n(V)}(g) = \text{det}_V(1 + u g) . \label{eqn:asymGendet}
\end{equation}
Note that the sum truncates since the antisymmetric product $\wedge^n(V)$ vanishes for $n > \text{dim}(V)$. In physics language, the difference compared to the symmetric case is a reflection of the statistics---it simply generalizes the familiar case of the fermionic/bosonic partition function given by $\left(1\pm u\right)^{\pm 1}$ respectively, with $u=e^{-\beta}$. This fermionic generating function can again be written into a (fermionic) plethystic exponential
\begin{equation}
\text{det}_V(1 + u g) = \exp \left( \sum_{m=1}^{\infty}\frac{(-1)^{m+1}}{m} u^m \text{Tr}_V(g^m) \right) \equiv \text{PE}_f[u \ch_V^{}(g)] .
\label{eqn:asymPEdef}
\end{equation}
Similarly to the definition of $\text{sym}^n[f(x)]$, one can define the antisymmetric product of functions $\wedge^n[f(x)]$ as the $u^n$ coefficient of $\text{PE}_f[uf(x)]$. With this definition, we obviously have $\wedge^n[\ch_V^{}(x)]=\ch_{\wedge^n(V)}^{}(x)$. Frequently we omit the subscript on \(\text{PE}_f\), as it is usually clear by context what is meant.

\subsubsection*{Short representations}

The character formulae are modified when a unitarity bound is saturated, \textit{i.e.} when $\D=\D_l$ in eq.~\eqref{eqn:unitaritybound}. Let us first look at some examples. Consider the short representation formed by the free scalar field $\ph$, with $\D=\D_0\equiv(d-2)/2$ and $l=\underline{0}\equiv(0,\cdots,0)$. The shortening condition comes from the EOM $\partial^2 \ph=0$, as a result of which there are only \textit{traceless} symmetric components \(\pd_{\{\m_1}\cdots \pd_{\m_n\}}\ph\) in the descendants. Thus the representation looks like
\begin{equation}
R_{[\D_0;\underline{0}]} = \begin{array}{cc|cccc} & &  & \text{scaling dim} & \mbox{ } & \text{spin} \\ \ph & & & \D_0 & & 0 \\ \pd_{\m_1}\ph & & &  \D_0 + 1 & & \Yfund \\  \pd_{\{\m_1}\pd_{\m_2\}} \ph & & & \D_0 + 2 & & \Ysymm \\  \pd_{\{\m_1}\pd^{}_{\m_2}\pd_{\m_3\}} \ph & & & \D_0 + 3 & & \Ythrees \\ \vdots & & &  \vdots & & \vdots \end{array},
\label{eq:free_scalar_rep_table}
\end{equation}
with \(\underbrace{\Box \cdots \Box}_{n}\) the Young diagram labeling the traceless symmetric representation of \(SO(d)\) with \(n\) indices, corresponding to \(l = (n,0,\cdots,0)\). The set of components $\partial_{\{\mu_1}\cdots\partial_{\mu_n\}}\ph$ is obtained from $\partial_{\mu_1}\cdots\partial_{\mu_n}\ph$ by contracting two of the indices while leaving the others fully symmetric. Therefore, we have
\begin{equation}
\ch_{(n,0,\cdots,0)}^{(d)}(x) = \left\{ \begin{array}{ll}
\ch_{\text{sym}^n(\Box)}^{(d)}(x) & \hspace{5mm} n<2 \\
& \\
\ch_{\text{sym}^n(\Box)}^{(d)}(x) - \ch_{\text{sym}^{n-2}(\Box)}^{(d)}(x) & \hspace{5mm} n\ge2
\end{array}
\right. .
\end{equation}
Using this, as well as eq.~\eqref{eq:symm_vec_P}, we obtain the character for the short representation in eq.~\eqref{eq:free_scalar_rep_table} as
\begin{equation}
\widetilde{\ch}^{(d)}_{[\D_0;\underline{0}]}(q;x) = \sum_{n=0}^{\infty} q^{\D_0 + n} \ch_{(n,0,\cdots,0)}^{(d)}(x) = q^{\D_0}(1-q^2) P^{(d)}(q;x) ,
\label{eq:chi_free_scalar_sum}
\end{equation}
where we have used a tilde to emphasize that this is a short representation. This result is rather simple to interpret. Using eq.~\eqref{eq:chi_ddim} we can write it as
\begin{equation}
\text{free }\ph: \quad \widetilde{\ch}^{(d)}_{[\D_0;\underline{0}]}= \ch^{(d)}_{[\D_0;\underline{0}]} - \ch^{(d)}_{[\D_0 + 2;\underline{0}]} \,,
\label{eq:chi_free_scalar}
\end{equation}
which clearly reflects subtracting off the states \((\pd^2\ph, \pd_{\m}\pd^2\ph, \pd_{\m_1}\pd_{\m_2}\pd^2\ph,\cdots)\) from the long multiplet \((\ph, \pd_{\m}\ph, \pd_{\m_1}\pd_{\m_2}\ph, \cdots)\).

Similar interpretations can be given to other short representations. For example, a conserved current \(j_{\m}\) has \(\D = d-1\) and \(l = (1,0,\cdots,0)\). Current conservation dictates \(\pd^{\m}j_{\m} = 0\), where \(\pd^{\m}j_{\m}\) itself is an operator with \(\D = d\), \(l = \underline{0}\). Thus the character is given by
\begin{equation}
\text{conserved }j_{\m}: \quad \widetilde{\ch}^{(d)}_{[d-1;(1,0,\cdots,0)]}= \ch^{(d)}_{[d-1;(1,0,\cdots,0)]} - \ch^{(d)}_{[d;\underline{0}]}.
\label{eq:chi_cons_j}
\end{equation}

In some instances, we must subtract off another short representation, as opposed to a long representation. An example of this is the left-handed field strength in four dimensions, \(F_{L\, \m \n} = F_{\m\n} + \tilde{F}_{\m\n}\) with \(\D = 2\) and \(l = (1,1)\). The EOM and Bianchi identity imply \(\pd^{\m}F_{L\, \m \n} = 0\), so that descendant states proportional to \((\pd\cdot F_{L})_{\n}\) should be removed. However, the operator \((\pd\cdot F_L)_{\n}\) has \(\D = 3\) and \(l = (1,0)\), which is the same as a conserved current, and therefore saturates a unitarity bound itself. Manifestly one can see this because \(\pd^{\m}(\pd\cdot F_L)_{\m}\)  vanishes automatically by anti-symmetry. Hence, the character is obtained by subtracting the \textit{short} representation \(\widetilde{\ch}^{(4)}_{[3;(1,0)]}\) from \(\ch^{(4)}_{[2;(1,1)]}\):
\begin{subequations}
\label{eqn:chiFL4d}
\begin{align}
4d\text{ field strength }F_{L\, \m\n}: \quad \widetilde{\ch}^{(4)}_{[2;(1,1)]} &= \ch^{(4)}_{[2;(1,1)]} - \widetilde{\ch}^{(4)}_{[3;(1,0)]} \label{eq:chi_F_L_4d_a}\\
&= \ch^{(4)}_{[2;(1,1)]} - \ch^{(4)}_{[3;(1,0)]} + \ch^{(4)}_{[4;(0,0)]}.
\label{eq:chi_F_L_4d}
\end{align}
\end{subequations}

\subsubsection*{The general shortening rule and notation for conformal characters}

Let us make some clarification about our notations on conformal characters. We will always use $\ch_{[\D;l]}^{(d)}(q;x)$, without a tilde, to denote the function in eq.~\eqref{eq:chi_ddim}. That is
\begin{equation}
\ch_{[\D;l]}^{(d)}(q;x) \equiv q^\D \ch_l^{(d)}(x) P^{(d)}(q;x) . \label{eqn:chidef}
\end{equation}
On the other hand, we will use $\widetilde{\ch}_{[\D;l]}^{(d)}(q;x)$, with the tilde, to denote the actual conformal character of the conformal representation labeled by $[\D;l]$. For long representations, $\widetilde{\ch}$ is just given by $\ch$:
\begin{equation}
\widetilde{\ch}_{[\D;l]}^{(d)}(q;x) = \ch_{[\D;l]}^{(d)}(q;x) \quad, \quad \text{for}\hspace{2mm} \D > \D_l \, . \label{eqn:chiTildeLong}
\end{equation}
For short representations ($\D=\D_l$), $\widetilde{\ch}$ is different from $\ch$. As in the above examples, $\widetilde{\ch}_{[\D_l;l]}^{(d)}$ is obtained from $\ch_{[\D_l;l]}^{(d)}$ by subtracting off another conformal character $\widetilde{\ch}$. The general shortening rule is\footnote{The characters for all short unitary conformal irreps are given in eqs.~(3.25)-(3.27) and (3.32)-(3.34) of~\cite{Dolan:2005wy}. Here we only summarize the results, and refer the reader to~\cite{Dolan:2005wy} for the proof.}
\renewcommand\arraystretch{1.5}
\begin{equation}
\widetilde{\ch}_{[\D_l;l]}^{(d)} = \left\{\begin{array}{ll}
\ch_{[\D_l;l]}^{(d)} - \widetilde{\ch}_{[\D_l+2;l^-]}^{(d)} & \hspace{5mm} \text{for}\hspace{2mm} l=\underline{0} \\
\ch_{[\D_l;l]}^{(d)} - \widetilde{\ch}_{[\D_l+1;l^-]}^{(d)} & \hspace{5mm} \text{for}\hspace{2mm} l \ne \underline{0}
\end{array}\right. \, , \label{eqn:chiTildeShort}
\end{equation}
\renewcommand\arraystretch{1.0}
with the lowered spin $l^-$ obtained from $l$ by a simple replacement:
\begin{equation}\def\arraystretch{1.5}
l^- \equiv \left\{\begin{array}{ll}
l & \hspace{5mm} \text{for } l=(\frac{1}{2},\cdots,\frac{1}{2}) \text{ in odd } d \\
l \text{ with } l_{p_l^{}} \to l_{p_l^{}} - \text{sgn}\left(l_{p_l^{}}\right) & \hspace{5mm} \text{for all other cases}
\end{array}\right. \, , \label{eqn:lminus}
\end{equation}
where we recall from eq.~\eqref{eqn:unitaritybound} that $l_{p_l^{}}$ is defined as the last component in $l$ that has the same absolute value as $l_1$, and sgn is the standard signum function:
\begin{equation}
\def\arraystretch{1.2}
\text{sgn}(x) \equiv \left\{\begin{array}{rl}
 1 & \hspace{5mm} \text{for}\hspace{2mm} x>0 \\
 0 & \hspace{5mm} \text{for}\hspace{2mm} x=0 \\
-1 & \hspace{5mm} \text{for}\hspace{2mm} x<0
\end{array}\right. \, .\nonumber
\end{equation}
Explicit examples of eq.~\eqref{eqn:chiTildeShort} are given in eqs.~\eqref{eq:chi_free_scalar}, ~\eqref{eq:chi_cons_j}, and~\eqref{eqn:chiFL4d}. For any $\widetilde{\ch}$, one can use eq.~\eqref{eqn:chiTildeShort} iteratively until the character being subtracted off corresponds to a long representation. So it is clear that any $\widetilde{\ch}$ is a linear combination of the $\ch$. Note that in this iteration, generically the $\widetilde{\ch}$ being subtracted in eq.~\eqref{eqn:chiTildeShort} is a short irrep itself. This is because generically eq.~\eqref{eqn:lminus} reduces $p_l^{}$ by 1, \textit{i.e.} $p_{l^-}^{}=p_l^{}-1$; since $\Delta_{l^-} = \Delta_{l} + 1$, the values for $\Delta_{l^-}$ and $p_{l^-}^{}$ generically saturate the bound in eq.~\eqref{eqn:unitaritybound}. The only exceptions are when $l=\underline{0}$ and $(\frac{1}{2},\cdots,\frac{1}{2},\pm\frac{1}{2})$, or when $p_l^{}=1$; in these cases the $\widetilde{\ch}$ being subtracted is a long representation and the iteration terminates.

\subsection{Character orthogonality}\label{subsec:char_ortho}

For a compact Lie group, the characters of the irreps are orthonormal with respect to integration over the group:
\begin{equation}
\int d\m_G^{} \, \ch_{R_1}^*(g) \ch_{R_2}^{}(g) = \d_{R_1R_2},
\label{eq:char_ortho_compact_G}
\end{equation}
where \(d\mu_G^{}\) is the invariant measure (Haar measure) on \(G\). When the integrand is a class function (conjugation invariant)---as clearly is the case for characters---the Haar measure can be restricted to the torus and is given by the Weyl integration formula (see appendix~\ref{app:Weyl}):
\begin{equation}
\int d\m_G^{} \to \frac{1}{\abs{W}} \oint \prod_i \frac{dx_i}{2\pi i x_i} \prod_{\a \in \text{rt}(G)}(1 - x^{\a})  \, . \label{eq:weyl_int}
\end{equation}
The product over $\alpha\in \text{rt}(G)$ is taken over all the roots of $G$, \(\abs{W}\) is the order of the Weyl group, and the contours are taken along \(\abs{x_i} = 1\). We use the shorthand notation $x^\alpha\equiv x_1^{\alpha_1}\cdots x_r^{\alpha_r}$ for $\alpha=(\alpha_1,\cdots,\alpha_r)$. The above can be simplified to a product over only the positive roots, see eq.~\eqref{eq:weyl_int_pos_roots}, which is quite useful for practical calculations. The Haar measures for the classical groups are tabulated in appendix~\ref{app:Weyl}.

The above results do not immediately generalize to non-compact groups, such as the conformal group. However, for the specific representations of \(SO(d+2,\mathbb{C})\) we use in this work, there is a notion of character orthogonality that we can make use of. As we work in a free field limit, the scaling dimension of operators coincides with the canonical dimension. In \(d\ge 2\) dimensions, this implies that all values of \(\D\) are positive, half-integer. In this case, the characters \(\ch^{(d)}_{[\D;l]}\) defined in eq.~\eqref{eqn:chidef} with \(\D \in \mathbb{Z}/2\) are orthonormal with respect to integration over the maximal compact subgroup of \(SO(d+2,\mathbb{C})\), namely \(SO(d+2)\):
\begin{equation}
\int d\m \ \ch^{(d)\,*}_{[\D;\, l]}(q;x)\ch^{(d)}_{[\D';\, l']}(q;x) = \int d\m \ \ch^{(d)}_{[\D;\, l]}\left(q^{-1};x^{-1}\right)\ch^{(d)}_{[\D';\, l']}(q;x) = \d_{\D\D'}\d_{ll'} \, , \label{eq:conf_char_ortho}
\end{equation}
where the Haar measure $d\mu$ (restricted to the torus) is given by\footnote{This measure is normalized so that eq.~\eqref{eq:conf_char_ortho} holds.  Up to a proportionality constant it is just the measure for \(SO(d+2)\):  \(d\m = (2r+2)d\m_{SO(d+2)}\) for both \(d=2r\) and \(d=2r+1\), with \(d\m_{SO(d+2)}\) given in eqs.~\eqref{eq:haar_SO_odd} and~\eqref{eq:haar_SO_even} (replace \(x_{r+1}\to q\) in these formulas to obtain eq.~\eqref{eq:haar_SO_d+2}).}
\begin{equation}
\int d\m = \oint \frac{dq}{2\pi i q} \int d\m_{SO(d)}^{}(x) \, \frac{1}{P^{(d)}(q;x)P^{(d)}(q^{-1};x^{-1})} \, .
\label{eq:haar_SO_d+2}
\end{equation}

With the explicit expression of the character in eq.~\eqref{eqn:chidef}, it is trivial to check that eq.~\eqref{eq:conf_char_ortho} holds (using eq.~\eqref{eq:char_ortho_compact_G} for the \(SO(d)\) characters). We emphasize that the modified characters \(\widetilde{\ch}\) for short irreps are \textit{not} orthonormal with respect to this measure, as can be seen from the fact that the \(\widetilde{\ch}\) are linear combinations of the \(\ch\). We will have to carefully account for this non-orthonormality.

As a technical matter, one has to keep in mind the covering group \(Spin(d+2,\mathbb{C})\) in applying the contour integrals of eqs.~\eqref{eq:weyl_int} and~\eqref{eq:haar_SO_d+2}. In particular, if the theory under consideration has either half-integer scaling dimensions or spinors, then we are in the double cover \(Spin(d+2,\mathbb{C})\) of \(SO(d+2,\mathbb{C})\). In the characters, this appears as possible square roots of the arguments \(q\) and \(x\). For simplicity of discussion, let us focus on half-integer scaling dimensions. Then the integral over \(q = e^{i\th_q}\), needs to be extended to the double cover:
\begin{equation}
\oint_{|q|=1} \frac{dq}{2\pi iq} f(q) = \int_0^{2\pi} \frac{d\theta_q}{2\pi} f(q) \quad \to \quad \int_0^{4\pi} \frac{d\theta_q}{4\pi} f(q) =  \oint_{|q|=1} \frac{dq}{2\pi iq} f(q^2) \,, \label{eqn:doublecover}
\end{equation}
 that is, one leaves the \(dq/(2\pi iq)\) piece alone and replaces \(q \to q^2\) everywhere else in the integrand.

\subsection{Summary}

Conformal representations are labeled by their scaling dimension \(\D\) and spin \(l = (l_1,\cdots,l_r)\). In \(d \ge 2\) dimensions, unitary irreps are infinite dimensional. Structurally, they consist of a primary operator \(\scO_{\D,l}\) of scaling dimension \(\D\) and spin \(l\) together with an infinite number of descendants consisting of derivatives acting on \(\scO_{\D,l}\).

A long irrep does not saturate a unitarity bound. Its character is given by
\begin{equation}
\ch^{}_{[\D;l]}(q,x) = q^{\D} \ch_l^{}(x) P(q;x) ,\quad \text{for}\hspace{2mm} \D > \D_l , \label{eqn:chiTildeLongRecap}
\end{equation}
where $\ch_l^{}(x)$ are the $SO(d)$ spin-$l$ characters with detailed expressions given in appendix~\ref{app:sodspin}, and the momentum generating function \(P(q;x)\) is
\begin{align}
P(q;x) \equiv \sum_{n=0}^{\infty} q^n \ch_{\text{sym}^n(\Box)}^{}(x) = \left\{\begin{array}{ll}
\displaystyle\prod_{i=1}^{r}\dfrac{1}{(1-q \,x_i)(1-q / x_i)} & \hspace{5mm} d=2r \\
& \\
\dfrac{1}{1-q}\displaystyle\prod_{i=1}^{r}\dfrac{1}{(1-q\,x_i)(1-q/x_i)} & \hspace{5mm} d=2r+1
\end{array}
\right. . \label{eqn:PRecap}
\end{align}

A short irrep saturates a unitarity bound, $\D=\D_l$. In this case, some descendant state becomes null; this state and its descendants are removed to form the short multiplet. The general shortening rule is summarized in eq.~\eqref{eqn:chiTildeShort}, from which we readily see that the \(\widetilde{\ch}\) are linear combinations of the \(\ch\). Some explicit examples of this are given in eqs.~\eqref{eq:chi_free_scalar}, ~\eqref{eq:chi_cons_j}, and~\eqref{eqn:chiFL4d}; we reproduce the free scalar character here:
\begin{equation}
\widetilde{\ch}^{}_{[\D_0;\underline{0}]} = \ch_{[\D_0;\underline{0}]}^{} - \ch_{[\D_0 + 2;\underline{0}]}^{} = q^{\D_0}(1-q^2) P(q;x) . \label{eqn:chiphiRecap}
\end{equation}

In the free field limit, where all scaling dimensions are positive half-integer, the characters $\ch_{[\D;l]}^{}(q;x)$ are orthonormal with respect to an integration over \(SO(d+2)\), the maximal compact subgroup of \(SO(d+2,\mathbb{C})\)
\begin{equation}
\int d\m \, \ch_{[\D;l]}^{*} \ch_{[\D';l']}^{} = \d_{\D\D'} \d_{ll'} \, , \label{eqn:chiORecap}
\end{equation}
with the Haar measure given by eq.~\eqref{eq:haar_SO_d+2}.

We  frequently make use of symmetric and anti-symmetric products of representations. The plethystic exponential serves as a generating function for them:
\begin{subequations}\label{eqn:GenPE}
\begin{align}
\sum_{n=0}^\infty u^n \ch_{\text{sym}^n(V)}^{}(g) &= \frac{1}{\det_V (1-u g)} = \text{PE} \left[ u \ch_V^{}(g) \right] , \label{eqn:symGenPE} \\
\sum_{n=0}^{\infty} u^n \ch_{\wedge^n(V)}(g) &= \text{det}_V(1 + u g) = \text{PE}_f \left[ u \ch_V^{}(g) \right], \label{eq:asymGenPE}
\end{align}
\end{subequations}

Finally, a comment on notation. In this section we have used a superscript to denote what spacetime dimensionality functions are in, \textit{e.g.} \(\ch^{(d)}_{[\D;l]}(q;x) = q^{\D}\ch_l^{(d)}(x)P^{(d)}(q;x)\). Such a notation is cumbersome. Since the meaning is generally clear in context, we will typically drop the superscript, as we have done in this summary.

\section{Counting operators: Hilbert series}
\label{sec:compute}

We now proceed to building the Hilbert series as a partition function on the operator basis \(\mathcal{K}\). As discussed in the Section~\ref{sec:overview}, the essential idea is to consider the enlarged operator space $\scJ$ and select out $\scK\subset \scJ$. $\scJ$ is built from tensor products of single particle modules $R_{\Phi_i}$. In the special case that the $R_{\Phi_i}$ are representations of the conformal group, then $\scJ$ itself forms a (reducible) representation of the conformal group; decomposing $\scJ$ back into conformal irreps, we identify $\scK$ as the scalar primaries. 

To explicitly obtain the Hilbert series, we associate characters $\chi_{\Phi_i}(q,x)$ to the $R_{\Phi_i}$, allowing us to define a generating function for $\scJ$. The piece corresponding to $\scK$ is then selected using character orthogonality, such that we ultimately arrive at a matrix integral expression for the Hilbert series.

We stress that the reduction from $\scJ$ to $\scK$ can also be addressed without recourse to conformal representation theory. We will show in sec.~\ref{subsec:nonlinearHilbert} how the cohomological nature of the IBP redundancy allows us to obtain the Hilbert series using the language of differential forms. 

We begin by deriving an explicit integral expression for the Hilbert series for a real scalar in \(d\) dimensions. Along the way, we show how the physical partition function can be obtained rather simply from our formalism. We next present the straightforward generalization to multiple fields and possible internal gauge and global symmetries. It is then explained how to include parity if one wishes to consider parity invariant EFTs; this will prove useful for the study of the real scalar field presented in section~\ref{sec:rsf}. Some of the derivation and details in these first sections are rather technical and bear little weight on practical computations of \(H\). For this reason we give a summary of the main expression for the Hilbert series in section~\ref{subsec:Hilbert_summary}, emphasizing its basic ingredients.

\subsection{Deriving a matrix integral formula for the Hilbert series}\label{subsec:Hderivation}

For clarity of presentation, we will derive the Hilbert series for a real scalar field $\phi$ in $d$-dimensions. The modifications to include multiple fields, symmetries {\it etc.} are described in the following subsection.

\subsection*{The character generating function $Z(\ph, q, x)$ for $\scJ$}

We start with the single particle module for a real scalar field, eq.~\eqref{eq:SPM_scalar},
\be
R_\ph = \left(
\begin{array}{cc}
\ph & \\
\pd_{\m_1}\ph & \\
\pd_{\{\m_1}\pd_{\m_2\}} \ph & \\
\pd_{\{\m_1}\pd^{}_{\m_2}\pd_{\m_3\}} \ph & \\
\vdots &
\end{array} \right) \,. \nonumber
\label{eqn:phitower}
\ee
The corresponding weighted character is (see eqs.~\eqref{eq:chi_free_scalar_sum} and~\eqref{eqn:chiphiRecap})
\be
\chi^{}_{R_\phi}(q,x)=\sum_{k=0}^\infty q^{\Delta_0+k} \ch_{(k,0,\ldots,0)}^{}(x) \,. \nonumber
\ee
Recall that the operator space $\mathcal{J}$ is constructed by taking symmetric tensor products of $R_{\phi}$, eq.~\eqref{eq:J_sym_prod}. Therefore, a generating function which labels states in $\mathcal{J}$ is readily constructed from $\chi_{R_{\phi}}^{}$ with the plethystic exponential (see eqs.~\eqref{eqn:symGendet} and \eqref{eqn:symPEdef}). We call this the character generating function and denote it by $Z$,
\be
Z(\ph, q, x)= \sum_{n=0}^\infty  \ph ^n \ch_{\text{sym}^n(R_\phi)}^{}(q,x) = \text{PE} \left[ \ph\, \chi^{}_{R_\phi}(q,x) \right] \label{eqn:ZphiSOd} \,.
\ee

\subsection*{IBP addressed by conformal representation theory}

The key insight is that the single particle module forms precisely the free scalar representation of the conformal group: $R_{\phi}=R_{[\Delta_0,\underline{0}]}$, and so $Z(\ph, q, x)= \text{PE} \left[ \ph\, \chi^{}_{R_\phi}\right] =\text{PE} \left[ \ph\,\widetilde{\chi}_{[\Delta_0,\underline{0}]}\right] $. Therefore, the operator space $\scJ$ generated by $R_\ph$ will also form a conformal representation. We just need to decompose $\scJ$ into conformal irreps, and then the highest weight state in each irrep, namely the primary operator, is what will survive the IBP redundancy. Then the operator basis $\scK$ is just the operator space formed by the scalar primaries. We now follow this strategy to define the Hilbert series $H(\ph,p)$ from the character generating function $Z(\ph,q,x)$. 

Characters nicely keep track of the decomposition into conformal irreps:
\be
\text{sym}^{n}\left(R_{[\D_0;\underline{0}]}\right) = \sum_{\D,l} b^{(n)}_{\D,l} \, R_{[\D;l]} \hspace{2mm} \implies \hspace{2mm} \widetilde{\ch}_{\text{sym}^n\left([\D_0;\underline{0}]\right)}^{} = \sum_{\D,l} b^{(n)}_{\D,l} \, \widetilde{\ch}_{[\D;l]}^{} \, ,
\label{eq:decomp}
\ee
with $b^{(n)}_{\D,l}$ some unknown multiplicities.\footnote{In general, the sum over \(\D\) on the rhs of eq.~\eqref{eq:decomp} should be an integral over \(\D\), as the scaling dimension is allowed to be continuous in the conformal group. However, as our starting point is to take tensor products of free field representations that have integer or half-integer scaling dimensions, the representations that show up in the decomposition have \(\D \in \mathbb{Z}/2\). In this sense, \(\D\) is effectively quantized for our purposes and the sum in eq.~\eqref{eq:decomp}, as opposed to an integral, is appropriate.} Hence, we can rewrite the character generating function as
\begin{equation}
Z(\ph,q,x) = 1 + \sum_{n=1}^{\infty} \ph^n \sum_{\D,l} b^{(n)}_{\D,l} \widetilde{\ch}_{[\D;l]} = 1 + \sum_{\D,l} C_{\D,l}(\ph) \widetilde{\ch}_{[\D;l]} ,
\label{eq:Z_decomp}
\end{equation}
where we have separated out the \(n=0\) piece for convenience, and performed the sum over \(n\) with the definition $C_{\D,l}(\ph) \equiv \sum_{n=1}^\infty \phi^nb^{(n)}_{\D,l}$. Keeping only the scalar primaries amounts to replacing each $\widetilde{\ch}_{[\D;l]}$ with $p^\Delta\delta_{l,\underline{0}}$. Therefore, the Hilbert series is defined as
\begin{align}
H(\ph,p) \equiv 1 + \sum_{\D,l} C_{\D,l}(\ph) p^\D \delta_{l,\underline{0}} = 1 + \sum_{\D} C_{\D,\underline{0}}(\ph) p^\D . \label{eq:H_define}
\end{align}
\(C_{\D,\underline{0}}(\ph)\) gives the number of operators of dimension \(\D\), weighted by the number of fields \(\ph\) in the a given operator. For example, in four dimensions at \(\D = 8\) we have two operators, \(\ph^8\) and \([(\pd_{\m}\ph)^2]^2\), so we anticipate that \(C_{8,\underline{0}}(\ph) = \ph^8 + \ph^4\).

\subsection*{Computing the multiplicities $C_{\D,\underline{0}}(\ph)$}

Now it is clear that to obtain the Hilbert series, our task is to compute the weighted multiplicities $C_{\D,\underline{0}}(\ph)$. They can be projected out from $Z(\phi,q,x)$ using character orthonormality (eq.~\eqref{eqn:chiORecap}), which we reproduce here:
\be
\int d\m \, \ch^{*}_{[\D;l]}\ch^{}_{[\D';l']} = \d_{\D\D'}\d_{ll'} \, , \nonumber
\ee
where the Haar measure \(d\m\) is given in eq.~\eqref{eq:haar_SO_d+2}. Note that this orthonormality relation is among the $\ch$, not the $\widetilde{\ch}$.\footnote{We remind the reader, per eqs.~\eqref{eqn:chidef}-\eqref{eqn:lminus}, that $\chi$ (no tilde) is defined as a function while $\widetilde{\chi}$ is the actual character for a conformal representation. This distinction is important for the next several equations to carefully derive $H(\phi,p)$; outside of this context we typically use $\chi$ and $\widetilde{\chi}$ for characters of long and short irreps, respectively.} However, the $\widetilde{\ch}$ are linear combinations of the $\ch$, so we first express eq.~\eqref{eq:Z_decomp} purely in terms of the $\ch$, and then make use of the above orthonormality relation to project out the coefficients. Carrying out this procedure carefully, we find
\begin{subequations}
\label{eq:extract_c}
\begin{alignat}{3}
\D \ne \D_0+2,d:&& \qquad C_{\D,\underline{0}}(\ph)&= \int d\m \, \ch^*_{[\D;\underline{0}]} \big( Z-1 \big) \, , \label{eq:c_Delta} \\
\D = \D_0+2:&& \qquad C_{\D_0+2,\underline{0}}(\ph)&=\int d\m \, \big(\ch^*_{[\D_0+2;\underline{0}]}+ \ch^*_{[\D_0;\underline{0}]} \big) \big( Z-1 \big) \, , \label{eq:c_Delta_2}\\
\D = d:&& \qquad C_{d,\underline{0}}(\ph)&=\int d\m \, \big(\ch^*_{[d;\underline{0}]}+ \ch^*_{[d-1;(1,0,\ldots,0)]} \big) \big( Z-1 \big) \, . \label{eq:c_d}
\end{alignat}
\end{subequations}

Eq.~\eqref{eq:c_Delta} gives the generic expression for $C_{\D,\underline{0}}(\ph)$, but there are two exceptional cases at \(\D = \D_0+2\) and \(\D = d\).\footnote{In two dimensions \(\D_0+2 = d = 2\) and instead of the two equations~\eqref{eq:c_Delta_2} and~\eqref{eq:c_d}, we have the single equation \(C_{2,\underline{0}}(\ph) = \int d\m \big(\ch^*_{[2;0]} + \ch^*_{[1;1]} + \ch^*_{[0;0]}\big)\big(Z-1\big)\).} They stem from the overlaps between the $\widetilde{\ch}$ and the $\ch$. For example, using
\begin{align}
\widetilde{\ch}_{[\D_0;\underline{0}]}^{} &= \ch_{[\D_0;\underline{0}]}^{} - \ch_{[\D_0+2;\underline{0}]}^{} , \nonumber
\end{align}
we have
\begin{align}
Z-1 &\supset C_{\D_0,\underline{0}}\widetilde{\ch}_{[\D_0;\underline{0}]} +C_{\D_0+2,\underline{0}} \widetilde{\ch}_{[\D_0+2;\underline{0}]} \nonumber \\
&= C_{\D_0,\underline{0}}\ch_{[\D_0;\underline{0}]} + (C_{\D_0+2,\underline{0}} - C_{\D_0,\underline{0}}) \ch_{[\D_0+2;\underline{0}]} .
\end{align}
Therefore, the coefficient $C_{\D_0+2,\underline{0}}$ is given by eq.~\eqref{eq:c_Delta_2}.

To properly compute the $C_{\D,\underline{0}}$, we need to pay attention to \textit{all} short irreps \(\widetilde{\ch}_{[\D;l]}^{}\) in eq.~\eqref{eq:Z_decomp} that include an \(l = \underline{0}\) component when written as linear combinations of the \(\ch_{[\D;l]}^{}\). From eq.~\eqref{eqn:chiTildeShort}, it is not hard to see that apart from the $\widetilde{\ch}_{[\D_0;\underline{0}]}^{}$ already discussed above, all possible such short irreps are
\begin{alignat}{3}\label{eqn:shortchis}
\widetilde{\ch}_{[\pm r]}^{} &\equiv \widetilde{\ch}_{[d-r;(1,1,\ldots,1,1,\pm 1)]}^{} &&= \ch_{[\pm r]}^{} - \widetilde{\ch}_{[r-1]}^{} , \nonumber \\
\widetilde{\ch}_{[r-1]}^{} &\equiv \widetilde{\ch}_{[d-(r-1);(1,1,\ldots,1,1,0)]}^{} &&= \ch_{[r-1]}^{} - \widetilde{\ch}_{[r-2]}^{} , \nonumber \\
\widetilde{\ch}_{[r-2]}^{} &\equiv \widetilde{\ch}_{[d-(r-2);(1,1,\ldots,1,0,0)]}^{} &&= \ch_{[r-2]}^{} - \widetilde{\ch}_{[r-3]}^{} , \\
&~\,\vdots &&~\,\vdots \nonumber \\
\widetilde{\ch}_{[1]}^{} &\equiv \widetilde{\ch}_{[d-1;(1,0,\ldots,0,0,0)]}^{} &&= \ch_{[1]}^{} - \ch_{[d;\underline{0}]}^{} ,\nonumber
\end{alignat}
where we have used obvious shorthand notations for the subscripts. Rewriting the $\widetilde{\ch}$ into the $\ch$, the recursive nature exhibited in eq.~\eqref{eqn:shortchis} allows us to absorb all the nontrivial overlaps into $\widetilde{\ch}_{[1]}^{}$:
\begin{align}
Z-1 &\supset C_{[\pm r]} \widetilde{\ch}_{[\pm r]}^{} + C_{[r-1]} \widetilde{\ch}_{[r-1]}^{} + C_{[r-2]} \widetilde{\ch}_{[r-2]}^{} + \cdots + C_{[1]} \widetilde{\ch}_{[1]}^{} + C_{d,\underline{0}}\ch_{[d;\underline{0}]}^{} \nonumber \\
 &= C'_{[1]} \widetilde{\ch}_{[1]}^{} + C_{d,\underline{0}}\ch_{[d;\underline{0}]}^{} + \cdots \nonumber \\
 &= C'_{[1]} \ch_{[1]}^{} + \left(C_{d,\underline{0}}-C'_{[1]}\right) \ch_{[d;\underline{0}]}^{} + \cdots \, ,
\end{align}
where the $+\cdots$ terms in the second and third lines contain $\ch_{[\pm r]}^{}, \ch_{[r-1]}^{}, \ldots, \ch_{[2]}^{}$, which are orthogonal to $\ch_{[1]}^{}$ and $\ch_{[d;\underline{0}]}^{}$. Therefore, the coefficient $C_{d,\underline{0}}$ is given by eq.~\eqref{eq:c_d}.

\subsection*{Computing $H(\ph,p)$ from the character generating function $Z(\ph,q,x)$}

To compute the Hilbert series, we plug the $C_{\D,\underline{0}}$ results in eq.~\eqref{eq:extract_c} into eq.~\eqref{eq:H_define}:
\begin{equation}
H(\ph, p) = 1 + \int d\mu \left[ \sum_\D p^\D \ch_{[\D;\underline{0}]}^* + p^{\D_0+2} \ch_{[\D_0;\underline{0}]}^* + p^d \ch_{[d-1;(1,0,\ldots,0)]}^* \right] \left(Z-1\right) . \label{eq:H_plug_c}
\end{equation}
Let us massage this into a more useful form. First focusing on the sum
\begin{align}
\sum_\D p^\D \ch_{[\D;\underline{0}]}^* = \sum_{n=0}^\infty \left(\frac{p}{q}\right)^{\D_0+\frac{n}{2}} P\left(q^{-1};x^{-1}\right) = \frac{1}{1-(p/q)^{1/2}} \left(\frac{p}{q}\right)^{\D_0} P\left(q^{-1};x^{-1}\right) , \nonumber
\end{align}
we get
\begin{align}
\int d\mu \left[ \sum_\D p^\D \ch_{[\D;\underline{0}]}^* \right] \left(Z-1\right)
&= \int d\m_{SO(d)}^{} \oint \frac{dq}{2\pi i} \frac{1}{q} \frac{1}{P(q;x)} \frac{1}{1-(p/q)^{1/2}} \left(\frac{p}{q}\right)^{\D_0} (Z-1) \nonumber \\
&= \int d\m_{SO(d)}^{} \frac{1}{P(p;x)} \left[ Z(\ph,p,x)-1 \right] \nonumber \\
&= \int d\m_{SO(d)}^{} \frac{1}{P(p;x)} Z(\ph,p,x) - 1 - (-p)^d .
\end{align}
In the first line above, the square root in the integrand indicates that the proper measure $d\mu$ is that for the double cover group. Practically, this means that we should send $q\to q^2$ in the integrand, as explained around eq.~\eqref{eqn:doublecover}. After doing so, in the second line, the integral evaluates to the residue at the single pole $q=\sqrt{p}$ inside the contour $|q|=1$.\footnote{Note that both $1/P(q;x)$ and $Z(\ph,q,x)$ are analytic functions inside the contour, since $\abs{x}=1$, $\abs{\ph}<1$, and $\abs{p}<1$. The expression $\left(p/q\right)^{\D_0} (Z-1)$ is regular at $q=0$ because the expansion of $(Z-1)$ starts with order $q^{\D_0}$, as is obvious from eq.~\eqref{eqn:ZphiSOd}.} Finally, in the last line, we have used the fact that
\be
\frac{1}{P(p;x)} = \text{det}_{\Box} \big(1- p \, h(x)\big) = \sum_{n=0}^d (-p)^n \ch_{\wedge^n(\Box)}^{}(x) , \label{eqn:Pasymgen}
\ee
which follows from eq.~\eqref{eqn:asymGendet}. 

With this result, we write the Hilbert series in eq.~\eqref{eq:H_plug_c} as
$H(\ph,p) = H_0(\ph,p) + \D H(\ph,p)$ with
\begin{align}
H_0(\ph,p) &\equiv \int d\m_{SO(d)}^{} \frac{1}{P(p;x)} Z(\ph,p,x) ,  \\
\D H(\ph,p) &\equiv (-1)^{d+1} p^d + p^{\D_0 + 2} \int d\m \, \ch^*_{[\D_0 ; \underline{0}]} \big( Z - 1\big) + p^{d} \int d\m \, \ch^*_{[d-1 ; (1,0,\ldots,0)]} \big( Z - 1\big).
\end{align}
The last two terms in $\D H(\ph,p)$ are evaluated in a similar way as above:
\begin{align}
p^{\D_0+2} \int d\m \, \ch^*_{[\D_0;\underline{0}]} \big(Z-1 \big) &= p^{\D_0+2}\int d\m_{SO(d)} \oint \frac{dq}{2\pi i} \frac{1}{q} \frac{1}{P(q^2;x)} \frac{1}{q^{2\D_0}} \Big( Z(\ph,q^2,x) - 1 \Big) \nonumber \\
&= p^{\D_0+2}\int d\m_{SO(d)}^{} \left. Z(\ph,q^2,x)\right|_{q^{2\D_0}}  \nonumber \\
&= p^{\D_0+2}\int d\m_{SO(d)}^{} Z(\ph,q,x)\big|_{q^{\D_0}} \, , \nonumber 
\end{align}
where in the first line we have sent $q\to q^2$ in the integrand as a consequence of using the double cover measure, and in going from the first to the second line we made use of the fact that the expansion of \(\big(Z(\ph,q^2,x) - 1\big)\) begins at \(q^{2\D_0}\). Similarly,
\begin{align}
p^{d} \int d\m \, \ch^*_{[d-1;(1,0,\ldots,0)]} \big(Z-1 \big) = p^d \int d\m_{SO(d)}^{} \ch_{\Box}^{}(x) \left. \left[\frac{1}{P(q;x)} \Big(Z(\ph,q,x) - 1 \Big)\right] \right|_{q^{(d-1)}} \, . \nonumber
\end{align}
Putting them together, we find
\begin{align}
\D H &= (-1)^{d+1}p^d + p^{\D_0+2}\int d\m_{SO(d)}^{} Z\big|_{q^{\D_0}} + p^d \int d\m_{SO(d)}^{} \ch_{\Box}^{}(x) \left. \left[\frac{1}{P} (Z-1)\right] \right|_{q^{(d-1)}} \nonumber \\
&= (-1)^{d+1}p^d + p^{\D_0+2}\ph , \label{eq:DH_single_scalar}
\end{align}
where in the second line, we have specified to the case of a single real scalar field $\ph$.

Let us summarize what we have derived. We split the Hilbert series into two parts
\begin{equation}
H(\ph,p) = H_0(\ph,p) + \D H(\ph,p) , \label{eq:split_H}
\end{equation}
and find that
\begin{subequations}
\label{eqn:Hresults}
\begin{align}
H_0 &= \int d\m_{SO(d)}^{} \frac{1}{P(p;x)} Z(\ph,p,x) , \label{eqn:H0result} \\
\D H &= (-1)^{d+1}p^d + p^{\D_0+2}\int d\m_{SO(d)}^{} Z\big|_{q^{\D_0}} + p^d \int d\m_{SO(d)}^{} \ch_{\Box}^{}(x) \left. \left[\frac{1}{P} (Z-1)\right] \right|_{q^{d-1}} . \label{eqn:DHresult}
\end{align}
\end{subequations}
The above expression for $H_0$ is the basic skeleton for any theory one wishes to consider and is a central result; $\Delta H$ is specific to the case where the single particle modules are conformal representations. Eq.~\eqref{eqn:Hresults} expresses the Hilbert series as an integral over \(SO(d)\) matrices, giving a concrete computational method for obtaining \(H\). From the above expression, it is manifest that $\Delta H$ only contains operators with mass dimension $\le d$. After explaining how to include multiple fields and possible symmetries (next subsection), the last two terms in \(\D H\) can be evaluated more explicitly, as shown in appendix~\ref{app:Delta_H}.

\subsection*{Relation of the generating function to the physical partition function}

As a side note, we mention that the character generating function \(Z(\ph,q,x)\) is closely related to the physical partition function \(\mathcal{Z}(q)\) of the free field theory. The partition function for the free theory is given by~\cite{Cardy:1991kr}
\begin{align}
\mathcal{Z}(q) &= \frac{1}{(1- q^{\D_0})(1-q^{\D_0+1})^{\text{dim}[(1,0,\ldots,0)]}(1-q^{\D_0+2})^{\text{dim}[(2,0,\ldots,0)]} \ldots} \nonumber \\
&= \prod_{n=0}^{\infty}\frac{1}{(1-q^{\D_0+n})^{\text{dim}[(n,0,\ldots,0)]}},
\label{eq:free_part_fxn}
\end{align}
where
\begin{equation}
\text{dim}[(n,0,\ldots,0)] = \binom{n+d-1}{n} - \binom{n+d-3}{n-2} ,
\label{eq:dim_sod_symm_trace}
\end{equation}
is the dimension of the representation \(l = (n,0,\ldots,0)\). The form of \(\mathcal{Z}(q)\) has a simple interpretation: the Hilbert space is a Fock space, built from single-particle states. By the state-operator correspondence, the single particle states correspond to $\partial^n \ph \equiv \partial_{\{\mu_1} \cdots \partial_{\mu_n\}}\ph$ with $n=0,1,2,\ldots$. These states have energy $q^{\D_0+n}$ with $q=e^{-\beta}$, which are eigenvalues of the Hamiltonian in radial quantization (the dilatation operator). As the theory is free, multi-particle states are simply built from products of the single particle states, hence we arrive at~\eqref{eq:free_part_fxn} for the partition function.

The determinant form of our character generating function $Z(\ph, q, x)$ can be obtained by combining eq.~\eqref{eqn:ZphiSOd} and the identity eq.~\eqref{eqn:GenPE}:
\be
Z(\ph, q, x) = \prod_{n=0}^\infty \frac{1}{\det_{(n,0,\ldots,0)} \left(1 - \ph q^{\Delta_0+n} h(x) \right) } , \label{eqn:ZphiDet}
\ee
with  $h\in SO(d)$. Comparing with eq.~\eqref{eq:free_part_fxn}, we see that it is very simple to obtain $\mathcal{Z}(q)$ from $Z(\ph, q, x)$: send \(\ph \to 1\) and \(x \to 1\), namely 
\begin{equation}
\mathcal{Z}(q) = Z(\ph=1,q,x=1) . \label{eq:partition_from_Z}
\end{equation}

The reason for this is that $\mathcal{Z}(q)$ only requires information on the number of generators of a given scaling dimension. In terms of characters, this is accomplished by setting the arguments of the character to unity, as this measures the dimension of the representation, \(\text{dim}(V) = \ch_V^{}(x=1)\). The relation~\eqref{eq:partition_from_Z} continues to hold when we add more fields and gauge symmetries. The use of plethysm and the character generating function $Z(\phi,q,x)$ provides a simple interpretation of expressions of the free-field partition function $\mathcal{Z}(q)$ found in~\cite{Aharony:2003sx,Polyakov:2001af} and related works.

\subsection{Multiple fields and internal symmetries}\label{sec:mult_fields_symmetries}

For a general EFT we can have multiple fields \(\{\Ph_i\}\) transforming under some internal symmetry group \(G\). After removing the EOM redundancy, the operator space $\scJ$ is generated by the single particle modules $R_{\Ph_i}$. In the following we assume:
\begin{itemize}
  \item The fields $\Ph_i$ transform linearly under \(G\), so that the single particle modules form linear representations $R_{G,\Ph_i}$ under $G$. The non-linear case is more subtle and we address it in section~\ref{sec:non-linear}.
  \item The $R_{\Phi_i}$ form representations $R_{SO(d+2,C),\Phi_i}$ of the conformal group. This assumption is easily relaxed, as the splitting $H = H_0 + \Delta H$ and the formula for $H_0$, eq.~\eqref{eqn:H0result}, can be derived using the differential form technique described in~\cite{Henning:2015alf} and sec.~\ref{subsec:nonlinearHilbert}. We presently stick to this assumption primarily for concreteness, where it includes (see sec.~\ref{subsec:unitary}) scalars $\ph_i$, spinors $\psi_i$, as well as $d/2$-form field strengths $F_i$ in even dimensions.
\end{itemize}

With these assumptions we have \(R_{\Ph_i} = R_{SO(d+2,\mathbb{C}),\Ph_i} \otimes R_{G,\Ph_i}\) with the associated character $\ch_{\Ph_i}^{}(q,x,y) = \widetilde{\ch}_{SO(d+2,\mathbb{C}),\Ph_i}^{}(q;x) \ch_{G,\Ph_i}^{}(y)$, where \(y = (y_1,\ldots, y_{\text{rank}(G)})\) collectively denotes the coordinates of the torus of \(G\). Specifically,
\begin{subequations}
\label{eq:generating_reps}
\begin{align}
\ch_{\ph_i}^{}(q,x,y) &= \widetilde{\ch}_{[\D_0;\underline{0}]}^{}(q;x) \ch_{G,\ph_i}^{}(y) \, ,\\
\ch_{\ps_i}^{}(q,x,y) &= \widetilde{\ch}_{[\D_0+\frac{1}{2};(\frac{1}{2},\ldots,\frac{1}{2},\pm\frac{1}{2})]}^{}(q;x) \ch_{G,\ps_i}^{}(y) \, ,\\
\ch_{F_i}^{}(q,x,y) &=\widetilde{\ch}_{[\D_0+1;(1,\ldots,1,\pm1)]}^{}(q;x) \ch_{G,F_i}^{}(y) \quad (d \text{ even}) ,
\end{align}
\end{subequations}

We next construct the character generating function \(Z(\{\Ph_i\},q,x,y)\) for the operator space $\scJ$. Letting
\begin{equation}
Z_{\Ph_i}(\Ph_i,q,x,y) = \left\{ \begin{array}{lcl} \dfrac{1}{\text{det}_{R_{\Ph_i}}\big(1- \Ph_i g \big)} = \text{PE}\big[ \Ph_i \ch_{\Ph_i}^{} \big] & & \text{Bosonic fields} \\ & & \\
\text{det}_{R_{\Ph_i}}\big(1+ \Ph_i g \big) = \text{PE}_f\big[ \Ph_i \ch_{\Ph_i}^{} \big] & & \text{Fermionic fields} \end{array} \right.
\end{equation}
we have
\begin{equation}
Z(\{\Ph_i\},q,x,y) = \prod_i Z_{\Ph_i}(\Ph_i,q,x,y) = \text{PE} \left[ \sum_i \Ph_i \ch_{\Ph_i}^{}(q,x,y) \right] ,
\end{equation}
where it is understood that proper care of statistics is taken in the plethystic exponential.

The operator basis $\scK$ then consists of \(G\)-invariant scalar conformal primaries in $\scJ$. Everything in the derivation of \(H(\ph,p)\) in section~\ref{subsec:Hderivation} continues to hold when we allow for multiple $R_{\Ph_i}$. To further project out the \(G\)-singlets, we simply integrate over \(G\). Thus, the results in eq.~\eqref{eqn:Hresults} are modified by replacing \(Z(\ph,q,x)\) with \(Z(\{\Ph_i\},q,x,y)\) and adding a $\int d\m_G^{}(y)$ in front, where $d\m_G^{}(y)$ is the Haar measure of $G$.

Two comments/caveats about our formalism are as follows. First, for scalar fields $\phi$ and spinor fields $\psi$, our Hilbert series does not count their kinetic terms $|D\phi|^2=\phi^\dagger (-D^2) \phi$ or $\bar\psi i\slashed D \psi$, as these terms are proportional to the EOM. This is not the case for gauge fields, as we work with $F_{\mu\nu}$ as opposed to $A_\mu$, as discussed in sec.~\ref{sec:2.5} (see also comments there on the treatment of the covariant derivative). 
Second, knowing the possible generating representations, eq.~\eqref{eq:generating_reps}, gives us the means to explicitly evaluate the last two terms in $\D H(\ph,p)$ of eq.~\eqref{eqn:DHresult}. This is worked out in app.~\ref{app:Delta_H}. We emphasize that while $\D H(\ph,p)$ is conceptually important, computing it is often of little practical relevance: it only contains a handful of terms, all related to operators with \(\D \le d\).

\subsection{Parity invariance: from $SO(d)$ to $O(d)$}\label{subsec:parity}

If the EFT under consideration is parity invariant, then the spacetime rotation group is \(O(d)\). Parity acts as a reflection in \(\mathbb{R}^d\), so that \(O(d)\) is a certain product of this \(Z_2\) action times \(SO(d)\). Therefore, as a group, \(O(d)\) consists of two connected components, the parity even component $O_+(d)=SO(d)$ and the parity odd component $O_-(d)$.

For a single real scalar field, when including parity invariance the main piece of the Hilbert series, \(H_0\), is given by
\begin{equation}
H_0(\ph,p) = \int d\m_{O(d)}^{} \, \text{det}_\Box^{}(1-p g) Z(\ph,p,x) , \label{eq:H_0_parity_start}
\end{equation}
where the integral is over all \(g \in O(d)\) with measure \(d\m_{O(d)}^{}\). The \(\text{det}_\Box^{}(1-p g)\) factor reduces to the \(1/P(p;x)\) factor of eq.~\eqref{eqn:H0result} on the parity even component. We recall from eq.~\eqref{eqn:ZphiDet} that \(Z\) is an infinite product of \(l = (n,0,\ldots,0)\) representations,
\begin{equation}
Z(\ph,p,x) = \prod_{n=0}^{\infty}\frac{1}{\text{det}_{(n,0,\ldots,0)}^{}(1 - \ph p^{\D_0+n} g)}. \nonumber
\end{equation}
The integral in eq.~\eqref{eq:H_0_parity_start} is split into the parity even and odd pieces of \(O(d)\),
\begin{align}
H_0(\ph,p) &= \frac{1}{2}\int d\m_{+} \text{det}_\Box^{}(1-p g_+) Z_+ \ + \  \frac{1}{2}\int d\m_{-} \text{det}_\Box^{}(1-p g_-) Z_- \nonumber \\
&\equiv \frac{1}{2}\Big[H_{0,+}(\ph,p) + H_{0,-}(\ph,p)\Big] , \label{eq:H_parity_split}
\end{align}
where \(g_{\pm} \in O_{\pm}(d)\), \(Z_{\pm}=Z(g_{\pm})\), \(d\m_{\pm}=d\mu_{O_\pm(d)}^{}\) is the Haar measure normalized as \(\int d\m_{\pm} = 1\), and the factors of \(1/2\) are a consequence of further normalizing \(\int d\m_{O(d)}^{} =1\). The parity even piece \(H_{0,+}\) is given by eq.~\eqref{eqn:H0result}. In this subsection, our focus is on bringing the parity odd piece,
\begin{equation}
H_{0,-}(\ph,p) = \int d\m_{-} \text{det}_\Box^{}(1-p g_-) Z_-(\ph,p,x) , \label{eqn:H0parityGeneral}
\end{equation}
into a more amenable form for explicit computation.

The group $O(d)$ is segmented into the cosets of its $SO(d)$ subgroup. Therefore, a general element \(g_- \in O_-(d)\) can be taken in the form \(g_- = g_+ \, \mathcal P\), with \(g_+ \in SO(d)\) and $\mathcal{P}$ denoting the parity element. The basic procedure is to make use of this relation to work out the determinants $\det_{(n,0,\ldots,0)}^{}(1-ag_-)$ and the measure \(d\m_-\) in eq.~\eqref{eqn:H0parityGeneral}. This is fairly simple in odd dimensions because we can take parity to commute with rotations so that $O(2r+1) = SO(2r+1) \times Z_2$. In even dimensions the parity element does not commute with general rotations so that the orthogonal group is a semidirect product \(O(2r) = SO(2r) \ltimes Z_2\). This complicates the procedure in even dimensions; here we summarize the result and work out the details in app.~\ref{app:parity}.

\subsubsection*{Odd dimensions: $d=2r+1$}

We can take the action of parity to flip the sign of all components of a vector, \(\mathcal{P}: v_{\m} \to  -v_{\m}\), \textit{i.e.} $\rho_\Box^{}(\mathcal{P})=-I$ where $\rho_l(\mathcal{P})$ denotes the representation matrix and $I$ is the identity matrix.. Then the action on general tensor representations is easy to deduce: a tensor of odd rank (odd number of indices) flips sign, while an even rank tensor is invariant:
\begin{equation}
\r_l^{}(\mathcal{P}) = (-1)^{\abs{l}} I,
\label{eq:parity_rep_odd_dim}
\end{equation}
where \(\abs{l} \equiv l_1 + \cdots + l_r\). The above is valid for tensor representations, \(l_i \in \mathbb{N}\) (see eq.~\eqref{eqn:spinortransformdodd} below for spinors).

For the Hilbert series in eq.~\eqref{eqn:H0parityGeneral}, all the determinants are over the symmetric tensor representations \(l = (n,0,\ldots,0)\). Using \(g_- = g_+ \mathcal{P}\) and eq.~\eqref{eq:parity_rep_odd_dim}, we get the determinants
\begin{equation}
\text{det}_{(n,0,\ldots,0)}\big(1- a g_-\big) = \text{det}_{(n,0,\ldots,0)}\big(1- (-1)^na g_+\big).
\end{equation}
The \((n,0,\ldots,0)\) representations arise from adding \(n\) derivatives on top of the field \(\ph\), so this captures the obvious statement that \(\pd^n\ph\) flips sign under parity if there are an odd number of derivatives.

The measure on \(O_-(2r+1)\) is the parity transformation of the \(SO(2r+1)\) measure; since the adjoint representation, \(l = (1,1,0,\ldots,0)\), doesn't transform under parity, neither does the measure. Therefore, we have $d\mu_-=d\mu_+$.

Let us put these together and reweight the Hilbert series in a momentum grading scheme by sending \(p \to t\) and \(\ph \to \ph/t^{\D_0}\), so that \(t\) encodes how many derivatives there are. In this scheme, eq.~\eqref{eqn:H0parityGeneral} gives
\begin{align}
H_{0,-}^{(2r+1)}\left(\frac{\ph}{t^{\D_0}},t\right) &= \int d\m_+ \, \text{det}_{(1,0,\ldots,0)}(1+tg_+) \prod_{n=0}^{\infty} \frac{1}{\text{det}_{(n,0,\ldots,0)}\left[1 - \ph (-t)^{n} g_+\right]} \nonumber \\
&= H_{0,+}^{(2r+1)}\left(\frac{\ph}{(-t)^{\D_0}},-t\right) . \label{eqn:H0mphidodd}
\end{align}

\subsubsection*{Even dimensions: $d=2r$}

In even dimensions, we take $\mathcal{P}$ as a reflection about the hyperplane orthogonal to the $d$th axis in $\mathbb{R}^d$, i.e. $\rho_{\Box}^{}(\mathcal{P}) = \text{diag}(1,\dots,1,-1)$. Since \(\mathcal{P}\) does not commute with general rotations, more effort is required to track the action of \(\mathcal{P}\) on general \(l = (l_1,\ldots,l_r)\) irreps and to work out the determinants and measure in eq.~\eqref{eqn:H0parityGeneral}. The details are worked out in app.~\ref{app:parity}. The end result is (see eq.~\eqref{eqn:H0mphidevenapp})
\begin{equation}
H_{0,-}^{(2r)}(\ph,p) = \int d\m_{Sp(2r-2)}^{} \frac{1-p^2}{P^{(2r-2)}(p;\tilde{x})} \text{PE}\Bigg[\ph p^{\D_0}P^{(2r-2)}(p;\tilde{x}) \, + \, \ph^2 \frac{p^{2\D_0+2}}{1-p^2} P^{(2r-2)}(p^2;\tilde{x}^2) \Bigg] , \label{eqn:H0mphideven}
\end{equation}
where the integral is over the symplectic group $Sp(2r-2)$, $P^{(2r-2)}$ is the usual $P$ function but in $2r-2$ dimensions, and $\tilde{x} \equiv (x_1,\dots,x_{r-1})$. 

\subsubsection*{Intrinsic parity}

So far we have focused on scalar fields with trivial intrinsic parity. To show how to address intrinsic parity, we discuss two typical field contents: pseudoscalars and spinors.

Let parity act on the spacetime coordinates (in Euclidean space) by switching the sign of the last coordinate, \((z_1,\ldots,z_d) \to (z_1,\ldots, z_{d-1},-z_d)\). Under this parity transformation a scalar field \(\ph(z)\) only has its arguments transform, \(\ph(z_1,\ldots,z_d) \to \ph(z_1,\ldots,z_{d-1},-z_d)\), while a pseudoscalar \(\varphi(z)\) also flips sign: \(\varphi(z_1,\ldots,z_d) \to - \varphi(z_1,\ldots,z_{d-1},-z_d)\). We have already explained in this section how to include parity for scalars: one splits the Hilbert series into its contribution from even and odd pieces, \(H(\ph,p) = \frac{1}{2}\big[H_+(\ph,p) + H_-(\ph,p)\big]\), and then proceeds to computing \(H_{\pm}(\ph,p)\). For a pseudoscalar, the procedure is essentially the same except that on the parity odd piece, the spurion \(\varphi\) also needs to flip sign. In other words,
\begin{equation}
H(\varphi,p) = \frac{1}{2} \Big[ H_+(\varphi,p) + H_-(-\varphi,p) \Big] . \label{eq:hilbert_pseudoscalar}
\end{equation}

For spinors, we need to split the cases of even and odd dimensions. In even dimensions the spinors are chiral, and parity interchanges the two chiralities $\psi_L \leftrightarrow \psi_R$. Their direct sum $\psi=\psi_L \oplus \psi_R$ forms a representation under $O(2r)$. We assign a single spurion $\psi$ for this representation and follow a procedure similar to the scalar case to compute $H_+(\psi,p)$ and $H_-(\psi,p)$ (see also the gauge field example in appendix~\ref{appsubsec:examples}). The full Hilbert series is again given by their average
\begin{equation}
H(\psi,p) = \frac{1}{2} \Big[ H_+(\psi,p) + H_-(\psi,p) \Big] . \label{eq:hilbert_spinorsdeven}
\end{equation}

In odd dimensions spinors are not chiral. In Minkowski signature, the two components $\psi$ and $\bar{\psi}$, transform under parity as
\begin{subequations}
\begin{eqnarray}
P \psi(z^0, \ldots, z^{2r-1 }, z^{2r}) P^{-1} &=& \pm i \gamma^{2r} \psi(z^0, \ldots, z^{2r-1}, -z^{2r}), \\
P \bar{\psi} (z^0, \ldots, z^{2r-1 }, z^{2r}) P^{-1} &=& \mp \bar{\psi}(z^0, \ldots, z^{2r-1 }, -z^{2r}) i \gamma^{2r} ,
\end{eqnarray}
\end{subequations}
where $i\gamma^{2r}$ is hermitian with our metric $(+,-,\ldots, -)$, and both signs $\pm$ ensure $P^2=1$. Note that the signs are opposite for $\psi$ and $\bar{\psi}$. After Wick rotation, we can combine it with 180$^\circ$ rotations in all other coordinates so that parity reverses all of the spacetime coordinates.  After switching to the $(+,+,\ldots, +)$ metric, we find
\begin{subequations}
\label{eqn:spinortransformdodd}
\begin{eqnarray}
P \psi(z) P^{-1} &=& \pm \psi(-z) , \\
P \bar{\psi}(z) P^{-1} &=& \mp \bar{\psi}(-z) .
\end{eqnarray}
\end{subequations}
We see that $\psi$ and $\bar{\psi}$ transform separately under $O(2r+1)$, namely each of them forms a representation of $O(2r+1)$, but they have opposite intrinsic parity. As a result, the mass term $\bar\psi \psi$ is odd under parity. This is expected because the little group for a massive fermion in odd dimensions is $SO(2r)$, whose spinors are chiral. For the Hilbert series calculation, one computes the parity odd piece $H_-(\psi,\bar\psi,p)$ again in a similar way with the scalar case. But to account for their intrinsic parities, we send the spurions $\psi \rightarrow \psi$, $\bar{\psi} \rightarrow -\bar{\psi}$ in accordance with eq.~\eqref{eqn:spinortransformdodd}. That is, the full Hilbert series is given by
\begin{equation}
H(\psi,\bar\psi,p) = \frac{1}{2} \Big[ H_+(\psi,\bar\psi,p) + H_-(\psi,-\bar\psi,p) \Big] . \label{eq:hilbert_spinorsdodd}
\end{equation}

\subsection{Reweighting and summary of main formulas}\label{subsec:Hilbert_summary}

Given an EFT with a set of fields \(\{\Ph_i\}\) and internal symmetry group \(G\), the Hilbert series counts the number of operators in the operator basis $\scK$. In this section we have shown how this counting function can be explicitly computed as a matrix integral. The final takes the form
\begin{equation}
H(\{\Ph_i\},p) = H_0(\{\Ph_i\},p) + \D H(\{\Ph_i\},p),
\label{eq:hil_split_summary}
\end{equation}
where the bulk of the Hilbert series is contained in the \(H_0\) term. The \(\D H\) piece, with expression given in eq.~\eqref{eqn:DHresult} (see also app.~\ref{app:Delta_H}), only contains operators of mass dimension less than or equal to the spacetime dimensionality $d$. Because almost all of \(H\) is contained in \(H_0\), we frequently refer to \(H_0\) as the Hilbert series.

This section made use of conformal representation theory to derive the Hilbert series. We stress, however, that this is not necessary. The splitting \(H = H_0 + \D H\) and the formula for \(H_0\) below can be derived by other means---in sec.~\ref{subsec:nonlinearHilbert} we do this using Hodge theory (see also sec.~\ref{sec:2.6} for a heuristic physical derivation of \(H_0\)). In particular, the formula for \(H_0\) \textit{is valid for all single particle modules}, regardless of whether or not they are unitary conformal representations. However, the explicit formula for \(\D H\) in eq.~\eqref{eqn:DHresult} requires the assumption that the single particle modules are also conformal irreps.

The starting point for a practical computation of \(H_0\) is
\begin{equation}
H_0(\{\Ph_i\},p) = \int d\m_G(y) \int d\m_{SO(d)}(x) \, \frac{1}{P(p;x)} \text{PE}\left[\sum_{i} \Ph_i \ch_{\Ph_i}^{}(p,x,y) \right].
\label{eq:H_0_summary}
\end{equation}
The ingredients in the above equation are
\begin{itemize}
\item The Haar measures for \(G\) and the Lorentz group \(SO(d)\). These measures are restricted to the torus using the Weyl integration formula. The torus of \(G\) is parameterized by \(y = (y_1,\ldots,y_{\text{rank}(G)})\), while that of \(SO(d)\) is parameterized by \(x= (x_1,\ldots,x_{\lfloor \frac{d}{2} \rfloor})\). Measures for the classical groups are given in app.~\ref{app:Weyl}.
\item The \(1/P(p;x)\) factor, with \(P(p;x)\) the function in eq.~\eqref{eqn:PRecap}, accounts for IBP redundancies.
\item The characters \(\ch_{\Ph_i}^{}\) for the single particle modules $R_{\Ph_i}$. They are a product of the weighted \(SO(d)\) character for the module times the character under \(G\). If \(R_{\Ph_i}\) is a conformal representation, then the weighted \(SO(d)\) character coincides with the conformal character, \textit{i.e.} (see eq.~\eqref{eq:generating_reps}) 
    \begin{equation}
    \ch_{\Ph_i}^{}(p,x,y) = \widetilde{\ch}_{SO(d+2,\mathbb{C}),\Ph_i}^{}(p;x) \, \ch_{G,\Ph_i}^{}(y) . \nonumber
    \end{equation}
    As the single particle modules already have the EOM removed, the EOM redundancies are accounted for in these characters.
\item The plethystic exponential is a generating function for symmetric or anti-symmetric products (appropriately chosen according to the statistics of each \(\Ph_i\)) of the generating representations.
\item The matrix integral and the use of characters automatically accounts for group relations that stem from finite rank conditions, such as Fierz identities or Gram determinant constraints.
\end{itemize}

To include parity as a symmetry of the EFT, the matrix integral for \(H\) includes an integral over \(O(d)\) instead of $SO(d)$. The integral splits into two pieces---the parity even and odd components---with the parity even component given by the \(SO(d)\) expression in~\eqref{eq:H_0_summary}. The expression for the parity odd component depends on whether \(d\) is even or odd. Relevant formulae can be found in sec.~\ref{subsec:parity} and app.~\ref{app:parity}.

One may wish to choose different weights to access certain information in the Hilbert series. In this section, we have derived the Hilbert series \(H(\{\Ph_i\},p)\) in the mass grading scheme as a function of the weights \(\{\Ph_i\}\) and \(p\). The momentum grading scheme---which we find frequently useful and will use in subsequent sections---is to weight operators by their field content as well as the number of derivatives in them. This scheme is readily obtained from the mass grading scheme by sending
\begin{equation}
\Ph_i \to \frac{\Ph_i}{t^{\D_{\Ph_i}}}~~,~~p \to t,
\label{eq:reweight}
\end{equation}
where \(\D_{\Ph_i}\) is the mass dimension of \(\Ph_i\) and $t$ counts powers of derivatives.

\section{Constructing operators: kinematic polynomial rings}
\label{sec:rsf}

The question of counting and constructing  operator bases can be translated into that of counting and constructing the kinematic quantities which form the basis of scattering amplitudes in QFTs. In this section, we map out this translation and explore the kinematic structure that underlies operator bases. As well as providing a concrete connection to the physical observables of a theory---scattering amplitudes---we emphasize that this momentum space picture is particularly powerful for the construction of the basis elements (\textit{i.e.} the forms of the operators), especially when used in conjunction with the machinery developed in the previous sections.

We start in sec.~\ref{sec:four_one} by passing to momentum space where operators are in one-to-one correspondence with polynomials in momenta. EOM and IBP redundancies become equivalence relations between polynomials, leading to a natural formulation of \(\scK\) in terms of polynomial rings with ideals. This generalizes our previous work in \(d=1\)~\cite{Henning:2015daa} to arbitrary \(d\). Section~\ref{sec:tour} exhaustively parades through the basis involving four scalar fields, exploring the consequences of EOM and IBP redundancy, spacetime dimensionality, and permutation symmetry. With an understanding of the ring formalism from the four point example, we pause in sec.~\ref{sec:four_three} to reflect on the physical connection to operators, contact terms, and amplitudes. Section~\ref{sec:general_algebraic} returns to the rings to tackle the general case: we establish that the rings are Cohen-Macaulay, which means we can find a finite set of generators of the ring which allow us to construct any other element uniquely. Here we also touch upon the case of spin, emphasizing that the scalar operators studied presently appear universally since they dictate the kinematics of momenta. Some comments about the relationship between elements of the ring and conformal primaries are made in sec.~\ref{sec:conf_prim_ring}. In sec.~\ref{subsec:n=5_basis} we explain a simple algorithm for constructing elements of the ring---\textit{i.e.} constructing operators in \(\scK\)---and use this to solve the basis for five scalar fields.

\subsection{Polynomial ring in particle momenta} \label{sec:four_one}

To begin, we pass to momentum space and consider the Fourier transform of the scalar fields, $\phi_i$, (where $i$ is a flavor index, such that $\phi_i$ and $\phi_j$ are indistinguishable if $i=j$),
\be
\phi_i(x) = \int d^dp \, \widetilde{\phi_i}(p) \, e^{i \,p^\mu x_\mu}\,.
\ee
An operator consisting of $n$ distinct $\phi_i$ fields and $k$ derivatives then takes the general form
\be
\phi_1 \cdots \phi_n \partial_{\mu_1} \cdots \partial_{\mu_k} \sim \int d^dp_1 \cdots d^dp_n \,\widetilde{\phi}_{1}(p_1)\cdots \widetilde{\phi}_{n}(p_n) F^{(n,k)}(p_1,\cdots, p_n) \exp\left(i\sum_{i=1}^n p_i^\mu x_\mu \right) \,,\nonumber\\
\label{eq:fourier}
\ee
where $F^{(n,k)}$ is a degree $k$ polynomial in the $n$ momenta $\{p_i^{\mu}\}$, $i=1,2,\cdots,n$, whose form depends on how the derivatives act in the operator; this is a one-to-one correspondence between operators and such polynomials (up to a constant, $F^{(n,k)}$ is the Feynman rule in momentum space).  Because we are interested in Lorentz scalar operators, we are concerned with polynomials invariant under an $SO(d)$ symmetry acting on the Lorentz indices of the  $\{p_i^\mu\}$. That is, we are interested in the polynomial ring,\footnote{We work in the field \(\mathbb{C}\), although any characteristic zero field, such as \(\mathbb{R}\), is equally valid. We refer the reader unfamiliar with commutative algebra to \cite{Henning:2015daa} for a primer that is relevant to our current application.}
\be
M_{n}^{SO(d)}={\mathbb C}[p_1^\mu,\cdots,p_n^\mu]^{ SO(d)} \, ,
\label{eq:ringwithnoting}
\ee
where the superscript indicates we are imposing invariance under $SO(d)$.

The effects of EOM and IBP lead to equivalence relations between polynomials.  The EOM $\partial^2 \phi=0$ translates to $p_i^2=0$\footnote{We continue to consider Euclidean spacetime and assume complex momentum (note that for real momenta, $p^2=0$ would force $p^{\mu}=0$).} and the vanishing of total derivatives $\partial_\mu(\ldots)=0$ is the statement of momentum conservation, $\sum_{i=1}^n p_i^\mu=0$. These define equivalence relations
\be
\mathcal{O}_1\sim \mathcal{O}_2+ \partial^2\phi \,\mathcal{O}_3 &\implies& F_1(\{p_i\})\sim  F_2(\{p_i\})+  p_i^2 F_3(\{p_i\}) \,, \label{eq:polyredunA}\\
\mathcal{O}_1\sim \mathcal{O}_2+  \partial_\mu\mathcal{O}^\mu_3 &\implies& F_1(\{p_i\})\sim  F_2(\{p_i\})+ \left( \sum_i p_i^\mu \right) F^\mu_3(\{p_i\})  \,.
\label{eq:polyredunB}
\ee
Such polynomial relations between elements of a ring are embodied in an ideal of the ring. The equivalence class of polynomials under these two redundancies lie in the quotient ring
\be
M_{n,\scK}^{SO(d)}=\bigg[{\mathbb C}[p_1^\mu,\cdots,p_n^\mu] / \langle p_1^\mu+\ldots+p_n^\mu,p_1^2,\cdots,p_n^2 \rangle\bigg] ^{SO(d)}\,,
\label{eq:componentring}
\ee
where $\langle\ldots\rangle$ is the standard notation for an ideal, and the subscript $\scK$ indicates EOM and IBP equivalences are accounted for.

For the case of indistinguishable scalars, the Fourier coefficients $\widetilde{\phi_i}$ in eq.~\eqref{eq:fourier} become identical, and the symmetry of the integral leads us to consider symmetric polynomials in the indices $1,2,\cdots,n$, namely polynomials invariant under $S_n$, where $S_n$ acts by permutations of the $n$ momenta. In this case, the result given in eq.~\eqref{eq:componentring} generalizes to
\be
M_{n,\scK}^{SO(d)\times S_n}=\bigg[{\mathbb C}[p_1^\mu,\cdots,p_n^\mu] / \langle p_1^\mu+\cdots+p_n^\mu,p_1^2,\cdots,p_n^2 \rangle\bigg] ^{SO(d) \times S_n}\,.
\label{eq:componentringID}
\ee

\noindent \textbf{A word on notation.} We denote the rings under consideration by \(M_{n,\scI}^{G}\). Here, \(n\) refers to the number of momenta in the ring. \(\scI\) denotes the operator space under consideration: \(\scI = \scK\) is the operator basis with both EOM and IBP imposed; \(\scI = \scJ\) is the set of operators with only EOM imposed (remove \(p_1^{\m}+ \cdots +p_n^{\m}\) from the ideals above); if \(\scI\) is empty we impose neither EOM nor IBP, as in eq.~\eqref{eq:ringwithnoting}. \(G\) denotes the symmetry group we impose invariance under. Generically, \(G = (S)O(d) \times \Sigma\) where \((S)O(d)\) means either \(SO(d)\) or \(O(d)\) and \(\Sigma \subseteq S_n\) is a permutation group acting on the momenta. We typically consider the extremes \(\Sigma = S_n\) (all indistinguishable fields) or \(\Sigma\) being trivial (all distinguishable fields). Finally, we work in the momentum grading scheme, where the \(p_i^{\m}\) are degree one and assigned a weight parameter \(t\). The Hilbert series of the rings are denoted as \(H_{n,\scI}^G(t)\) or \(H(M_{n,\scI}^G,t)\).

\subsection*{Translation to a ring of Lorentz invariants}

A natural way to take Lorentz invariance into account is to work with a ring generated by the invariants, which include the symmetric invariants $s_{ij}\equiv p_{i\,\mu} p_j^\mu$, and the antisymmetric invariants $\epsilon_{p_{i_1} p_{i_2}\cdots p_{i_d}}\equiv\epsilon_{\mu_1\mu_2\cdots\mu_d} \,p_{i_1}^{\mu_1} p_{i_2}^{\mu_2} \cdots p_{i_d}^{\mu_d}$. Then any element in this ring is automatically Lorentz invariant. However, this feature does not come without a cost. To simplify discussion, we impose parity---$O(d)$ invariance, as opposed to $SO(d)$---which eliminates all the antisymmetric invariants involving the $\epsilon$ tensor.

In $d$ dimensions only $d$ momenta can be linearly independent, which leads to nontrivial relations among the $s_{ij}$. In particular, the Gram matrix
\be
\left( \begin{array}{cccc}
s_{11} &  \,\, s_{12} & \cdots &  s_{1n} \\
s_{12} & \,\, s_{22} & \cdots & s_{2n} \\
\vdots & \vdots & \ddots & \vdots \\
s_{1n} &  \,\, s_{2n} & \cdots & s_{nn} \\
  \end{array} \right)  \, ,
  \label{eq:gram}
\ee
is at most rank $d$. This dictates that any $(d+1) \times (d+1)$ sub-matrix has vanishing determinant, which gives a set of nontrivial relations among the $s_{ij}$, called rank conditions or Gram conditions. Denoting $\{\D\}$  as the set of all $(d+1) \times (d+1)$ minors of the Gram matrix, we have the ring isomorphism\footnote{In the math literature, this result is known as the first and second fundamental theorems of invariant theory.}
\be
M_{n}^{O(d)}={\mathbb C}[p_1^\mu,\cdots,p_n^\mu]^{O(d)} ={\mathbb C}[\{ s_{ij} \}]/\langle \{ \Delta \} \rangle \, .
\ee
Clearly, $\{\D\}$ is nontrivial only for the case $n\ge d+1$.

To study the effects of EOM and IBP, we begin with the case $ n\le d$, such that there are no Gram conditions to impose. Our initial ring has $n(n+1)/2$ invariants $s_{ij}$. After imposing EOM, $s_{ii}=0$, and we are left with $n(n-1)/2$ Mandelstam invariants $\{s_{ij}, i<j\}$. As we will always impose EOM, in the rest of this paper it is understood that the notation $s_{ij}$ has $i\ne j$, unless explicitly stated otherwise. Now our ring becomes
\be
M^{O(d)}_{n,\scJ}={\mathbb C}[\{s_{ij}\}]  \,,~~~n\le d\,.
\ee

IBP is the statement of momentum conservation: $\sum_{i=1}^{n} p_i^\m=0$. Contracting  this  with each of the $p_i^\mu$ gives $n$ equations of the form
\be
X_i \equiv \sum_{j\ne i} s_{ij} =0 \,.
\label{eq:xvars}
\ee
These $n$ equations reduce the number of independent Mandelstam invariants  to $n(n-1)/2-n=n(n-3)/2$. Alternatively, this reduction can be seen as the following: imposing IBP conditions means we can entirely eliminate one momenta, say \(p_n^{\m}\), such that we can consider only $(n-1)(n-2)/2$ Mandelstam invariants. Having done this, one relation between these invariants remains---from \(p_n^2 =0\)---such that the number of generators of the ring after IBP becomes $(n-1)(n-2)/2-1=n(n-3)/2$.

The $n$ equations in eq.~\eqref{eq:xvars} form the part of an ideal which accounts for IBP,
\be
M^{O(d)}_{n\le d,\scK}=\bigg[{\mathbb C}[\{s_{ij}\}] / \langle X_1,\ldots,X_n \rangle&&\!\!\!\!\bigg] ~~~~~~\text{(distinguishable)}\,,  \\
M^{O(d)\times S_n}_{n\le d,\scK}=\bigg[{\mathbb C}[\{s_{ij}\}] / \langle X_1,\ldots,X_n\rangle&&\!\!\!\!\bigg]^{S_{n}} ~~~\text{(indistinguishable)}\,,
\ee
where in the indistinguishable case, permutation invariance is imposed where $S_n$ acts on $s_{ij}$ via $\s \in S_n: s_{ij}\to s_{\sigma(i)\sigma(j)}$. For the distinguishable case, this module can be written as a ring freely generated by $n(n-3)/2$ Mandelstam invariants. For example, we can choose $M^{O(d)}_{n\le d,\scK}={\mathbb C}[\{s_{ij}\}]$ with $1\le i<j\le n-1$, $(i,j)\ne(n-2,n-1)$.

Let us now take \(n>d\) and consider the Gram conditions. Nominally we should expect to see their effects when $n=d+1$. However, when $n = d+1$, the Gram condition that the full $(d+1)\times(d+1)$ determinant vanishes follows from momentum conservation. Thus, when considering IBP, Gram conditions only affect the case $n\ge d+2$.

When $n\ge d+2$, EOM, IBP and Gram conditions together imply the number of independent generators is given by $\nobreak{(d-1) n - d(d+1)/2}$. In this formula $(d-1) n$ is the number of independent components of the $n$ momenta after requiring them to be on-shell (EOM), and  $ d(d+1)/2$ is the number of generators of the Poincar\'e group, each one causing a relation between the components: $d$ translations (IBP) and $d(d-1)/2$ Lorentz transformations. In summary, the dimensions for the rings are (note that \(n(n-3)/2 = n(d-1)-d(d+1)/2\) at both \(n=d\) and \(n=d+1\).)
\begin{subequations}\def\arraystretch{1.8}
\label{eq:dim_modules}
\begin{align}
\text{dim}(M_{n,\mathcal{K}}^G) &= \left\{ \begin{array}{ccl} \frac{1}{2}n(n-3) & & \text{for } n \le d+1 \\ (d-1)n - \frac{1}{2}d(d+1) & & \text{for }n\ge d \end{array} \right. \\
\textcolor{white}{3} \nonumber \\
\text{dim}(M_{n,\mathcal{J}}^G) &= \left\{ \begin{array}{ccl} \frac{1}{2}n(n-1) & & \text{for } n \le d \\ (d-1)n - \frac{1}{2}d(d-1) & & \text{for }n\ge d-1 \end{array} \right.
\end{align}
\end{subequations}
where we have also included the dimensions for \(M_{n,\scJ}^G\), obtained by similar considerations.

Phrased in terms of our initial ring and ideal, the final quotient ring we are interested in is,
\be
M^{O(d)\times \Sigma}_{n,\scK}=\bigg[{\mathbb C}[\{s_{ij}\}] / \langle X_1,\ldots,X_n , \{\Delta\}\rangle&&\!\!\!\!\bigg]^{\Sigma} \,,
\label{eq:ddimring}
\ee
where $\Sigma$ is a subgroup of $S_n$.
As before, for the distinguishable case, the $X_i$ in the ideal can be used to eliminate $n$ of the $s_{ij}$ from the ring. 

For the indistinguishable case, a useful result is that the $s_{ij}$ decompose into \(S_n\) irreducible representations as
\begin{equation}
s_{ij} = s_{ji}, i \ne j \leftrightarrow V_{(n)} \oplus V_{(n-1,1)} \oplus V_{(n-2,2)},
\label{eq:sij_decomp_Sn}
\end{equation}
where the notation refers to the partition associated with a Young diagram. Together the  $(n)\oplus (n-1,1)$ form the natural representation of $S_n$, with a basis given by the variables $X_i$  defined in eq.~\eqref{eq:xvars}. IBP completely removes these components, by eq.~\eqref{eq:ddimring}. Thus, (up to dealing with the additional Gram conditions),  understanding the structure of this ring requires an understanding of the invariants of the $(n-2,2)$ representation of $S_n$; this is, unfortunately, an open and difficult mathematical problem. 

\subsection{Tour of the four point ring} \label{sec:tour}

A lot of the subtleties in the above can already be seen in the ring when \(n=4\), where it is easy to identify the generators and relations between them. Thus, we consider the four point ring \(M_{n=4,\scI}^G\) with and without IBP, in all space-time dimensions, for distinguishable and indistinguishable scalar particles, and imposing $O(d)$ and $SO(d)$ invariance. We pay particular attention to the generators of the polynomial ring, the relations between them, and how this leads to the Hilbert series.

\subsubsection*{(i) $\scI = \scJ, \ G = O(d), \ d\ge 4$}
We begin with $d\ge4$ such that Gram conditions do not play a role. Consider first the case we impose parity, $O(d)$, and where the four scalars are distinguishable. Our initial ring has $n(n+1)/2=10$ invariants $p_i\cdot p_j$; after imposing EOM we have $n(n-1)/2=6$ Mandelstam invariants $s_{ij}$. Before imposing IBP, the ring is freely generated by these invariants, and the Hilbert series reflects this
\begin{subequations}
\be
M^{O(d\ge 4)}_{4,\scJ}&=&{\mathbb C}[s_{12},s_{13},s_{14},s_{23},s_{24},s_{34}] \, \label{eq:rdistEOMring} \,, \\
H^{O(d\ge 4)}_{4,\scJ}&=&\frac{1}{(1-q_{s_{12}}) (1-q_{s_{13}}) (1-q_{s_{14}}) (1-q_{s_{23}}) (1-q_{s_{24}}) (1-q_{s_{34}})} \to \frac{1}{(1-t^2)^6} \,, \nonumber \label{eq:hdistEOMring} \\
\ee
\end{subequations}
where $q_{s_{ij}}$ are gradings introduced to count powers of the $s_{ij}$ generators, and in the final expression for the Hilbert series we grade by $t$, counting powers of derivatives as per the previous section.

\subsubsection*{(ii) $\scI = \scK, \ G = O(d), \ d\ge 3$}
After imposing IBP conditions we have $n(n-3)/2=2$ Mandelstam invariants. We use familiar notations $\{s_{12}=s_{34},s_{13}=s_{24},s_{23}=s_{14}\}\equiv\{s,t,u\}$, and then use the final IBP relation $s+t+u=0$ to eliminate $u$. That is,
\begin{subequations}
\be
M^{O(d\ge 3)}_{4,\scK}&=&{\mathbb C}[s,t] \,,\label{eq:reomibp}\\
H^{O(d\ge 3)}_{4,\scK}&=&\frac{1}{(1-q_{s}) (1-q_{t}) } \to \frac{1}{(1-t^2)^2} \,.
\label{eq:heomibp}
\ee
\end{subequations}

\subsubsection*{(iii) $\scI = \scK, \ G = O(d), \ d=2$}
In $d=2$, we need to account for Gram conditions. Requiring that every $3\times3$ sub-determinant of the Gram matrix,
\be
\begin{pmatrix}0&s&t&u\\s&0&u&t\\t&u&0&s\\u&t&s&0\\\end{pmatrix} \,, \nonumber
\ee
vanishes results in the condition $stu=st(-s-t)=0$. That is,
\begin{subequations}
\be
M^{O(2)}_{4,\scK}&=&{\mathbb C}[s,t]/\langle s^2t+t^2 s\rangle \,, \label{eq:hil_n4_O2_K}\\
H^{O(2)}_{4,\scK}&=&\frac{1-q_s^2q_t}{(1-q_{s}) (1-q_{t}) } \to \frac{1+t^2+t^4}{1-t^2} \,.
\ee
\end{subequations}
(A choice of ordering scheme appears in the first equality for the Hilbert series, see \textit{e.g.} the appendix in~\cite{Henning:2015daa}.)

The ring in eq.~\eqref{eq:hil_n4_O2_K} can be explicitly ``solved'' as follows. Change variables to \(\theta \equiv s-t\) and \(\eta \equiv s+t\), then the ideal \(\avg{\et^3 - \et\th^2}\) implies we can always eliminate powers of \(\et\) cubic or higher, \textit{e.g.} \(\et^5 \to \et^3\th^2 \to \et\th^4\). In this way, the monomials \(\th^n\), \(\et \th^n\), and \(\et^2\th^n\) freely span the ring, giving the direct sum decomposition
\begin{equation}
M^{O(2)}_{4,\scK}={\mathbb C}[s,t]/\langle s^2t+t^2 s\rangle = \big(1 \oplus \et \oplus \et^2\big)\mathbb{C}[\th]\, . \label{eq:hil_n4_O2_K_2}
\end{equation}
The existence of such a direct sum decomposition is a special property of so-called Cohen-Macaulay rings, which we discuss in sec.~\ref{sec:general_algebraic}. In the language of sec.~\ref{sec:general_algebraic}, we say that \(M^{O(2)}_{4,\scK}\) is a freely generated module over \(\mathbb{C}[\th]\), with a basis for the module given by \(\{1,\et,\et^2\}\).

\subsubsection*{(iv) $\scI = \scJ, \ G = O(d)\times S_4, \ d\ge 4$}
Still imposing parity, we now turn to the indistinguishable cases. Imposing only EOM, the ring is
\begin{subequations}
\be
M^{O(d\ge 4)\times S_4}_{4,\scJ}&=&{\mathbb C}[s_{12},s_{13},s_{14},s_{23},s_{24},s_{34}]^{S_4} \,, \label{eq:diffring}\\
H^{O(d\ge 4)\times S_4}_{4,\scJ}&=&\frac{1+t^6+t^8+t^{10}+t^{12}+t^{18}} {\left(1-t^2\right) \left(1-t^4\right)^2 \left(1-t^6\right)^2 \left(1-t^8\right)} \,.
\label{eq:diffhilb}\ee
\end{subequations}
We give the details on the consequences of the  $S_4$ invariance of the ring in eq.~\eqref{eq:diffring} (which is not straightforward) and how it leads to the above Hilbert series in appendix~\ref{app:EOMS4ring} (obtaining just the Hilbert series is, however, relatively simple). 

\subsubsection*{(v) $\scI = \scK, \ G = O(d)\times S_4, \ d\ge 3$}
Interestingly, further imposing IBP greatly simplifies the generator problem for indistinguishable particles. We again work with $\{s,t,u\}$, subject to $s+t+u=0$.
Special to $n=4$, permutations in $S_4$ of the momenta indices become simple permutations in $S_3$ of the Mandelstam invariants $\{s,t,u\}$. Hence, we construct polynomials invariant under $S_3$ permutations of $s,t,u$. Such a ring is freely generated by the symmetric polynomials, $e_1\equiv s+t+u$, $e_2\equiv st+su+tu$, and $e_3\equiv stu$. Momentum conservation simply removes the generator $e_1$. That is,
\begin{subequations}
 \be
M^{O(d\ge 3)\times S_4}_{4,\scK}&=&\bigg[{\mathbb C}[s,t,u]/\langle s+t+u\rangle\bigg]^{S_3(s,t,u)} ={\mathbb C}[e_1,e_2,e_3]/\langle e_1\rangle = {\mathbb C}[e_2,e_3]  \,, \label{eq:reomibpIN}\\
H^{O(d\ge 3)\times S_4}_{4,\scK}&=&\frac{1}{(1-q_{e_2}) (1-q_{e_3}) } \to \frac{1}{(1-t^4)(1-t^6)} \,.
\label{eq:heomibpIN}
\ee
\end{subequations}
where, similar to above, $q_{e_2}$ and $q_{e_3}$ are gradings introduced to count powers of the $e_2=st+su+tu$ and $e_3=stu$ generators.

\subsubsection*{(vi) $\scI = \scK, \ G = O(2)\times S_4, \ d=2$}
In $d=2$, the Gram constraint $stu=0$ simply sets $e_3=0$. That is, 
\begin{subequations}
\be
M^{O(2)\times S_4}_{4,\scK}&=&{\mathbb C}[e_2] \,,\label{eq:deq2indistingA}
\\
H^{O(2)\times S_4}_{4,\scK}&=&\frac{1}{(1-q_{e_2}) } \to \frac{1}{(1-t^4)} \,.
\label{eq:deq2indisting}
\ee
\end{subequations}

\subsubsection*{(vii) $\scI = \scK, \ G = SO(d)\times \Sigma$}
Finally, we omit parity and impose \(SO(d)\), rather than \(O(d)\), invariance. This means invariants involving the epsilon tensor are now available,
\begin{equation}
\epsilon_{p_{i_1}p_{i_2}\cdots p_{i_d}} \equiv \epsilon_{\mu_1\mu_2\ldots \mu_d}\,p_{i_1}^{\mu_1}p_{i_2}^{\mu_2}\ldots p_{i_d}^{\mu_d}
\end{equation}
However, only one power of the epsilon tensor can appear in any element of the ring, because of the relation
\be
\epsilon_{\mu_1\ldots \mu_d}\epsilon_{\nu_1\ldots \nu_d} = \left | \begin{array}{ccc} g_{\mu_1 \nu_1}&\ldots& g_{\mu_1 \nu_d}\\ \vdots &\ddots& \vdots \\   g_{\mu_d \nu_1}&\ldots& g_{\mu_d \nu_d} \end{array}\right| \,.
\label{eq:epsilontog}
\ee
Note that in $d\ge4$, any epsilon contraction with the $p_i^{\m}$ will be identically zero as we only have \(n=4\) momenta (in $d=4$, momentum conservation implies $\epsilon_{p_1p_2p_3p_4}=0$).

For the four point ring then, $SO(d)$ and $O(d)$ invariance are only different for $d\le 3$. For distinguishable particles in \(d=3\) there is only one independent epsilon invariant, say $\epsilon_{p_1p_2p_3}$, hence
\begin{subequations}
\be
M^{SO(3)}_{4,\scK}&=&{\mathbb C}[s,t] \oplus \epsilon_{p_1p_2p_3}{\mathbb C}[s,t]  \,\\
H^{SO(3)}_{4,\scK}&=& \frac{1+t^3}{(1-t^2)^2} \,.
\ee
\end{subequations}
For indistinguishable particles invariance under $S_4$ requires the additional invariant to be $\epsilon_{\text{sym}}\equiv \epsilon_{p_1 p_2 p_3} (s_{12}-s_{13})(s_{12}-s_{23})(s_{13}-s_{23})$, and hence
\begin{subequations}
\be
M^{SO(3)\times S_4}_{4,\scK}&=&{\mathbb C}[e_2,e_3] \oplus \epsilon_{\text{sym}}{\mathbb C}[e_2,e_3]  \, , \label{eq:n4_d3_ident_ring} \\
H^{SO(3)\times S_4}_{4,\scK}&=& \frac{1+t^9}{(1-t^4)(1-t^6)} \,.
\label{eq:so3indist}
\ee
\end{subequations}

In two dimensions, for distinguishable particles, the \(SO(2)\) invariants are \((s,t,u)\) and six epsilon invariants \(\epsilon_{p_ip_j}\), \(i<j\). Momentum conservation allows us to eliminate \(u\) as well as three of the \(\epsilon_{p_ip_j}\), say the \(\epsilon_{p_ip_4}\) with \(i=1,2,3\). Then, analogous to the \(O(2)\) ring in eq.~\eqref{eq:hil_n4_O2_K_2} we have
\begin{subequations}
\be
M^{SO(2)}_{4,\scK}&=&\Big(1 \oplus (s+t) \oplus \epsilon_{p_1p_2} \oplus \epsilon_{p_1p_3} \oplus \epsilon_{p_2p_3} \oplus (s+t)^2\Big){\mathbb C}[s-t]  \,\\
H^{SO(2)}_{4,\scK}&=& \frac{1+4t^2+t^4}{1-t^2} \,,
\ee
\end{subequations}
Finally, there is no difference between $SO(2)$ and $O(2)$ for \(n=4\) indistinguishable particles since no non-vanishing $S_4$ symmetric invariant can be formed with one power of $\epsilon_{p_ip_j}$.

\subsection{Interpreting the ring in terms of operators and amplitudes}\label{sec:four_three}
We were able to completely solve the operator basis at $n=4$ in section~\ref{sec:tour}, by which we mean that we found a set of generators, and understood the relations among them (or lack of relations). This enables us to easily construct the independent polynomials---{\it i.e.} the independent operators---$F^{(4,k)}_\alpha$ in the basis (here $\alpha$ labels the independent operators at the given $n$ and $k$). Before proceeding to the more general case---where things get significantly more involved---let us use the relatively simple and concrete formulas of the \(n=4\) case to reflect on what this means.

The association of a polynomial equation in the momenta of the particles with a given operator is exactly what one obtains when deriving the Feynman rule in momentum space for an operator. That is, an operator $\mathcal{O}^{(n,k)}_\alpha$ involving $n$ powers of $\phi$ and $k$ derivatives gives rise to a momentum space Feynman rule $F^{(n,k)}_\alpha$. For example, for \(n=4\) distinguishable scalars the ring in eq.~\eqref{eq:reomibp} corresponds to operators as
\begin{equation}
s^nt^m \in \mathbb{C}[s,t] \ \Leftrightarrow \ \pd_{\m_1}\dots\pd_{\m_n}\pd_{\n_1}\dots\pd_{\n_m}\ph_1\, \pd^{\m_1}\dots \pd^{\m_n}\ph_2 \, \pd^{\n_1}\dots\pd^{\n_m}\ph_3 \, \ph_4.
\end{equation}
Actually, to be precise, the above should have the derivatives traceless---such a construction of operators is generally cumbersome, and this is one way in which the momentum space picture is advantageous. Another advantage is in the permutation symmetry: elements \((st+su+tu)^n(stu)^m \in \mathbb{C}[e_2,e_3]\) of eq.~\eqref{eq:reomibpIN} essentially correspond to the same type of operator as above (with right number of derivatives, of course) but with the flavor indices dropped, all \(\ph_i =\ph\). In this case, the operator may look simpler, but it is masking the fact that one has to go through the exercise of symmetrization (taking all possible Wick contractions) for any quantity computed. 

\begin{figure}
\centering
\includegraphics[width=10cm]{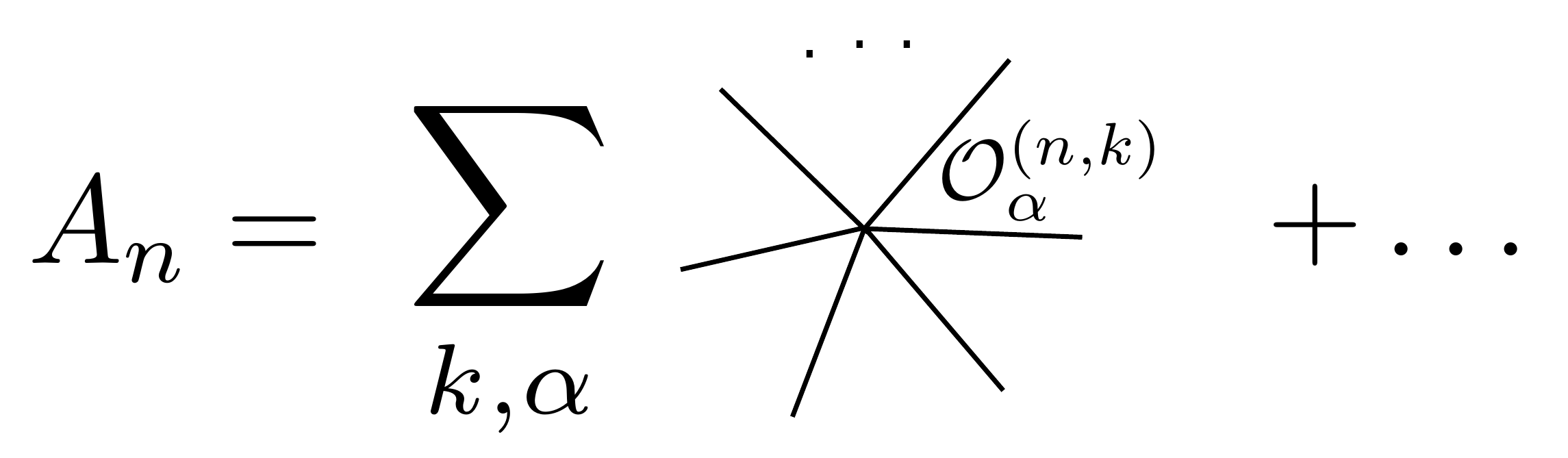}
\caption{Contact interaction contributions to the $n$-point scattering amplitude from  Feynman diagrams associated with the operators $\mathcal{O}^{(n,k)}_\alpha$.}
\label{fig:feyndiag}
\end{figure}

The polynomials $F^{(n,k)}_\alpha$ form the Feynman rules. Consider the $n$-point tree-level amplitude in the EFT of a real scalar field. The operators $\mathcal{O}^{(n,k)}_\alpha$ contribute as contact interactions, as depicted in Fig.~\ref{fig:feyndiag}. Further contributions involving propagators are indicated by the $+\ldots$ in the figure. 

More explicitly, we can write the general $n$ point, tree-level amplitude, $A_n$, as
\be
A_{n}(p_1,\ldots,p_n)  &=& \sum_k\sum_{\alpha} c_{\alpha}\,F^{(n,k)}_{\alpha}(p_1,\ldots,p_n) +\ldots \, .
\label{eq:amplitude}
\ee
The $c_{\alpha}$ are the Wilson coefficients of the operators $\mathcal{O}^{(n,k)}_{\alpha}$ in the Lagrangian. For example, if we impose a \(Z_2\) symmetry \(\ph \to -\ph\) (which eliminates the \(\ph^3\) vertex), the \(n=4\) point amplitude only involves contact interactions at tree-level,
\begin{equation}
A_{4}(p_1,\ldots,p_4)  = \sum_{n,m} c_{n,m} (st+su+tu)^n(stu)^m.
\label{eq:4pt_amp}
\end{equation}
A well-known result is that unitarity and analyticity constrain the Wilson coefficient associated to \((\pd\ph)^4\) to be positive, \(c_{1,0}>0\)~\cite{Adams:2006sv}. This result is obtained in the forward scattering regime where \(st + su + tu \to -s^2\) and \(stu \to 0\). The structure of the operator basis in terms of the two \textit{generators} \(e_2 = st + su + tu\) and \(e_3 = stu\) suggests \(c_{n,0} > 0\) for all \(n\ge 1\); indeed, this is true~\cite{Adams:2006sv}.

\subsection{Constructing the basis at higher $n$: generators and algebraic properties}\label{sec:general_algebraic}
With a healthy understanding of the four-point operator basis, we turn to the general \(n\) case. Although the basis becomes notably more involved, a rich structure underlies this complexity. A unifying theme in our discussion is how these properties are reflected in the Hilbert series of the ring. Useful references for background and further information are~\cite{Sturmfels:inv,Stanley79,Stanley78}. 

Intuitively, we would like to find generators for the operator basis, \textit{i.e.} some finite set of operators (Feynman rules) from which all others can be obtained, just as we did for the \(n=4\) ring in sec.~\ref{sec:tour}. It turns out that we can always find a particularly nice set of generators for the rings \(\Mgen\) such that any \(g \in \Mgen\) can be expressed \textit{uniquely} in terms of the generators (the choice of generators is not unique, however). This is because the \(\Mgen\) are so-called Cohen-Macaulay (CM) rings.

In sec.~\ref{subsubsec:CM_property} we prove that \(\Mgen\) is a Cohen-Macaulay integral domain. A reader less interested in mathematical details needs only to note the decomposition in eq.~\eqref{eq:hironaka} together with the associated definitions of ``primary'' and ``secondary'' invariants. In some instances, such as when \(G = SO(d)\), the secondary invariants exhibit a pairing as a result of \(\Mgen\) being not only CM, but even Gorenstein. Useful in this regard are the Hilbert series for \(\Mgen\), where a necessary and sufficient condition is that the Hilbert series exhibits a certain palindromic form~\cite{Stanley78}.

We discuss a conjecture~\cite{Thiery:2000:AIG:377604.377612} on a set of primary invariants for \(M_{n,\scI}^{O(d)\times S_n}\) in sec.~\ref{subsubsec:Thiery_conjecture}. In sec.~\ref{subsubsec:relation_bw_J_and_K} we touch upon the relationship between the operator spaces \(\scJ\) and \(\scK\).

In subsec.~\ref{subsubsec:spin_generalization} we briefly consider the generalization to the case that operators are composed of fields with spin, as well as the case that the operator itself carries spin. An important point here is that the rings for scalar fields, \(\Mgen\), are also important for understanding the spin case, due to the universal feature of kinematics which is encoded in the \(\Mgen\). However, the Cohen-Macaulay property generally breaks down when spin is involved, as a result of polarization tensors. 

\subsubsection{The Cohen-Macaulay property}\label{subsubsec:CM_property}
We begin by reviewing some relevant results from invariant theory~\cite{Sturmfels:inv,Stanley79,Stanley78}. Let \(R\) be a graded ring whose coordinates form a representation of some symmetry group \(G\) and let \(R^G\) denote the subring consisting of \(G\)-invariant elements of \(R\). For the groups \(G\) under consideration, a fundamental result is that \(R^G\) is finitely generated, meaning all \(f \in R^G\) can be expressed as polynomials in some finite set of generators. In general there may be relations among the generators, so that we are not guaranteed that \(f\) can be expressed uniquely from the generators. 

Within the generators, it is always possible to identify a set \(\{\th_1,\dots,\th_m\}\), \(m = \text{dim}(R^G)\), of homogeneous elements which are algebraically independent and their freely generated ring \(\Cbb[\th_1,\dots,\th_m]\) forms a subring of \(R^G\). \(R^G\) can therefore be taken as a finitely generated module over \(\Cbb[\th_1,\dots,\th_m]\). The set \(\{\th_1,\dots,\th_m\}\) is called a \textit{homogeneous system of parameters} (HSOP). Note that a choice of HSOP is by no means unique (a simple example: we could equally well have chosen \(\{\th_1^2,\th_2,\dots,\th_m\}\) as a HSOP).

The basic problem is to identify some HSOP and then determine the structure of \(R^G\) as a module over \(\Cbb[\th_1,\dots,\th_m]\). Broadly speaking, this entails (1) finding the rest of the generators and (2) classifying the algebraic relations between generators (including relations among relations, called syzygies). If \(R^G\) is a \textit{free} module over \(\Cbb[\th_1,\dots,\th_m]\) task (2) is greatly simplified, because it implies that \(R^G\) has the decomposition
\begin{equation}
R^G = \bigoplus_{i=1}^s \et_i \Cbb[\th_1,\dots,\th_m].
\label{eq:hironaka}
\end{equation}
In other words, every \(g \in R^G\) can be expressed \textit{uniquely} as a linear combination of the \(\et_i\) with coefficients in \(\Cbb[\th_1,\dots,\th_m]\): \(g = \sum_{i=1}^sf_i(\th_1,\dots,\th_m) \et_i\). In such a case, \(R^G\) is said to be \textit{Cohen-Macaulay}, and the decomposition in eq.~\eqref{eq:hironaka} is called a Hironaka decomposition~\cite{Sturmfels:inv,Stanley79}. This decomposition shows that CM rings very much resemble vector spaces.

In the decomposition in eq.~\eqref{eq:hironaka} the \(\th_1,\dots,\th_m\) are called \textit{primary invariants} and the the \(\et_1,\dots,\et_s\) \textit{secondary invariants}.\footnote{Note that \(\et_1\) can always be taken as \(\et_1 = 1\). Moreover, \(\{\et_1,\dots,\et_s\}\) may be polynomials of some smaller set \(\{\et_1',\dots,\et_r'\}\), \(r\le s\), sometimes called irreducible secondary invariants.} In terms of operators, the secondary invariants behave like seed operators---like \(\ph^n\) or a parity violating term such as \(\e^{\m_1\m_2\m_3}\pd_{\m_1\a_1\a_2}\ph \, \pd_{\m_2\a_3}\ph \, \pd_{\m_3}\ph \, \pd^{\a_1\a_2\a_3}\ph\) in \(d=3\) (\(\epsilon_{\text{sym}}\) in eq.~\eqref{eq:n4_d3_ident_ring})---upon which we can add momenta in the form of \(n\)-point Mandelstam variables (the primary invariants).

Given a Hironaka decomposition, it is trivial to write down the Hilbert series. Defining \(d_i \equiv \text{deg}(\th_i)\) and \(e_j \equiv \text{deg}(\et_j)\), we have
\begin{equation}
H(R^G,t) = \frac{\sum_{j=1}^st^{e_j}}{\prod_{i=1}^m(1-t^{d_i})}.
\label{eq:hil_hiro}
\end{equation}
Importantly, the numerator is a sum of strictly positive terms. For a ring which is not CM, syzygies can lead to negative terms in the numerator (with the denominator chosen in the form to reflect the HSOP, as in eq.~\eqref{eq:hil_hiro}). If we compute some Hilbert series and are able to bring it to the form in eq.~\eqref{eq:hil_hiro}, it is a good indication (although not sufficient) that the ring is CM.

In some cases one notices a remarkable property about the numerator of the Hilbert series in a CM ring: it is palindromic. That is, the numerator \(N(t) = \sum_{k=0}^r a_kt^k\) obeys \(a_k = a_{r-k}\) (equivalently, \(N(t) = t^r N(1/t)\) where \(r\) is the maximal degree of \(t\) in \(N(t)\)). This pairing, indicative of a duality, is enough to tell us that the ring is \textit{Gorenstein}: A theorem of Stanley, theorem 4.4 in~\cite{Stanley78}, says that a Cohen-Macaulay integral domain is Gorenstein if and only if its Hilbert series is palindromic.\footnote{\(R\) is an integral domain if for all non-zero \(x,y\in R\) then \(xy \ne 0\).}

With the above preparatory remarks we now come to the main point: the rings \(\Mgen\) are Cohen-Macaulay. This essentially follows from the fundamental theorem of Hochster and Roberts which asserts that the invariant ring of a reductive group is Cohen-Macaulay~\cite{Hochster74}. Omitting EOM and IBP constraints, it is clear that \(\Cbb[p_1^{\m},\dots,p_n^{\m}]^G\) is CM since \(G = (S)O(d)\times \Sigma\) is reductive. We need to check that the CM property holds when we add back in EOM and IBP.

As \(G\) is a direct product, we can apply invariance under individual groups one-by-one and see the CM property at each step along the way. For simplicity of discussion, we will include parity and work with \(O(d)\), although this is non-essential. Applying \(O(d)\) invariance, per the discussion in sec.~\ref{sec:four_one}, we end up with
\begin{subequations}
\begin{align}
 R_0^{\Sigma} &\equiv \mathbb{C}[p_1^{\m},\dots,p_n^{\m}]^{O(d)\times \Sigma} = \left[\bigslant{\Cbb[\mathbf{s}]}{\avg{\{\D\}}}\right]^{\Sigma}, \\
 R_1^{\Sigma} &\equiv \left[\bigslant{\mathbb{C}[p_1^{\m},\dots,p_n^{\m}]}{\avg{\{p_i^2\}}}\right]^{O(d)\times \Sigma} = \left[\bigslant{\Cbb[\mathbf{s}]}{\avg{\{s_{ii}\},\{\D\}}}\right]^{\Sigma}, \\
 R_2^{\Sigma} &\equiv \left[\bigslant{\mathbb{C}[p_1^{\m},\dots,p_n^{\m}]}{\avg{\{p_i^2\},\sum_i p_i^{\m}}}\right]^{O(d)\times \Sigma} = \left[\bigslant{\Cbb[\mathbf{s}]}{\avg{\{s_{ii}\},\{X_i\},\{\D\}}}\right]^{\Sigma} ,
\end{align}
\end{subequations}
where \(\mathbf{s}\) collectively denotes the \(s_{ij}\) including diagonal components \(s_{ii}\). When \(n \le d\) there are no Gram constraints, \(\{\D\} = \varnothing\), and all the above rings are isomorphic to \(\Cbb[V]^{\Sigma}\) with \(V\) some representation of \(\Sigma\) (\textit{e.g.} eq.~\eqref{eq:sij_decomp_Sn}). In this case it is clear that the resulting rings are CM as they are just invariant rings of a finite group, which are CM~\cite{Stanley79,Sturmfels:inv}.

We proceed to the case \(n > d\), where \(\avg{\{\D\}}\) is non-trivial. We will show that \(\Cbb[\mathbf{s}]/I\) is CM, which will subsequently be used to show that \(\big[\Cbb[\mathbf{s}]/I\big]^{\Sigma}\) is CM. We first need two other characterizations of CM rings: (1) The fact that \(R^G\) is a free \(\Cbb[\th_1,\dots,\th_m]\)-module implies that for \textit{every} HSOP \(\{\th_1',\dots,\th_m'\}\) then \(R^G\) is a free module over \(\Cbb[\th_1',\dots,\th_m']\). (2) As a free \(\Cbb[\th_1,\dots,\th_m]\)-module, this implies \(\{\th_1,\dots,\th_m\}\) is a regular sequence in \(R^G\); an equivalent definition of a CM ring is that it possesses a regular sequence of algebraically independent elements and length equal to the dimension of the ring.\footnote{Given a ring \(R\), an element \(f\in R\) is a non-zero divisor, or \textit{regular}, if for all non-zero \(g \in R\) then \(f\cdot g \ne 0\). A sequence \(\{f_1,\dots,f_n\}\) is called a \textit{regular sequence} if \(f_1\) is a non-zero divisor in \(R\) and \(f_i\) is a non-zero divisor in \(R/\avg{f_1,\dots,f_{i-1}}\). From these definitions, it is clear that \(R\) is particularly well behaved when quotiented by regular elements/sequences, \textit{i.e.} \(R\) is a free \(\Cbb[f_1,\dots,f_n]\)-module if \(f_1,\dots,f_n\) are algebraically independent.}

First, note that \(\{s_{11},\dots,s_{nn}\}\) is a regular sequence in \(R_0\): corollary 3.2 of~\cite{Stanley78} tells us that for a sequence \(\{\th_1,\dots,\th_n\}\) in \(R_0\) we have the following inequality of Hilbert series,
\begin{equation}
H(R_0,t) \le \frac{H\big(\bigslant{R_0}{\avg{\th_1,\dots,\th_n}},t\big)}{\prod_{i=1}^n(1-t^{\text{deg}{\th_i}})},
\label{eq:hil_reg_seq_inequality}
\end{equation}
with equality holding if and only if \(\{\th_1,\dots,\th_n\}\) is a regular sequence. For the sequence \(\{s_{11},\dots,s_{nn}\}\) we have
\begin{equation}
\bigslant{R_0}{\avg{s_{11},\dots,s_{nn}}} = \bigslant{\Cbb[\mathbf{s}]}{\avg{s_{11},\dots,s_{nn},\{\D\}}} = R_1,
\end{equation}
and by the matrix integral formula for the Hilbert series,
\begin{equation}
H(R_1,t) = \int d\m_{O(d)}(g)\left[\frac{(1-t^2)}{\text{det}_{\Box}(1-t g)}\right]^n = (1-t^2)^n H(R_0,t).
\end{equation}
Since eq.~\eqref{eq:hil_reg_seq_inequality} is saturated, \(\{s_{11},\dots,s_{nn}\}\) is a regular sequence in \(R_0\). We further know that \(R_0\) is CM by Hochster-Roberts' theorem, and hence we can find a regular sequence of length \(\text{dim}(R_0)\). Let \(\{s_{11},\dots,s_{nn},\th_1,\dots,\th_{N_1}\}\) be such a sequence, where \(N_1 = \text{dim}(R_0) - n = \text{dim}(R_1)\). By definition, since \(\{s_{11},\dots,s_{nn},\th_1,\dots,\th_{N_1}\}\) is a \(R_0\)-sequence, then \(\{\th_1,\dots,\th_{N_1}\}\) is a regular sequence on \(R_0/\avg{s_{11},\dots,s_{nn}}\). But \(R_1 = R_0/\avg{s_{11},\dots,s_{nn}}\), so \(\{\th_1,\dots,\th_{N_1}\}\) is a \(R_1\)-regular sequence of length \(N_1 = \text{dim}(R_1)\), and hence \(R_1\) is Cohen-Macaulay. Similar reasoning establishes \(R_2\) as CM.

Given that \(R_{1,2}\) are CM, we can show that \(R_{1,2}^{\Sigma}\) is CM via a straightforward modification of the proof that the invariant ring of a finite group is CM (\textit{e.g.}~\cite{Stanley79} theorem 3.2 or~\cite{Sturmfels:inv} section 2.3). The basic sketch is as follows. Let \(\{\th_1,\dots,\th_m\} \in R_i^{\Sigma}\), \(m = \text{dim}(R_i^{\Sigma})\), be a HSOP for \(R_i^{\Sigma}\). Since \(\text{dim}(R_i) = \text{dim}(R_i^{\Sigma})\) (finite groups do not have enough symmetry to remove continuous degrees of freedom), \(\{\th_1,\dots,\th_m\}\) is also a HSOP for \(R_i\). Let \(U_i\) be the set of \(\Sigma\) non-invariant elements of \(R_i\). As a module over \(\Cbb[\th_1,\dots,\th_m]\), \(R_i\) has the direct sum decomposition \(R_i = R_i^{\Sigma} \oplus U_i\). Now, as established above, \(R_i\) is CM. This means that \(R_i\) is a free module over \textit{any} HSOP; in particular, \(R_i\) is a free \(\Cbb[\th_1,\dots,\th_m]\)-module. Then the direct sum decomposition of \(R_i\) as \(\Cbb[\th_1,\dots,\th_m]\)-modules implies that \(R_i^{\Sigma}\) is a free module over \(\Cbb[\th_1,\dots,\th_m]\) and hence CM.

The rings \(\Mgen\) are also integral domains. Therefore, \(\Mgen\) is Gorenstein if its Hilbert series exhibits the palindromic form \(H(\Mgen,t) = (-1)^m t^p H(\Mgen,1/t)\) for some integer \(p\). This is always the case when \(G = SO(d)\) (no permutation group), as can easily be seen from the matrix integral formula
\begin{equation}
H(M_{n,\scI}^{SO(d)},1/t) \propto \int d\m_{SO(d)}(x)\big[P^{(d)}(1/t;x)\big]^n = \pm t^k H(M_{n,\scI}^{SO(d)},t),
\label{eq:hil_goren_1overt}
\end{equation}
for some \(k\in \mathbb{Z}\) (some care needs to be taken with the contours as we send \(t \to 1/t\)). For disconnected groups, such as \(G = O(d)\) or \(G = (S)O(d) \times \Sigma\), we do not have a completely general statement. When \(G = O(d) = SO(d) \ltimes Z_2\), \(M_{n,\scI}^{O(d)}\) is Gorenstein for either \(n\)-even or for \(n\)-odd, depending on \(d=2r\) versus \(d = 2r+1\) and whether \(\scI = \scJ\) or \(\scK\) (if \(\scM_{n,\scJ}^{O(d)}\) is Gorenstein, then \(M_{n+1,\scK}^{O(d)}\) is as well since momentum conservation effectively removes one momentum). Including full permutations, \(G = (S)O(d) \times S_n\), we have the following. There are no Gram conditions in \(\scJ\) when \(n \le d\) and none in \(\scK\) when \(n \le d+1\); in such cases, after applying \((S)O(d)\) invariance, we are left with the ring \(\Cbb[V]^{S_n}\) with \(V\) a representation of \(S_n\) (for \(\scJ\), \(V = V_{(n)}\oplus V_{(n-1,1)}\oplus V_{(n-2,2)}\); for \(\scK\), \(V = V_{(n-2,2)}\)). It is straightforward to see,~\cite{Thiery:2000:AIG:377604.377612} theorem 6.2, that \(\Cbb[V_{(n)}\oplus V_{(n-1,1)}\oplus V_{(n-2,2)}]^{S_n}\) is Gorenstein for \(n\) even, while \(\Cbb[V_{(n-2,2)}]^{S_n}\) is Gorenstein for \(n\) odd. For \(G = (S)O(d) \times S_n\) with \(n > d\), explicit calculations of the Hilbert series, \textit{e.g.} see sec.~\ref{sec:case}, indicate that the ring is usually \textit{not} Gorenstein, and we suspect this is generally the case for \(n\) sufficiently larger than \(d\).

\subsubsection{A set of primary invariants in the absence of Gram conditions}\label{subsubsec:Thiery_conjecture}
When \(\scI = \scJ\), \(G = O(d) \times S_n\), and \(n \le d\), determining the elements of \(M_{n\le d,\scJ}^{O(d)\times S_n} = \Cbb[s_{ij}]^{S_n} = \Cbb[V_{(n)}\oplus V_{(n-1,1)}\oplus V_{(n-2,2)}]^{S_n} \) is equivalent to a graph isomorphism problem, \textit{e.g.}~\cite{Thiery:2000:AIG:377604.377612}. As a graph, we take \(n\) vertices so that \(s_{ij}\) corresponds to an edge between vertices \(i\) and \(j\). Using intuition from the fundamental theorem of symmetric polynomials,~\cite{Thiery:2000:AIG:377604.377612} conjectured the following to be a HSOP for \(\Cbb[s_{ij}]^{S_n}\):
\begin{subequations}
\label{eq:Thiery_prim}
\begin{align}
X_1 + \cdots + X_n, \ &\ldots, \ X_1^n + \cdots +X_n^n, \\
s_{12}^2 + \cdots + s_{n-1\, n}^2, \ &\ldots, \ s_{12}^{n(n-3)/2 +1}+ \cdots s_{n-1\, n}^{n(n-3)/2 +1},\label{eq:Thiery_sij_prim}
\end{align}
\end{subequations}
where we recall \(X_i \equiv \sum_{j\ne i} s_{ij}\). Some evidence for this conjecture is that the Hilbert series suggests a HSOP of minimal degree has degrees coinciding with the above. The polynomials in eq.~\eqref{eq:Thiery_sij_prim} could also serve as a HSOP when the \(X_i\) have been eliminated by momentum conservation (\textit{i.e.} when \(\scI = \scK\) and we work with \(M_{n\le d+1,\scK}^{O(d) \times S_n} = \Cbb[V_{(n-2,2)}]^{S_n}\)). The power sum polynomials in eq.~\eqref{eq:Thiery_sij_prim} are obviously \(S_n\) invariant: by construction, they are invariant under the larger group \(S_{n(n-1)/2}\) which permutes the \(s_{ij}\) as the defining representation.

Having knowledge of a HSOP helps to algorithmically construct the operator basis, see subsection~\ref{subsec:n=5_basis}. Additionally, in the absence of Gram constraints, the above can serve as a HSOP for some generalizations discussed below.

\subsubsection{Relationship between $\scJ$ and $\scK$}\label{subsubsec:relation_bw_J_and_K}
The rings \(M_{n,\scJ}^G\) and \(M_{n+1,\scK}^G\) are closely related since momentum conservation effectively allows us to eliminate one momentum in \(M_{n+1,\scK}^G\). Eliminating, say, \(p_{n+1}^{\m}\) we are left with the \(n\) momenta \(p_1^{\m},\dots,p_n^{\m}\) subject to the on-shell conditions \(p_i^2 = 0\) plus the additional constraint \( p_{n+1}^2 = (p_1 + \dots + p_n)^2 = 0\),
\begin{align}
\left[\bigslant{\Cbb[p_1^{\m},\dots,p_{n+1}^{\m}]}{\avg{\{p_i^2\},\sum_{i=1}^{n+1} p_i^{\m}}}\right]^{G} &\sim \left[\bigslant{\Cbb[p_1^{\m},\dots,p_{n}^{\m}]}{\avg{\{p_i^2\},(\sum_{i=1}^{n} p_i^{\m})^2}}\right]^{G} \nonumber \\
 &\sim \left[\bigslant{M_{n,\scJ}}{\avg{p_1\cdot p_2 + \dots p_{n-1}\cdot p_n}}\right]^G.
\end{align}
To wit, in the absence of permutation symmetry, this relationship is clearly reflected in the Hilbert series
\begin{equation}
H(M_{n+1,\scK}^{(S)O(d)},t) = \int d\m \frac{1}{P(t;x)} \big[(1-t^2)P(t;x)\big]^{n+1} = (1-t^2) H(M_{n,\scJ}^{(S)O(d)},t).
\end{equation}

In one-dimensional field theories we observed~\cite{Henning:2015daa} a similar relationship and were able to use this to arrive at consistency conditions and recursion relations. It would be interesting to explore this relationship in more depth in the \(d\)-dimensional case studied here.

\subsubsection{Generalizations with spin}\label{subsubsec:spin_generalization}
Here we make a few comments when spin enters the problem. Moving beyond scalar fields, what algebraic properties do we expect if we have fermions, gauge fields, or other particles with spin? Physically, each field that composes an operator \(\scO \in \scK\) has an associated momentum (continuous degree of freedom) and a polarization tensor (or spinor) parameterizing a \textit{finite} number of polarization states. Hence, the appropriate algebraic formulation will still consist of polynomials in the \(s_{ij}\), but will also include a finite number of terms accounting for how the polarization tensors can be contracted amongst themselves or with the \(p_i^{\m}\) to form Lorentz invariants.

The above indicates that, for fixed field content and arbitrary numbers of derivatives, the relevant rings involving fields with spin (whose precise description we have not yet formulated) will be finitely generated and of the same dimension as the rings involving only scalar fields. In particular, the \(\Mgen\) describe kinematics of the momenta---universal to all operators---so that the primary and secondary invariants of \(\Mgen\) will still appear and the primary invariants \(\{\th_1,\dots,\th_m\}\) can serve as a HSOP in this case as well. However, the rings for operators composed of spinning particles in general will \textit{not} be free as \(\Cbb[\th_1,\dots,\th_m]\)-modules, \textit{i.e.} they will not be Cohen-Macaulay. This happens because Gram conditions now involve the polarization tensors in addition to the momenta. In keeping with our theme, these properties are reflected in the Hilbert series---see sec.~\ref{sec:case} for examples.

Although this paper is focused on operators belonging to the operator basis \(\scK\), which are necessarily Lorentz scalars, a natural generalization (for example, in studying operator content of free CFTs) is to look at operators carrying spin. The Hilbert series formula eq.~\eqref{eq:H_0_summary} is trivially generalized to count primary operators in the free theory of a given spin. What about explicit construction of these operators? Let \(M_{n,\scI,l}^G\) denote the set of elements in \(\Cbb[p_1^{\m},\dots,p_n^{\m}]/\avg{\{p_i^2\},\sum_ip_i^{\m}}\) which transform as the \(l = (l_1,\dots,l_{\lfloor\frac{d}{2}\rfloor})\) representation of \((S)O(d)\). Note that the spin does not change when we multiply a polynomial \(f\in M_{n,\scI,l}^G\) by an \((S)O(d)\) invariant polynomial. Moreover, \(\text{dim}(M_{n,\scI,l}^G) = \text{dim}(\Mgen)\).  We thus see that the \(M_{n,\scI,l}^G\) are finitely generated modules over \(M_{n,\scI}^G\) (but, in general, not free, and hence not CM)~\cite{Stanley79b}. In particular, any HSOP for \(\Mgen\) will also be a HSOP for \(M_{n,\scI,l}^G\).

While scalar fields may seem like a special case in field theory, the fact is that the operator bases for scalar fields is actually encoding the kinematics of momenta. For this broad reason, and more pointed ones hinted at above, understanding the rings \(\Mgen\) is essential for understanding operator spectra in more general cases.

\subsection{Conformal primaries and elements of $M_{n,\mathcal{K}}^G$} \label{sec:conf_prim_ring}
Recall from sec.~\ref{sec:overview} that the single particle module for scalars, \(R_{\ph}\), is an irrep of the conformal group. This leads to the fact that the operator basis \(\scK\) is spanned by scalar conformal primaries appearing in the tensor products of \(R_{\ph}\) (equivalently, in the OPE of \(\ph(x)\)). How does this connect with the momentum space picture in terms of the rings \(M_{n,\scK}^G\)?

By construction, elements of \(M_{n,\scK}^G\) correspond to operators which are not total derivatives. However, this does not mean they directly correspond to conformal primaries; that is, an element \(f \in M_{n,\scK}^G\) may not transform as a conformal primary when acted upon by the generator for special conformal transformations, \(K_{\m}f\). As a quotient ring, \(M_{n,\scK}^G\) consists of equivalence classes, and in this way any representative element \(f \in M_{n,\scK}^G\) is in an equivalence class containing a conformal primary.

If one wants explicit conformal primaries, one option is to construct \(K_{\m}\) in momentum space and then enforce the correct transformation behavior. An equivalent option is to construct the conformal Casimir in momentum space and find polynomials which are eigenstates of this operator. See~\cite{Katz:2016hxp} for related discussions. 

A related point is the question of endowing an inner product on \(M_{n,\scK}^G\).  There is a natural inner product inherited from field theory, namely the two-point function. With this we can orthogonalize the elements of \(M_{n,\scK}^G\) since the free scalar field theory is conformal so that primary operators obey \(\avg{O_i(x)O_j(y)} \sim \d_{ij}/\abs{x-y}^{\D_i + \D_j}\). Note that elements of different degrees in \(M_{n,\scK}^G\)---regardless of whether they are actually primary or not---are automatically orthogonal under the inner product since they are eigenvectors of the dilatation operator with different eigenvalues.

\subsection{Algorithms for operator construction and the $n=5$ basis}
\label{subsec:n=5_basis}
In this section we explain a simple algorithm for explicitly constructing elements of the operator basis with the goal of determining a set of primary and secondary invariants. We then apply this algorithm to the case of \(n=5\) identical scalars in \(d \ge 4\) dimensions, \textit{i.e.} for \(M_{5,\scK}^{(S)O(d\ge 4)\times S_5}\), determining the relevant symmetrized Mandelstam variables appropriate for 5-point amplitudes.

We are unaware of an explicit construction of these variables in the physics literature, and note that such a set could be of technical use, for example, to investigate the generalization of~\cite{Heemskerk:2009pn} beyond 4-point correlation functions. In the math literature, the invariant ring \(\Cbb[V_{(3,2)}]^{S_5} \big(= M_{5,\scK}^{(S)O(d\ge 4)\times S_5}\big)\) was solved in~\cite{dixmier}. Beyond \(n>5\) we are unaware of an existing solution. In these cases, algorithmic approaches~\cite{Sturmfels:inv,Thiery:2000:AIG:377604.377612} appear to be the most straightforward line of attack. Freely available computer programs, like  \texttt{Singular}~\cite{singular,King2013} and \texttt{Macaulay2}~\cite{m2} have dedicated packages for these problems; see the end of this subsection for an example using \texttt{Singular}. The algorithm we outline is lower-level, although it benefits from its conceptual simplicity and flexibility to handle the varying cases of interest to us.

For concreteness, we take \(\scI = \scK\) and \(G = O(d)\times S_n\) and work with \(M_{n,\scK}^{O(d) \times S_n}\). As representations of \(S_n\) the \(s_{ij}\) decompose into \(V_{(n)}\oplus V_{(n-1,1)}\oplus V_{(n-2,2)}\) with the basis for \(V_{(n)}\oplus V_{(n-1,1)}\) given by \(X_i =\sum_{j\ne i}s_{ij}\) and a basis \(x_1,\dots,x_m\), \(m = n(n-3)/2\), for \(V_{(n-2,2)}\) given by the orthogonal linear combinations of \(s_{ij}\). We can take this basis to coincide with the usual one built from tabloids of the standard tableaux (\textit{e.g.}~\cite{Sagan:sym}),
\begin{equation}
\ytableausetup{boxsize=1em,tabloids}
\begin{ytableau}
i & j & & \cdots & \\
k & l
\end{ytableau} \quad \rightarrow \quad s_{ij} - s_{il} - s_{jk} + s_{kl}.
\label{eq:xi_basis}
\end{equation}
Fixing an order \(\vec{s} = (s_{12},s_{13},\dots,s_{1n},s_{23},\dots,s_{n-1\,n})\) and letting \(S\) denote the change of basis matrix we have, for example, at \(n = 4\)
\begin{equation}
\begin{pmatrix} X \\ x \end{pmatrix} = S \vec{s} \ \rightarrow \ \begin{pmatrix} X_1 \\ X_2 \\ X_3 \\ X_4 \\ x_1 \\ x_2 \end{pmatrix} = \begin{pmatrix} 1 & 1 & 1 & 0 & 0 & 0 \\ 1 & 0 & 0 & 1 & 1 & 0 \\ 0 & 1 & 0 & 1 & 0 & 1 \\ 0 & 0 & 1 & 0 & 1 & 1 \\ 1 & 0 & -1 & -1 & 0 & 1 \\ 0 & 1 & -1 & -1 & 1 & 0 \end{pmatrix} \begin{pmatrix} s_{12} \\ s_{13} \\ s_{14} \\ s_{23} \\ s_{24} \\ s_{34} \end{pmatrix}.\label{eq:sij_convert_basis}
\end{equation}
For an element \(\s \in S_n\), we obtain the representation matrices \(\r(\s)\) as
\begin{equation}
S \r_{s_{ij}}(\s) S^{-1} = \begin{pmatrix} \r_{(n)\oplus(n-1,1)}(\s) & 0 \\ 0 & \r_{(n-2,2)}(\s) \end{pmatrix}.
\end{equation}
In terms of the coordinates \(x_1,\dots,x_m\) we have
\begin{equation}
M_{n,\scK}^{O(d)\times S_n} = \left[\bigslant{\Cbb[x_1,\dots,x_m]}{\avg{\D(x_a)}}\right]^{S_n},
\end{equation}
where the \(X_i\) have been removed by momentum conservation and the Gram conditions are expressed in terms of the \(x_a\).

We first outline how to determine a basis for elements of fixed degree in \(M_{n,\scK}^{O(d)\times S_n}\). At a fixed degree \(k\), this is done by first symmetrizing over the degree \(k\) elements of \(\Cbb[x_1,\dots,x_m]\). This produces an overcomplete basis for degree \(k\) elements of \(M_{n,\scK}^{O(d)\times S_n}\). If \(n \le d+1\), one can find the linearly independent elements by numerically evaluating the elements on enough randomly chosen points \((x_1,\dots,x_m) \in \mathbb{R}^m\). If \(n > d+1\) we account for Gram conditions by instead picking random values for \(p_1^{\m},\dots,p_n^{\m}\) and then evaluate the \(x_a\) as functions of the \(p_i^{\m}\) per eq.~\eqref{eq:xi_basis}.

Now let's explain in more detail. For simplicity, we set \(\text{deg}(s_{ij}) = \text{deg}(x_a) = 1\). First, one computes the Hilbert series and brings it to the form as in eq.~\eqref{eq:hil_hiro} to reflect the degrees of a set of primary and secondary invariants,
\begin{equation}
H_{n,\scK}^{O(d)\times S_n}(t) = \frac{\sum_j^{s}t^{e_j}}{\prod_{i=1}^m(1-t^{d_i})} = \sum_{k=0} c_k t^k.
\label{eq:hil_alg_exp}
\end{equation}
The Taylor expansion tells us that there are \(c_k\) independent degree \(k\) elements in the ring.

Let \(*\) denote the Reynolds operator which symmetrizes a polynomial:
\begin{equation}
f(\mathbf{x}) \mapsto f^*(\mathbf{x}) = \frac{1}{n!} \sum_{\s \in S_n} f\big(\r_{(n-2,2)}(\s)\mathbf{x}\big).
\label{eq:reyn}
\end{equation}
At a fixed degree \(k\), \(\Cbb[x_1,\dots,x_m]\) is spanned by the monomials \(\mathbf{x}^{\a} \equiv x_1^{\a_1}\dots x_m^{\a_m}\) with \(\abs{\a} = \a_1 + \dots + \a_m = k\). Let \(A^{(k)}\) denote the set of polynomials obtained by symmetrizing the monomials. Since we are symmetrizing, we can restrict \(\a\) to be in the set of partitions of \(k\) into at most \(m\) parts,
\begin{equation}
A^{(k)} = \big\{ (\mathbf{x}^{\a})^* \ | \ \a_1 \ge \dots \ge \a_m \ge 0, \ \abs{\a} = k \big\}. 
\label{eq:A_k}
\end{equation}

The set \(A^{(k)}\) is an overcomplete span of the degree \(k\) elements in \(\Cbb[x_1,\dots,x_m]^{S_n}\). To find a set of linear independent elements in \(A^{(k)}\) is just a linear algebra problem. A simple method is to evaluate the elements numerically on (at least) \(c_k\) random points. If there are no Gram constraints (\(n \le d+1\) \(\Leftrightarrow\) the ideal \(\avg{\{\D\}}\) is trivial) then these can be \(c_k\) random points \((x_1,\dots,x_m) \in R^{m}\). When \(\{\D\}\) is non-trivial (\(n > d+1\)), one instead picks \(c_k\) random values for the momenta \(p_1^{\m},\dots,p_n^{\m}\) subject to \(p_i^2\) and \(\sum_i p_i^{\m} = 0\). This in turn provides \(c_k\) points \((x_1,\dots,x_m)\) via eq~\eqref{eq:xi_basis} and \(s_{ij} = p_i \cdot p_j\). By working directly with the momenta, finite rank conditions are automatically accounted for. Following this procedure, we let \(B^{(k)}\) denote a set of degree \(k\), linearly independent elements which span the degree \(k\) elements in \(M_{n,\scK}^{O(d)\times S_n}\). Note \(\abs{B^{(k)}} = c_k\) with the \(c_k\) given in eq.~\eqref{eq:hil_alg_exp}.

To determine primary and secondary invariants, we go degree-by-degree. Let \(k_{\text{min}}\) be the lowest degree for which \(B^{(k)}\) is non-empty. The Hilbert series will indicate how many elements in \(B^{(k_{\text{min}})}\) should be considered primary 
invariants and how many secondary, and one then splits up the elements of \(B^{(k_{\text{min}})}\) accordingly. We then increment the degree and repeat. At degree \(k\), the Hilbert series may indicate that \(B^{(k)}\) contains new primary and/or secondary invariants: one identifies these by finding the elements of \(B^{(k)}\) which are linearly independent from degree \(k\) polynomials built from already identified primary and secondary invariants (see below for an example). The Hilbert series provides much information on where to look for primary and secondary invariants. Finally, one needs to make sure that a candidate set of primary invariants is algebraically independent. This is easily done by computing the Jacobian determinant, which is non-vanishing if and only if the polynomials are algebraically independent. This algorithm terminates when the Hilbert series indicates no further primary or secondary invariants remain.

\subsubsection{The $n=5$ operator basis}
We now use the above algorithm to obtain the operator basis for \(n=5\) identical scalars in \(d\ge 4\) dimensions, \textit{i.e.} find a set of primary and secondary invariants for \(M_{5,\scK}^{(S)O(d\ge4)\times S_5}\). This ring is isomorphic to the invariant ring \(\Cbb[x_1,\dots,x_5]^{S_5}\) where the \(x_i\) are a basis for the \(S_5\) representation \(V_{(3,2)}\), eq.~\eqref{eq:xi_basis}.\footnote{Explicitly, the change of basis is \begin{equation}
\begin{pmatrix}X_1 \\ X_2 \\ X_3 \\ X_4 \\ X_5 \\ x_1 \\ x_2 \\ x_3 \\ x_4 \\ x_5 \end{pmatrix} = 
\left(\begin{array}{cccccccccc}
 1 & 1 & 1 & 1 &   &   & \textcolor{white}{-1} & \textcolor{white}{-1}  & \textcolor{white}{-1}  & \textcolor{white}{-1}  \\
 1 &  \textcolor{white}{-1}  &   &   & 1 & 1 & 1 &   &   &   \\
 \textcolor{white}{-1}  & 1 &   &   & 1 &   &   & 1 & 1 &   \\
   &   & 1 &   &   & 1 &   & 1 &   & 1 \\
   &   &   & 1 &   &   & 1 &   & 1 & 1 \\
 1 &   &   & -1 &   & -1 &   &   &   & 1 \\
 1 &   &   & -1 & -1 &   &   &   & 1 &   \\
 1 &   & -1 &   & -1 &   &   & 1 &   &   \\
   & 1 &   & -1 & -1 &   & 1 &   &   &   \\
   & 1 & -1 &   & -1 & 1 &   &   &   &   \\
\end{array}\right) 
\begin{pmatrix} s_{12} \\ s_{13}\\ s_{14}\\s_{15}\\s_{23}\\s_{24}\\s_{25}\\s_{34}\\s_{35}\\s_{45} \end{pmatrix}\label{eq:n=5_change_basis}
\end{equation}}

The Hilbert series is readily calculated as
\begin{subequations}\label{eq:hil_n=5_alg}
\begin{align}
H_{5,\scK}^{(S)O(d\ge 4)\times S_5} &= \frac{1 + t^6 + t^7 + t^8 + t^9 + t^{15}}{(1-t^2)(1-t^3)(1-t^4)(1-t^5)(1-t^6)} \qquad \ (\text{deg}(s_{ij}) =1)\label{eq:hil_n=5_full} \\
&=1 + t^2 + t^3 + 2 t^4 + 2 t^5 + 5 t^6 + 4 t^7 + 8 t^8 + 9 t^9 + 
 13 t^{10} + \dots, \label{eq:hil_n=5_expand}
\end{align}
\end{subequations}
where, for simplicity, we set \(\text{deg}(s_{ij}) = \text{deg}(x_a) = 1\) (send \(t \to t^2\) to obtain our conventional grading \(\text{deg}(s_{ij}) = 2\)). The Hilbert series indicates that a minimal set of generators consists of five primary invariants \(\{\th_2,\th_3,\th_4,\th_5,\th_6\}\) and six secondary invariants \(\{\et_0,\et_6,\et_7,\et_8,\et_9,\et_{15}\}\) where the subscript indicates the degree. The degree-zero secondary is just the trivial element, \(\et_0 = 1\). We note in passing the palindromic nature of the numerator, which indicates the ring is Gorenstein.

Eq.~\eqref{eq:hil_n=5_expand} indicates there are no degree one elements. Indeed, all degree one elements of \(\Cbb[x_1,\dots,x_5]\) vanish under symmetrization, \textit{e.g.} \((x_1)^* = \frac{1}{5!}\sum_{\s \in S_5} \r_{(3,2)}(\s)x_1 = 0\).

The degree two elements of \(\Cbb[x_1,\dots,x_5]\) are spanned by the monomials \(x_ax_b\); we symmetrize over these to obtain elements of \(\Cbb[x_1,\dots,x_5]^{S_5}\). We only need to symmetrize over \(x_1^2\) and \(x_1x_2\), so that the set in eq.~\eqref{eq:A_k} is
\begin{equation}
A^{(2)} = \{(x_1^2)^*,(x_1x_2)^*\}, \nonumber
\end{equation}
where the symmetrization gives\footnote{Throughout this section, we take the symmetrization operation up to a constant. We choose to normalize \(f^*\) so that all terms in \(f^*\) have the smallest possible integer coefficients, \textit{e.g.} if \((x_1^2)^* = \frac{1}{5!}\sum_{\s \in S_5}(\r_{(3,2)}(\s)x_1)^2 = \frac{2}{15}(3x_1^2 + \dots)\) we drop the \(\frac{2}{15}\) factor and take \((x_1^2)^* = 3x_1^2 + \dots\). In particular, equalities like eqs.~\eqref{eq:reyn_n=5_deg2} or~\eqref{eq:deg4_rel} should be understood under this convention.}
\small{\begin{align}
(x_1^2)^* = (x_1x_2)^* = &\ 3 x_1^2-2 x_1 x_2 + 3x_2^2 - 2x_1x_3 -2x_2x_3 +3x_3^2 - 2x_1x_4 \nonumber\\ 
&-2x_2x_4 +2x_3x_4 + 3x_4^2 +4x_1x_5 -4x_3x_5 -4x_4x_5 +4x_5^2.
\label{eq:reyn_n=5_deg2}
\end{align}}\normalsize
The linear dependence between \((x_1^2)^*\) and \((x_1x_2)^*\) is trivial to solve and we may simply take \(B^{(2)} = \{(x_1^2)^*\}\). The Hilbert series tells us this element is a primary invariant, so we set
\begin{equation}
\th_2 = (x_1^2)^*.
\end{equation}
We note that \(\th_2\), as given in eq.~\eqref{eq:reyn_n=5_deg2}, is precisely what is obtained by taking \(s_{12}^2 + s_{13}^2 + \cdots + s_{45}^2\) and converting the \(s_{ij} \to (X_i,x_a)\) (using eq.~\eqref{eq:n=5_change_basis}) and setting \(X_i = 0\).

At degree three we consider \(A^{(3)} = \{(x_1^3)^*,(x_1^2x_2)^*,(x_1x_2x_3)^*\}\). The first two elements vanish; the last one does not and we can take it to be the primary invariant \(\th_3\),
\small{\begin{align}
\th_3 = (x_1x_2x_3)^* = 
&\ x_1^3-x_1^2 x_2-x_1 x_2^2+x_2^3-x_1^2 x_3+4 x_1 x_2 x_3-x_2^2 x_3-x_1x_3^2-x_2 x_3^2 \nonumber \\
&+x_3^3-x_1^2 x_4+2 x_1 x_2 x_4-x_2^2 x_4-x_1 x_3x_4-x_2 x_3 x_4+x_3^2 x_4 -x_1 x_4^2 \nonumber \\
&-x_2 x_4^2+x_3 x_4^2+x_4^3+2 x_1^2x_5-3 x_1 x_2 x_5+3 x_2 x_3 x_5 -2 x_3^2 x_5+2 x_1 x_4 x_5 \nonumber\\
&+x_2 x_4x_5-2 x_3 x_4 x_5-2 x_4^2 x_5-x_2 x_5^2+2 x_4 x_5^2
\end{align}}\normalsize

At degree four \(A^{(4)} = \{(x_1^4)^*, (x_1^3x_2)^*, (x_1^2x_2^2)^*, (x_1^2x_2x_3)^*, (x_1x_2x_3x_4)^*\}\). By direct calculation, the first four elements are equal to each other; therefore, we can take \(B^{(4)} = \{(x_1^4)^*,(x_1x_2x_3x_4)^*\}\). The Hilbert series indicates there is a new primary invariant at degree four, while the other degree four element comes from \(\th_2^2\). Therefore there must be some linear combination such that \(a_1 (x_1^4)^* +a_2 (x_1x_2x_3x_4)^* = \th_2^2\).

Let us use the numerical evaluation described above to find the relationship between \(\th_2^2\) and the elements of \(B^{(4)}\) (although the relationship is not hard to find by hand in this case). Since \(\abs{B^{(4)}}=2\), we need two random points \(\mathbf{x}^{(i)} \equiv (x_1^{(i)},\dots,x_5^{(i)}) \in \mathbb{R}^5\), \(i=1,2\), to evaluate the polynomials on:
\begin{equation}
\texttt{RowReduce}\left[\left(\begin{array}{ccc}
\left.(x_1^4)^*\right|_{\mathbf{x}^{(1)}} & \left.(x_1x_2x_3x_4)^*\right|_{\mathbf{x}^{(1)}} & \left.\th_2^2\right|_{\mathbf{x}^{(1)}} \\
\left.(x_1^4)^*\right|_{\mathbf{x}^{(2)}} & \left.(x_1x_2x_3x_4)^*\right|_{\mathbf{x}^{(2)}} & \left.\th_2^2\right|_{\mathbf{x}^{(2)}}\end{array}\right)\right] = \bigg(\begin{array}{ccc}1&0&3\\ 0&1&2\end{array}\bigg).\nonumber
\end{equation}
The relationship lies in the kernel of this matrix,
\begin{equation}
\texttt{NullSpace}\left[\bigg(\begin{array}{ccc}1&0&3\\ 0&1&2\end{array}\bigg)\right] = \big(\begin{array}{ccc} 3 & 2 & -1 \end{array}\big) \nonumber \ \Rightarrow \ 0 = 3(x_1^4)^* + 2(x_1x_2x_3x_4)^* - \th_2^2. \label{eq:deg4_rel}
\end{equation}
(We have cast the above equations as \texttt{Mathematica} commands.) 
We then choose some other linear combination to serve as the primary invariant \(\th_4\), \textit{e.g.}
\begin{equation}
\th_4 \equiv 5(x_1^4)^* - 2(x_1x_2x_3x_4)^*.
\end{equation}
(This is the linear combination that comes from converting \(s_{12}^4 + \cdots + s_{45}^4\) into a polynomial in the \(x_a\).)

This algorithm proceeds in a straightforward manner. We make a few final comments
\begin{description}
\item[Primary invariants: ] Per the discussion around eq.~\eqref{eq:Thiery_sij_prim}, the primary invariants can be chosen to be the polynomials \(s_{12}^i + \cdots + s_{45}^i\), \(i = 2,\dots,6\). This will always be true when there are no Gram conditions to consider. Of course, the above algorithm can always be used. For example, we found a different degree six invariant that has a more compact form (in terms of the \(x_a\)).
\begin{equation}
\left(x_1 x_3 x_4-x_2 x_3 x_4-x_1 x_2 x_5+x_2 x_3 x_5+x_2 x_4 x_5-x_2
   x_5^2\right)^2. \nonumber
\end{equation}
(We found this by noting the degree three polynomial in parentheses is invariant under the alternating group \(A_5\) but odd under \(S_5\), so that its square is invariant under \(S_5\).) Expanding \(s_{12}^6 + \cdots + s_{45}^6\) into \(x_1,\dots,x_5\) results in an expression over half a page long, so the above is quite a simplification.
\item[Algebraic independence: ] One needs to ensure that the candidate primary invariants are algebraically independent, \textit{e.g.} by checking the rank of the Jacobian matrix. Another option is to run an elimination routine using, for example, Gr\"obner bases. 
\item[Secondary invariants: ] There are six secondaries \(\{\et_0 = 1,\et_6,\et_7,\et_8,\et_9,\et_{15}\}\). It turns out we can take \(\et_{15} = \et_6\et_9\) or \(\et_7\et_8\), so a set of irreducible secondary invariants can be given by \(\{\et_6,\et_7,\et_8,\et_9\}\). The expressions for the secondary invariants are, unfortunately, too long to include in this paper.
\item[Computer packages: ] The software \texttt{Singular} contains algorithms devoted to describing invariant rings~\cite{singular,King2013}. With a few lines of input, it returns a set of primary and secondary invariants:
\begin{description}
\item \texttt{>LIB ``finvar.lib'';}
\item \texttt{>ring R=0,(x1,x2,x3,x4,x5),dp;}
\item \texttt{>matrix A[5][5] = ...} (input matrices which generate the group, \textit{e.g.} \(A,B,C,D\) corresponding to the permutations \((12),(23),(34),(45)\))
\item \texttt{>matrix P,S,IS = invariant\_ring (A,B,C,D);}
\end{description}
The output matrices P, S, and IS give a set of primary, secondary, and irreducible secondary invariants, respectively.
\end{description}

\subsection{Summary}
Operators \(\scO^{(n,k)}\) composed of \(n\) scalar fields and \(k\) derivatives have a one-to-one correspondence with homogeneous degree \(k\) polynomials \(F^{(n,k)}\) in the momenta \(p_1^{\m},\dots,p_n^{\m}\). This leads to a formulation of the operator basis in terms of a polynomial ring quotiented by an ideal to account for EOM (on-shell, \(p_i^2=0\)) and IBP (momentum conservation, \(\sum_i p_i^{\m} = 0\)) redundancies. Phrased in terms of Lorentz invariants \(s_{ij} = p_i \cdot p_j\), this quotient ring is
\begin{equation}
M_{n,\scK}^{O(d) \times \Sigma} = \left[\bigslant{\mathbb{C}[\{s_{ij}\}]}{\avg{X_1,\ldots,X_n,\{\D\}}}\right]^{\Sigma},
\end{equation}
where the \(s_{ii} = p_i^2 = 0\) are removed by EOM, the \(X_i = \sum_{j \ne i}s_{ij}\) are the Lorentz invariant consequences of momentum conservation, \(\{\D\}\) are the set of Gram conditions (vanishing of \((d+1)\times (d+1)\) minors in \(s_{ij}\)), and \(\Sigma \subseteq S_n\) is a possible permutation group in the case the particles are indistinguishable.

In the \(n=4\) case, we obtained explicit solutions to these rings with relative ease in section~\ref{sec:tour}, showing that the rings are generated by the familiar Mandelstam variables \(\{s,t,u\}\) (or symmetric combinations thereof). The insight gained here provides the foundation for understanding the rings at higher \(n\), where the structure of the rings \(\Mgen\) become significantly more involved.

We showed that there is a precise notion of generators for the rings \(\Mgen\) in section~\ref{sec:general_algebraic}: we can always find a direct sum decomposition of the rings into primary invariants and secondary invariants,
\begin{equation}
\Mgen = \bigoplus_{i=1}^{s}\et_i \mathbb{C}[\th_1,\ldots,\th_m].
\label{eq:summary_hiro}
\end{equation}
In other words, any \(f \in \Mgen\) can be expressed \textit{uniquely} as a linear combination of the secondary invariants \(\et_i\) whose coefficients are polynomials in the primary invariants \(\th_j\).

The kinematic nature of the rings \(\Mgen\) imply that they are fundamental to understanding general operator spectra, including when operators are composed of fields with spin. Here we expect the \textit{same} primary and secondary invariants of the \(\Mgen\) to show up, supplemented by some additional generators physically connected to polarization tensors. In the case of spin, however, we generally will not have a clean direct sum decomposition like that in eq.~\eqref{eq:summary_hiro}.

The description of the operator basis in terms of quotient rings immediately lends itself to algorithmic approaches for constructing the basis. We outlined a simple algorithm in section~\ref{subsec:n=5_basis}, and then used this to find a set of primary and secondary invariants for the ring \(M_{5,\scK}^{O(d\ge 4)\times S_5}\).

\section{Applications and examples of Hilbert series}
\label{sec:case}
In this section we compute the Hilbert series in several cases of interest using the matrix integral formula derived in sec.~\ref{sec:compute}. The case studies we examine are as follows.

In sec.~\ref{subsec:H_rsf_d=4} we detail the case of a single real scalar field in \(d=4\), and tabulate the Hilbert series for the rings \(M_{n,\scK}^{(S)O(4)\times S_n}\) of sec.~\ref{sec:rsf} up to \(n=8\).

In sec.~\ref{subsec:H_d=2} we compute a closed form expression for the Hilbert series for \(n\) \textit{distinguishable} scalars in \(d=2,3\) dimensions. Both the above examples highlight how the Hilbert series reflects properties of the operator bases---such as dimensionality, primary and secondary invariants, the role of gram conditions, \textit{etc.}---as well as the marked increase in difficulty when particles are identical.

In sec.~\ref{subsec:hilbert_spinning} we examine the Hilbert series when the particles carry spin. The number of tensor structures in the amplitude decomposition eq.~\eqref{eq:amp_tens} is encoded in the Hilbert series. This reproduces the recent determination of these numbers~\cite{Kravchuk:2016qvl}, which were obtained using a different technique. (For completeness, we re-derive the results of~\cite{Kravchuk:2016qvl} in app.~\ref{app:helicity_count}.)

In sec.~\ref{subsec:quant_redund} we examine the number of operators as a function of mass dimension as we turn on various constraints. The main (and, perhaps at first, surprising) result is that EOM and IBP redundancies give a polynomial suppression in the total number of operators, while Gram conditions give an \textit{exponential} suppression.

\subsection{Hilbert series for a real scalar field in $d=4$}
\label{subsec:H_rsf_d=4}
We work in \(d=4\) dimensions with the EFT of a single real scalar field and compute the Hilbert series for fixed powers of \(\ph\) and arbitrary numbers of derivatives. This corresponds to computing the Hilbert series for the rings \(M_{n,\scK}^{(S)O(4)\times S_n}\) discussed in sec.~\ref{sec:rsf}. Although the computation is straightforward, the permutation symmetry makes it somewhat difficult; operationally, the plethystic exponential accounts for this symmetry in the integrand, but leads to more complicated contour integrals (compare to the next subsection). 

We first discuss the \(SO(4)\) case, and subsequently include parity. As in sec.~\ref{sec:rsf}, we use the momentum weighting scheme eq.~\eqref{eq:reweight}. In this scheme, eq.~\eqref{eq:H_0_summary} takes the form
\begin{equation}
H(\ph,t) = \int d\m_{SO(4)} \frac{1}{P(t;x)} \text{PE}\left[\ph (1-t^2) P(t;x)\right] + \D H,
\label{eq:H_0_scalar_4d_example}
\end{equation}
with
\begin{equation}
P(t;x) = \frac{1}{(1-tx_1)(1-t/x_1)(1-tx_2)(1-t/x_2)}, \nonumber
\end{equation}
the measure given by (see eq.~\eqref{eq:haar_SO_even})
\begin{equation}
d\m_{SO(4)} = \frac{1}{4}\frac{dx_1}{2\pi i x_1}\frac{dx_2}{2\pi i x_2} \big(1-x_1x_2\big)\big(1-\frac{1}{x_1x_2}\big)\big(1-\frac{x_1}{x_2}\big)\big(1-\frac{x_2}{x_1}\big), \nonumber
\end{equation}
and, as per eq.~\eqref{eq:DH_single_scalar},
\begin{equation}
\D H = -t^4 + \ph t^2, \nonumber
\end{equation}
\textit{i.e.} \(\D H\) only contributes to the \(\ph^0\) and \(\ph^1\) terms in \(H(\ph,t)\). The basic idea to enumerate all terms for a fixed power of \(\phi\) is to expand the plethystic exponential to order \(\ph^n\) and then evaluate the \(SO(4)\) contour integrals using the residue theorem.

As a concrete example, take \(n = 4\).\footnote{The \(n=1,\ 2,\) and 3 cases are, of course, easier to compute. The results end up being trivial: \(H(\ph,t)|_{\ph} = \ph\), \(H(\ph,t)|_{\ph^2} = \ph^2\), and \(H(\ph,t)|_{\ph^3} = \ph^3\), which is related to the fact that the 1-, 2-, and 3-point amplitudes are trivial in massless theories.} Expanding \(\text{PE}\left[\ph (1-t^2) P(t;x)\right]\) to \(\scO(\ph^4)\) gives the fourth symmetric product of the argument of the plethystic exponential,
\begin{equation}
\text{sym}^4[f(z)] = \frac{1}{4!}\big[f(z)^4 + 6 f(z)^2f(z^2) + 3f(z^2)^2+ 8 f(z)f(z^3) + 6f(z^4) \big], 
\label{eq:sym4}
\end{equation}
where in place of \(f(z)\) we use \(\ph (1-t^2)P(t;x)\); \textit{e.g.} the \(f(z^4)\) term corresponds to
\begin{equation}
\ph^4(1-t^8)P(t^4;x^4) = \ph^4\frac{(1-t^8)x_1^4x_2^4}{(1-t^4x_1^4) (x_1^4- t^4) (1-t^4x_2^4) (x_2^4-t^4)}. \nonumber
\end{equation}
These factors are then inserted into eq.~\eqref{eq:H_0_scalar_4d_example} and the contour integrals, taken around \(\abs{x_i}=1\), are evaluated using the residue theorem recalling that \(\abs{t} < 1\).

Carrying out the above procedure, we obtain
\begin{equation}
\left.H(\ph,t)\right|_{\ph^4} = \ph^4 \frac{1}{(1-t^4)(1-t^6)} = \ph^4(1+ t^4 + t^6 + t^{10} + 2 t^{12} + \dots) \,,
\end{equation}
in agreement with eq.~\eqref{eq:heomibpIN} (recall that $SO(4)$ and $O(4)$ give the same result for $n=4$). In the expansion above we can begin to read off the operators in the basis. For example, the \(\ph^4\), \(\ph^4t^4\), and \(\ph^4t^6\) terms respectively correspond to the representative operators \(\ph^4\), \([(\pd_{\m}\ph)^2]^2\), and \((\pd_{\{\m}\pd_{\n\}}\ph)^2(\pd_{\s}\ph)^2\). They have momentum space Feynman rules \(1\), \(st+su+tu\), and \(stu\) (up to \(s+t+u=0\))---\textit{i.e.} they are the first three elements of the ring in eq.~\eqref{eq:reomibpIN}. 

For the \(O(4)\) case we also need to include parity (see sec.~\ref{subsec:parity} and app.~\ref{app:parity}). Under parity, we assume $\phi$ is scalar in the strict sense, namely not a pseudo-scalar. The Hilbert series is
\be
H^{O(4)}(\phi,t)= \frac{1}{2}\left(H_{+}(\phi,t)+H_{-}(\phi,t)\right) \,, \label{eq:H_O4_avg}
\ee
where \(H_+\) is the parity even contribution given in eq.~\eqref{eq:H_0_scalar_4d_example}. \(H_-\) takes the form (eq.~\eqref{eqn:H0mphideven} in the momentum weighting scheme)
\begin{equation}
H_{-}(\ph,t) = \int d\m_{Sp(2)}^{} \frac{1-t^2}{P^{(2)}(t;x)} \text{PE}\Bigg[\ph P^{(2)}(t;x) \, + \, \ph^2 \frac{t^{2}}{1-t^2} P^{(2)}(t^2;x^2) \Bigg] + \D H_-, \label{eq:H0_scalar_O4_example}
\end{equation}
with
\begin{equation}
P^{(2)}(t,x) = \frac{1}{(1-t x)(1-t/x)} \ , \  d\m_{Sp(2)}^{} = \frac{1}{2} \frac{dx}{2\pi i x} \big(1-x^2\big)\big(1-\frac{1}{x^2}\big) \ , \ \D H_- = t^4 + \ph t^2.\nonumber
\end{equation}
It is straightforward to check that \(H_-|_{\ph^n} = H_+|_{\ph^n}\) for \(n\le 4\), so that in the average eq.~\eqref{eq:H_O4_avg} we obtain the expected result \(H^{O(4)}|_{\ph^n} = H^{SO(4)}|_{\ph^n}\) for \(n \le 4\).

\begin{table}
\small
\centering
\scalebox{0.8}{\begin{tabular}{|r c l |}
\hline
$n=5$ & &\\
\hline && \\
$D$& $=$ & $\left(1-t^4\right)\left(1-t^6\right) \left(1-t^8\right) \left(1-t^{10}\right) \left(1-t^{12}\right)$\\ && \\
$N_{SO(4)}$&$=$&$1+t^{10}+t^{12}+2 t^{14}+2 t^{16}+ t^{18}+ t^{22}+t^{24} + t^{28}+t^{30}$\\ && \\
 $N_{O(4)}$&$=$&$1+t^{12}+t^{14}+ t^{16} + t^{18}+t^{30}$ \\ &&\\
   \hline
   $n=6$ & &\\
\hline && \\
$D$& $=$ & $\left(1-t^4\right) \left(1-t^6\right)^2 \left(1-t^8\right)^3 \left(1-t^{10}\right) \left(1-t^{12}\right)$\\ && \\
$N_{SO(4)}$&$=$&$1+3 t^{10}+6 t^{12}+11 t^{14}+17 t^{16}+22 t^{18}+31 t^{20}+36 t^{22}+48 t^{24}+53 t^{26}+58 t^{28}+58
   t^{30}+48 t^{32}+38 t^{34}+23 t^{36}$\\&&$+14 t^{38}+6 t^{40}+4 t^{42}+2 t^{44}+t^{46}$\\ && \\
 $N_{O(4)}$&$=$&$1+2 t^{10}+5 t^{12}+7 t^{14}+9 t^{16}+11 t^{18}+13 t^{20}+14 t^{22}+21 t^{24}+24 t^{26}+28 t^{28}+32
   t^{30}+26 t^{32}+22 t^{34}+13 t^{36}$\\&&$+7 t^{38}+3 t^{40}+t^{42}+t^{44}$ \\ &&\\
   \hline
$n=7$ && \\
\hline && \\
$D$&$=$&$\left(1-t^4\right) \left(1-t^6\right)^2 \left(1-t^8\right)^3 \left(1-t^{10}\right) \left(1-t^{12}\right)
   \left(1-t^{14}\right) \left(1-t^{20}\right) \left(1-t^{24}\right) $ \\ &&\\
   $N_{SO(4)}$&$=$&$1+5 t^{10}+14 t^{12}+29 t^{14}+68 t^{16}+131 t^{18}+254 t^{20}+464 t^{22}+820 t^{24}+1332 t^{26}+2115
   t^{28}+3136 t^{30}+4485 t^{32}$\\&&$+6134 t^{34}+8108 t^{36}+10309 t^{38}+12778 t^{40}+15297 t^{42}+17841
   t^{44}+20258 t^{46}+22387 t^{48}+24111 t^{50}+25356 t^{52}$\\&&$+25974 t^{54}+25975 t^{56}+25389 t^{58}+24151
   t^{60}+22454 t^{62}+20336 t^{64}+17933 t^{66}+15385 t^{68}+12855 t^{70}+10365 t^{72}$\\&&$+8140 t^{74}+6150
   t^{76}+4479 t^{78}+3130 t^{80}+2096 t^{82}+1317 t^{84}+798 t^{86}+442 t^{88}+229 t^{90}+109 t^{92}+44
   t^{94}+13 t^{96}+3 t^{98}$ \\ &&\\
   $N_{O(4)}$&$=$&$1+4 t^{10}+11 t^{12}+19 t^{14}+41 t^{16}+73 t^{18}+134 t^{20}+237 t^{22}+413 t^{24}+664 t^{26}+1052
   t^{28}+1563 t^{30}+2231 t^{32}+3055 t^{34}$\\&&$+4039 t^{36}+5136 t^{38}+6372 t^{40}+7637 t^{42}+8909
   t^{44}+10125 t^{46}+11193 t^{48}+12058 t^{50}+12687 t^{52}+13000 t^{54}+13005 t^{56}$\\&&$+12711 t^{58}+12095
   t^{60}+11240 t^{62}+10178 t^{64}+8975 t^{66}+7696 t^{68}+6429 t^{70}+5184 t^{72}+4064 t^{74}+3068
   t^{76}+2232 t^{78}$\\&&$+1555 t^{80}+1043 t^{82}+654 t^{84}+396 t^{86}+220 t^{88}+114 t^{90}+55 t^{92}+23
   t^{94}+7 t^{96}+2 t^{98}$
    \\ &&\\
   \hline
$n=8$ && \\
\hline && \\
$D$&$=$&$\left(1-t^4\right) \left(1-t^6\right)^2 \left(1-t^8\right)^4 \left(1-t^{10}\right) \left(1-t^{12}\right)
   \left(1-t^{14}\right) \left(1-t^{16}\right) \left(1-t^{20}\right) \left(1-t^{24}\right)
   \left(1-t^{30}\right)$ \\&&\\
   $N_{SO(4)}$&$=$&$1+7 t^{10}+20 t^{12}+49 t^{14}+134 t^{16}+319 t^{18}+775 t^{20}+1741 t^{22}+3743 t^{24}+7525 t^{26}+14516
   t^{28}+26494 t^{30}+46454 t^{32}$\\&&$+78002 t^{34}+126172 t^{36}+196794 t^{38}+297183 t^{40}+434786
   t^{42}+618293 t^{44}+855582 t^{46}+1154256 t^{48}+1520246 t^{50}$\\&&$+1957689 t^{52}+2467159 t^{54}+3046857
   t^{56}+3690302 t^{58}+4387179 t^{60}+5123633 t^{62}+5881714 t^{64}+6640351 t^{66}$\\&&$+7377142 t^{68}+8067728
   t^{70}+8688015 t^{72}+9216173 t^{74}+9631884 t^{76}+9919148 t^{78}+10067172 t^{80}+10069631
   t^{82}$\\&&$+9926275 t^{84}+9643336 t^{86}+9231191 t^{88}+8705813 t^{90}+8087214 t^{92}+7397387 t^{94}+6660141
   t^{96}+5900232 t^{98}$\\&&$+5139908 t^{100}+4400687 t^{102}+3700557 t^{104}+3053776 t^{106}+2470701
   t^{108}+1958189 t^{110}+1518091 t^{112}+1149962 t^{114}$\\&&$+849837 t^{116}+611610 t^{118}+427825
   t^{120}+290332 t^{122}+190531 t^{124}+120626 t^{126}+73466 t^{128}+42807 t^{130}+23806 t^{132}$\\&&$+12559
   t^{134}+6229 t^{136}+2886 t^{138}+1246 t^{140}+476 t^{142}+171 t^{144}+50 t^{146}+12 t^{148}+2 t^{150}$ \\ && \\
   $N_{O(4)}$&$=$&$1+6 t^{10}+17 t^{12}+35 t^{14}+84 t^{16}+184 t^{18}+419 t^{20}+911 t^{22}+1924 t^{24}+3816 t^{26}+7309
   t^{28}+13298 t^{30}+23251 t^{32}$\\&&$+39007 t^{34}+63068 t^{36}+98330 t^{38}+148496 t^{40}+217271
   t^{42}+308982 t^{44}+427618 t^{46}+576946 t^{48}+759929 t^{50}$\\&&$+978683 t^{52}+1233454 t^{54}+1523340
   t^{56}+1845114 t^{58}+2193616 t^{60}+2561887 t^{62}+2940999 t^{64}+3320380 t^{66}$\\&&$+3688794 t^{68}+4034121
   t^{70}+4344269 t^{72}+4608338 t^{74}+4816179 t^{76}+4959773 t^{78}+5033731 t^{80}+5034902 t^{82}$\\&&$+4963167
   t^{84}+4821641 t^{86}+4615506 t^{88}+4352773 t^{90}+4043446 t^{92}+3698498 t^{94}+3329885 t^{96}+2949931
   t^{98}$\\&&$+2569776 t^{100}+2200207 t^{102}+1850175 t^{104}+1526820 t^{106}+1235322 t^{108}+979101
   t^{110}+759067 t^{112}+575035 t^{114}$\\&&$+424988 t^{116}+305868 t^{118}+213980 t^{120}+145221 t^{122}+95301
   t^{124}+60345 t^{126}+36750 t^{128}+21406 t^{130}+11906 t^{132}$\\&&$+6274 t^{134}+3110 t^{136}+1441
   t^{138}+620 t^{140}+237 t^{142}+86 t^{144}+26 t^{146}+7 t^{148}+2 t^{150}$ \\ & &\\
   \hline
\end{tabular}}
\caption{Hilbert series of the form $N/D$ (see eq.~\eqref{eq:numdenom}) at fixed order $\phi^n$, and to all-orders in the power of derivatives. Results are for both $SO(4)$ and $O(4)$ spacetime groups. }
\label{tab:fixed_u}
\end{table}

For \(n>4\) the Hilbert series becomes increasingly more intricate. We write 
\be
H(\phi,t)\bigg|_{\mathcal{O}(\phi^n)} = \phi^n \frac{N^{(n)}(t)}{D^{(n)}(t)} \,,
\label{eq:numdenom}
\ee
for some numerator $N^{(n)}(t)$ and denominator $D^{(n)}(t)$. These functions are, of course, not unique since one can always multiply and divide eq.~\eqref{eq:numdenom} by any function of $t$. A useful choice (which can always be done) is to bring the denominator to the form \(D = \prod_{i=1}^{m} (1-t^{d_i})\) so that it reflects some set of algebraically independent generators (primary invariants) of degree \(d_i\) with \(m\) the dimension of the ring.\footnote{The dimension of the ring is unambiguously determined from the Hilbert series as the order of the pole as \(t\to 1\), \textit{i.e.} the number of \((1-t)\) factors in the denominator when the Hilbert series is maximally factorized; in the math literature this is properly called the \textit{Krull dimension}.} Since the rings under consideration are Cohen-Macaulay, the numerator can be written as a strictly positive sum whose terms reflect the degrees of the secondary invariants, see eq.~\eqref{eq:hil_hiro}.

Table~\ref{tab:fixed_u} gives the Hilbert series for $n=5,6,7,8$ (for which a straightforward Mathematica implementation of the evaluation of eqs.~\eqref{eq:H_0_scalar_4d_example} and \eqref{eq:H0_scalar_O4_example} is tractable). We leave a detailed study of the invariants to a future work, but point out a few pieces of information reflected in the Hilbert series.

The denominator is a product of \(\text{dim}(M_{n,\scK}^{(S)O(4)\times S_n}) = 3n - 10\) (\(n \ge 4\)) terms. In the absence of Gram conditions, the dimension of the ring would be \(n(n-3)/2\) with primary invariants of degree \(4,6,8,\dots, n(n-3)+2\)~\cite{Thiery:2000:AIG:377604.377612}, see eq.~\eqref{eq:Thiery_sij_prim}; the Hilbert series sheds light on how Gram conditions cut down these invariants. Per the numerators, only \(N^{(5)}_{O(4)}\) is palindromic, so we can conclude \(M_{5,\scK}^{O(4)\times S_5}\) is Gorenstein while the rest are not (see the discussion around eq.~\eqref{eq:hil_goren_1overt}; also compare to the next subsection).\footnote{A similar calculation in \(d = 2\) and \(3\) dimensions reveals that at \(n=5\) both the \(SO(\cdot)\) and \(O(\cdot)\) cases are Gorenstein; at \(n=6\), \(O(2)\) and \(SO(3)\) are Gorenstein, while \(SO(2)\) and \(O(3)\) are not.}

As an example of getting information on secondary invariants, we see that the Hilbert series always contains a single parity non-invariant operator with ten derivatives (the \(t^{10}\) term present in the \(SO(4)\) but not \(O(4)\) Hilbert series numerators)---it corresponds to \(\ph^{n-4}\e^{\m\n\r\s}(\pd_{\m}\ph)( \pd_{\n}\pd^{\a}\ph) (\pd_{\r}\pd^{\b}\pd^{\g}\ph) (\pd_{\s}\pd_{\a}\pd_{\b}\pd_{\g}\ph)\). As seen in the table, results for $n\ge5$ become very lengthy; the marked increase in complexity over the $\mathcal{O}(\phi^4)$ Hilbert series reflects the increased complexity in moving beyond the simple kinematics of four-point scattering amplitudes.

\subsection{Closed form Hilbert series for $n$ distinguishable scalars in $d=2,3$}
\label{subsec:H_d=2}

In order to examine the interplay of EOM, IBP, and Gram conditions, it is illuminating to omit the permutation symmetry and consider the case of distinguishable particles, \textit{i.e.} study the rings \(M_{n,\mathcal{K}}^{(S)O(d)}\). In this case, it is straightforward to evaluate the matrix integral and obtain a closed form expression for the Hilbert series for arbitrary \(n\). We will look at \(d=2\) and 3 dimensions---the techniques extend in an obvious way, although massaging the equations gets more difficult.

The integrand of the matrix integral eq.~\eqref{eq:H_0_summary} (in the momentum weighting scheme eq.~\eqref{eq:reweight}) contains the plethystic exponential \(\text{PE}[\sum_i \ph_i (1-t^2)P(t;x)]\); to get the Hilbert series for distinguishable scalars we take the term linear in each \(\ph_i\), namely\footnote{Strictly speaking, we need to include the \(\D H \supset t^2 \sum_i \ph_i + t^{d-2\D_0}\sum_{i<j}\ph_i\ph_j\) terms, eqs.~\eqref{eqn:DH1} and~\eqref{eq:DH_term_ii_2r_phi_and_F}, which adds a \(+t^2\) to the \(n=1\) and \(n=2\) cases. We will not be careful to consistently denote these exceptional cases in this subsection.}
\begin{equation}
H\big(M_{n,\scK}^{SO(d)}; t\big) = \int d\m_{SO(d)} \frac{1}{P(t;x)} \Big[(1-t^2)P(t;x)\Big]^n.
\label{eq:H_n_disting}
\end{equation}

In two dimensions the above takes the form
\begin{equation}
H^{SO(2)}_{n,\scK} = (1-t^2)^n\oint \frac{dx}{2\pi i x} \left[ \frac{1}{(1-tx)(1-t/x)} \right]^{n-1},
\label{eq:H_n_disting_SO2}
\end{equation}
namely it has a pole of order \(n-1\) at \(x = t\). Define
\begin{equation}
f_{k,l}(x,t) \equiv \frac{x^k}{(1-t x)^l},
\label{eq:def_f_kl}
\end{equation}
and let \(f^{(m)}_{k,l}\) denote the \(m\)th derivative with respect to \(x\). Then, by the residue theorem, the above integral is equal to
\begin{equation}
\oint \frac{dx}{2\pi i} f_{n-2,n-1}\frac{1}{(x-t)^{n-1}} = \frac{1}{(n-2)!}\left. f^{(n-2)}_{n-2,n-1}\right|_{x=t}.
\end{equation}

To evaluate \(f^{(m)}_{k,l}\), note that the first derivative is given by
\begin{equation}
\frac{\pd}{\pd x}f_{k,l} = k \, f_{k-1,l+1} + (l-k)\, t\, f_{k,l+1}.
\label{eq:recurse_f}
\end{equation}
Making use of this recursively, one readily finds
\begin{equation}
\frac{1}{m!}f^{(m)}_{k,l} = \sum_{i=0}^{m} \binom{k}{m-i} \binom{l-k+m-1}{i}\, t^i \, f_{k-m+i, l+m},
\label{eq:f_kl_deriv_eval}
\end{equation}
with \(\binom{a}{b}\) denoting a binomial coefficient.

With this, the Hilbert series in eq.~\eqref{eq:H_n_disting_SO2} is
\begin{equation}
H_{n,\scK}^{SO(2)} = \frac{1}{(1-t^2)^{n-3}}\sum_{i=0}^{n-2}{\binom{n-2}{i}}^2 t^{2i}. 
\label{eq:H_n_disting_SO2_eval}
\end{equation}
The denominator reflects the dimensionality of the ring and indicates the anticipated result that some choice of \(n(d-1) - d(d+1)/2 = n-3\) (see eq.~\eqref{eq:dim_modules}) linear combinations of the \(s_{ij}\) serve as primary invariants. The reflection property of the binomial coefficient tells us that the numerator is palindromic; in fact, we see Pascal's triangle:
\begin{equation}
\begin{array}{clc}
n=3 & : & \ \ \ 1 + t^2 \\
n=4 & : & \ \ \ 1 + 4t^2 + t^4 \\
n=5 & : & \ \ \ 1 + 9t^2 + 9 t^4 + t^6\\
n=6 & : & \ \ \ 1 +16 t^2 + 36t^4+ 16 t^6 + t^8\\
n=7 & : & \ \ \ 1 +25 t^2 + 100t^4+ 100 t^6 + 25t^8 + t^{10}\\
\end{array}
\label{eq:pascal_SO2}
\end{equation}
The total number of secondary invariants is given by \(\sum_{i=0}^{n-2}{\binom{n-2}{i}}^2 = \binom{2(n-2)}{n-2}\).

We can include parity and work out the \(O(2)\) case. While this follows from the machinery worked out in App.~\ref{app:parity}, it is easiest to recognize \(P = 1/\text{det}_{\Box}(1- t g)\) and use the fact that on the parity odd component \(g\) is conjugate to \( \big(\begin{smallmatrix} 1 & \\ & -1 \end{smallmatrix}\big)\) so \(P(g_-) = 1/(1-t^2)\);  hence,
\begin{equation}
H_{n,\scK}^{O(2)} = \frac{1}{2}\big[ H_{n,\scK}^{SO(2)} + (1-t^2) \big]. \nonumber
\end{equation}
Using the binomial expansion \((1-t^2)^{n-2} = \sum_{i=0}\binom{n-2}{i}(-t^2)^i\) we have
\begin{equation}
H_{n,\scK}^{O(2)} = \frac{1}{(1-t^2)^{n-3}}\sum_{i=0}^{n-2}\frac{1}{2}{\binom{n-2}{i}}\left[\binom{n-2}{i} + (-1)^i \right] t^{2i}.
\label{eq:H_n_disting_O2_eval}
\end{equation}
The first few numerators are (they are palindromic for even \(n\))
\begin{equation}
\begin{array}{clc}
n=3 & : & \ \ \ 1  \\
n=4 & : & \ \ \ 1 + t^2 + t^4 \\
n=5 & : & \ \ \ 1 + 3t^2 + 6 t^4\\
n=6 & : & \ \ \ 1 +6 t^2 + 21t^4+ 6 t^6 + t^8\\
n=7 & : & \ \ \ 1 +10 t^2 +55t^4+ 45 t^6 + 15t^8 \\
\end{array}
\label{eq:numer_O2}
\end{equation}
The total number of secondary invariants is \( \frac{1}{2} \binom{2(n-2)}{n-2}\), \textit{i.e.} half of the secondary invariants in the \(SO(2)\) ring are parity invariant, while the other half are parity odd (implying they are proportional to the \(\e\)-tensor).

In two dimensions, the Gram constraints are the vanishing of the \(3\times 3\) minors of \(s_{ij}\). They reduce the dimensionality of the ring from \(n(n-3)/2\) to \(n-3\), \textit{i.e.} \((n-2)(n-3)/2\) primary invariants are removed due to the Gram conditions. However, since the Gram constraints are \(\scO(s_{ij}^3)\), these primary invariants are not completely removed from the ring---they should still be present in some ways as secondary invariants. Indeed, the \(t^2\) term in the sum in eq.~\eqref{eq:H_n_disting_O2_eval} is equal to \(\binom{n-2}{2}\); moreover, the \(t^4\) term is equal to \(\binom{\binom{n-2}{2}+1}{2}\), as we can use each of these terms twice without hitting a Gram constraint. The Gram constraints kick in at \(\scO(t^6)\) in the numerator, making it more difficult to unravel what is going on.

The same basic technique---making use of eq.~\eqref{eq:f_kl_deriv_eval}---gives a straightforward way to evaluate eq.~\eqref{eq:H_n_disting} in higher dimensions. Let us work out the \(d=3\) case. Using the \(SO(3)\) measure in eq.~\eqref{eq:haar_classical_pos_roots}, the Hilbert series is given by
\begin{align}
H^{SO(3)}_{n,\scK} &= (1-t^2)^n\oint \frac{dx}{2\pi i x}(1-x) \left[ \frac{1}{(1-tx)(1-t)(1-t/x)} \right]^{n-1} \nonumber \\
&= \frac{(1-t^2)^n}{(1-t)^{n-1}} \oint \frac{dx}{2\pi i} \left[f_{n-2,n-1}\frac{1}{(x-t)^{n-1}} - f_{n-1,n-1}\frac{1}{(x-t)^{n-1}}\right].
\label{eq:H_n_disting_SO3}
\end{align}
Making use of the residue theorem and eq.~\eqref{eq:f_kl_deriv_eval}, one readily finds
\begin{equation}
H^{SO(3)}_{n,\scK} = \frac{1}{(1-t^2)^{2n-6}} \frac{(1+t)^{n-3}}{(1-t)^2} \sum_{i=0}^{n-2}\left[{\binom{n-2}{i}}^2 - t\,\binom{n-1}{i+1}\binom{n-3}{i}\right]t^{2i}.
\label{eq:H_n_disting_SO3_eval1}
\end{equation}
We stripped off the factor \((1-t^2)^{-(2n-6)}\), which corresponds to the dimension of the ring, eq.~\eqref{eq:dim_modules}. The polynomial in the sum vanishes at \(t = 1\), as does its first derivative; this indicates that we can factor a \((1-t)^2\) term out of this sum. After some manipulation we obtain,
\begin{equation}
H^{SO(3)}_{n,\scK} = \frac{N^{SO(3)}_{n,\scK}(t)}{(1-t^2)^{2n-6}} ,
\label{eq:H_n_disting_SO3_eval2}
\end{equation}
where the polynomial in the numerator is
\begin{equation}
N^{SO(3)}_{n,\scK}(t) \equiv (1+t)^{n-3} \sum_{i=0}^{n-3}\left[{\binom{n-3}{i}}^2 - t\,\binom{n-3}{i+1}\binom{n-3}{i}\right]t^{2i}\, .
\label{eq:H_n_disting_SO3_numpoly}
\end{equation}

The first few terms for the numerator are
\begin{equation}
\begin{array}{clc}
n=3 & : & \ \ \ 1  \\
n=4 & : & \ \ \ 1 + t^3 \\
n=5 & : & \ \ \ 1 + t^2 +4t^3+  t^4 + t^6\\
n=6 & : & \ \ \ 1+3 t^2+10 t^3+6 t^4+6 t^5+10 t^6+3 t^7+t^9\\
n=7 & : & \ \ \ 1+6 t^2+20 t^3+21 t^4+36 t^5+56 t^6+36 t^7+21 t^8+20 t^9+6 t^{10}+t^{12}
\end{array}
\label{eq:numer_SO3}
\end{equation}
We see that the numerators are palindromic, indicating these rings are Gorenstein. The total number of secondary invariants is proportional to the Catalan number \(C_{n-3}\),\footnote{\(C_m = \frac{1}{m+1}\binom{2m}{m} = \binom{2m}{m}-\binom{2m}{m+1}\). The appearance of the Catalan numbers appears common in any dimensionality: in \(d=2\) we found above \(\binom{2(n-2)}{n-2}= (n-1)C_{n-2}\), while in \(d=4\) one finds \(C_{n-3}C_{n-4}\).}
\begin{equation}
N^{SO(3)}_{n,\scK}(t\to1) = 2^{n-3} \left[ \binom{2(n-3)}{n-3} - \binom{2(n-3)}{n-2}\right] = 2^{n-3} C_{n-3}.
\label{eq:num_second_SO3_catalan}
\end{equation}

In the case of \(O(3)\), the parity odd piece of the Hilbert series is obtained simply by sending \(t \to -t\) in \(SO(3)\) Hilbert series (see sec.~\ref{subsec:parity}). That is,
\begin{equation}
H^{O(3)}_{n,\scK}(t) = \frac{1}{2}\Big( H^{SO(3)}_{n,\scK}(t) + H^{SO(3)}_{n,\scK}(-t) \Big).
\end{equation}
We note that the numerators are palindromic for \(n\)-odd.

\subsection{Hilbert series for spinning particles}
\label{subsec:hilbert_spinning}

Consider an \(n\)-point amplitude involving particles with spin. As discussed in section~\ref{sec:overview}---see eq.~\eqref{eq:amp_tens}---the amplitude can be decomposed into a finite number of tensor structures (which depend on the polarization tensors) multiplying scalar functions of the Mandelstam invariants. In this helicity amplitude decomposition~\cite{Jacob:1959at} there is a tensor structure for each independent helicity configuration.\footnote{For our purposes, the word ``helicity'' refers to the spin states under the relevant little group, \(SO(d-1)\) for massive particles or \(SO(d-2)\) for massless particles.}

\begin{table}
\centering
\footnotesize{
\textbf{Hilbert series for \(n=4\) gauge fields in \(d = 4,5\)}
\[\def\arraystretch{1.5}
\begin{array}{|r|clc|c|}\hline
G \ \ \ \ \ ~~ & \textcolor{white}{G} & \text{Numerator} & \textcolor{white}{G}  & \# \text{ tensor} \\
\hline
SO(4) &&  4(3+t^2) & &  16\\ 
O(4)   &&   2(3+t^2)  & & 8\\ 
SO(5) && 6 +9t+4t^2  & & 19 \\ 
O(5)   && 6 + 4t^2  & & 10 \\  
\hline
SO(4) \times S_4 & &  3+5t^2+t^4-2t^6& & 7\\ 
O(4)  \times S_4  & & 2+3t^2+t^4-t^6 & & 5\\ 
SO(5)  \times S_4  & & 2+3t^2+2t^4 & & 7\\ 
O(5) \times S_4   & & 2+3t^2+2t^4 & & 7 \\ 
\hline
\end{array}
\]
}
\vspace{-15pt}
\caption{Hilbert series for \(n=4\) gauge fields in \(d=4,5\) dimensions. The top half of the table is for distinguishable gauge fields; the bottom identical (accounts for \(S_4\) permutations). The second column lists the numerators of the Hilbert series; the denominators are given by \((1-t^2)^2\) and \((1-t^4)(1-t^6)\) for the distinguishable and identical cases, respectively. The last column lists the number of tensor structures in the 4-point amplitude for gauge fields, see eq.~\eqref{eq:num_tens_hil_ratio}.}
\label{tab:F_4pt}
\end{table}

In this section we explore how the Hilbert series reflects this amplitude decomposition. Recently,~\cite{Kravchuk:2016qvl} developed a general procedure for counting these tensor structures. We will show how the Hilbert series also computes this number, as well as provides information about the underlying algebra, in connection with section~\ref{subsubsec:spin_generalization}. The authors of~\cite{Kravchuk:2016qvl} obtain their results primarily in the language of decomposing conformal correlation functions (the correspondence is between \(\text{CFT}_{d-1}\) and \(\text{QFT}_d\)~\cite{Heemskerk:2009pn,Fitzpatrick:2010zm,Penedones:2010ue,Fitzpatrick:2011ia,Kravchuk:2016qvl}); to aid in accessibility, in appendix~\ref{app:helicity_count} we re-derive the results of~\cite{Kravchuk:2016qvl} from the scattering viewpoint.

We ask what operators belong to the operator basis \(\scK\) for a fixed number \(n\) of fields \(\Ph_{1},\dots,\Ph_{n}\) (possibly with spin, possibly identical, \textit{etc}.) and arbitrary numbers of derivatives. Let \(H_{\text{spin}}^G(t)\) be the Hilbert series for this set, \textit{i.e.} \(H_{\text{spin}}^G(t) = \sum_k c_kt^k\) where \(c_k\) is the number of independent operators schematically of the form \(\Ph_{1}\cdots \Ph_{n}\pd^k\). As usual, \(G = (S)O(d)\times \Sigma\), \(\Sigma \subseteq S_n\), is the symmetry group we impose invariance under. Further, let \(H_{\text{scalar}}^G(t)\) denote the Hilbert series corresponding to \(n\) scalar fields.

The number of tensor structures is read off by taking the ratio of spin to scalar Hilbert series and sending \(t \to 1\):
\begin{equation}
\text{\# tensor structures} =\lim_{t\to1}\frac{H_{\text{spin}}^G(t)}{H_{\text{scalar}}^G(t)}
\label{eq:num_tens_hil_ratio}
\end{equation}
This formula follows from the notion---discussed in sec.~\ref{subsubsec:spin_generalization}---that the appropriate algebra when spin is involved is finitely generated over the algebra for scalars.

Table~\ref{tab:F_4pt} lists the Hilbert series for \(n=4\) gauge fields in \(d=4,5\) dimensions, while Table~\ref{tab:F_5pt} does the same for \(n=5\) gauge fields. In these tables we arranged the denominators of the Hilbert series to be equal to the denominators of the corresponding scalar Hilbert series. In every case, the number of tensor structures agrees precisely with~\cite{Kravchuk:2016qvl} (the corresponding problem in \(\text{CFT}_{d-1}\) is for conserved currents).

As an example, take \(n=4\) identical photons in \(d=4\). The Hilbert series is\footnote{In terms of chiral fields \(F_{L,R}\)---special to \(4d\)---we have \(H_{F_L^4}=H_{F_R^4} = (1+t^2-t^6)/[(1-t^4)(1-t^6)]\), \(H_{F_L^3F_R}=H_{F_LF_R^3}=t^2/[(1-t^4)(1-t^6)]\), and \(H_{F_L^2F_R^2} = 1/[(1-t^2)(1-t^4)]\).}
\begin{equation}
H_{F^4}^{SO(4)\times S_4}(t) = \frac{3+5t^2+t^4-2t^6}{(1-t^4)(1-t^6)}.
\label{eq:HF4_iden}
\end{equation}
The \(d=4\) Hilbert series for \(n=4\) identical scalars is \(1/(1-t^4)(1-t^6)\). Hence,  the number of \(SO(4)\) tensor structures for \(n=4\) identical gauge fields is equal to 7,
\begin{equation}
\left.\big(3+5t^2+t^4-2t^6\big)\right|_{t\to 1} = 7 \nonumber
\end{equation}
Note that the Hilbert series contains a negative sign in the numerator; in particular, it is impossible to bring eq.~\eqref{eq:HF4_iden} to a form like eq.~\eqref{eq:hil_hiro} with a numerator of strictly positive terms. This implies the underlying algebra is \textit{not} Cohen-Macaulay.

\begin{table}
\centering
\footnotesize{
\textbf{Hilbert series for \(n=5\) gauge fields in \(d=4,5\)}
\[\def\arraystretch{1.0}
\begin{array}{|r|clc|c|}\hline
G \ \ \ \ \ ~~ & \textcolor{white}{G} & \text{Numerator} & \textcolor{white}{G} & \# \text{ tensor} \\
\hline
& & & & \\
SO(4) && 32(1+5t^2-4t^4)            & & 32\\ & & & & \\
O(4)   && 16(1+5t^2-4t^4)          & & 32\\ & & & & \\
SO(5) && 22 + 95 t + 145 t^2 + 65 t^3 - 35 t^4 - 39 t^5 - 10 t^6  & & 243 \\ & & & & \\
O(5)   && 22 + 145 t^2 - 35 t^4 - 10 t^6  & & 122 \\ & & & & \\
\hline
& & & & \\
SO(4) \times S_5 & &  2\big(2t^4 +7 t^6+17t^8 + 28 t^{10} + 35 t^{12} + 42 t^{14} + 39 t^{16} + 28 t^{18}& & 32\\
 & &  + 18 t^{20} + 4 t^{22} - 7 t^{24} - 8 t^{26} - 7 t^{28} - 7 t^{30} - t^{32} + 2 t^{34} \big) & & \\
& & & & \\
O(4)  \times S_5  & & 2t^4 +7 t^6+17t^8 + 28 t^{10} + 35 t^{12} + 42 t^{14} + 39 t^{16} + 28 t^{18} & & 32\\
 & &  + 18 t^{20} + 4 t^{22} - 7 t^{24} - 8 t^{26} - 7 t^{28} - 7 t^{30} - t^{32} + 2 t^{34} & & \\
& & & & \\
SO(5)  \times S_5  & & t + 5 t^3 + 4 t^4 + 16 t^5 + 16 t^6 + 30 t^7 + 36 t^8 + 51 t^9 +  63 t^{10} + 73 t^{11} & & 243\\
& &  + 89 t^{12} + 92 t^{13} + 110 t^{14} + 103 t^{15} + 117 t^{16} + 103 t^{17} + 108 t^{18} & & \\
& & + 91 t^{19} + 88 t^{20} + 71 t^{21} + 59 t^{22} + 49 t^{23} + 32 t^{24} + 27 t^{25} + 13 t^{26} & & \\ 
& & + 12 t^{27} + 2 t^{28} + 3 t^{29} - 3 t^{30} - 2 t^{32} - t^{33}& & \\
& & & & \\
O(5) \times S_5 & & 4 t^4 + 16 t^6 + 36 t^8 + 63 t^{10} + 89 t^{12} + 110 t^{14} + 117 t^{16} + 108 t^{18} & & 122 \\
 & & + 88 t^{20} + 59 t^{22} + 32 t^{24} + 13 t^{26} + 2 t^{28} - 3 t^{30} - 2 t^{32} & &\\
& & & & \\
\hline 
\end{array}
\]
}
\vspace{-15pt}
\caption{Same as Table~\ref{tab:F_4pt}, but for \(n=5\) gauge fields. The denominators of the Hilbert series are \((1-t^2)^5\) for distinguishable fields and \((1-t^4)(1-t^6)(1-t^8)(1-t^{10})(1-t^{12})\) for identical. The corresponding \(H_{\text{scalar}}^G\) have numerators given by \(SO(4): 1+q^4\); \(O(4),(S)O(5): 1\); \((S)O(4)\times S_5:\) table~\ref{tab:fixed_u}; \((S)O(5)\times S_5:\) exercise for the reader. For the tensor structures, we point out that \(32 = 2^5\), \(243 = 3^5\), and \(122 = (3^5+1)/2\) (see text).}
\label{tab:F_5pt}
\end{table}

As a second example, the \(O(4)\times S_5\) Hilbert series for \(n=5\) identical gauge fields is shown in table~\ref{tab:F_5pt}. The number of terms in the numerator is
\[
\left. \big(O(4),\ n=5 \ \text{gauge numerator}\big)\right|_{t\to 1} = 192
\]
The \(O(4)\) Hilbert series for \(n=5\) identical scalars is given in Table~\ref{tab:fixed_u}; in particular, the numerator indicates there are 6 secondary invariants. Therefore, the number of tensor structures is equal to \(192/6 = 32 = 2^5\).

The number of tensor structures is simply the number of independent helicity amplitudes.  Intuitively, when setting up a scattering experiment one picks a helicity state for each external particle; therefore, the number of independent helicity configurations is bounded by \(N^{(n)}_{h,\text{max}} \equiv \prod_{i=1}^n h_i\), where \(h_i\) is the number of possible helicity states for the \(i\)th particle. Symmetry may relate configurations and reduce this number, as long as the symmetry operation preserves the kinematics of the scattering configuration. 

An explicit counting of the independent helicity amplitudes is straightforward to obtain by using momentum conservation and Lorentz transformations to fix a scattering configuration~\cite{Kravchuk:2016qvl}---see appendix~\ref{app:helicity_count} where we re-derive the results of~\cite{Kravchuk:2016qvl}. If our interest is in low-dimensional field theories, then the most applicable result is the intuitive one:
\begin{equation}
\text{If } n>d>4: \ \ \# \text{tensor structures} = N^{(n)}_{h,\text{max}} = \prod_{i=1}^n h_i.
\end{equation}
The reason every helicity configuration is independent in this case is because we exhaust all of the symmetry by rotating the momenta into some configuration. When \(n=d>4\), parity can still act on the polarization tensors, and this will leave us with \(N^{(n)}_{h,\text{max}}/2\) or \((N^{(n)}_{h,\text{max}}+1)/2\), for \(N^{(n)}_{h,\text{max}}\) even or odd, independent tensor structures in \(O(d)\). These results and the other cases are explained in appendix~\ref{app:helicity_count}.

\subsection{Quantifying the effects of EOM, IBP, and Gram redundancy}
\label{subsec:quant_redund}

It is interesting to explore some basic quantification of the effects of EOM, IBP, and Gram redundancy. For concreteness, we do this in the context of the EFT of a real scalar field.

Asymptotic behavior of the full partition function of scalar field theory is known---for a free scalar in \(d\)-dimensions the number of \textit{all} operators of mass dimension \(\D\) grows exponentially at large \(\D\)~\cite{Cardy:1991kr},
\begin{equation}
\r^{(d)}_{\text{all ops}}(\D) \approx a_d(\D) \exp\big(\b_d \D^{1-1/d}\big).
\label{eq:cardy}
\end{equation}
One can show that the density of scalar operators, as well as scalar primaries, have the same exponential growth but differ in the prefactor (which is a power law in \(\D\)). This implies that enforcing Lorentz invariance (scalar operators) and then momentum conservation (primaries) leads to power law suppression of the overall number of operators. However, note that enforcing finite rank conditions leads to \textit{exponential} suppression: not accounting for linear dependencies amounts to taking \(d \to \infty\) in the above equation.

\begin{figure}
\centering
\includegraphics[width=12cm]{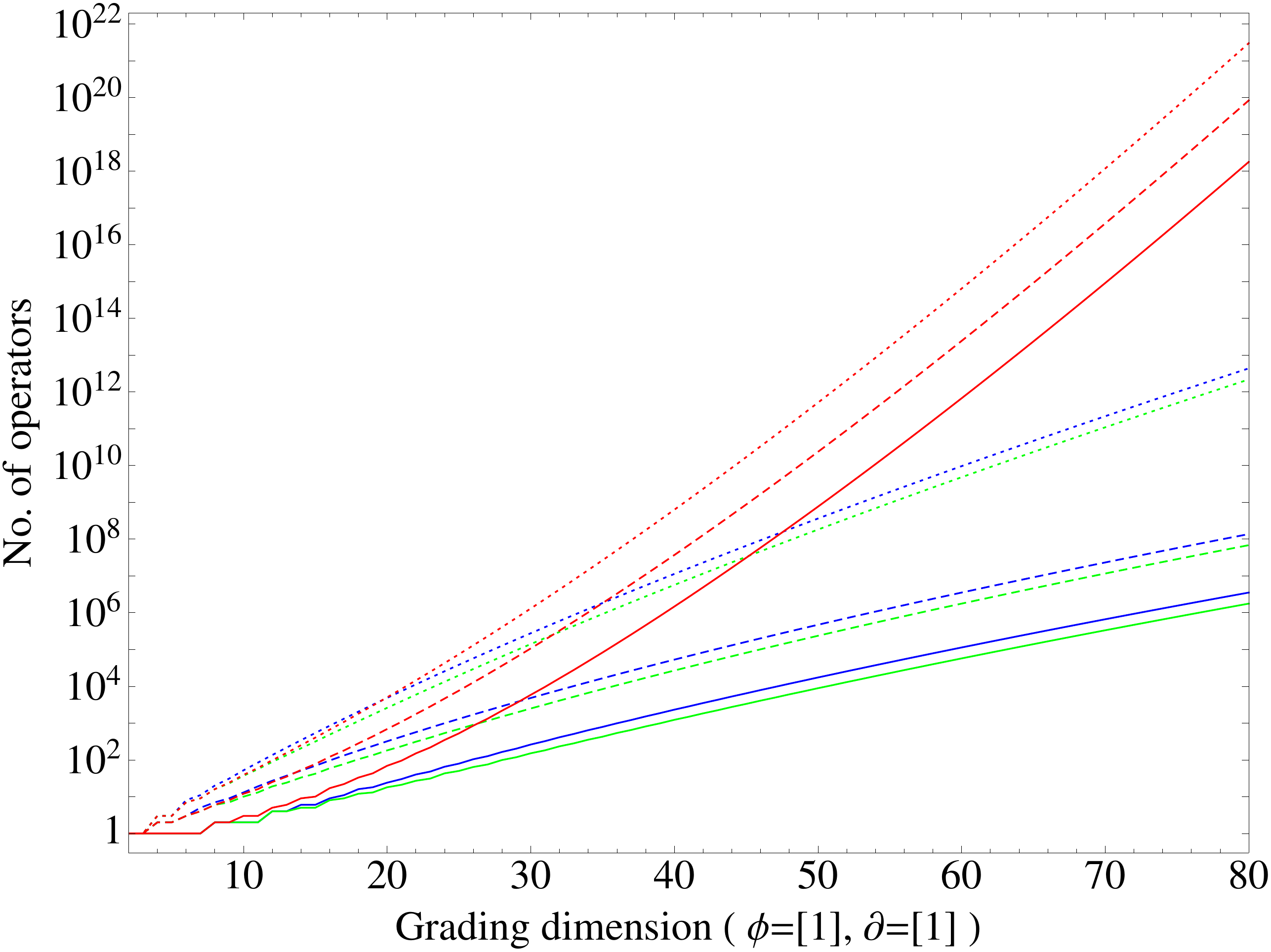}
\caption{Growth of the number of independent scalar operators with grading dimension ($[\phi]=1$ and $[\partial]=1$) in the EFT of a real scalar field, up to dimension 80. Red curves are when spacetime rank conditions (Gram conditions) are not enforced, which is equivalent to considering spacetime dimension $d\gg1$;  green and blue curves include these rank conditions in $d=2$ dimensions, with and without  parity ($O(2)$ and $SO(2)$) imposed on the operator basis, respectively. The solid curves are after imposing both EOM and IBP redundancy; the dashed curves are after only imposing EOM; the dotted curves are without EOM and IBP imposed.}
\label{fig:one}
\end{figure}

We can already illustrate the comparison between the polynomial suppression of EOM and IBP  with the exponential suppression  of Gram constraints by explicit evaluation of the Hilbert series up to moderately high orders. To illustrate the Gram constraints, we take two extreme cases: $d=2$ and  $d\to \infty$.  For the $d=2$ case we evaluate the formulas we presented above, including or omitting EOM and IBP redundancies through inclusion/omission of the shortening of the character of the scalar field and inclusion/omission of the momentum ($1/P$) factor in the integrand. For the $d\to \infty$ we apply Molien's formula as in~\cite{Thiery:2000:AIG:377604.377612, tagkey1973iii}. 

In order to make a reasonable comparison, we choose to work away from the canonical scaling dimension \([\phi]=(d-2)/2\) and grade the Hilbert series with $[\phi]=1$ and $[\partial]=1$ for both \(d=2\) and \(d\to \infty\)  and evaluate up to grading dimension 80. For the $d=2$ case we do this with and without parity invariance; for the $d\to \infty$ case, both \(SO(d)\) and \(O(d)\) give equivalent results.

The results of this are plotted in Fig.~\ref{fig:one}, and exhibit the expected exponential suppression between the $d\to \infty$  (red) results and the $d=2$ (green ($O(2)$), blue ($SO(2)$)) results. This is in contrast with the polynomial suppression that is apparent between the curves enforcing neither EOM or IBP (dotted), EOM  only (dashed), and EOM and IBP (solid).\footnote{That the \(d=2\) dotted curve looks exponentially bigger than the dashed and solid curves is an artifact of taking \([\ph] = 1\) in our comparison. The canonical dimension \([\ph] = 0\) leads to an infinity of operators at any mass dimension; removing the \((1-\ph)^{-1}\) zero mode in the Hilbert series and then using \([\ph]=0\) gives \(d=2\) curves with the same exponential behavior and relative polynomial suppression.} We leave a more detailed analysis of the precise scaling ({\it i.e.} coefficients $a_d$ and $\beta_d$ in eq.~\eqref{eq:cardy}) to future work.

\section{Non-linear realizations}
\label{sec:non-linear}

The formalism for computing the Hilbert series described in sec.~\ref{sec:compute} was based on the assumption that the fields $\Phi_i$ transform linearly under the internal symmetry group $G$. However, many phenomenological theories are built from a non-linear realization of $G$, most famously chiral perturbation theory~\cite{Pagels:1974se,Weinberg:1978kz,Gasser:1983yg,Gasser:1984gg,Fearing:1994ga,Bijnens:1999sh}, in which the building blocks of the Lagrangian transform in a more complicated way under $G$. 

As mentioned in sec.~\ref{sec:2.4}, for theories of pions, scattering in the soft limit places an additional requirement on the fields interpolating the single particle states such that they vanish with momentum. We expect these considerations to lead directly to the single particle module $R_u$ in eq.~\eqref{eq:SPM_non-linear}.  In sec.~\ref{subsec:blocks} we give a derivation of $R_u$ using the CCWZ~\cite{Coleman:1969sm,Callan:1969sn} description of non-linearly realized symmetries, providing a concrete justification for the form of this single particle module.

This module forms the starting point for building the Hilbert series; sec.~\ref{subsec:nonlinearHilbert} presents the derivation mirroring the approach taken in  sec.~\ref{sec:compute}. We first construct the enlarged operator space $\scJ$ by taking tensor products of the modules. In going from $\scJ$ to $\scK$, however, we can no longer straightforwardly appeal to conformal representation theory, as in this case $R_u$ does not correspond to a conformal representation. Instead we use the differential form technique described in~\cite{Henning:2015alf} as an alternative approach to compute the Hilbert series.

We have applied our non-linear realization counting formalism to the chiral Lagrangian under the symmetry breaking scenario of $G = SU(N)_L \times SU(N)_R$ broken to $H = SU(N)_V$.  We found a smaller number of $O(p^6)$ operators compared with the results in~\cite{Bijnens:1999sh}. It turns out that there is a redundancy in the list of operators presented in~\cite{Bijnens:1999sh}. We will report the details elsewhere.

\subsection{Linearly transforming building blocks}\label{subsec:blocks}

Consider a spontaneous symmetry breaking $G\to H \subset G$, with $X^i\in \mathfrak{g}/\mathfrak{h}$ denoting the broken generators. We are interested in the EFT of the Goldstone bosons $\pi^i(x)$. In a non-linear realization, the building block of the EFT Lagrangian is the unitary representation matrix~\cite{Coleman:1969sm,Callan:1969sn}
\be
\xi(x) = e^{i \pi^i(x) X^i/f_\pi} ,
\ee
and its derivatives; namely, the Lagrangian is a $G$-invariant polynomial of $\xi, \xi^{-1}$, and their derivatives.

As a non-linear realization, under a transformation $g\in G$, $\xi$ is stipulated to transform as
\be
g \xi = \xi ' h(g,\xi) , \hspace{5mm} \mbox{\textit{i.e.}} \hspace{5mm} \xi \to \xi ' = g \xi h^{-1}(g,\xi) , \label{eqn:xitransform}
\ee
where we require $h(g,\xi)\in H$. In this section, we only consider \textit{global} transformations $g$. However, observe that $h(g,\xi)$ is \textit{local}, as it depends on $\xi(x)$. To better see what polynomials are invariant under the transformation eq.~\eqref{eqn:xitransform}, it is helpful to define the Maurer-Cartan form
\be
w_\mu \equiv \xi^{-1}\partial_\mu \xi = u_\mu^i X^i + v_\mu^a T^a = u_\mu + v_\mu ,
\label{eq:def_maurer_cartan}
\ee
with $T^a\in \mathfrak{h}$ denoting the unbroken generators.\footnote{The typical convention includes a factor of \(i\) in the definition, \( w_\mu = -i u_\mu - i v_\mu\), so that \(u_{\m}\) and \(v_{\m}\) are Hermitian. Formulas that follow are easily modified to adhere to the standard convention.}  Note in particular that $w_\mu$ is valued in the Lie algebra $\mathfrak{g}$, with $v_\mu \in \mathfrak{h}$ belonging to the unbroken algebra and $u_\mu \in \mathfrak{g}/\mathfrak{h}$ in the coset space. Following eq.~\eqref{eqn:xitransform}, it is easy to see that the Maurer-Cartan form transforms as
\be
w_\mu \to h w_\mu h^{-1}+h(\partial_\mu h^{-1}) .
\ee
Since $h(\partial_\mu h^{-1})\in \mathfrak{h}$, the components of $w_\mu$ must transform as
\begin{subequations}
\begin{align}
u_\mu &\to h u_\mu h^{-1} , \\
v_\mu &\to h v_\mu h^{-1} + h(\partial_\mu h^{-1}) .
\end{align}
\end{subequations}
We see that $u_\mu$ transforms homogeneously while the transformation of $v_\mu$ is inhomogeneous, similar to a gauge field. If we put $v_\mu$ together with the derivative $\partial_\mu$, namely if we define
\begin{equation}
D_\mu \equiv\partial_\mu + v_\mu , \nonumber
\end{equation}
then the transformation of $D_\mu$ is homogeneous:
\begin{equation}
D_\mu \to h D_\mu h^{-1} . \nonumber
\end{equation}
This can be understood from a different perspective by regarding the transformation of $\xi$ in eq.~\eqref{eqn:xitransform} as a bi-linear transformation under a larger group $G\times H$, with $G$ global and $H$ local. From this point of view, $v_\mu$ is the gauge field for the local group $H$, and $D_\mu$ the covariant derivative. As usual, the covariant derivative also leads to a field strength $F_{\mu\nu}\equiv[D_\mu, D_\nu]$ which transforms homogeneously.

Now we see that the linearly transforming building blocks of the Lagrangian are $\xi$, $u_\mu$, $D_\mu$, and $F_{\mu\nu}$, with their transformation properties under the group $G\times H$ as
\begin{subequations}
\begin{align}
\xi &\to g\xi h^{-1} , \\
u_\mu &\to h u_\mu h^{-1} , \\
D_\mu &\to h D_\mu h^{-1} , \\
F_{\mu\nu} &\to h F_{\mu\nu} h^{-1} .
\end{align}
\end{subequations}
Note that $\xi$ is the only component that retains explicit dependence on $g\in G$ under a transformation. However, eq.~\eqref{eq:def_maurer_cartan} implies we can always trade derivatives of \(\xi\) for products of \(\xi\) and \(u_{\mu}\),
\begin{align}
D_\mu \xi &= \partial_\mu \xi -\xi v_\mu = \xi u_\mu , \nonumber \\
D_\mu \xi^{-1} &= \partial_\mu \xi^{-1} + v_\mu \xi^{-1} = - u_\mu \xi^{-1} , \nonumber
\end{align}
\textit{i.e.} $\xi$ only enters without derivatives. Then to make invariant terms under $G$, we have to form the combination $\xi^{-1}\xi$. But since $\xi^{-1}\xi=1$, $\xi$ drops out of the Lagrangian. Therefore, we are left with the building blocks $u_\mu$, $D_\mu$, and $F_{\mu\nu}$. Because these building blocks only transform under $H$, imposing $H$-invariance is equivalent to the original $G$-invariance.

It seems that we have converted the original non-linear realization theory of the global symmetry $G$ into a linearly realized theory of the local group $H$ with a ``matter field'' $u_\mu$, covariant derivative $D_\mu$, and field strength $F_{\mu\nu}$. However, a crucial feature is that---contrary to a usual gauge theory---the gauge field $v_\mu$ (and hence the field strength $F_{\mu\nu}$) is actually \textit{not} an independent degree of freedom from the field $u_\mu$. To see this, consider the identity of the Maurer-Cartan form
\be
\partial_\mu w_\nu - \partial_\nu w_\mu + [w_\mu,w_\nu] = 0 ,
\ee
which trivially follows from $\partial_\mu(\xi^{-1}\xi)=0$. In terms of $u_\mu$, $D_\mu$, and $F_{\mu\nu}$ this reads
\be
D_\mu u_\nu - D_\nu u_\mu + [u_\mu, u_\nu]+F_{\mu\nu}=0 . \label{eqn:MaurerCartanIdentity}
\ee
Recalling that
\begin{subequations}
\begin{align}
[T^a, T^b] &\in \mathfrak{h} , \nonumber \\
[T^a, X^i] &\in \mathfrak{g}/\mathfrak{h} , \nonumber
\end{align}
\end{subequations}
we clearly have $F_{\mu\nu} \in \mathfrak{h}$ and $D_\mu u_\nu - D_\nu u_\mu \in \mathfrak{g}/\mathfrak{h}$. But $[u_\mu, u_\nu]$ could have both components:
\begin{equation}
[u_\mu, u_\nu] = [u_\mu, u_\nu]_\mathfrak{h} + [u_\mu, u_\nu]_{\mathfrak{g}/\mathfrak{h}} \nonumber .
\end{equation}
Splitting into components, eq.~\eqref{eqn:MaurerCartanIdentity} reads\footnote{If $G/H$ is a symmetric space, \([X,X] \sim T \in \mathfrak{h}\), then $[u_\mu,u_\nu]_{\mathfrak{g}/\mathfrak{h}}=0$, and this further simplifies to
\begin{subequations}
\begin{align}
F_{\mu\nu} &= -[u_\mu, u_\nu] , \nonumber\\
D_\mu u_\nu - D_\nu u_{\mu} &= 0 . \nonumber
\end{align}
\end{subequations}
Note that the usual chiral symmetry breaking patterns are symmetric spaces.}
\begin{subequations}\label{eqn:MCidentity}
\begin{align}
F_{\mu\nu} &= - [u_\mu, u_\nu]_\mathfrak{h} , \label{eqn:Ffromp} \\
D_\mu u_\nu - D_\nu u_\mu &= - [u_\mu, u_\nu]_{\mathfrak{g}/\mathfrak{h}} . \label{eqn:pcurlfromp}
\end{align}
\end{subequations}

Eq.~\eqref{eqn:Ffromp} shows that the field strength $F_{\mu\nu}$ is fully determined by the field $u_\mu$. This result is expected. Recall that $H$ is considered as a local group because in the transformation eq.~\eqref{eqn:xitransform}, $h(g,\xi)$ is \textit{local}, as it depends on $\xi(x)$. However, this also means that while the group $H$ is considered as ``gauged'', its local transformation parameter is not introduced as a free parameter, but instead fully fixed by the field $\xi(x)$. Therefore, the gauge field should not be expected as an independent field from the ``matter'' field $u_\mu$.

To sum up, in a non-linear realization of the global symmetry $G$, all local operators can be built from just \textit{two} building blocks, \(u_{\m}\) and the covariant derivative \(D_{\m}\). These objects transform linearly under the unbroken group $H$. By forming operators invariant under the unbroken group $H$, we are guaranteed that these are \(G\)-invariant as well.

\subsection{Computing the Hilbert series}\label{subsec:nonlinearHilbert}

We identified above the building blocks of operators to be the linearly transforming objects \(u_{\m}\) and \(D_{\m}\). By virtue of eq.~\eqref{eqn:Ffromp}, antisymmetric combinations of the covariant derivatives can always be eliminated for polynomials in the \(u_{\m}\). Moreover,~\eqref{eqn:pcurlfromp} implies a similar result for the curl of \(u_{\m}\). So, up to equation of motion, all operators are built from \(u_{\m}\) and symmetrized derivatives acting on \(u_{\m}\), \textit{i.e.} out of $u_{\m}, \ D_{\left( \m_1 \right.}u_{\left. \m_2\right )}, \ D_{\left( \m_1 \right.}D_{\m_2}u_{\left. \m_3\right)}$, \textit{etc}.

The equation of motion for \(u_{\m}\) is
\be
D^\mu u_\mu =0,  \label{eqn:pEOM}
\ee
which can obtained from variation of the action with respect to \(\xi\) (see appendix~\ref{app:eom_nonlin}). Given that the divergence and curl (eq.~\eqref{eqn:MCidentity}) of the field \(u_{\mu}\) are constrained, we can expect its harmonic behavior to be determined. Concretely, the EOM and eq.~\eqref{eqn:MCidentity} imply that \(D^2u_{\m}\) can always be eliminated in favor of polynomials of \(u_{\m}\) and \(D_{\{\mu}u_{\nu\}}\).\footnote{The general expression is straightforward to obtain, although the final result is not very enlightening. However, for a symmetric space it takes a simple form:
\begin{equation}
\text{symmetric coset: }D^2 u_{\m} = \big[u^{\n}\big[u_{\n},u_{\m}\big]\big].
\label{eq:laplacian_p}
\end{equation}
} Below we will use cohomology to address IBP redundancy; we point out here that we can discuss cohomology with the covariant derivative $D$ instead of the usual derivatives $\partial$ because $D\wedge D$ can be rewritten using $u_{\mu}$ and therefore effectively be treated as zero in our analysis.

\subsubsection*{The character generating function $Z(u,q,x,y)$ for $\scJ$}

The above considerations lead us to the following single particle module
\begin{equation}\def\arraystretch{1.1}
R_u = \left(\begin{array}{c}u_{\m} \\ D_{\{\m_1}u_{\m_2\}} \\ D_{\{\m_1}D^{}_{\m_2\,}u_{\m_3\}} \\ \vdots \end{array}\right) . \label{eqn:Ru}
\end{equation}
The corresponding weighted character can be constructed as follows.
First consider the block $D^k u_\mu \equiv D_{\{\mu_1} \cdots D_{\mu_k\}} u_\mu$. As $[u_\mu]=1$, this block possesses mass dimension $k+1$. If we introduce the weight variable $q$ to keep track of the mass dimension, the appropriately $q$-weighted $SO(d)$ character of this block should be
\begin{equation}
\chi_{u,k}^{} (q,x,y) = q^{k+1} \chi_{(k+1,0,\cdots,0)}^{}(x) \chi_{H,u}^{}(y) ,
\end{equation}
where following the notation in sec.~\ref{sec:compute}, we have used $\chi_{H,u}^{}(y)$ to denote the character of $u_\mu$ under the internal symmetry group $H$. Then, the $q$-weighted $SO(d)$ character for the full generating tower $R_u$ can be obtained by summing over $k$:
\begin{align}
\chi_u^{} (q, x, y) &= \sum_{k=0}^\infty \chi_{u,k}^{} (q, x, y) = \left[(1-q^2)P(q;x)-1\right] \chi_{H,u}^{}(y) , \label{eqn:chiu}
\end{align}
where we have used eq.~\eqref{eq:chi_free_scalar_sum}. Therefore, the generating function for the operator space \(\scJ = \bigoplus_{n=0}^{\infty} \text{sym}^n(R_u)\) is
\begin{align}
Z(u,q,x,y) &= \sum_{n=0}^{\infty} u^n \ch_{\text{sym}^n(R_u)}^{}(q,x,y) = \text{PE} \left[ u \chi_u^{}(q,x,y) \right] , \label{eqn:Zu}
\end{align}
where we have assigned a weight \(u\) to the field \(u_{\m}\).

\subsubsection*{IBP addressed by Hodge theory}

The operator basis $\scK$ consists of operators in $\scJ$ that are (1) Lorentz scalars, (2) invariant under the unbroken group $H$, and (3) independent under integration by parts. With the $SO(d)$ character generating function $Z(u,q,x,y)$ at hand, the first two conditions are straightforward to impose (by integrating over \(SO(d)\) and \(H\)). Now we explain how to address IBP redundancy by counting differential form operators in $\scJ$. This method has its footing in Hodge theory.

IBP redundancy imposes an equivalence relation among the scalar operators in $\scJ$,
\begin{equation}
\scO^a_0 \sim \scO^b_0 \quad \text{if} \quad \scO^a_0 = \scO_0^b + \pd \cdot \scO^c_1 , \label{eq:equiv_IBP_0form}
\end{equation}
where, anticipating our language, the subscript denotes that scalars are 0-forms, taken equivalent up to the divergence of a 1-form. The wording of this equivalence relation becomes more familiar by taking the Hodge dual.\footnote{A \(k\)-form \(\o_k\) on a \(d\)-dimensional manifold in coordinates is \(\o_k = \frac{1}{k!} \o_{\m_1\dots \m_k}\de x^{\m_1}\wedge \dots \wedge \de x^{\m_k}\). The Hodge dual \(*\o_k\) is a \((d-k)\)-form,
\begin{align}
*\o_k &= \frac{1}{k!} \o_{\m_1\dots \m_k} *\big(\de x^{\m_1}\wedge \dots \wedge \de x^{\m_k}\big) \nonumber = \frac{1}{k!(d-k)!} \o_{\m_1\dots \m_k} \e^{\m_1 \dots \m_k}_{\textcolor{white}{\m_1 \dots \m_k}\m_{k+1} \dots \m_d} \de x^{\m_{k+1}} \wedge \dots \wedge \de x^{\m_d}, \nonumber
\end{align}
\textit{i.e.} the \((d-k)\)-form \(*\o_k\) is essentially obtained by contraction with the epsilon tensor. Minus signs, tensor densities, raised/lowered indices, \textit{etc.} are not important for our discussion, so we will ignore these.} The dual of any 0-form is a \(d\)-form. The dual of a scalar which is the divergence of a 1-form is an \textit{exact} \(d\)-form, \textit{i.e.} it is given by the exterior derivative \(\de\) acting on a \((d-1)\)-form. In the dual picture, the equivalence relation~\eqref{eq:equiv_IBP_0form} reads
\begin{equation}
\widetilde{\scO}^a_d \sim \widetilde{\scO}^b_d \quad \text{if} \quad \widetilde{\scO}^a_d = \widetilde{\scO}^b_d + \de \widetilde{\scO}^c_{d-1},
\label{eq:equiv_IBP_dform}
\end{equation}
where \(\widetilde{\o}_{d-k} = *\o_k\) denotes the Hodge dual. In words: two \(d\)-forms are taken equivalent if they differ by an exact \(d\)-form. So the precise statement about IBP redundancy (formulated in the dual picture) is that \textit{operators in the basis are closed but not exact \(d\)-forms}. 

To proceed, we need a little more terminology. The exterior derivative takes a \(k\)-form to a \((k+1)\)-form. Hodge duality gives a natural adjoint to \(\de\): the codifferential \(\d = *\de *\) (we are ignoring minus signs in definitions), which takes a \(k\)-form to a \((k-1)\)-form. Intuitively, one thinks of \(\d\) as taking a divergence, similar to the intuition that \(\de\) takes a curl. Unsurprisingly, \(\d^2 = 0\) (the double divergence of some form vanishes by antisymmetry). If \(\d \o_k = 0\), then \(\o_k\) is said to be \textit{co-closed}. If \(\o_k = \d \b_{k+1}\), in which case \(\d \o_k =0\) follows trivially, \(\o_k\) is said to be \textit{co-exact}. 

With this language, eq.~\eqref{eq:equiv_IBP_0form} reads \(\scO_0^a \sim \scO_0^b\) if \(\scO_0^a = \scO_0^b + \d \scO_1^c\). Now we can formulate the IBP redundancy in the original picture (as opposed to the dual picture): \textit{the operator basis $\scK$ consists of all 0-forms that are not co-exact}. Namely, for counting we have
\begin{equation}
\#\big(\text{operators}\big) = \#\big(0\text{-forms}\big) - \#\big(\text{co-exact } 0\text{-forms}\big) . \label{eqn:0form}
\end{equation}
Co-exact \(0\)-forms come from \(1\)-forms that do not vanish when acting with \(\d\), \textit{i.e.} \(1\)-forms that are not co-closed:
\begin{align}
\small
\# \left(\text{co-exact }0\text{-forms}\right) &= \# \left(1\text{-forms}\right) - \# \left(\text{co-closed } 1\text{-forms}\right) \label{eqn:peeled} \\
 &= \# \left(1\text{-forms}\right) - \# \left(\text{co-closed but not co-exact }1\text{-forms}\right) \nonumber \\
 & \quad - \# \left(\text{co-exact }1\text{-forms}\right) .
\end{align}
We can iterate the logic for $\# \left(\text{co-exact }1\text{-forms}\right)$, ultimately arriving at the sequence:
\small{
\begin{equation}
\#\big(\text{operators}\big) = \left\{ \sum_{k=0}^d (-1)^k\#\big(k\text{-forms}\big) \right\} + \left\{\sum_{k=1}^d (-1)^{k+1}\#\big(\text{co-closed but not co-exact }k\text{-forms}\big) \right\} . \nonumber
\end{equation}
}\normalsize
As indicated in the equation, the counting is naturally divided into two sets: one which counts all possible forms (with appropriate signs) and another which corrects exceptional cases misidentified in the first set. This leads to splitting the Hilbert series into two pieces:
\begin{equation}
H(u,p) = H_0(u,p) + \D H(u,p), \nonumber
\end{equation}
exactly as in eq.~\eqref{eq:hil_split_summary}. We additionally gain the interpretation of \(\D H\) as counting co-closed but not co-exact \(k\)-form operators.

For the $H_0$ piece, it is straightforward to count all \(k\)-forms which appear in \(Z(u,q,x,y)\) using \(SO(d)\) character orthogonality. Note that we also need to include a factor $p^k$ in front of the number of $k$-forms to properly weight the mass dimension. Therefore, while integrating over the $SO(d)$ Haar measure, we should multiply $Z(u,q,x,y)$ by the factor
\begin{equation}
\sum_{k=0}^{d} p^k (-1)^k \ch_{\wedge^k(\Box)}^{}(x) = \frac{1}{P(p;x)} , \label{eqn:pfactor}
\end{equation}
where we have used eq.~\eqref{eqn:Pasymgen}. To further project out invariants of the unbroken group $H$, we integrate over the Haar measure $d\mu_H^{}(y)$. So in the end, $H_0$ is given by
\begin{equation}
H_0(u,p) = \int d\m^{}_H(y) \int d\m_{SO(d)}^{}(x)\frac{1}{P(p;x)}Z(u,p,x,y) , \label{eq:H_0_non-lin}
\end{equation}
which is structurally identical to eq.~\eqref{eq:H_0_summary}.

To compute the $\Delta H$ piece, we enumerate all the co-closed but not co-exact forms. For each rank $k$, the type of these forms and their contributions to $\D H$ are listed in the table below:
\[\def\arraystretch{1.3}
\begin{array}{c|c|c}
\text{rank} & \text{operators} & \text{contributions to }\D H \\
\hline
1\text{-forms} & \hspace{2mm} u \text{ and } *\left(\wedge^{d-1} u\right) \hspace{2mm} & \hspace{2mm} p^2 u \ch_{H,u}^{} \text{ and }p^d u^{d-1}\wedge^{d-1}(\ch_{H,u}^{}) \\
2\text{-forms} &  *\left(\wedge^{d-2} u\right) & (-1)^3 p^d u^{d-2}\wedge^{d-2}(\ch_{H,u}^{}) \\
3\text{-forms} &  *\left(\wedge^{d-3} u\right) & (-1)^4 p^d u^{d-3}\wedge^{d-3}(\ch_{H,u}^{}) \\
\vdots & \vdots & \vdots \\
k\text{-forms  } &  *\left(\wedge^{d-k} u\right) & (-1)^{k+1} p^d u^{d-k}\wedge^{d-k}(\ch_{H,u}^{}) \\
\vdots & \vdots & \vdots \\
d\text{-forms} & *1 & (-1)^{d+1} p^d
\end{array}
\]
The 1-form $u$ is co-closed (\textit{i.e.} its divergence is zero) due to the equation of motion $D^\mu u_\mu=0$. The Hodge dual of the wedge products, $*\left(\wedge^{d-k} u\right)$, are co-closed essentially because the curl of $u_\mu$ vanishes:
\begin{equation}
D_{\mu_1} \left( \epsilon^{\mu_1 \cdots \mu_{d-k} \nu_1 \cdots \nu_k} u_{\nu_1} \cdots u_{\nu_k} \right) = \sum\limits_{i=1}^k \epsilon^{\mu_1 \cdots \mu_{d-k} \nu_1 \cdots \nu_k} u_{\nu_1} \cdots u_{\nu_{i-1}} \left(D_{\mu_1} u_{\nu_i}\right) u_{\nu_{i+1}} \cdots u_{\nu_k} = 0 . \nonumber
\end{equation}
Finally, $*1$ is co-closed as it is a constant whose divergence must be zero. Clearly, none of the operators in the list can be written as a divergence of another operator, simply because there is no derivative involved in any of these operators. So they are all co-closed but not co-exact forms. We do not have a proof that this list exhausts all the possibilities, but we conjecture this is the case.

In the contributions to $\Delta H$ column, the antisymmetric product of the character function $\wedge^k \big[\ch_{H,u}^{}(y)\big]$ gives the character of the representation $\wedge^k (R_{H,u})$. The power of $u$ is given by the power of $u_\mu$ in the corresponding operator, while the power of $p$ is given by the mass dimension of the operator plus the rank of the form (the number of divergences needed to bring it into a $0$-form). Putting all the listed contributions together, we get
\begin{equation}
\D H (u, p) = \int d\m^{}_H(y) \left[ p^2 u \ch_{H,u}^{}(y) + p^d\sum_{k=1}^{d}(-1)^{k+1} u^{d-k} \wedge^{d-k}\left[\ch_{H,u}^{}(y)\right] \right] .
\end{equation}
Clearly, $\D H$ only contains operators with mass dimension \(\le d\).

\subsection{Summary}

For a non-linearly realized theory of a global symmetry $G$ broken to a subgroup $H$, the single particle module takes the form
\begin{equation}\def\arraystretch{1.1}
R_u=\left(\begin{array}{c}u_{\m} \\ D_{\{\m_1}u_{\m_2\}} \\ D_{\{\m_1}D^{}_{\m_2\,}u_{\m_3\}} \\ \vdots \end{array} \right) \,,
\end{equation}
where $u_\mu$ transforms linearly under the subgroup $H$ and $D_\mu$  is the covariant derivative. $R_u$ is structurally identical to the single particle module of a scalar field, $R_{\phi}$, but without the "top" component $\phi \in R_{\phi}$. This structure follows from the CCWZ formalism and is physically related to soft limits of pion amplitudes. 

The character $\chi_u^{}(q,x,y)$ for $R_u$ and the generating function $Z(u,q,x,y)$ for the operator space $\scJ$ are
\begin{align}
\chi_u^{} (q, x, y) &= \left[(1-q^2)P(q;x)-1\right] \chi_{H,u}^{}(y) , \\
Z(u,q,x,y) &= \text{PE} \left[ u \chi_u^{}(q,x,y) \right] ,
\end{align}
where $\chi_{H,u}^{}(y)$ is the character of $u_\mu$ under the unbroken group $H$.

Making use of differential forms, the IBP redundancy is recast as an equivalence of scalar (0-form) operators up to a co-exact 0-form. Counting the number of such equivalence classes leads to a natural splitting of the Hilbert series as
\begin{equation}
H(u,p) = H_0(u,p) + \D H(u,p) , \label{eqn:nonlinearHsplitRecap}
\end{equation}
where
\begin{subequations}
\begin{align}
H_0(u,p) &= \int d\m^{}_H(y) \int d\m^{}_{SO(d)}(x) \frac{1}{P(p;x)} Z(u,p,x,y) , \\
\D H(u,p) &= \int d\m^{}_H(y) \left[p^2 u \ch_{H,u}^{}(y) + p^d \sum_{k=1}^{d}(-1)^{k+1} u^{d-k}\wedge^{d-k}\left[\ch_{H,u}^{}(y)\right] \right] . \label{eq:DH_pion}
\end{align}
\end{subequations}
The splitting $H = H_0 + \Delta H$ and the formula for $H_0$ are universally valid, arising from a natural inclusion/exclusion accounting of IBP redundancy. In the general case, the derivation of these results also tells us that $\Delta H$ counts co-closed by not co-exact $k$-form operators, $k=1,\dots,d$. The formula for $\Delta H$ in eq.~\eqref{eq:DH_pion} is specific to a theory consisting solely of pions which non-linearly realize $G$. We explicitly identified the operators in $\Delta H$ associated to each term in the integrand of eq.~\eqref{eq:DH_pion}, and we conjecture these are the only such operators.

\section{Discussion}
\label{sec:discuss}

In this paper we have introduced a number of systematic techniques that allowed us to study and eliminate redundancies in the operator bases of relativistic EFTs. 
Most particularly, we have shown how to treat the redundancies associated with EOM and IBP identities; in this way we detailed the construction of operator bases that properly account for the independence of all (in-principle) physical measurements one can make in a relativistic quantum theory.  This can be seen as a final squeeze of the EFT approach---making full use of all {\it a priori} kinematic selection rules arising from the full Poincar\'e symmetry of the $S$-matrix.

We presented quite general results for EFTs in $d$ dimensions, with scalars, fermions, and $d/2$-form fields, and detailed the inclusion of parity invariance on the theory. Our results easily include invariances under linear internal/gauge groups, and we included a discussion of how to deal with non-linearly realized internal symmetries. Throughout the paper we made use of character  theory; we found a direct and useful connection to conformal representation theory in the case of linearly realized symmetries, and a slight modification of these ideas for the non-linear story.

After making use of Poincar\'e invariance of the \(S\)-matrix to construct the operator basis, the natural question is: what do unitarity and analyticity of the \(S\)-matrix imply? This transitions us beyond pure kinematics into the realm of dynamics. In this context, the constraint of causality/analyticity has been shown to enforce positivity constraints on Wilson coefficients, see~\cite{Adams:2006sv} and \textit{e.g.}~\cite{Bellazzini:2014waa,deRham:2017avq}. 

Through requiring locality, and by further considering the behavior of scattering amplitudes in the soft limit, a systematic classification of Lorentz invariant single scalar EFTs in $d<6$ dimensions has also been recently obtained \cite{Cheung:2014dqa,Cheung:2016drk}. This suggests that it should be possible to impose such behavior at the level of the single particle module, as discussed in sec.~\ref{sec:overview}. This would be in the vein of imposing {\it a priori} requirements of shift symmetries on theories. We did not follow this line of thought here, instead identifying the building blocks that form the single particle modules for non-linear theories via a CCWZ construction, but it would be interesting to make this connection.

Of course, a whole wealth of new structure appears in going from considering the operator basis to  the full, dynamical EFT. It would, for instance, be interesting to study  the phenomena of the holomorphy of the SM EFT \cite{Alonso:2013hga}, and general non-renormalization theorems \cite{Cheung:2015aba} in our operator basis language. 

The formulation in terms of a polynomial ring of kinematic variables that we developed for scalar fields was very useful for explicitly constructing the operator basis (or equivalently, constructing the independent Feynman rules). We believe that developing a similar ring picture for spinning particles would be very useful to understand more about the structure of these amplitudes; this would go beyond the more basic counting of the tensor and Mandelstam structures that was presented in sec.~\ref{sec:case}.

Finally, as far as the enumeration of operators in bases is concerned, much remains to study about the Hilbert series. Can we obtain a fully closed form for a relativistic theory in \(d>1\) dimensions? What can we learn from asymptotics? One motivation to explore along these lines is to capture analytic properties that are not evident in any truncated EFT expansion, similar to the recursion relations and properties found in $d=1$ dimensions in our previous work~\cite{Henning:2015daa}. As with the other facets of the operator basis evidenced in this paper, here too one expects a richer structure as we move from quantum mechanics to quantum field theory. 
\newline

%\newline 
%\newline 

\noindent \textbf{Note added: } 
During the finalization of this manuscript, the preprints~\cite{deMelloKoch:2017caf,deMelloKoch:2017dgi} appeared, with some overlap with the ideas of this paper.
\newline 

\section*{Acknowledgments}
We would like to thank Hsin-Chia Cheng, Ben Gripaios, Zuhair Khandker, Aneesh Manohar, Adam Martin, David Poland, Kyoji Saito, Andy Stergiou, Jed Thompson, and Junpu Wang for useful discussions. 
BH is grateful to Walter Goldberger, Siddarth Prabhu, and Witek Skiba for conversations. And we thank Francis Dolan, whose off the beaten path ingenuity we sadly never got to know personally, for lifting us from one to $d$-dimensions. XL is supported by DOE grant DE-SC0009999; TM is supported by U.S. DOE grant DE-AC02-05CH11231, and by WPI, MEXT, Japan; TM also acknowledges computational resources provided through ERC grant number 291377: LHCtheory. 
HM was supported by the U.S. DOE under Contract DE-AC02-05CH11231, and
by the NSF under grants PHY-1316783 and PHY-1638509.  HM was also
supported by the JSPS Grant-in-Aid for Scientific Research (C)
(No.~26400241 and 17K05409), MEXT Grant-in-Aid for Scientific Research on Innovative Areas (No. 15H05887, 15K21733), and by WPI, MEXT, Japan.

\newpage
\appendix

\section{Character formulae for classical Lie groups}
\label{app:sodspin}

The Weyl character formula (WCF) provides an explicit formula to obtain the characters of irreducible representations. The WCF is covered in most group theory textbooks, for example~\cite{FultonHarris,Brocker:2003}; here, we only review the bare bones of the formula and then specify directly to \(SO(d)\) characters.

Let \(G\) be a compact, connected Lie group of rank \(r\). We denote the coordinates of the torus by \(x = (x_1,\dots,x_r)\). Let \(l = (l_1,\dots,l_r)\) be the highest weight vector---in the orthonormal basis---of an irreducible representation. Let \(\r\) be the half-sum of positive roots,
\begin{equation}
\r = \frac{1}{2} \sum_{\a \in \text{rt}_+(G)}\a.
\label{eq:def_rho}
\end{equation}
The WCF then gives the character for the irreducible representation to be
\begin{equation}
\ch_l(x) = \frac{A_{\r + l}}{A_{\r}},
\label{eq:WCF}
\end{equation}
where \(A_{\l}\) is an antisymmetric sum of the Weyl group \(W\) acting on the vector \(\l\),
\begin{equation}
A_{\l} = \sum_{w \in W}(-1)^w x^{w(\l)},
\label{eq:def_A_lambda}
\end{equation}
where \((-1)^{w} = sgn(w)\) is the sign of group element \(w \in W\).

\(A_{\r}\) in eq.~\eqref{eq:WCF} can be shown to be equal to a certain product over the positive roots (this is sometimes referred to as Weyl's denominator formula):
\begin{equation}
A_{\r} = \prod_{\a \in \text{rt}_+(G)}(x^{\a/2} - x^{-\a/2}) = x^{\r} \prod_{\a \in \text{rt}_+(G)}(1 - x^{-\a}),
\label{eq:Weyl_denom_form}
\end{equation}
where the second equality follows from the definition of \(\rho\), eq.~\eqref{eq:def_rho}. In particular, the product \(A^*_{\r}A_{\r} = A_{-\r}A_{\r}\) is equal to the Jacobian factor that shows up in the Weyl integration formula,
\begin{equation}
A^*_{\r}A_{\r} = \prod_{\a \in \text{rt}_+(G)} (1- x^{\a})(1-x^{-\a}) = \prod_{\a \in \text{rt}(G)}(1-x^{\a}).
\end{equation}

The WCF for the classical groups can be cast into a more ``user-friendly'' expression involving determinants. The basic identity is that the anti-symmetric sum of permutations gives a determinant,
\begin{equation}
\sum_{\s \in S_r}(-1)^{\s}x^{\l_1}_{\s(1)}\dots x^{\l_r}_{\s(r)} = \text{det}( x_i^{\l_j}).
\label{eq:det_S_r}
\end{equation}
Since the the Weyl group contains the permutation group, this allows us to rewrite the sum over \(W\) in eq.~\eqref{eq:def_A_lambda} in terms of various determinants. For further details on how to obtain these, as well as other useful formulas, see~\cite{FultonHarris}.

\subsection*{WCF for $SO(d)$}

For even dimensions, \(d= 2r\), the Weyl group consists of permutations together with an even number of sign flips: \(W_{SO(2r)} = S_r \ltimes (Z_2^r/Z_2)\) where the \(Z_2^r\) are the \(r\) different sign flips we can perform on \(\l = (\l_1,\dots,\l_r)\) and the modding out by \(Z_2\) is requiring only even numbers of sign flips. The half-sum on positive roots is \(\r = (r-1,r-2,\dots,0)\).

The numerator $A_{\rho+l}^{SO(2r)}$ can be computed from eq.~\eqref{eq:def_A_lambda} by first summing over the sign flips and then the permutations in the Weyl group~\cite{FultonHarris}:
\begin{small}
\begin{align}
A_{\l}^{SO(2r)} &= \sum_{\sigma\in S_r} (-1)^\sigma \sum_\text{even flips} x_{\sigma(1)}^{\lambda_1} \cdots x_{\sigma(r)}^{\lambda_r} \nonumber \\
&= \sum_{\sigma\in S_r} (-1)^\sigma \frac{1}{2} \left[ \sum_\text{all flips} x_{\sigma(1)}^{\lambda_1} \cdots x_{\sigma(r)}^{\lambda_r} + \left(\sum_\text{even flips} - \sum_\text{odd flips}\right) x_{\sigma(1)}^{\lambda_1} \cdots x_{\sigma(r)}^{\lambda_r} \right] \nonumber \\
&= \frac{1}{2} \sum_{\sigma\in S_r} (-1)^\sigma \left[ \left(x_{\sigma(1)}^{\lambda_1} + x_{\sigma(1)}^{-\lambda_1}\right) \cdots \left( x_{\sigma(r)}^{\lambda_r} + x_{\sigma(r)}^{-\lambda_r} \right) + \left(x_{\sigma(1)}^{\lambda_1} - x_{\sigma(1)}^{-\lambda_1}\right) \cdots \left( x_{\sigma(r)}^{\lambda_r} - x_{\sigma(r)}^{-\lambda_r} \right) \right] \nonumber \\
&= \frac{1}{2}\left( \text{det}\big[x_i^{\l_j}+x_{i}^{-\l_j}\big] +\text{det}\big[x_i^{\l_j}-x_{i}^{-\l_j}\big] \right) ,
\end{align}
\end{small}where $\l=\rho+l$, and we have applied the determinant formula eq.~\eqref{eq:det_S_r}. Note that the second determinant above vanishes if \(l_r = 0\). Using the special case $l=(0,\cdots,0)$ in the above, we obtain the denominator:
\begin{align}
A_{\rho}^{SO(2r)} &= \frac{1}{2} \text{det}\big[x_i^{\rho_j}+x_{i}^{-\rho_j}\big] =
\left| {\begin{array}{*{20}{c}}
{x_1^{r - 1} + x_1^{ - \left( {r - 1} \right)}}& \cdots &{{x_1} + x_1^{ - 1}}&1\\
 \vdots &{}& \vdots & \vdots \\
{x_r^{r - 1} + x_r^{ - \left( {r - 1} \right)}}& \cdots &{{x_r} + x_r^{ - 1}}&1
\end{array}} \right| \nonumber \\
&= \left| {\begin{array}{*{20}{c}}
{{{\left( {{x_1} + x_1^{ - 1}} \right)}^{r - 1}}}& \cdots &{{x_1} + x_1^{ - 1}}&1\\
 \vdots &{}& \vdots & \vdots \\
{{{\left( {{x_r} + x_r^{ - 1}} \right)}^{r - 1}}}& \cdots &{{x_r} + x_r^{ - 1}}&1
\end{array}} \right| = V(x_1+x_1^{-1},\ldots,x_{r}+x_{r}^{-1}) ,
\end{align}
where \(V(y_1,\dots,y_r)\) is the Vandermonde determinant
\begin{equation}
V(y_1,\dots,y_r) = \prod_{1 \le i < j \le r}(y_i - y_j) .
\end{equation}
Using $y_i-y_j=x_i\left(1-x_i^{-1}x_j\right)\left(1-x_i^{-1}x_j^{-1}\right)$, one gets
\begin{equation}
A_{\r}^{SO(2r)} = x_1^{r-1}x_2^{r-2}\dots x_{r-1} \prod_{1 \le i < j \le r}\big(1-x_i^{-1}x_j^{-1}\big)\big(1-x_i^{-1}x_j\big) ,
\end{equation}
which manifestly agrees with eq.~\eqref{eq:Weyl_denom_form}. Putting it all together, the character for an irreducible representation of \(SO(2r)\) labeled by \(l\) is given by
\begin{equation}
\ch_l^{(2r)}(x) = \frac{\frac{1}{2}\left( \text{det}\big[x_i^{\l_j}+x_{i}^{-\l_j}\big] +\text{det}\big[x_i^{\l_j}-x_{i}^{-\l_j}\big] \right) }{V(x_1+x_1^{-1},\ldots,x_{r}+x_{r}^{-1})} ,
\end{equation}
with \(\l_j=\rho_j+l_j=r-j+l_j\).

For odd dimensions, \(d = 2r+1\), the Weyl group consists of permutations together with any number of sign flips: \(W_{SO(2r+1)} = S_r \ltimes Z_2^r\). The half-sum on positive roots is \(\r = (r - \frac{1}{2},r- \frac{3}{2},\dots, \frac{1}{2})\). Following a similar procedure to the above, we get~\cite{FultonHarris}
\begin{equation}
A_{\r + l}^{SO(2r+1)} = \text{det}\big[x_i^{\l_j}-x_{i}^{-\l_j}\big] ,
\end{equation}
where \(\l=\r+l\) so that \(\l_j=r-j+\frac{1}{2}+l_j\). When specifying to $l=(0,\cdots,0)$, we get
\begin{equation}
A_{\r}^{SO(2r+1)} = V(x_1+x_1^{-1},\ldots,x_{r}+x_{r}^{-1})\big(x_1^{\frac{1}{2}}-x_1^{-\frac{1}{2}}\big)\ldots\big(x_r^{\frac{1}{2}}-x_r^{-\frac{1}{2}}\big)
\end{equation}
Combining it all together, the character is given by
\begin{equation}
\ch_l^{(2r+1)}(x) = \frac{\text{det}\big[x_i^{\l_j}-x_{i}^{-\l_j}\big] }{V(x_1+x_1^{-1},\ldots,x_{r}+x_{r}^{-1}) \prod_{i=1}^r\big(x_i^{\frac{1}{2}} - x_i^{-\frac{1}{2}}\big)},
\end{equation}
with \(\l_j=r-j+\frac{1}{2}+l_j\).

\subsection*{Explicit formulas for certain representations}

For reference, here we record the specific characters of \(SO(d)\) irreps that are either utilized frequently in this work or may be useful for Hilbert series calculations.

The character of the vector representation, \(l = (1,0,\dots,0)\), in even and odd dimensions is
\begin{subequations}
\begin{align}
\ch_{(1,0,\dots,0)}^{(2r)}(x) &= \sum_{i=1}^r\Big( x_i + \frac{1}{x_i} \Big),\\
\ch_{(1,0,\dots,0)}^{(2r+1)}(x) &= 1+ \sum_{i=1}^r\Big( x_i + \frac{1}{x_i} \Big).
\end{align}
\end{subequations}

The properties of spinors depends on \(d\) mod 8. In odd dimensions, spinors are either real or pseudo-real; therefore, the characters of spinors in odd dimensions are always self-conjugate. The \(d = 3\), 5, 7, and 9 spin \(1/2\) characters are
\small{\begin{equation}
\def\arraystretch{1.8} \begin{array}{lrcl} d=3: \ & \ch_{(\frac{1}{2})}^{(3)} & = & \sqrt{x} \ + \ \text{c.c.}, \\
d=5: \ & \ch_{(\frac{1}{2},\frac{1}{2})}^{(5)} & = & \sqrt{x_1x_2} + \sqrt{\dfrac{x_1}{x_2}} \ + \ \text{c.c.}, \\
d=7: \ & \ch_{(\frac{1}{2},\frac{1}{2},\frac{1}{2})}^{(7)} & = & \sqrt{x_1x_2x_3} + \sqrt{\dfrac{x_1x_2}{x_3}} + \sqrt{\dfrac{x_1x_3}{x_2}} + \sqrt{\dfrac{x_2x_3}{x_1}} \ + \ \text{c.c.}, \\
d=9: \ & \ch_{(\frac{1}{2},\frac{1}{2},\frac{1}{2},\frac{1}{2})}^{(9)} & = & \sqrt{x_1x_2x_3x_4} + \sqrt{\dfrac{x_1x_2x_3}{x_4}} + \sqrt{\dfrac{x_1x_2x_4}{x_3}}+ \sqrt{\dfrac{x_1x_3x_4}{x_2}} \  \\
\ & \ & \ &+ \sqrt{\dfrac{x_2x_3x_4}{x_1}}+ \sqrt{\dfrac{x_1x_2}{x_3x_4}} + \sqrt{\dfrac{x_1x_3}{x_2x_4}}  + \sqrt{\dfrac{x_1x_4}{x_2x_3}}+ \ \text{c.c.},
\end{array}
\end{equation}}\normalsize
where \(+~\text{c.c.}\) means \(+~\text{complex conjugate}\), \(x_i^* = x_i^{-1}\). In even dimensions, spinors are chiral. These spinors may be real, pseudo-real, or complex. They are complex in \(d=2 \text{ mod } 8\) and \(d = 6 \text{ mod } 8\) (\textit{i.e.} \(d=2 \text{ mod } 4\)); in this case the characters for the chiral spinors are conjugate: \(\ch^{(4k+2)}_{(l_1,\dots,-l_{2k+1})} = \big(\ch^{(4k+2)}_{(l_1,\dots,l_{2k+1})}\big)^*\). In \(d = 4 \text{ mod } 8\) spinors are pseudo-real, while they are real in \(d = 8 \text{ mod } 8\); in both cases, the characters for the chiral representations are self-conjugate: \(\ch^{(4k+4)}_{(l_1,\dots,\pm l_{2k+2})} = \big(\ch^{(4k+4)}_{(l_1,\dots,\pm l_{2k+2})}\big)^*\). One readily sees these properties in the equation below, which gives the characters for spinors in \(d = 2\), 4, 6, and 8:
\small{\begin{equation}
\def\arraystretch{1.8} \begin{array}{ll}
d= 2: \ & ~~~~~~ \begin{array}{rcl} \ch_{(\frac{1}{2})}^{(2)} & = & \sqrt{x} \\ \ch_{(-\frac{1}{2})}^{(2)} & = & \dfrac{1}{\sqrt{x}} \end{array} \\
d= 4: \ & ~~~~ \begin{array}{rcl} \ch_{(\frac{1}{2},\frac{1}{2})}^{(4)} & = & \sqrt{x_1x_2} \ + \ \text{c.c.} \\ \ch_{(\frac{1}{2},-\frac{1}{2})}^{(4)} & = & \sqrt{\dfrac{x_1}{x_2}} \ + \ \text{c.c.} \end{array} \\
d= 6: \ & ~~ \begin{array}{rcl} \ch_{(\frac{1}{2},\frac{1}{2},\frac{1}{2})}^{(6)} & = & \sqrt{x_1x_2x_3} + \sqrt{\dfrac{x_1}{x_2x_3}} + \sqrt{\dfrac{x_2}{x_1x_3}}  + \sqrt{\dfrac{x_3}{x_1x_2}} \\  \ch_{(\frac{1}{2},\frac{1}{2},-\frac{1}{2})}^{(6)} & = & \dfrac{1}{\sqrt{x_1x_2x_3}} + \sqrt{\dfrac{x_2x_3}{x_1}} + \sqrt{\dfrac{x_1x_3}{x_2}}  + \sqrt{\dfrac{x_1x_2}{x_3}}  \end{array} \\
d= 8: \ & \begin{array}{rcl} \ch_{(\frac{1}{2},\frac{1}{2},\frac{1}{2},\frac{1}{2})}^{(8)} & = & \sqrt{x_1x_2x_3x_4} + \sqrt{\dfrac{x_1x_2}{x_3x_4}} + \sqrt{\dfrac{x_1x_3}{x_2x_4}}  + \sqrt{\dfrac{x_1x_4}{x_2x_3}} \ + \ \text{c.c.}\\  \ch_{(\frac{1}{2},\frac{1}{2},\frac{1}{2},-\frac{1}{2})}^{(8)} & = &  \sqrt{\dfrac{x_1}{x_2x_3x_4}} + \sqrt{\dfrac{x_2}{x_1x_3x_4}} + \sqrt{\dfrac{x_3}{x_1x_2x_4}} + \sqrt{\dfrac{x_4}{x_1x_2x_3}} \ + \ \text{c.c.} \end{array}
\end{array}
\end{equation}}\normalsize
As a side comment we remind the reader that if we include fermions we are actually working with the covering group \(Spin(d)\) of \(SO(d)\). Computations in this paper frequently involve integrating over the group using contour integrals; as discussed at the end of section~\ref{subsec:char_ortho}, we need to make sure to do this properly for the covering group. In practice, this is simply achieved by sending \(x_i \to x_i^2\), so that all the square roots in the above characters disappear.

Finally, we mention that the fundamental weights in the orthonormal basis have highest weight vectors \(l=(1,0,\dots,0),\ (1,1,0,\dots,0), \ (1,1,1,0,\dots,0),\) \textit{etc}. They correspond to the vector and anti-symmetric representations. For even \(d\), the $\frac{d}{2}$-forms have \(l=(1,1,1,\dots,\pm1)\).

\subsection*{Character formulae for $SU(r+1)$ and $Sp(2r)$}

We reproduce results found in \textit{e.g.}~\cite{FultonHarris} for character formulae in the orthogonal basis for the remaining classical groups.

The character for the representation of $SU(r+1)$ with highest weight vector $l=(l_1,\ldots,l_r)$ can be written as a ratio determinants of $(r+1)\times(r+1)$ matrices,
\be
\chi_l^{SU(r+1)}(x)=\frac{\text{det}\left[ x_j^{\overline{l}_i+r+1-i}\right]}{\text{det}\left[ x_j^{r+1-i}\right]} \,,
\ee
where $\overline{l}=(l_1,\ldots,l_r,0)$, and $x_{r+1}=\prod_{i=1}^r x_i^{-1}$. Leaving \(x_{r+1}\) independent, this is the character formula for \(U(r+1)\) and is a Schur polynomial.

The character for a representation of $Sp(2r)$ with highest weight vector $l=(l_1,\ldots, l_r)$  can be written as a ratio of determinants of $r\times r$ matrices,\be
\chi_l^{Sp(2r)}(x)=\frac{\text{det}\left[ x_j^{(l_i+r+1-i)}- x_j^{-(l_i+r+1-i)}\right]}{\text{det}\left[ x_j^{(r+1-i)}-x_j^{-(r+1-i)}\right]} \,.
\ee

\newpage
\section{Weyl integration formula}
\label{app:Weyl}

In this appendix we explain the Weyl integration formula, focusing especially on how the factor \(\abs{W}^{-1}\prod_{\a \in \text{rt}(G)}(1 - x^{\a})\) in eq.~\eqref{eq:weyl_int} appears. The Weyl integration formula is covered in many group theory texts; chapter IV of~\cite{Brocker:2003} gives a very thorough explanation. We hope to summarize and pseudo-derive some of the main features in a way that is more easily read by physicists. After doing this, we then explain how, for the cases that interest us, the usual Weyl integration formula essentially calculates the same thing \(\abs{W}\) times; we can remove this redundancy, which can significantly ease computations. Finally, we record the Haar measures, when restricted to the torus, for the classical groups \(SU(N)\), \(SO(N)\), and \(Sp(2N)\).

Let \(G\) be a compact, connected Lie group with \(\text{dim}(G) = n\) and \(\text{rank}(G) = r\). It is a fact, which we will not prove, that every element \(g \in G\) can be conjugated into a maximal torus \(T \subset G\), \textit{i.e.} \(g\) can be diagonalized by another group element: \(h^{-1}g h \in T\).

In fact, \(g\) is conjugate to multiple elements in \(T\). The different elements are basically related by permutations of the eigenvalues of \(g\). The discrete subgroup in \(G\) which carries out these permutations is called the Weyl group \(W\).\footnote{Precisely, let \(N(T)\) be the normalizer of \(T\) in \(G\),
\begin{equation}
N(T) = \{ g \in G \ | \ g T g^{-1} \in T \}, \nonumber
\end{equation}
\textit{i.e.} \(N(T)\) consists of all group elements which, acting by conjugation, leave elements of the torus within the torus. Obviously, \(N(T)\) contains \(T\) since the torus acts trivially on itself (as all elements of \(T\) commute). The elements which act non-trivially make up the Weyl group, \textit{i.e.} \(W = N(T)/T\). As is obvious by the definition, the elements of \(W\) belong to \(G/T\).} Thus, \(g\) is conjugate to \(\abs{W}\) elements in \(T\), where \(\abs{W}\) is the order of the Weyl group (the number of elements in \(W\)). For example, for \(G = SU(N)\) the Weyl group is the permutation group \(S_N\) with \(\abs{S_N} = N!\).

Let \(f\) be a function on the group. We consider averaging over this function,
\begin{equation}
\int_G d\m(g) \, f(g), \nonumber
\end{equation}
where \(d\m\) is the Haar measure on \(G\) normalized as \(\int_G d\m = 1\). We will elaborate on this measure below. Since every \(g\in G\) can be written as \(g = h t h^{-1}\), for \(t \in T\) and \(h \in G/T\), we can rewrite the above integral to be over \(T\) and \(G/T\),
\begin{equation}
\int_G d\m(g) \, f(g) \propto \int_T d\m(t) \int_{G/T}d\m(h) \, (\text{Jacobian}) f(h t h^{-1}),
\end{equation}
where there is a Jacobian factor from using \(g = h t h^{-1}\) and switching domains of integration from \(G\) to \(G/T \times T\). The proportionality constant is fixed by noting that the mapping \(G/T \times T \to G\), \((h,t) \mapsto h t h^{-1} \in G\) covers \(G\) \(\abs{W}\) times. Below we will explicitly compute the Jacobian. Upon doing so, we arrive at the Weyl integration formula,
\begin{equation}
\int_G d\m(g)\, f(g) = \frac{1}{\abs{W}} \int_T d\m(t) \det(1- \text{Ad}_t)|_{\mathfrak{g}/\mathfrak{t}} \int_{G/T}d\m(h) \, f(h t h^{-1}).
\label{eq:app_Weyl_int_form}
\end{equation}
In the Jacobian Ad is the adjoint map, which acts on the Lie algebra \(\mathfrak{g}\) by \(\text{Ad}_t(\mathfrak{g}) = t \mathfrak{g} t^{-1}\). Note that the determinant is restricted to the \(\mathfrak{g}/\mathfrak{t}\) subspace, where \(\mathfrak{t}\) is the Cartan subalgebra.\footnote{In plainer language, this is the determinant in the adjoint representation with zero's omitted.}

The Weyl integration formula is particularly useful when \(f\) is a class function, so that \(f(h t h^{-1}) = f(t)\). In this case, the integral over \(G/H\) is trivial and we have
\begin{equation}
\int_G d\m(g)\, f(g) = \frac{1}{\abs{W}} \int_T d\m(t) \left.\det(1- \text{Ad}_t)\right|_{\mathfrak{g}/\mathfrak{t}} \, f(t).
\label{eq:weyl_class_fxn}
\end{equation}
As \(T = U(1)^r\), the measure of the torus is simply
\begin{equation}
\int_T d\m(t) = \int_0^{2\pi} \prod_{i=1}^r d\th_i = \oint_{\abs{x_i}=1} \prod_{i=1}^r \frac{dx_i}{2\pi i x_i}.
\label{eq:measure_on_torus}
\end{equation}
The determinant is equal to the product over the roots of \(G\), \(\det(1- \text{Ad}_t)|_{\mathfrak{g}/\mathfrak{t}} = \Pi_{\a \in \text{rt}(G)} (1- x^{\a})\), so that the above coincides with the expression given in eq.~\eqref{eq:weyl_int}.

Let us now compute the Jacobian from the change of variables \(g = h t h^{-1}\). First, it is helpful to recall some facts about the Haar measure, which is the unique left and right invariant measure on \(G\). It can be constructed from the Maurer-Cartan forms
\begin{equation}
i \o \equiv g^{-1}d g \equiv \d g \, . \nonumber
\end{equation}
Note that \(\o = \o^A t^A\) is valued in the Lie algebra \(\mathfrak{g}\), where \(t^{A}\) are the generators of \(\mathfrak{g}\). The metric
\begin{equation}
ds^2 = \text{Tr}\big( \d g \, \d g^{-1} \big), \nonumber
\label{eq:G_metric}
\end{equation}
where \(\d g^{-1} = d g^{-1} g\), is invariant under left and right multiplication and therefore provides an invariant measure on the group.\footnote{\(\d g\) is invariant under left multiplication, \(g \to g' g\): \(\d g \to \d g\), while it is conjugated under right multiplication, \(g \to g g'\): \(\d g \to g'^{-1} \d g g'\). The trace ensures invariance under both left and right multiplication.} From \(0 = d(g^{-1}g)\) we have \(\d g^{-1} = - \d g\), so that the metric is
\begin{equation}
ds^2 = - \text{Tr}\big( \d g\, \d g \big) = \d^{AB} \o^A \o^B,
\end{equation}
where we have normalized the generators as \(\text{Tr}(t^{A}t^{B}) = \d^{AB}\). Just as the invariant volume on a metric space \(ds^2 = \et_{\m\n}dx^{\m}dx^{\n}\) is \(\sqrt{\abs{\det \et}} d^nx\), we have the invariant volume on the group
\begin{equation}
d\m_{G} = d^n\o = \o^1 \wedge \dots \wedge \o^n = \frac{1}{n!}\e_{A_1\dots A_n}\o^{A_1}\wedge \dots \wedge \o^{A_n} \, . \nonumber
\end{equation}

With the explicit parameterization of the measure on \(G\) by \(\d g\) it is straightforward to find the induced measure on \(G/T \times T\) when we change variables to \(g = h t h^{-1}\). We rewrite the measure as
\begin{equation}
ds^2 = \text{Tr} \big( \d g \, \d g^{-1} \big) = \text{Tr}\big( dg \, dg^{-1} \big) = \text{Tr}\big( h^{-1} dg h \, h^{-1} dg^{-1} h \big). \nonumber
\end{equation}
Inserting \(g = h t h^{-1}\) we have
\begin{align}
h^{-1} \, dg\, h &= h^{-1}\big( dh\, t\, h^{-1} + h \, dt\, h^{-1} + h\, t\, dh^{-1}\big) h \nonumber \\
&= \d h \,t + t\, \d h^{-1} + dt \nonumber \\
&= [\d h,t] + dt, \nonumber
\end{align}
where we defined \(\d h \equiv h^{-1} dh\) and used \(\d h^{-1} = - \d h\) in the last line. Similarly,
\begin{equation}
h^{-1} \, \d g^{-1}\, h = [\d h, t^{-1}] + dt^{-1}, \nonumber
\end{equation}
so that
\begin{equation}
ds^2 = \text{Tr}\big( [\d h, t][\d h, t^{-1}]\big) + \text{Tr}\big(dt \, dt^{-1} \big) + \text{Tr}\big([\d h,t] dt^{-1} \big) + \text{Tr}\big(dt [\d h, t^{-1}]\big) \,.
\label{eq:metric_four_terms}
\end{equation}
Consider the commutator
\begin{equation}
[\d h, t] = \big( \d h - t \, \d h \, t^{-1} \big) t \, . \nonumber
\end{equation}
While \(\d h \in \mathfrak{g}\) , the subtraction \(\d h - t \, \d h \, t^{-1}\) is valued in \(\mathfrak{g}/\mathfrak{t}\). This means the third trace in~\eqref{eq:metric_four_terms} vanishes since \(t\, dt^{-1} \in \mathfrak{t}\). Similarly, \(\text{Tr}\big( [\d h,t^{-1}] dt \big) = 0\).

We write
\begin{equation}
\d h - t \, \d h \, t^{-1} = [1 - \text{Ad}_t](\d h) \, , \nonumber
\end{equation}
where \(\text{Ad}_t(\d h) = t \, \d h \, t^{-1}\) is the adjoint action on the Lie algebra. Then the metric is given by
\begin{equation}
ds^2 = \text{Tr} \Big[ [1 - \text{Ad}_t](\d h) \, [1 - \text{Ad}_t](\d h^{-1}) \Big] + \text{Tr}\Big[ \d t \, \d t^{-1} \Big],
\end{equation}
where \(\d t \equiv t^{-1}dt\). Since \(\d t \in \mathfrak{t}\) while \([1-\text{Ad}_t](\d h) \in \mathfrak{g}/\mathfrak{t}\), we have achieved a parameterization of the algebra as the direct sum \(\mathfrak{g} = \mathfrak{t} \oplus \mathfrak{g}/\mathfrak{t}\). In matrix form the metric reads,
\begin{equation}
\et_{AB} = \begin{pmatrix} [1- \text{Ad}_t]^2 & \\ & 1 \end{pmatrix}, \nonumber
\end{equation}
and hence
\begin{equation}
\sqrt{\abs{\det \et}} = \det( 1- \text{Ad}_t ) |_{\mathfrak{g}/\mathfrak{t}} ,
\end{equation}
where the determinant is restricted to the \(\mathfrak{g}/\mathfrak{t}\) subspace.

With this Jacobian factor, we arrive at the Weyl integration formula
\begin{equation}
\int_G d\m(g)\, f(g) = \frac{1}{\abs{W}} \int_T d\m(t) \det(1- \text{Ad}_t)|_{\mathfrak{g}/\mathfrak{t}} \int_{G/T}d\m(h) \, f(h t h^{-1}), \nonumber
\end{equation}
where we recall that the \(\abs{W}^{-1}\) factor arises because the map \(G/T \times T \to G\): \((h,t) \mapsto h t h^{-1}\) is a \(\abs{W}\)-fold covering of \(G\). The Jacobian factor is the determinant in the adjoint representation, restricted to the \(\mathfrak{g}/\mathfrak{t}\) subspace. This means it is picking up the roots of \(G\), so that
\begin{equation}
\det(1- \text{Ad}_t)|_{\mathfrak{g}/\mathfrak{t}} = \prod_{\a \in \text{rt}(G)}(1- x^{\a}),
\end{equation}
where the product is over the roots of \(G\). Upon inserting this for the Weyl integration formula when \(f(g)\) is a class function, eq.~\eqref{eq:weyl_class_fxn}, we arrive at the formula quoted in the main text.

\subsection{A simplification when $f(g)$ is Weyl invariant}

Let us assume that the function we integrate over is Weyl invariant: \(f(h^{-1}gh) = f(g)\) for \(h \in W \subset G/T\). In this case the Weyl integration formula is, in some sense, redundant: since the \(G/T \times T \to G\) map is a \(\abs{W}\)-fold covering of \(G\), with each covering related by a Weyl transformation, we are picking up the same contribution \(\abs{W}\) times. To only pick up the contribution once, we can replace
\begin{equation}
\frac{1}{\abs{W}}\prod_{\a \in \text{rt}(G)}(1- x^{\a}) \to \prod_{\a \in \text{rt}_+(G)}(1- x^{\a})
\end{equation}
in the Weyl integration formula. Note that the product on the right hand side is over only the positive roots.

The key to proving the above replacement is to take the Jacobian factor, \(\prod_{\a \in \text{rt}(G)}(1- x^{\a}) = \prod_{\a \in \text{rt}_+(G)}(1- x^{\a})(1-x^{-\a})\), and show that it is equal to a sum over Weyl transformations of \(\prod_{\a \in \text{rt}_+(G)}(1-x^{\a})\),\footnote{To show this, use the Weyl denominator formula (see appendix~\ref{app:sodspin}) for the \(\prod_{\a \in \text{rt}_+(G)}(1-x^{-\a})\) factor:
\begin{align*}
\prod_{\a \in \text{rt}_+(G)}(1-x^{\a})(1-x^{-\a}) &= \sum_{w\in W}x^{-\r}(-1)^wx^{w(\r)} \prod_{\a \in \text{rt}_+(G)}(1-x^{\a}) \\
&= \sum_{w\in W} x^{w(\r)}\Big[(-1)^w\prod_{\a \in \text{rt}_+} \big(x^{-\frac{1}{2}\a} - x^{\frac{1}{2} \a} \big)\Big] \\
&= \sum_{w\in W} x^{w(\r)} \prod_{\a \in \text{rt}_+(G)}\big(x^{-\frac{1}{2}w(\a)} - x^{\frac{1}{2}w(\a)} \big) \\
&= \sum_{w\in W}\prod_{\a \in \text{rt}_+(G)}(1-x^{w(\a)}).
\end{align*}
In the first line we used the Weyl denominator formula; in the second line we used the definition eq.~\eqref{eq:def_rho} that \(\r\) is the half-sum of positive roots to bring \(x^{-\r}\) into the product; the third line follows from the action of the Weyl group on the root system (algebraically, the Weyl group is generated by reflections of the simple roots and it maps the root system into itself); to arrive at the fourth line we again used the definition of \(\r\) to bring a factor of \(x^{-w(\r)}\) out of the product.}
\begin{equation}
\prod_{\a \in \text{rt}_+(G)}(1- x^{\a})(1-x^{-\a}) = \sum_{w \in W}\prod_{\a \in \text{rt}_+(G)}(1- x^{w(\a)}). \nonumber
\end{equation}
If \(f(g)\) is Weyl invariant, then for each term in the sum above we can perform the inverse Weyl transformation so that we end up with \(\abs{W}\) copies of \(\prod_{\a \in \text{rt}_+(G)}(1- x^{\a})\).

Note that all class-functions are Weyl invariant; in particular, characters obey this property. Therefore, for all computations described in the main text we can do this replacement. Explicitly, if the function we integrate over is Weyl invariant, we can take the Haar measures to be
\begin{equation}
\int d\m_G =  \oint_{\abs{x_i}=1}\left[\prod_{i=1}^r \frac{dx_i}{2\pi i x_i} \right] \prod_{\a \in \text{rt}_+(G)}\big(1-x^{\a}\big).
\label{eq:weyl_int_pos_roots}
\end{equation}
This simplification can significantly ease computations, especially for large rank groups (on a computer, it essentially reduces computation time by a factor of \(\abs{W}\), which grows factorially with the rank of the group).

\subsection{The measures for the classical Lie groups}

The root systems for classical Lie groups are contained in any group theory textbook, so it is very simple to write down the Haar measures when restricted to the torus. For reference, we record these measures here. First we record them as the the product over all roots (with an overall \(\abs{W}^{-1}\) factor), and then as a product only over the positive roots (without the \(\abs{W}^{-1}\) factor). We use that the measure on the torus, \(d\m_T\), is as in eq.~\eqref{eq:measure_on_torus}; we omit the measure \(d\m_{G/T}\) below since it is never necessary for our analyses (since we only ever deal with class functions).

When restricted to the torus, the group measure is given as the measure on the torus times a product of the roots of the group.
\begin{equation}
\int d\m^{}_G =  \frac{1}{\abs{W}}\oint_{\abs{x_i}=1}\prod_{i=1}^r \frac{dx_i}{2\pi i x_i} ~ \prod_{\a \in \text{rt}(G)}\big(1-x^{\a}\big)
\end{equation}
\begin{description}
\item{\(SU(r + 1)\):} The Weyl group is \(S_{r+1}\) and the measure is
\begin{align}
d\m^{}_{SU(r+1)} &= \frac{1}{(r+1)!} \prod_{i=1}^r \frac{dx_i}{2\pi i x_i}~ \prod_{1\le i<j \le r+1} \Big(1- \frac{x_i}{x_j}\Big) \Big(1 - \frac{x_j}{x_i}\Big) \\
\text{where} ~~ x_{r+1} &\equiv \prod_{i=1}^r \frac{1}{x_i}. \nonumber
\end{align}
\item{\(SO(2r + 1)\):} The Weyl group is \(S_r\ltimes Z_2^r\) and the measure is
\begin{align}
d\m^{}_{SO(2r+1)} = \frac{1}{r!2^{r}} &\prod_{i=1}^r  \frac{dx_i}{2\pi i x_i}~\prod_{i=1}^r \Big(1-x_i\Big)\Big( 1- \frac{1}{x_i}\Big) \nonumber \\
\times& \prod_{1\le i<j\le r} \Big(1- x_ix_j\Big) \Big(1-\frac{x_i}{x_j}\Big) \Big(1- \frac{1}{x_ix_j}\Big) \Big(1 - \frac{x_j}{x_i}\Big)
\label{eq:haar_SO_odd}
\end{align}
\item{\(Sp(2r)\):} The Weyl group is \(S_r\ltimes Z_2^r\) and the measure is
\begin{align}
d\m^{}_{Sp(2r)} = \frac{1}{r!2^{r}} &\prod_{i=1}^r \frac{dx_i}{2\pi i x_i}~\prod_{i=1}^r \Big(1-x_i^2\Big)\Big( 1- \frac{1}{x_i^2}\Big) \nonumber \\
\times & \prod_{1\le i<j\le r} \Big(1- x_ix_j\Big) \Big(1-\frac{x_i}{x_j}\Big) \Big(1- \frac{1}{x_ix_j}\Big) \Big(1 - \frac{x_j}{x_i}\Big)
\label{eq:haar_Sp2r}
\end{align}
\item{\(SO(2r)\):} The Weyl group is \(S_r\ltimes (Z_2^r/Z_2)\) and the measure is
\begin{align}
d\m^{}_{SO(2r)} = \frac{1}{r!2^{r-1}} &\prod_{i=1}^r \frac{dx_i}{2\pi i x_i} \nonumber \\
\times & \prod_{1\le i<j\le r} \Big(1- x_ix_j\Big) \Big(1-\frac{x_i}{x_j}\Big) \Big(1- \frac{1}{x_ix_j}\Big) \Big(1 - \frac{x_j}{x_i}\Big)
\label{eq:haar_SO_even}
\end{align}
\end{description}

Restricting only to the positive roots, as in eq.~\eqref{eq:weyl_int_pos_roots}, the measures for the classical groups are
\begin{subequations}
\label{eq:haar_classical_pos_roots}
\begin{align}
d\m^{}_{SU(r+1)} &=  \prod_{i=1}^r \frac{dx_i}{2\pi i x_i} ~ \prod_{1 \le i<j \le r+1} \Big(1- \frac{x_i}{x_j}\Big),  \\
d\m^{}_{SO(2r+1)} &= \prod_{i=1}^r \frac{dx_i}{2\pi i x_i}~\prod_{i=1}^r \Big(1-x_i\Big) ~\prod_{1\le i<j \le r} \Big(1- x_ix_j\Big) \Big(1-\frac{x_i}{x_j}\Big) ,\\
d\m^{}_{Sp(2r)} &= \prod_{i=1}^r \frac{dx_i}{2\pi i x_i}~\prod_{i=1}^r \Big(1-x_i^2\Big)~\prod_{1\le i<j\le r} \Big(1- x_ix_j\Big) \Big(1-\frac{x_i}{x_j}\Big) ,\\
d\m^{}_{SO(2r)} &= \prod_{i=1}^r \frac{dx_i}{2\pi i x_i} ~ \prod_{1\le i<j\le r} \Big(1- x_ix_j\Big) \Big(1-\frac{x_i}{x_j}\Big),
\end{align}
\end{subequations}
where, as above, for \(SU(r+1)\) we define \(x_{r+1} \equiv \prod_{i=1}^r x_i^{-1}\).

\newpage
\section{Parity}
\label{app:parity}

In this appendix, we show how to address parity in even dimensions $d=2r$.\footnote{In odd dimensions, parity is simpler to deal with since it can be taken to commute with rotations; the relevant details were explained in section~\ref{subsec:parity}.} We start in section~\ref{appsubsec:preliminary} by discussing the general irreducible representations (irreps) and characters of $O(2r)$~\cite{Weyl:classical}. Parity acts as an outer automorphism of the Lie algebra $\mathfrak{so}_{2r}$; we show in section~\ref{appsubsec:folding} how the parity odd characters arise by \textit{folding} $\mathfrak{so}_{2r}$ by the outer automorphism. We also discuss two notions of folding which are not typically made clear in the literature. In section~\ref{appsubsec:HilbertParity}, we describe how to compute the Hilbert series when including parity as a symmetry. A central result is a plethystic exponential formula for the determinant \(\nobreak{[\text{det}_l^{}(1-a g)]^{-1}}\) in an arbitrary representation \(l=(l_1,\ldots,l_r)\), eq.~\eqref{eqn:parityDetapp}. Finally, we give two explicit computation examples in section~\ref{appsubsec:examples}---the real scalar field in $d=2r$ and the gauge field in $d=4$.

\subsection{Representations and characters of $O(2r)$}\label{appsubsec:preliminary}

We adopt the convention that the parity element $\mathcal{P}$ flips the last component of the vector, \textit{i.e.} its representation matrix is $\r_\Box^{}(\mathcal{P}) = \text{diag}(1,\ldots,1,-1)$. \(\mathcal{P}\) does not commute with generic rotations, and hence the orthogonal group $O(2r)$ is a semidirect product of its two subgroups $SO(2r)$ and $Z_2=\{e,\mathcal{P}\}$, \textit{i.e.} $O(2r) = SO(2r) \ltimes Z_2$. One can segment $O(2r)$ by the cosets of its subgroup $SO(2r)$, which yields two connected components $O(2r)=\{O_+(2r)\equiv SO(2r), O_-(2r)\equiv SO(2r)\mathcal{P}\}$. In other words, \(g_- \in O_-(2r)\) can be taken in the form \(g_- = g_+\mathcal P\) with \(g_+ \in SO(2r)\).\footnote{We note that the results in this appendix apply for spinor representations as well. We will not be careful to distinguish between \(O(2r)\) and its covering group \(Pin(2r)\) (or the fact that there is a choice in this covering). For the character theory required here, it is enough to take the \(Spin(2r)\) characters together with the definition of parity as an outer-automorphism of the Lie algebra.}

\subsubsection*{The defining representation and the torus of the parity odd component}

The defining/vector representation matrix $\r_\Box^{}(g_-)$ for any parity odd element $g_-$ can be brought to the form
\begin{equation}
  \left(
    \begin{array}{ccccc}
      \left( \begin{array}{cc}
        c_1 & s_1\\
        -s_1 & c_1
      \end{array} \right)
      & 0 & \cdots & 0 & 0\\
      0 & \left( \begin{array}{cc}
        c_2 & s_2\\
        -s_2 & c_2
      \end{array} \right) & \cdots & 0 & 0\\
      \vdots & \vdots & \ddots & \vdots & \vdots \\
      0 & 0 & \cdots & \left( \begin{array}{cc}
        c_{r-1} & s_{r-1}\\
        -s_{r-1} & c_{r-1}
      \end{array} \right) & 0 \\
      0 & 0 & \cdots & 0 & \left( \begin{array}{cc}
        1 & 0\\
        0 & -1
      \end{array} \right)
    \end{array}
  \right)\label{eqn:skewdiagonal}
\end{equation}
by an orthogonal transformation, \textit{i.e.} by a group conjugation, where $c_i = \cos\theta_i$, $s_i = \sin\theta_i$. The corresponding eigenvalues are
\begin{equation}
x_1, x_1^{-1}, x_2, x_2^{-1}, \ldots, x_{r-1}, x_{r-1}^{-1}, +1, -1 \, , \label{eqn:parityEigenvaluesapp}
\end{equation}
with $x_i = e^{i\theta_i}$. An immediate consequence is that the character $\ch_l^-\equiv\text{Tr}_l^{}(g_-)$ only depends on these $(r-1)$ arguments, which we collectively denote by
\begin{equation}
\tilde{x} \equiv (x_1,\ldots, x_{r-1}) . \label{eqn:xtildedefapp}
\end{equation}

\subsubsection*{General irreps and characters}

The irreps of $O(2r)$ can be induced from the irreps of its subgroup $SO(2r)$ (\textit{e.g.}~\cite{Brocker:2003}). Instead of following this standard technical procedure, let us understand it in a more heuristic way.

Consider the $SO(2r)$ representation space $R_l$ labeled by $l = (l_1, \ldots, l_r)$. When $l_r \ne 0$, parity exchanges the two chiral spaces $R_{(l_1, \ldots, l_{r-1}, l_r)}$ and $R_{(l_1, \ldots, l_{r-1},-l_r)}$, and hence the direct sum of the two is an irrep of $O(2r)$. In this case, the parity even character $\ch_l^+$ is a sum of the two $SO(2r)$ characters, while the parity odd character $\ch_l^-$ vanishes as its representation matrix is off block diagonal.

When $l_r=0$, $R_l$ itself forms an irrep of $O(2r)$, with parity exchanging chiral subspaces within $R_l$. We can assign an intrinsic parity to these representations---as an \(SO(2r)\) representation, $R_l$ actually induces two inequivalent irreps of $O(2r)$, \textit{e.g.} scalar vs pseudo-scalar, vector vs pseudo-vector, \textit{etc}. In this case, the parity even character $\ch_l^+$ is the $SO(2r)$ character, while the parity odd character $\ch_l^-$ coincides with a character of \(Sp(2r-2)\) (up to an overall sign reflecting the intrinsic parity assignment), a result worked out a Weyl ago~\cite{Weyl:classical}. This coincidence might be a bit mysterious; in the next subsection, we explain how it arises from folding the Lie algebra $\mathfrak{so}_{2r}$.

In summary, the general irreducible representations of $O(2r)$ are labeled by $l=(l_1,\ldots,l_r)$ with $l_1\ge \cdots \ge l_r\ge 0$,
\begin{subequations}
\label{eqn:Odirreps}
\begin{alignat}{2}
l_r>0 : && \quad R_{(l_1,\ldots,l_{r-1},l_r)}^{O(2r)} &= R_{(l_1,\ldots,l_{r-1},l_r)}^{SO(2r)} \oplus R_{(l_1,\ldots,l_{r-1},-l_r)}^{SO(2r)} , \\
l_r=0 : && \quad R_{(l_1,\ldots,l_{r-1},  0)}^{O(2r)} &= R_{(l_1,\ldots,l_{r-1},0)}^{SO(2r)} \hspace{3mm}\text{with $\pm$ intrinsic parity} ,
\end{alignat}
\end{subequations}
with corresponding characters
\begin{subequations}
\label{eqn:Odcharacters}
\begin{alignat}{4}
l_r>0 : && \quad \ch_l^+(x) &=\ch_{(l_1,\cdots,l_r)}^{}(x)+\ch_{(l_1,\ldots,-l_r)}^{}(x) &\quad,\quad&& \ch_l^-(\tilde{x}) &=0 , \\
l_r=0 : && \quad \ch_l^+(x) &=\ch_l^{}(x) &\quad,\quad&& \ch_l^-(\tilde{x}) &=\pm \ch_{(l_1,\ldots,l_{r-1})}^{Sp(2r-2)} (\tilde{x}) .
\end{alignat}
\end{subequations}
Throughout this appendix, we use the label $l$ to denote the $O(2r)$ irrep with positive intrinsic parity assignment. It is straightforward to modify formulae for the case of a negative intrinsic parity assignment (see the discussion in section~\ref{subsec:parity}).

\subsection{Understanding the parity odd character: folding $\mathfrak{so}_{2r}$} \label{appsubsec:folding}

At first glance it is rather surprising that the symplectic group shows up in the character formulae for $O_-(2r)$, especially considering that \(Sp(2r-2)\) is not a subgroup of \(O(2r)\) except for \(r\leq 2\). In this subsection, we provide an understanding of this.

\subsubsection*{Computing characters from the weights}

Consider the parity even character $\ch_l^+(x)$, which is the trace of $g_+$ in the $O(2r)$ representation $R_l$. In the basis with the Cartan generators $H_k$ ($k=1,2,\ldots,r$) diagonal, each state $\ket{\mu}$ is labeled by its eigenvalues $\mu_k$ under $H_k$, called the \textit{weight} $\mu=(\mu_1,\mu_2,\ldots,\mu_r)$, namely
\begin{equation}
H_k \ket{\mu} = \mu_k \ket{\mu} .
\end{equation}
The character is given by a sum over all the weights $\mu$ in the representation $R_l$:
\begin{equation}
\ch_l^+(x) = \text{Tr}_l^{}(g_+) = \sum_{\mu\in R_l} \bra{\mu} e^{i\sum_{k=1}^r \theta_k H_k} \ket{\mu} = \sum_{\mu\in R_l} x^\mu ,
\end{equation}
with $x^\mu \equiv x_1^{\mu_1}\cdots x_r^{\mu_r}$ and $x_k=e^{i\theta_k}$ as usual. Using $g_-=g_+\mathcal{P}$, the parity odd character $\ch_l^-(\tilde{x})$ is
\begin{equation}
\ch_l^-(\tilde{x}) =\text{Tr}_l^{}(g_+\mathcal{P}) = \sum_{\m \in R_l} \bra{\m} e^{i\sum_{k=1}^r \th_k H_k} \mathcal{P} \ket{\m} .
\end{equation}
As a $Z_2$ action, parity either acts trivially on a state, or exchanges two different $\ket{\mu}$. Since $\left\langle {\mu} \mathrel{\left | {\vphantom {\mu{\mu'}}} \right. \kern-\nulldelimiterspace}{{\mu'}} \right\rangle=\delta_{\mu\mu'}$, only states invariant under \(\mathcal{P}\) contribute to the character $\ch_l^-(\tilde{x})$:
\begin{equation}
\ch_l^-(\tilde{x}) = \sum_{\m \in R_l^{\mathcal{P}}} \bra{\m} e^{i\sum_{k=1}^r \th_k H_k} \ket{\m} = \sum_{\m \in R_l^{\mathcal{P}}} x^\m ,
\end{equation}
where $R_l^{\mathcal{P}}$ denotes the set of states invariant under $\mathcal{P}$.

In order to compute $\ch_l^-(\tilde{x})$, we therefore need to identify all the states invariant under parity. For $l_r>0$ this is fairly easy: because $R_l$ is a direct sum of two chiral spaces (see eq.~\eqref{eqn:Odirreps}) that get exchanged under parity, there is simply no state invariant under parity. Hence, $\ch_l^-(\tilde{x})=0$ for $O(2r)$ irreps with $l_r>0$. The case of $l_r=0$ is more complicated and requires a closer check on how the states in $R_l$ transform under parity. This is achieved by studying the action of parity on the root system of $\mathfrak{so}_{2r}$, which we explain below.

\subsubsection*{Root system of $\mathfrak{so}_{2r}$ and the parity outer automorphism}

\begin{figure}
\centering
\subfigure[]{
\begin{tikzpicture}\label{fig:ff}
	\draw (0,0) -- (2.7,0);
	\draw (3.3, 0) -- (4,0);
	\draw (4,0) -- (5,-.5);
	\draw (4,0) -- (5,.5);
	\draw[fill=white] (0,0) circle(.1);
	\draw[fill=white] (1,0) circle(.1);
	\draw[fill=white] (2,0) circle(.1);
	\draw[fill=white] (4,0) circle(.1);
	\draw[fill=white] (5,-.5) circle(.1);
	\draw[fill=white] (5,.5) circle(.1);
	\node at (3,0) {$\cdots$};
	\node at (-1,0) {$D_r$};
\end{tikzpicture}
\label{fig:Dn}
}\\
\subfigure[]{
\begin{tikzpicture}
	\draw (0,0) -- (2.7,0);
	\draw (3.3, 0) -- (4,0);
	\draw (4,0.1) -- (5,0.1);
	\draw (4,-0.1) -- (5,-0.1);
	\draw (4.4,0.2) -- (4.6,0);
	\draw (4.4,-0.2) -- (4.6,0);		
	\draw[fill=white] (0,0) circle(.1);
	\draw[fill=white] (1,0) circle(.1);
	\draw[fill=white] (2,0) circle(.1);
	\draw[fill=white] (4,0) circle(.1);
	\draw[fill=white] (5,0) circle(.1);	
	\node at (3,0) {$\cdots$};
	\node at (-1,0) {$B_{r-1}$};	
\end{tikzpicture}
\label{fig:Bn}
}\\
\subfigure[]{
\begin{tikzpicture}
	\draw (0,0) -- (2.7,0);
	\draw (3.3, 0) -- (4,0);
	\draw (4,0.1) -- (5,0.1);
	\draw (4,-0.1) -- (5,-0.1);
	\draw (4.4,0) -- (4.6,0.2);
	\draw (4.4,0) -- (4.6,-0.2);		
	\draw[fill=white] (0,0) circle(.1);
	\draw[fill=white] (1,0) circle(.1);
	\draw[fill=white] (2,0) circle(.1);
	\draw[fill=white] (4,0) circle(.1);
	\draw[fill=white] (5,0) circle(.1);	
	\node at (3,0) {$\cdots$};
	\node at (-1,0) {$C_{r-1}$};
\end{tikzpicture}
\label{fig:Cn}
}
\caption{Relevant Dynkin diagrams: (a) The $D_r$ Dynkin diagram possesses a reflection symmetry about the horizontal axis. This symmetry corresponds to the outer automorphism of the $\mathfrak{so}_{2r}$ Lie algebra under the parity transformation. (b) The Dynkin diagram $B_{r-1}$ (corresponding to Lie algebra $\mathfrak{so}_{2r-1}$) obtained by folding the \textit{roots} of $\mathfrak{so}_{2r}$ by the outer automorphism. (c) The Dynkin diagram of $C_{r-1}$ (corresponding to Lie algebra $\mathfrak{sp}_{2r-2}$) obtained by folding the \textit{co-roots} of $\mathfrak{so}_{2r}$ by the outer automorphism.}
\end{figure}
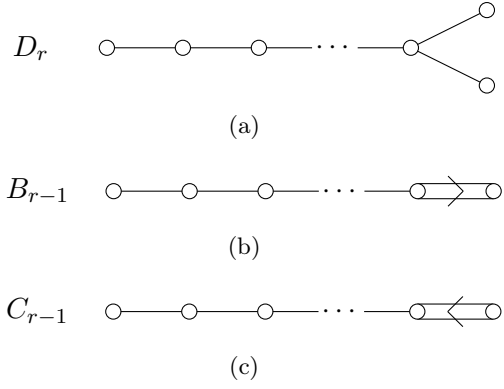

We adopt the following convention. The simple roots for $\mathfrak{so}_{2r}$ are $\alpha_i=e_i-e_{i+1}$ for $i=1,\ldots,r-1$ and $\alpha_r=e_{r-1}+e_r$, \textit{i.e.}
\begin{eqnarray}
  \alpha_1 &=& (1, -1, 0, \ldots, 0, 0, 0)\nonumber\\
  \alpha_2 &=& (0, 1, -1, \ldots, 0, 0, 0)\nonumber\\
  & & \vdots \nonumber\\
  \alpha_{r-2} &=& (0, 0, 0, \ldots, 1, -1, 0)\nonumber\\
  \alpha_{r-1} &=& (0, 0, 0, \ldots, 0, 1, -1)\nonumber\\
  \alpha_{r} &=& (0, 0, 0, \ldots, 0, 1, +1) ,
\label{eq:simpleroots}
\end{eqnarray}
with the complete set of roots given by $\pm e_i \pm e_j$ and $\pm e_i \mp e_j$, $i\ne j$. Parity exchanges the last two simple roots, which corresponds to the reflection of the Dynkin diagram $D_r$ around the horizontal axis (see fig.~\ref{fig:Dn}), an outer automorphism of the Lie algebra $\mathfrak{so}_{2r}$:
\begin{equation}
\mathcal{P}: \alpha_{r-1} \leftrightarrow \alpha_r .
\end{equation}

\subsubsection*{Strings of lowering operators and folding $\mathfrak{so}_{2r}$}

In an $O(2r)$ irrep with $l_r=0$, the highest weight state $l=(l_1,\ldots,l_{r-1},0)$ is invariant under parity, $\mathcal{P}\ket{l}=\ket{l}$. Starting from this highest weight state, all the other states are obtained by applying a string of lowering operators $E_i$ associated with the simple roots $\alpha_i$:
\begin{equation}
\ket{\mu} = E_1^{m_1} E_2^{m_2} \cdots E_r^{m_r} \ket{l} .
\end{equation}
We wish to determine if $\mathcal{P}\ket{\mu}=\ket{\mu}$. Since $\mathcal{P}\ket{l}=\ket{l}$, this amounts to studying if the string of lowering operator is invariant under the conjugation:
\begin{equation}
\mathcal{P} E_1^{m_1} E_2^{m_2} \cdots E_r^{m_r} \mathcal{P}^{-1} .
\end{equation}
From the action of parity on the simple roots, we have
\begin{subequations}
\begin{align}
\mathcal{P} E_i \mathcal{P}^{-1} &= E_i , \quad i=1,2,\ldots,r-2 , \\
\mathcal{P} E_{r-1} \mathcal{P}^{-1} &= E_r , \\
\mathcal{P} E_r \mathcal{P}^{-1} &= E_{r-1} .
\end{align}
\end{subequations}
Note that $E_{r-1}$ and $E_r$ commute with each other, and they both commute with all $E_i$ except $E_{r-2}$, as is obvious from the Dynkin diagram. It then follows that a parity invariant string must contain an equal number of $E_{r-1}$ and $E_r$ in between $E_{r-2}$. Namely $E_{r-1}$ and $E_r$ \textit{pair up}
\begin{equation}
\cdots \left(E_{r-1}E_r\right)^{s_1} E_{r-2}^{t_1} \cdots \left(E_{r-1}E_r\right)^{s_2} E_{r-2}^{t_2} \cdots .
\label{eq:string_pair_up}
\end{equation}
The crucial feature is that $E_{r-1}$ and $E_r$ only ever enter in the combination $\left(E_{r-1}E_r\right)$. They therefore act together, effectively giving the combined lowering operator $\tilde{E}_{r-1}$ associated to the new simple root
\begin{equation}
\tilde{\alpha}_{r-1} = \alpha_{r-1} + \alpha_r = (0, \ldots, 0, 2, 0) .
\end{equation}
So in the procedure of finding the parity invariant states from the highest weight state, we effectively work with a new set of \textit{parity symmetrized} simple roots $\alpha_1, \ldots, \alpha_{r-2}, \tilde{\alpha}_{r-1}$. This is precisely the root system for \(\mathfrak{sp}_{2r-2}\) shown in fig.~\ref{fig:Cn} (note that $\tilde{\alpha}_{r-1}$ is \textit{long} relative to the other roots). Pictorially, we ``fold'' the Dynkin diagram for \(\mathfrak{so}_{2r}\) (fig.~\ref{fig:Dn}) and obtain the diagram for \(\mathfrak{sp}_{2r-2}\). This is why we have $\ch_{(l_1,\ldots,l_{r-1},0)}^-(\tilde{x})=\ch_{(l_1,\ldots,l_{r-1})}^{Sp(2r-2)}(\tilde{x})$.

\subsubsection*{Two definitions of folding}

The folding of a Dynkin diagram is a procedure to use an outer automorphism of a Dynkin diagram and hence its root system to obtain a new root system. It is discussed, for example, in the \href{https://en.wikipedia.org/wiki/Dynkin_diagram#Folding}{Wikipedia page}. However, it is probably less known in the community that there are two kinds of folding one can define~\cite{saito1985extended}.\footnote{We additionally found some math.stackexchange posts helpful (\href{http://mathoverflow.net/questions/3888/folding-by-automorphisms}{one}, \href{http://mathoverflow.net/questions/111469/dual-versions-of-folding-symmetric-ade-dynkin-diagrams}{two}) as well as a set of notes~\cite{Stembridge}. We point out that Kostant~\cite{Kostant:1961} generalized Weyl's character formula to disconnected groups.} 
Operationally, the two versions correspond to averaging over an outer automorphism versus taking orbits of the automorphism. The two procedures are dual to each other: doing one on the roots is equivalent to doing the other on the co-roots.

The version in Wikipedia is to find a Lie subalgebra that is invariant under the outer automorphism. Therefore we are looking for a smaller group invariant under the outer automorphism. For the root system, it corresponds to taking the \textit{average} of the orbit of the roots under the outer automorphism,
\begin{equation}
\tilde{\alpha}_i = \frac{1}{n} \sum_{i'=1}^n \alpha_{i'} \, ,
\end{equation}
where $\alpha_{i'}$ denote all the simple roots that $\alpha_i$ can transform into under the outer automorphism, and $n$ is the total number of them. Note that root vectors that are transformed among each other by the outer automorphism are supposed to be orthogonal to each other. The nontrivial new simple root has $n=2$ for most cases ($A_n$, $D_{n\ne4}$, $E_6$), and $n=3$ for $D_4$.

In our case of $\mathfrak{so}_{2r}$, the subalgebra invariant under parity is obviously $\mathfrak{so}_{2r-1}$. The outer automorphism flips the sign of the last component of the root vectors. Indeed, the averaging affects the simple roots $\alpha_{r-1}$ and $\alpha_r$ in eq.~(\ref{eq:simpleroots}) and we find a new simple root
\begin{equation}
\tilde{\alpha}_{r-1} = \frac{1}{2} (\alpha_{r-1}+\alpha_r) = (0, \ldots, 0, 1, 0) .
\end{equation}
This is a short root. \(\tilde{\alpha}_{r-1}\) together with \(\a_1,\ldots,\a_{r-2}\) form the basis for the root lattice of \(\mathfrak{so}_{2r-1}\), namely $B_{r-1}$ shown in fig.~\ref{fig:Bn}. Note that this is a {\it new}\/ root system, not a sublattice of the original root lattice because of the non-integer coefficients.

There is an alternative way to define the folded Dynkin diagram and the corresponding root (weight) lattice. We take the sublattice of the root and weight lattices that consist of vectors \textit{invariant} under the outer automorphism. This is what we need if we want to include the outer automorphism as a part of the group, namely the group is extended by the outer automorphism, and we want to identify the weights that contribute to the characters. For this purpose, we introduce the new set of simple roots by the sum, without averaging:
\begin{equation}
\tilde{\alpha}_i = \sum_{i'=1}^n \alpha_{i'} \, .
\end{equation}
Obviously they are on the original root lattice because the coefficients are integers. Namely all vectors generated by the new set of simple roots $\tilde{\alpha}$ are subset of the original root lattice, and these vectors are invariant under the outer automorphism. Correspondingly, all the weights in the new sublattice are invariant.

This second procedure picks up all the weights in a given representation that are invariant under the outer automorphism and hence contribute to the character.  Therefore, this is the correct version of folding relevant to our discussions.  The new simple root is then a \textit{long} root, and $D_{r}$ folds to $C_{r-1}$ shown in fig.~\ref{fig:Cn}, namely \(\mathfrak{sp}_{2r-2}\). Note that \(\mathfrak{sp}_{2r-2}\) is not a subalgebra of \(\mathfrak{so}_{2r}\). Yet the characters are given by the weights of the \(\mathfrak{sp}_{2r-2}\) weight lattice. This is why $\ch_{(l_1,\ldots,l_{r-1},0)}^-(\tilde{x})=\ch_{(l_1,\ldots,l_{r-1})}^{Sp(2r-2)}(\tilde{x})$.

It is easy to see that two definitions of folding lead to dual lattices.  Using the convention that every roots are normalized to length two for simply-laced root lattices $A_r$, $D_r$ and $E_r$, the first definition of folding leads to a new root with squared length $2/n$, hence short roots, while the second definition leads to a long root with squared length $2n$. They are related by roots and co-roots. Therefore, it is natural that the first folding of $D_{r}$ leads to $B_{r-1}$, while the second to $C_{r-1}$.

\subsection{Hilbert series with parity}\label{appsubsec:HilbertParity}

We now show how to compute the Hilbert series when including parity as a symmetry. For a general field $\Ph$, the main piece of the Hilbert series, $H_0$, is given by
\begin{equation}
H_0(\Ph,p) = \int d\m_{O(d)}^{} \, \text{det}_\Box^{}(1-p g) Z(\Ph,p,x) , \label{eqn:H0Odapp1}
\end{equation}
with the character generating function
\begin{equation}
Z(\Ph,p,x) = \prod_{n=0}^{\infty}\frac{1}{\text{det}_{l(n)}^{}(1 - \Ph p^{\D_\Ph+n} g)} \, , \label{eqn:ZPhiapp}
\end{equation}
where $l(n)$ denotes the $O(d)$ representation formed by $\partial^n\Ph$. As the group $O(d)$ is segmented into the parity even and odd pieces $O_\pm(d)$, this integral can be split accordingly:
\begin{subequations}\label{eqn:H0Odapp}
\begin{align}
H_0(\Ph,p) &= \frac{1}{2}\Big[H_{0,+}(\Ph,p) + H_{0,-}(\Ph,p)\Big] , \label{eqn:H0Odsplitapp} \\
H_{0,\pm}(\Ph,p) &= \int d\m_{\pm} \text{det}_\Box^{}(1-p g_\pm) Z_\pm , \label{eqn:H0pmapp}
\end{align}
\end{subequations}
where \(g_{\pm} \in O_{\pm}(d)\), \(Z_{\pm}=Z(g_{\pm})\), \(d\m_{\pm}=d\mu_{O_\pm(d)}^{}\) is the Haar measure normalized as \(\int d\m_{O(d)}^{}=\int d\m_{\pm} = 1\). The key ingredients we need to discuss are how to evaluate a determinant of the form $\text{det}_l^{}(1-a g_-)$ as well as the integration measure $d\mu_\pm$.

\subsubsection*{Evaluating $\text{det}_l^{}(1-a g_-)$}

The evaluation of this determinant is a bit subtle, because the usual plethystic exponential identity does not apply:
\begin{align}
\frac{1}{\text{det}_l(1-a g_-)} = \exp\left[ \sum_{n=1}^{\infty}\frac{a^n}{n} \text{Tr}_l^{}(g_-^n) \right] \ne \text{PE} \left[ a \ch_l^-(\tilde{x}) \right] . \label{eqn:naivePE}
\end{align}
The reason is that the trace evaluation for the parity odd element is nontrivial:
\begin{align}
\text{Tr}_l^{}(g_-^n) \ne \ch_l^-(\tilde{x}^n) , \label{eqn:naiveTrace}
\end{align}
as opposed to the usual paradigm, \textit{e.g.} $\text{Tr}_l^{}(g_+^n) = \ch_l^+(x^n)$. To see this, let us take a close look at the structure of $g_-=g_+\mathcal{P}$.

As explained before, in a general representation, the eigenvalues of $g_+$ are $x^\mu = x_1^{\mu_1}\cdots x_r^{\mu_r}$. The parity action exchanges the coordinate $x_r \leftrightarrow x_r^{-1}$, which corresponds to flipping the last component of the weights $\mu_r \leftrightarrow -\mu_r$. Therefore, all the weights $\mu$ fall into two categories, the ones invariant under parity $\mu^I$, and the ones paired up under parity $\mu_\pm^P$. The invariant weights must have $\mu_r^I=0$, while the paired up weights have the structure $\mu_\pm^P = (\mu_1^P,\ldots,\mu_{r-1}^P,\pm\mu_r^P)$. However, note that $\mu_r^P$ does not have to be nonzero.

For notational convenience, let\footnote{Although cumbersome, we are trying to be careful with our notation. Introducing $\overline{x}=(x_1,\ldots,x_{r-1},1)$ allows us to compare equations depending on the $r$ variables $x=(x_1,\ldots,x_r)$ with those depending on the $r-1$ variables $\tilde{x}=(x_1,\ldots,x_{r-1})$, \textit{e.g.} eq.~\eqref{eqn:GammaDefapp}. Note that $\overline{x}$ implements a restriction, \textit{i.e.} $\chi_l^+(\overline{x})=\text{Tr}(\text{Res}_{SO(2r-1)}^{SO(2r)} g_+)$.}
\begin{equation}
\overline{x} \equiv (x_1, \ldots, x_{r-1}, 1) ,
\end{equation}
then in the basis with $g_+$ diagonal, $g_-$ takes the form
\begin{align}\def\arraystretch{0.8}
g_- = g_+ \mathcal{P} &= \left(
\begin{array}{c:c}
\begin{array}{ccc}
\ddots & & \\
& \overline{x}^{\mu^I} & \\
& & \ddots
\end{array} &  \\ \hdashline
 & \begin{array}{ccc}
 \ddots & & \\
 & \overline{x}^{\mu_+^P} \left(\begin{array}{cc}
   x_r^{\mu_r^P} & 0 \\
   0 & x_r^{-\mu_r^P}
   \end{array}\right) & \\
 & & \ddots
   \end{array}
\end{array}\right)\left(
\begin{array}{c:c}
\begin{array}{ccc}
\ddots & & \\
& 1 & \\
& & \ddots
\end{array} &  \\ \hdashline
 & \begin{array}{ccc}
 \ddots & & \\
 & \left(\begin{array}{cc}
   0 & 1 \\
   1 & 0
   \end{array}\right) & \\
 & & \ddots
   \end{array}
\end{array}\right) \nonumber \\
&\mapsto \left(
\begin{array}{c:c}
\begin{array}{ccc}
\ddots & & \\
& \overline{x}^{\mu^I} & \\
& & \ddots
\end{array} &  \\ \hdashline
 & \begin{array}{ccc}
 \ddots & & \\
 & \overline{x}^{\mu_+^P} \left(\begin{array}{cc}
   1 & 0 \\
   0 & -1
   \end{array}\right) & \\
 & & \ddots
   \end{array}
\end{array}\right) \, , \label{eqn:gminusmatrix}
\end{align}
where in the second line we made a further diagonalization. Note that the dependence on \(x_r\) in $g_-$ is completely washed out, as expected. It is now clear that the mismatch in eq.~\eqref{eqn:naiveTrace} happens only for even powers:
\begin{align}\def\arraystretch{1.2}
\text{Tr}_l^{}(g_-^n) = \left\{\begin{array}{ll}
\ch_l^+(\overline{x}^n)=\ch_l^-(\tilde{x}^n)+2\G_l^{}(\tilde{x}^n) & \quad \text{for} \hspace{2mm} n=2k \\
\ch_l^-(\tilde{x}^n) & \quad \text{for} \hspace{2mm} n=2k+1
\end{array}\right. \, ,
\end{align}
where we have defined
\begin{equation}
\Gamma_l^{}(\tilde{x}) \equiv \frac{1}{2}\Big[\ch_l^+(\overline{x}) - \ch_l^-(\tilde{x}) \Big] . \label{eqn:GammaDefapp}
\end{equation}

The determinant then follows:
\begin{align}
\frac{1}{\text{det}_l^{}(1-a g_-)} &= \exp\left[ \sum_{n=1}^\infty \frac{a^n}{n} \text{Tr}_l^{}(g_-^n) \right] = \exp\left[\sum_{n=1}^\infty \frac{a^n}{n} \ch_l^-(\tilde{x}^n) + \sum_{k=1}^\infty \frac{a^{2k}}{2k} 2\G_l^{}(\tilde{x}^{2k}) \right] \nonumber \\
&= \text{PE} \Big[ a \ch_l^-(\tilde{x}) + a^2 \G_l^{}(\tilde{x}^2) \Big] . \label{eqn:parityDetapp}
\end{align}
Equation~\eqref{eqn:parityDetapp} is the central result for handling determinants on the parity odd component of \(O(2r)\). It is easily generalized to the fermionic case:
\begin{align}
\text{det}_l^{}(1+a g_-) &= \exp\left[\sum_{n=0}^\infty \frac{-(-a)^n}{n} \text{Tr}_l^{}(g_-^n) \right] = \exp\left[ -\sum_{n=1}^\infty \frac{(-a)^n}{n} \ch_l^-(\tilde{x}^n) - \sum_{k=1}^\infty \frac{a^{2k}}{2k} 2\G_l(\tilde{x}^{2k}) \right] \nonumber \\
&= \text{PE}_f\Big[ a \ch_l^-(\tilde{x})\Big]\text{PE}\Big[-a^2 \G_l^{}(\tilde{x}^2)\Big] . \label{eqn:parityDetFapp}
\end{align}

The vector representation determinant in eq.~\eqref{eqn:H0pmapp} can be easily obtained from eq.~\eqref{eqn:parityDetapp}, or even more straightforwardly from eq.~\eqref{eqn:parityEigenvaluesapp}:
\begin{equation}
\text{det}_\Box (1-pg_-) = (1-p^2) \prod_{i=1}^{r-1} (1-px_i)(1-p/x_i) = \frac{1-p^2}{P^{(2r-2)}(p;\tilde{x})} . \label{eqn:parityDetVectorapp}
\end{equation}

\subsubsection*{Integration measure}

The integral over \(O(2r)\) splits into two separate integrals over the components \(O_{\pm}(2r)\),
\begin{align}
\int d\m_{O(2r)}^{} f(g) &= \frac{1}{2}\int d\mu_+ f(g_+) + \frac{1}{2} \int d\mu_- f(g_-) \nonumber \\
&= \frac{1}{2} \int d\m_{SO(2r)}^{} f(g_+) + \frac{1}{2} \int d\m_{Sp(2r-2)}^{} f(g_-) \, ,
\end{align}
where the factors of \(1/2\) are from normalizing \(\int d\m = \int d\m_{\pm} = 1\). The measure on \(O_+(2r)\) is obviously that of \(SO(2r)\). The measure of \(O_-(2r)\) can be computed using the Weyl integration formula given in appendix~\ref{app:Weyl}. As the folded root system is that of $Sp(2r-2)$, the end result is simply $d\mu_-=d\mu_{Sp(2r-2)}$.

\subsection{Examples}\label{appsubsec:examples}

To compute the Hilbert series, one follows eq.~\eqref{eqn:H0Odapp}. From above, we know that
\begin{subequations}
\label{eqn:measuredet}
\begin{alignat}{3}
d\mu_+ &= d\mu_{SO(2r)}^{}(x) \quad&,&\quad  &\text{det}_\Box^{}(1-p g_+) &= \frac{1}{P^{(2r)}(p;x)} \, , \\
d\mu_- &= d\mu_{Sp(2r-2)}^{}(\tilde{x}) \quad&,&\quad  &\text{det}_\Box^{}(1-p g_-) &= \frac{1-p^2}{P^{(2r-2)}(p;\tilde{x})} \, .
\end{alignat}
\end{subequations}
Our task is to compute the functions $Z_\pm$ for some given field content. In the following, we show two explicit examples---the real scalar field in $d=2r$ and the gauge field in $d=4$ dimensions.

\subsubsection*{Real scalar field in $d=2r$}

Our first example is the real scalar field $\ph$, whose derivative $\partial^n\ph$ forms the representation $l(n)=(n,0,\ldots,0)$ of $SO(2r)$ and hence is also a representation of $O(2r)$. According to eq.~\eqref{eqn:Odcharacters}, the characters are
\begin{align}
\ch_{(n,0,\ldots,0)}^+(x) &= \ch_{(n,0,\ldots,0)}^{}(x) , \\
\ch_{(n,0,\ldots,0)}^-(\tilde{x}) &= \ch_{(n,0,\ldots,0)}^{Sp(2r-2)}(\tilde{x}) .
\end{align}
The $Z_+(\ph,p,x)$ function is given in eq.~\eqref{eqn:ZphiSOd}. Focusing on the $Z_-$ function, we use eq.~\eqref{eqn:parityDetapp}:
\begin{align}
Z_-(\ph,p,\tilde{x}) &= \prod_{n=0}^{\infty}\frac{1}{\text{det}_{(n,0,\ldots,0)}(1 - \ph p^{n+\D_0} g_-)} \nonumber \\
&= \text{PE}\left[\ph p^{\D_0} \sum_{n=0}^{\infty}p^n \ch_{(n,0,\ldots,0)}^-(\tilde{x}) + \ph^2p^{2\D_0}\sum_{n=0}^{\infty}p^{2n} \G_{(n,0,\ldots,0)}^{}(\tilde{x}^2) \right] . \label{eqn:parityZhalf}
\end{align}
The two sums are evaluated as follows:
\begin{subequations}
\begin{align}
\sum_{n=0}^{\infty}p^n \ch_{(n,0,\ldots,0)}^-(\tilde{x}) &= \sum_{n=0}^{\infty}p^n \ch_{(n,0,\ldots,0)}^{Sp(2r-2)}(\tilde{x})
= \sum_{n=0}^{\infty}p^n \ch_{\text{sym}^n(1,0,\ldots,0)}^{Sp(2r-2)}(\tilde{x}) = \text{PE}\left[p\ch_{(1,0,\ldots,0)}^{Sp(2r-2)}(\tilde{x})\right] \nonumber \\
&= \prod_{i=1}^{r-1}\frac{1}{(1-p x_i)(1-p/x_i)} = P^{(2r-2)}(p;\tilde{x}) \, , \label{eqn:sum1} \\
\sum_{n=0}^{\infty}p^n\G_{(n,0,\dots,0)}(\tilde{x}) &= \frac{1}{2} \left[ \sum_{n=0}^\infty p^n\ch_{(n,0,\dots,0)}^+(\overline{x}) - \sum_{n=0}^\infty p^n \ch_{(n,0,\dots,0)}^-(\tilde{x}) \right] \nonumber \\
&=\frac{1}{2} \left[ (1-p^2)P^{(2r)}(p;\overline{x}) - P^{(2r-2)}(p;\tilde{x}) \right] = \frac{p}{1-p}P^{(2r-2)}(p;\tilde{x}) . \label{eqn:sum2}
\end{align}
\end{subequations}
Eq.~\eqref{eqn:sum1} uses the $Sp(2r-2)$ representation relation $R_{(n,0,\ldots,0)}^{}=R_{\text{sym}^n(1,0,\ldots,0)}$ as well as $\ch_{(1,0,\ldots,0)}^{Sp(2r-2)}(\tilde{x}) = \sum_{i=1}^{r-1} (x_i+x_i^{-1})$. In the second equation, the first sum in the brackets is the scalar conformal character, eq.~\eqref{eq:chi_free_scalar_sum}, evaluated on $\overline{x}$. In the end we obtain
\begin{align}
Z_-(\ph,p,\tilde{x})= \text{PE}\Bigg[\ph p^{\D_0}P^{(2r-2)}(p;\tilde{x}) \, + \, \ph^2 \frac{p^{2\D_0+2}}{1-p^2} P^{(2r-2)}(p^2;\tilde{x}^2) \Bigg] . \label{eqn:parityZ}
\end{align}
Gathering it all together yields the final result for the parity odd piece of the Hilbert series
\begin{equation}
H_{0,-}^{(2r)}(\ph,p) = \int d\m_{Sp(2r-2)}^{} \frac{1-p^2}{P^{(2r-2)}(p;\tilde{x})} \text{PE}\Bigg[\ph p^{\D_0}P^{(2r-2)}(p;\tilde{x}) \, + \, \ph^2 \frac{p^{2\D_0+2}}{1-p^2} P^{(2r-2)}(p^2;\tilde{x}^2) \Bigg] . \label{eqn:H0mphidevenapp}
\end{equation}

\subsubsection*{Gauge fields in $d=4$}

As a second example, we consider gauge fields in four dimensions. In addition to its physical interest, this case is an example of how to address representations with $l_r \ne 0$.

Gauge fields are handled by working with their field strengths \(F_{\m\n}\) which are the rank 2 antisymmetric representation of \(SO(2r)\) corresponding to \(l = (1,1,0,\dots,0)\). In four dimensions we can split \(F_{\m\n}\) into chiral components by defining the chiral components $F_{L,R} = F \pm \widetilde{F}$,\footnote{In Minkowski signature there is a factor of \(i\), \(F_{L,R} = F \pm i\widetilde{F}\).} which transform as the $l=(1,\pm 1)$ representations of $SO(4)$. The conformal representations for $F_{L,R}$ contain derivatives $\partial^n F_{L,R}$ in the $(n+1,\pm 1)$ representation of $SO(4)$, whose direct sum gives the $O(4)$ representation $l(n)=(n+1,1)$. By eq.~\eqref{eqn:Odcharacters}, the characters are
\begin{align}
\ch_{(n+1,1)}^+(x_1,x_2) &= \ch_{(n+1,1)}^{}(x_1,x_2) + \ch_{(n+1,-1)}^{}(x_1,x_2) , \\
\ch_{(n+1,1)}^-(x) &= 0 .
\end{align}
Using these, we obtain for the parity even piece
\begin{align}
Z_+(F,p,x_1,x_2) = \text{PE} \left[ \sum_{n=0}^\infty F p^{2+n} \ch_{(n+1,1)}^+(x_1,x_2) \right] = \text{PE} \left[ F \left(\widetilde{\ch}_{[2;(1,1)]}^{} + \widetilde{\ch}_{[2;(1,-1)]}^{} \right) \right] , \nonumber
\end{align}
with the conformal characters
\begin{align}
\widetilde{\ch}^{}_{[2;(1,1)]} &= \ch_{[2;(1,1)]}^{} - \ch_{[3;(1,0)]}^{} + \ch_{[4;(0,0)]}^{} \nonumber \\
&= p^2 \left[ \left( x_1x_2 + 1 + \frac{1}{x_1x_2} \right) - p \left( x_1 + \frac{1}{x_1} + x_2 + \frac{1}{x_2} \right) + p^2 \right] P^{(4)}(p;x_1,x_2) , \nonumber \\
\widetilde{\ch}^{}_{[2;(1,-1)]} &= \ch_{[2;(1,-1)]}^{} - \ch_{[3;(1,0)]}^{} + \ch_{[4;(0,0)]}^{} \nonumber \\
&= p^2 \left[ \left( \frac{x_1}{x_2} + 1 + \frac{x_2}{x_1} \right) - p \left( x_1 + \frac{1}{x_1} + x_2 + \frac{1}{x_2} \right) + p^2 \right] P^{(4)}(p;x_1,x_2) . \nonumber
\end{align}
The $Z_-$ factor is computed using eq.~\eqref{eqn:parityDetapp}. Noting that
\begin{equation}
\G_{(n+1,1)}^{}(x) = \frac{1}{2} \left[ \ch_{(n+1,1)}^+(x,1) - \ch_{(n+1,1)}^-(x) \right] = \ch_{(n+1,1)}^{}(x,1) ,
\end{equation}
we have
\begin{align}
Z_-(F,p,x) &= \text{PE} \left\{ \sum_{n=0}^\infty \left[ F p^{2+n} \ch_{(n+1,1)}^-(x) + F^2 p^{2(2+n)} \G_{(n+1,1)}^{}(x^2) \right] \right\} \nonumber \\
&= \text{PE} \left[ F^2 \sum_{n=0}^\infty p^{2(2+n)} \ch_{(n+1,1)}^{}(x^2,1) \right] = \text{PE} \left[ F^2 \widetilde{\ch}_{[2;(1,1)]}^{}(p^2;x^2,1) \right] .
\end{align}
Combining the $Z_\pm$ results with eq.~\eqref{eqn:measuredet}, the Hilbert series for gauge fields in \(4d\) looks like
\begin{equation}
H_0(F,p) = \frac{1}{2}\Big[ H_{0,+}(F,p) + H_{0,-}(F,p) \Big]
\end{equation}
with\footnote{This is somewhat schematic; for example, if gauge field is non-abelian there are other pieces pertaining to the gauge group. See section~\ref{sec:mult_fields_symmetries}.}
\begin{align}
H_{0,+} &= \int d\m_{SO(4)}^{}(x_1,x_2)\,\frac{1}{P^{(4)}(p;x_1,x_2)}\text{PE}\Big[F \widetilde{\ch}_{[2;(1,1)]}^{}(p;x_1,x_2) + F \widetilde{\ch}_{[2;(1,-1)]}^{}(p;x_1,x_2)\Big] \, , \nonumber \\
H_{0,-} &= \int d\m_{Sp(2)}^{}(x)\,\frac{1-p^2}{P^{(2)}(p;x)} \text{PE}\Big[F^2\, \widetilde{\ch}_{[2;(1,1)]}^{}(p^2;x^2,1)\Big] \, . \nonumber
\end{align}

\newpage
\section{Computation of $\Delta H$}
\label{app:Delta_H}

In this appendix, we explicitly evaluate \(\D H(\{\Ph_i\},p)\) for general field content, \textit{i.e.} based on general generating representations of eq.~\eqref{eq:generating_reps}. Generalizing \(\D H\) from eq.~\eqref{eqn:DHresult} to the general field content, we have
\begin{subequations}
\label{eq:DHapp}
\begin{align}
\D H &= (-1)^{d+1}p^d +  \D H_1 + \D H_2 , \\
\D H_1 &= p^{\D_0+2} \int d\m_G^{}(y) \int d\m_{SO(d)}^{}(x) \left. \Big[Z(\{\Ph_i\},q,x,y)\Big]\right|_{q^{\D_0}} , \\
\D H_2 &= p^d \int d\m_G^{}(y) \int d\m_{SO(d)}^{}(x) \, \ch_{\Box}^{}(x) \left. \left[\frac{1}{P(q;x)} \Big(Z(\{\Ph_i\},q,x,y)-1\Big)\right] \right|_{q^{d-1}} ,
\end{align}
\end{subequations}
where our focus is obviously on $\D H_1$ and $\D H_2$.

For $\D H_1$, we need the coefficient of \(q^{\D_0}\) in \(Z(\{\Ph_i\},q,x,y)\). As \(\D_0\) is the minimum allowed scaling dimension, saturated by the scalars \(\ph_i\), a non-zero coefficient of \(q^{\D_0}\) in \(Z\) can only come from scalar fields. It is easy to see that
\begin{equation}
Z_{\ph_i}(\ph_i,q,x,y) = 1 + q^{\D_0}\ph_i \ch_{G,\ph_i}(y) + \scO(q^{\D_0+\frac{1}{2}}), \nonumber
\end{equation}
and hence
\begin{equation}
\D H_1 = p^{\D_0+2} \int d\m^{}_G \sum_i \ph_i \ch^{}_{G,\ph_i}(y) . \label{eqn:DH1}
\end{equation}
This is non-zero only if there are scalars which are singlets under \(G\).

The other term \(\D H_2\) can be evaluated by similar considerations. However, as the field content depends on the spacetime dimensionality---whether \(d\) is even or odd---so does the result. Furthermore, the results pertaining to fermions depend on \(r\) mod 4, due to various properties of spinors in \(SO(d)\).

For even dimensions, \(d = 2r\), the contribution of scalars \(\ph_i\), and the \(\frac{d}{2}\)-form field strengths \(F_a\) and \(\bar{F}_a\), are independent of \(r\). And we find
\begin{equation}
\D H_2^{(2r)} \supset p^d \int d\m^{}_G \Bigg[ \sum_{i<j} \ph_i \ph_j \ch^{}_{G,\ph_i}\ch^{}_{G,\ph_j} + \sum_i \ph_i^2 \wedge^2(\ch^{}_{G,\ph_i}) + (-1)^{r-1} \sum_a (F_a + \bar{F}_a) \ch^{}_{G,F_a} \Bigg] , \label{eq:DH_term_ii_2r_phi_and_F}
\end{equation}
In many instances, the scalar terms in~\eqref{eq:DH_term_ii_2r_phi_and_F} will survive the integral over \(G\): if \(\ph\) is in a complex or pseudo-real representation, then the theory necessarily contains \(\ph^{\dag}\). In this case it is guaranteed that one of the \(\ph_i\) in the first term of~\eqref{eq:DH_term_ii_2r_phi_and_F} is conjugate to \(\ph_j\), and hence the product \(\ch^{}_{G,\ph_i}\ch^{}_{G,\ph_j}\) will contain the singlet representation, \(\ch^{}_{G,\ph_i}\ch^{}_{G,\ph_j} = \ch^*_{G,\ph_j}\ch^{}_{G,\ph_j} \supset 1 + \cdots\).

Spinors in \(SO(2r)\) are chiral, and the properties of these irreps depend on \(r\) mod 4. Accounting for this we find:
\begin{equation}
\D H_2^{(2r)} \supset \left\{   \begin{array}{rc>{\displaystyle}{l}}
r \text{ mod }4 = 2,4: & ~~ & p^d \int d\m^{}_G \sum_{i,j} \ps_i\bar{\ps}_j\ch^{}_{G,\ps_i}\ch_{G,\bar{\ps}_j}, \\
r \text{ mod }4 = 3 ~~~: & & p^d \int d\m^{}_G \bigg[\sum_{i<j} \ps_i\ps_j\ch^{}_{G,\ps_i}\ch^{}_{G,\ps_j} + \sum_{i<j} \bar{\ps}_i\bar{\ps}_j\ch^{}_{G,\bar{\ps}_i}\ch^{}_{G,\bar{\ps}_j} \\
& & ~~~~+ \sum_i \ps_i^2 \text{sym}^2\big(\ch_{G,\ps_i}\big) + \sum_i \bar{\ps}_i^2 \text{sym}^2\big(\ch_{G,\bar{\ps}_i}\big) \bigg],\\
r \text{ mod }4 = 1 ~~~: & & p^d \int d\m^{}_G \bigg[\sum_{i<j} \ps_i\ps_j\ch^{}_{G,\ps_i}\ch^{}_{G,\ps_j} + \sum_{i<j} \bar{\ps}_i\bar{\ps}_j\ch^{}_{G,\bar{\ps}_i}\ch^{}_{G,\bar{\ps}_j} \\
& & ~~~~+ \sum_i \ps_i^2 \wedge^2\big(\ch_{G,\ps_i}\big) + \sum_i \bar{\ps}_i^2 \wedge^2\big(\ch_{G,\bar{\ps}_i}\big) \bigg].
\end{array} \right.
\end{equation}
Note that the difference between \(r \text{ mod } 4=3\) and \(r \text{ mod } 4=1\) is only in the last terms, whether they are symmetric or antisymmetric combinations of the \(\ch_{G,\ps_i}^{}\).

The situation for odd dimensions $d=2r+1$ is simpler and we find:
\begin{equation}
\D H_2^{(2r+1)} \supset \left\{   \begin{array}{rc>{\displaystyle}{l}}
r \text{ mod }4 = 1, 4: & ~~ & p^d \int d\m^{}_G \bigg[\sum_{i<j} \ps_i \ps_j \ch^{}_{G,\ps_i} \ch^{}_{G,\ps_j} + \sum_{i} \ps_i^2 \wedge^2\ch^{}_{G,\ps_i} \bigg], \\
r \text{ mod }4 = 2, 3: & ~~ & p^d \int d\m^{}_G \bigg[\sum_{i<j} \ps_i \ps_j \ch^{}_{G,\ps_i} \ch^{}_{G,\ps_j} + \sum_{i} \ps_i^2 \text{sym}^2\ch^{}_{G,\ps_i} \bigg]. \\
\end{array} \right.
\end{equation}

Each piece in $\D H_2$ has an intuitive meaning in terms of conserved currents. The $SO(d)$ integral weighted by $\chi_{(1,0,..,0)}$ projects out terms which correspond to the exterior derivative of a current that vanishes through EOM. For example, consider two real scalars $\phi_1$ and $\phi_2$ transforming trivially under $G$: from eq.~\eqref{eq:DH_term_ii_2r_phi_and_F} we find a contribution $\Delta H\supset p^d \phi_1\phi_2$. This corresponds to the operator $\partial^\mu j_\mu$, where $j_\mu=(\phi_1\partial_\mu\phi_2-\phi_2 \partial_\mu \phi_1)$ is a conserved current after use of the EOM. This operator is over-subtracted when accounting for integration by parts in $H_0$---it is precisely a term which, in the language of section~\ref{sec:non-linear}, comes from a non-exact form yet is zero through the EOM when an exterior derivative acts upon it.

\newpage
\section{Hilbert series for ${\mathbb C}[s_{12},s_{13},s_{14},s_{23},s_{24},s_{34}]^{S_4}$}
\label{app:EOMS4ring}

The consequences of $S_4$ invariance (acting on the indices of the Mandelstam invariants) on the ring in eq.~\eqref{eq:diffring},
\be
R^{\text{disting.}}_{\text{EOM}}&=&{\mathbb C}[s_{12},s_{13},s_{14},s_{23},s_{24},s_{34}]^{S_4} \,,
\label{eq:appring}
\ee
have been explored in Ref.~\cite{Aslaksen1996}, where a Hironaka decomposition (the form given below in eq.~\eqref{eq:hiro}) of the ring is presented, from which the Hilbert series eq.~\eqref{eq:diffhilb},
\be
H(t)=\frac{1+t^6+t^8+t^{10}+t^{12}+t^{18}} {\left(1-t^2\right) \left(1-t^4\right)^2 \left(1-t^6\right)^2 \left(1-t^8\right)} \,,
\ee
follows trivially.

Before reproducing this result, however, we mention two ways in which just the Hilbert series can be obtained straightforwardly without finding such a decomposition (to our knowledge, such decompositions are unknown for $n\ge 5$). The first way is to simply use our formalism for the Hilbert series, for the parity even case (in $d\ge n$ dimensions, to avoid Gram conditions), but exclude the `$1/P$' factor in the integrand that accounts for IBP/momentum conservation. A second way would be to compute matrices $M^x_{ab}$, $a,b=1,\ldots,n(n-1)/2$, which describe the action of each element  $x \in S_n$ on the $s_{ij}$, and then use Molien's formula,
\be
H(t)=\frac{1}{n!}\sum_{x=1}^{n!} \frac{1}{\text{det}(1-t^2 M^x)} \,.
\ee

We now turn to the Hironaka decomposition of eq.~\eqref{eq:appring}.
Defining
\be
g_2 &=& s_{12}s_{34}+s_{13}s_{24}+s_{14}s_{23 } \,,\nonumber \\
g_3 &=& s_{12}s_{23}s_{31}+s_{12}s_{24}s_{41}+s_{13}s_{34}s_{41}+s_{23}s_{34}s_{42} \,, \nonumber \\
g_4 &=& s_{12}s_{23}s_{34}s_{41}+s_{12}s_{24}s_{43}s_{31}+s_{13}s_{32}s_{24}s_{41}  \,,   \nonumber \\
h_3 &=& s_{12}s_{13}s_{14}+s_{12}s_{23}s_{24}+s_{23}s_{13}s_{34}+s_{24}s_{34}s_{14}  \,, \nonumber
\ee
and setting
\be
t_2=e_2-g_2\,,~~~t_3=e_3-g_3-h_3\,,~~~t_4=e_4-g_4 \,, \nonumber
\ee
where $e_i$ denotes the $i$th elementary symmetric polynomial in the variables $s_{ij}$,
a Hironaka decomposition of eq.~\eqref{eq:appring} is,
\be
{\mathbb C}[s_{12},s_{13},s_{14},s_{23},s_{24},s_{34}]^{S_4} = \bigoplus_{i=1}^6 f_i {\mathbb C}[y_1,\ldots,y_6] \,,
\label{eq:hiro}
\ee
with $y_i$ being $e_1$, $g_2$, $t_2$, $g_3$, $h_3$, $g_4$, and $f_i$ being $1$, $t_3$, $t_4$, $e_5$, $t_3^2$, $t_3^3$.

\newpage
\section{Counting helicity amplitudes}
\label{app:helicity_count}

In this appendix we re-derive the results of reference~\cite{Kravchuk:2016qvl} for determining the number of independent tensor structures in the amplitude decomposition eq.~\eqref{eq:amp_tens} (= number of helicity amplitudes~\cite{Jacob:1959at}). 

The basic setup is described in section~\ref{subsec:hilbert_spinning}, where we consider \(n\)-point configurations from fields \(\Ph_1,\dots,\Ph_n\). We recall that polarization tensors are transverse and are specified by the little group \(SO(d-2)\) for massless particles and \(SO(d-1)\) for massive particles. For a given set of fields \(\Ph_1,\dots,\Ph_n\), we let \(h_i\) be the number of allowed helicity states for the \(i\)th particle and use \(N_h^{(n)}\) to denote the number of independent tensor structures.

Following the logic of~\cite{Kravchuk:2016qvl}, we first fix a scattering frame by using the Poincar\'e symmetry to bring the momenta \(p_1^{\m},\dots,p_n^{\m}\) to some configuration, and then examine the helicity configurations in this frame. If there is some remaining symmetry which does not change the momenta, \textit{i.e.} stabilizes the scattering frame, then this may relate various helicity configurations. If no such symmetry exists, then the number of independent helicity amplitudes is simply \(N_h^{(n)} = N^{(n)}_{h,\text{max}} \equiv \prod_{i=1}^n h_i\). Otherwise, this number may be reduced.

We go case-by-case, beginning with the situation where there is no remaining symmetry and work our way up in complexity. For \(n\ge d\) we can obtain explicit formulas for \(N_h^{(n)}\) in terms of the \(h_i\). For \(n<d\) there is no general expression in terms of the \(h_i\), but \(N_h^{(n)}\) can be easily computed from a group integral, see eq.~\eqref{eq:Nh_SOm}.

\subsubsection*{$n>d>4$: no symmetry}
We begin with the case \(n > d > 4\). Because \(n>d\), one uses all the Poincar\'e generators to adjust the momenta of particles. For example, if we consider \(2 \to n-2\) scattering, we can rotate to the following scattering configuration\footnote{In words the steps are (\(M_{\m\n}\) are Lorentz generators): (1) use translation generators to set momentum conservation (\(p_1^{\m} + p_2^{\m} = p_3^{\m} + \cdots + p_n^{\m}\)); (2) use boosts \(M_{0\m}\) to get to the COM frame (\(p_1^{\m} + p_2^{\m} = (p_1^0+p_2^0,0,\cdots,0)\)); (3) use \(M_{1i}\), \(i>1\), to rotate \(p_3^{\m}\) into (12) plane; (4) use \(M_{2i}\), \(i>2\), to rotate \(p_4^{\m}\) into (123) plane; \textit{etc}. Note that this procedure leaves us with \(n(d-1) - d(d+1)/2\) independent variables, eq.~\eqref{eq:dim_modules}.}
\begin{equation}
\begin{pmatrix} p_1^0 \\ p \\ 0 \\\vdots\\ \\ 0\end{pmatrix} + \begin{pmatrix} p_2^0 \\ -p \\ 0 \\\vdots\\ \\ 0\end{pmatrix} = \begin{pmatrix} p_3^0 \\ p_3^1 \\ p_3^2\\ 0 \\\vdots \\ 0\end{pmatrix} + \cdots + \begin{pmatrix} p_{d-1}^0 \\ p_{d-1}^1 \\\vdots \\ \\ p_{d-1}^{d-1}\\ 0\end{pmatrix} + \begin{pmatrix} p_d^0 \\ p_d^1  \\ \\ \vdots \\ \\ p_d^d\end{pmatrix} + \cdots + \begin{pmatrix} p_n^0 \\ p_n^1 \\ \\ \vdots \\ \\ p_n^d\end{pmatrix}.
\label{eq:scat_config}
\end{equation}
Because we have used all of the Poincar\'e symmetry to achieve this configuration, there are no further operations that can be done to relate the various external polarization tensors. Hence, the number of helicity amplitudes is just \(N_{h,\text{max}}^{(n)} = \prod_{i=1}^nh_i\).
\begin{equation}
n>d>4: \ \ N_h^{(n)} = \prod_{i=1}^nh_i
\label{eq:num_hel_ngtd}
\end{equation}

\subsubsection*{$n=d>4$: parity}
When parity is a symmetry of the theory it can relate helicity configurations when \(n\le d\). Here we explain the case for \(n=d\), where parity is the only spacetime symmetry which stabilizes the scattering frame. The case of \(n<d\) is covered below (when \(n<d\) parity is part of a larger spacetime symmetry \(O(m)\) that stabilizes the scattering frame).

When \(n=d>4\), the last component of \(p_d^{\m}\) in eq.~\eqref{eq:scat_config} vanishes by momentum conservation. If the theory possesses \(O(d)\) symmetry, the parity element \(\text{diag}(1,\dots,1,-1)\) acts trivially on the configuration,\footnote{We are working back in Euclidean space, as we have done throughout this paper. In Minkowski space, the following results hold for applying both parity and time-reversal symmetry.} and therefore can be used to relate different polarization states. In the invariant description, we can form \(\e_{\m_1\dots\m_d}p_{i_1}^{\m_1}\cdots p_{i_d}^{\m_d}\) only for \(n>d\) (when \(n=d\) these vanish by momentum conservation). Therefore, for \(n\le d\), contractions with the \(\e\)-tensor necessarily involve polarization tensors. 

To determine independent amplitudes, we take orbits of the polarization configurations under the action of parity. As parity is a \(Z_2\) action, these orbits involve at most two different helicity configurations, so the number of \(O(d)\) invariant tensor structures is essentially \(N_{h,\text{max}}^{(n)}/2\).

More precisely, parity acts trivially only if the configuration is all spin-0 helicity states; moreover, there is at most one such configuration, and it exists if and only if \(N_{h,\text{max}}^{(n)}\) is odd (\textit{i.e.} all \(h_i\) odd). Hence, for \(n=d > 4\), the number of \(O(d)\) invariant tensor structures is \(N_{h,\text{max}}^{(n)}/2\) or \((N_{h,\text{max}}^{(n)} + 1)/2\) if \(N_{h,\text{max}}^{(n)}\) is even or odd, respectively.

\subsubsection*{$n\le 4$: permutation symmetry}
If we have a permutation group \(\Sigma_n \subseteq S_n\), there can be permutations that stabilize the momenta configuration and therefore can be used to relate polarization configurations. Call this permutation subgroup \(\Sigma_{n}^{\text{kin}} \subseteq \Sigma_n\). Recall that the Mandelstam invariants \(s_{ij}\), when subject to on-shell and momentum conservation, fill out the \(V_{(n-2,2)}\) representation of \(S_n\), eq.~\eqref{eq:sij_decomp_Sn}. The various cases to consider are:
\begin{itemize}
\item[\(n\le 3\):] On-shell and momentum conservation fix all Mandelstam invariants (\(V_{(n-2,2)}\) does not exist for \(n\le 3\)), so all \(\s \in S_{n}\) stabilize the configuration. Hence, \(\nobreak{\Sigma_n^{\text{kin}} = \Sigma_n \subseteq S_n}\) for \(n\le 3\).
\item[\(n=4\):] The \(V_{(2,2)}\) representation of \(S_4\) is isomorphic to the \(V_{(2,1)}\) representation of \(S_3\); this is just a fancy way of saying the usual \((s,t,u)\), \(s+t+u = \sum_im_i^2\), are permuted according to \(S_3\). In particular, the permutations 
\begin{equation}
S_4^{\text{kin}} \equiv \{(e),\ (12)(34),\ (13)(24),\ (14)(23)\} \in S_4
\label{eq:S4_kin_def}
\end{equation}
stabilize the kinematics. Any nontrivial subgroup of \(S_4^{\text{kin}}\) is a \(Z_2\) group, \textit{e.g.} \(\{(e),(13)(24)\}\). Therefore, \(\Sigma_4^{\text{kin}}\) matters only if all four particles are identical, \(\Sigma_4^{\text{kin}} = S_4^{\text{kin}}\), or if we have two pairs of identical particles.
\item[\(n>4\):] After fixing a momentum configuration, there are no \(\s \in S_n\) which stabilize this configuration. In other words, all \(\s \in S_n\) act non-trivially on \(V_{(n-2,2)}\), and therefore \(\Sigma_n^{\text{kin}} = 1\) is trivial for \(n>4\).
\end{itemize}
We determine the independent helicity configurations from the distinct orbits under \(\Sigma_n^{\text{kin}}\). For example, for \(n=4\) identical gauge bosons in \(d=4\), the orbit of \((++--)\) is \(\nobreak{\sum_{\s \in S_4^{\text{kin}}} \s\circ (++--) = 2(++--) + 2(--++)}\); in this way one finds seven distinct orbits for \(SO(4)\) and five orbits for \(O(4)\) when we include parity, as in table~\ref{tab:F_4pt}. Below we give a concrete formula for counting these orbits in general.

\subsubsection*{$n<d$: leftover rotational symmetry}
When \(n < d\), we can bring all the momenta into some hyperplane, leaving rotations in the orthogonal submanifold. It is not hard to see this is a \((d-n+1)\)-dimensional submanifold for \(n\ge 3\), \textit{e.g.} from fixing a scattering frame as in eq.~\eqref{eq:scat_config}. Hence, there is a remaining \(SO(m)\) symmetry, \(m \equiv d-n+1\).\footnote{For \(n=2\) it is a remaining \(SO(d-2)\) symmetry; for \(n=1\) it is the little group for the particle.}  

As polarizations are transverse to momenta, this \(SO(m)\) symmetry acts non-trivially on some of the polarization states. For example, picking a scattering frame for \(n=4\) the remaining \(SO(d-3)\) acts as:
\begin{equation}\label{eq:scat_polar_config}
\begin{tikzpicture}[baseline=(current  bounding  box.center)]
\node at (0,0) {\(\def\arraystretch{1.25}
\left(\begin{array}{c} p_1^0 \\ p \\ 0 \\ 0 \\ \vdots \\ 0\end{array}\right) \ + \ \left(\begin{array}{c} p_2^0 \\ -p \\ 0 \\ 0 \\\vdots\\ 0\end{array}\right) \ = \ \left(\begin{array}{c} p_3^0 \\ k_1 \\ k_2\\ 0 \\\vdots \\ 0\end{array}\right) \ + \ \left(\begin{array}{c} p_4^0 \\ -k_1 \\ -k_2\\ 0 \\ \vdots \\ 0\end{array}\right)
\)};
\draw[blue,dashed,opacity=.7] (-3.8,-.13) -- (3.8,-.13) -- (3.8,-2.) -- (-3.8,-2.) -- (-3.8,-.13);
\draw[magenta,dashed,opacity=.7] (-3.8,.15) -- (3.8,.15) -- (3.8,1.18) -- (-3.8,1.18) -- (-3.8,.15);
\draw[decoration={brace,mirror,amplitude=5pt},decorate,blue]
  (-4.,-.13) -- (-4.,-2);
\node[blue] at (-5.5,-.8) {\footnotesize \(SO(d-3)\) acts};
\node[blue] at (-5.5,-1.2) {\footnotesize on polarizations};
\node[blue] at (-5.5,-1.6) {\footnotesize living here};
\draw[decoration={brace,amplitude=5pt},decorate,magenta]
  (4.,1.18) -- (4.,.15);
\node[magenta] at (5.5,1.1) {\footnotesize polarizations};
\node[magenta] at (5.5,0.7) {\footnotesize living here are};
\node[magenta] at (5.5,0.3) {\footnotesize \(SO(d-3)\) singlets};
\end{tikzpicture}
\end{equation}
In order for the amplitude to be Lorentz invariant, the polarization configurations need to be invariant under \(SO(m)\) (and a possible \(\Sigma_n^{\text{kin}}\)). 

The number of such invariants is counted with our favorite routine: character orthogonality. The idea is to take the representation of a particle under its little group, restrict to the subgroup of \(SO(m)\) rotations---as visualized in the above equation---and impose \(SO(m)\) invariance for the \(n\) particles tensored together.

Let \(G_i^{\text{lit}}\) be the relevant little group for the \(i\)th particle, \(V_i\) its representation under \(G_i^{\text{lit}}\), and \(\ze_i \equiv \text{Tr}_{V_i}^{}(g)\) the character of \(V_i\). We will typically denote this as \(\ze_i(x)\) where \(x = (x_1,\dots,x_s)\) parameterize the torus of \(G_i^{\text{lit}}\), \(s = \text{rank}(G_i^{\text{lit}})\). Note that \(h_i = \ze_i(1)\).

For example, in \(d=5\) the characters for a graviton and a massive vector---little groups \(SO(3)\) and \(SO(4)\), respectively---are 
\begin{subequations}\label{eq:lit_group_char_exam}
\begin{align}
d=5 \text{ graviton: ~~~~~} \ze(x_1) &= ~~~\ch_{(2)}^{(3)}(x_1) ~~~= x_1^2 + x_1^{} + 1 + x_1^{-1} + x_1^{-2}, \\
d=5 \text{ massive vector: ~} \ze(x_1,x_2) &= \ch_{(1,0)}^{(4)}(x_1,x_2) = x_1^{} + x_1^{-1} + x_2^{} + x_2^{-1}.
\end{align}
\end{subequations}
We remind the reader that \(\ch_l^{(n)}(x)\) is our usual notation for \(SO(n)\) characters, app.~\ref{app:sodspin} (also, if needed, \(O(n)\) characters are explained in app.~\ref{app:parity} and sec.~\ref{subsec:parity}).

\(SO(m)\) only acts on a subspace of \(V_i\), \textit{e.g.} eq.~\eqref{eq:scat_polar_config}. Let \(\xoverline{V}_i\) denote the \(SO(m)\) representation obtained by restricting the group action to this subspace,
\begin{equation}
\xoverline{V}_i \equiv \text{Res}^{G_i^{\text{lit}}}_{SO(m)}V_i \, .
\label{eq:restrict_Vi}
\end{equation}
The character \(\text{Tr}_{\overline{V}_i}(g)\) is readily obtained from \(\ze_i(x_1,\dots,x_s)\) by setting the appropriate \(x_i = 1\). Specifically, embedding \(SO(m)\) into \(G_i^{\text{lit}}\) so that 
\(x_1,\dots,x_r\) are coordinates for the torus of \(SO(m)\), then the restriction is obtained by setting \(x_{r+1}=\cdots = x_s = 1\) (\textit{i.e.} setting \(\th_i = 0\) in \(x_i = e^{i\th_i}\), \(i = r+1,\dots, s\)). Introducing the shorthand notation \(\xoverline{x} \equiv (x_1,\dots,x_r,1,\dots,1)\), the relevant character is given by
\begin{equation}
\ze_i(\xoverline{x}) \equiv \ze_i(x_1,\dots,x_r,1,\dots,1).
\label{eq:restrict_char}
\end{equation}
For example, if \(d=5\) and \(n=4\) then restricting the characters in eq.~\eqref{eq:lit_group_char_exam} to \(SO(2)\) gives
\begin{subequations}\label{eq:restrict_char_exam}
\begin{align}
\text{Restrict }d=5 \text{ graviton to }SO(2):\  \ze(\xoverline{x}) &= x_1^2 + x_1^{} + 1 + x_1^{-1} + x_1^{-2}, \\
\text{Restrict }d=5 \text{ massive vector to }SO(2): \ \ze(\xoverline{x}) &=  x_1^{} + 2 + x_1^{-1}.
\end{align}
\end{subequations}

We are now in a position to write down the number of independent polarization configurations. If the kinematic permutation group \(\Sigma_n^{\text{kin}}\) is trivial, then we simply multiply all the restricted characters, eq.~\eqref{eq:restrict_char}, together and average over \(SO(m)\):
\begin{equation}
\boxed{N_h^{(n)} = \int d\m_{SO(d-n+1)}^{} \ze_1(\xoverline{x}) \cdots \ze_n(\xoverline{x}).} 
\label{eq:Nh_SOm}
\end{equation}
(See eq. 5.8 in~\cite{Kravchuk:2016qvl}.) If parity is a symmetry, then the group integral is over \(O(m)\).

In fact, as is hopefully clear, the present discussion actually applies to all cases, not just \(n<d\). When \(n\ge d\) there is no remaining rotational symmetry, \textit{i.e.} \(SO(m)\) is trivial.\footnote{In a very particle physics language, the momentum configuration in eq.~\eqref{eq:scat_config} ``higgses'' or ``breaks'' the entire Poincar\'e symmetry, while when \(n<d\) the scattering frame, \textit{e.g.} eq~\eqref{eq:scat_polar_config}, breaks \(\text{Poincar\'e} = SO(d-1,1)\ltimes T^d \to SO(d-n+1)\).} In particular, the restricted characters in eq.~\eqref{eq:restrict_char} become \(\ze_i(1) = h_i\) and the integral in eq.~\eqref{eq:Nh_SOm} becomes trivial,
\begin{equation}
\int d\m_{SO(m)}^{}\ze_1(\xoverline{x})\cdots \ze_n(\xoverline{x}) \to \ze_1(1) \cdots \ze_n(1) = \prod_{i=1}^nh_i \, ,\nonumber
\end{equation}
up to a possible average over the parity action when \(n=d\).

If \(\Sigma_n^{\text{kin}}\) is non-trivial we do the following. Recall that taking the symmetric product of a representation, \(\text{sym}^nV\), is a recipe for symmetrizing indices. For example, if \(\r(g)\), \(g \in G\), is the representation matrix acting on \(V\) as \(v_i \to \r_{ij}(g)v_j\), then the appropriate representation acting on \(\text{sym}^2V\) is \(\frac{1}{2!} \sum_{\s \in S_2} \r_{i_1j_{\s(1)}}\r_{i_2j_{\s(2)}} = \frac{1}{2}(\r_{i_1j_1}\r_{i_2j_2} + \r_{i_1j_2}\r_{i_2j_1})\). Taking the trace by contracting with \(\d^{i_1j_1}\d^{i_2j_2}\), we obtain the familiar formula \(\text{Tr}_{\text{sym}^2V}(g) = \frac{1}{2}\big(\text{Tr}_V(g)^2 + \text{Tr}_V(g^2)\big)\), which has usually been phrased in this work as an action on the character \(\text{sym}^2[\chi(x)] = \frac{1}{2}\big(\chi(x)^2 + \chi(x^2)\big)\).

The point of reviewing the symmetric products is that it is the exact same procedure to take tensor products with any type of permutation symmetry. In particular, for \(n=4\) the possible non-trivial permutation groups are \(S_4^{\text{kin}}\) in eq.~\eqref{eq:S4_kin_def} if all particles are identical or \(\Sigma_4^{\text{kin,pairs}} \equiv \{(e),(13)(24)\}\) if two pairs of particles are identical (here we picked the pairs to be 1,3 and 2,4). Taking the tensor product with, for example, \(S_4^{\text{kin}}\) gives the character
\begin{equation}
\d^{i_1j_1}\cdots\d^{i_4j_4} \frac{1}{4} \sum_{\s \in S_4^{\text{kin}}} \r_{i_1j_{\s(1)}}\cdots \r_{i_4j_{\s(4)}} = \frac{1}{4}\Big( \text{Tr}(\r)^4 + 3 \text{Tr}(\r^2)^2\Big).
\end{equation}
Extending this to an action on functions, \(S_4^{\text{kin}}[f(x)]\), we have for the characters
\begin{equation}
S_4^{\text{kin}}\big[\ze(\xoverline{x})\big] = \frac{1}{4}\Big(\ze(\xoverline{x})^4 + 3\ze(\xoverline{x}^2)^2\Big).
\label{eq:S4_kin_char1}
\end{equation}
(One may wish to compare this to formula for the fourth symmetric product, eq.~\eqref{eq:sym4}.) Similarly, for \(\Sigma_4^{\text{kin,pairs}}\):
\begin{equation}
\Sigma_4^{\text{kin,pairs}}\big[\ze_1(\xoverline{x}),\ze_2(\xoverline{x})\big] = \frac{1}{2}\Big(\ze_1(\xoverline{x})^2\ze_2(\xoverline{x})^2 + \ze_1(\xoverline{x}^2)\ze_2(\xoverline{x}^2)\Big).
\label{eq:S4_kin_char2}
\end{equation}

To obtain the appropriate counting for these cases, we average eqs.~\eqref{eq:S4_kin_char1} and~\eqref{eq:S4_kin_char2} over \(SO(d-3)\), as in eq.~\eqref{eq:Nh_SOm}. That is,
\begin{equation}
\boxed{n=4, \text{ all identical: } N_4^{(h)} = \int d\m_{SO(d-3)}^{} \frac{1}{4}\Big(\ze(\xoverline{x})^4 + 3\ze(\xoverline{x}^2)^2\Big)\,}\, ,
\label{eq:n4_all_ident}
\end{equation}
and
\begin{equation}
\boxed{n=4, \text{two pairs identical: } N_4^{(h)} = \int d\m_{SO(d-3)}^{} \frac{1}{2}\Big(\ze_1(\xoverline{x})^2\ze_2(\xoverline{x})^2 + \ze_1(\xoverline{x}^2)\ze_2(\xoverline{x}^2)\Big)\,} \, .
\label{eq:n4_two_pair}
\end{equation}

We have been a bit cavalier in the above by ignoring the underlying statistics of the particles. For \(n=4\) this turns out to okay and the results in eqs.~\eqref{eq:n4_all_ident} and~\eqref{eq:n4_two_pair} hold whether the particles are bosons or fermions, but for \(n=3\) one has to be more careful. We refer to~\cite{Kravchuk:2016qvl} for details.

\subsection*{$d=4$ cases}
Let us enumerate the cases in \(d = 4\). We assume that \(\Ph_1,\dots,\Ph_n\) contains at least one particle with spin (if they are all scalars, then trivially the number of tensor structures is just one). The little group for massless particles is \(SO(2)\), so that all massless \(\Ph_i\) have two polarization states.\footnote{Really, it is the covering group \(U(1)\), but CPT ensures that all one-dimensional representations come in pairs.} The little group for massive particles is \(SO(3)\), so that a massive \(\Ph_i\) with spin \(l_i\) has \(h_i = 2l_i +1\). If \(n>4\) we simply have
\begin{equation}
N_{n>4}^{(h)} = \prod_{i=1}^n h_i \,.
\end{equation}
If \(n=4\), we need to distinguish the cases when parity and/or \(\Sigma_4^{\text{kin}}\) is non-trivial:
\begin{description}
\item[(1) No parity, \(\Sigma_4^{\text{kin}} = 1\):] \(N_4^{(h)} = N_{4,\text{max}}^{(h)} = \prod_{i=1}^4h_i\). 
\item[(2) With parity, \(\Sigma_4^{\text{kin}} = 1\):] As discussed above, the answer is simply \(N_{4,\text{max}}^{(h)}/2\) or \((N_{4,\text{max}}^{(h)}+1)/2\) depending on whether \(N_{4,\text{max}}^{(h)}\) is even or odd. We can also derive this result from eq.~\eqref{eq:Nh_SOm} with \(O(n-d+1)  = O(1) = \{1,\mathcal{P}\}\). Here, \(\{1,\mathcal{P}\}\) are the identity and parity elements restricted from their representations in the little groups \(O(2)\) or \(O(3)\). The integral in eq.~\eqref{eq:Nh_SOm} simply becomes a sum over these two elements,
\begin{equation}
\frac{1}{2}\sum_{g \in \{1,\mathcal{P}\}}\ze_1(g)\cdots \ze_4(g) = \frac{1}{2}\left(\prod_{i=1}^4h_i + \ze_1(\mathcal{P})\cdots\ze_4(\mathcal{P}) \right).
\label{eq:n=4_d=4_parity}
\end{equation}
For all massless particles with spin, parity in \(O(2)\) represents as \(\text{diag}(1,-1)\), which is remains the same upon restricting to \(O(1)\). For \(O(3)\) we can take parity to represent on the vector as \(\text{diag}(-1,-1,-1)\) (whose restriction to \(O(1)\) is \(\text{diag}(1,1,-1)\)). It is easy to see that for any integer spin \(\ze_i(\mathcal{P}) = 1\), while for any half-integer spin \(\ze_i(\mathcal{P}) = 0\). Therefore, the second term in eq.~\eqref{eq:n=4_d=4_parity} is non-zero only if all particles with spin are massive and have integer spin, in which case \(\ze_1(\mathcal{P})\cdots\ze_4(\mathcal{P}) = 1\) and we recover the stated result.
\item[(3) No parity, \(S_4^{\text{kin}}\):] We use eq.~\eqref{eq:n4_all_ident} with the integral being trivial. We have \(\ze(1) = \ze(1^2) = h\) (recall that \(\ze_i(\xoverline{x}^2)\) means \(\text{Tr}_{\overline{V}_i}(g^2)\)). Therefore we obtain \(N_4^{(h)}= \frac{h^2}{4}(h^2 + 3)\). If the four particles are massless this gives \(N_4^{(h)}= 7\), as in table~\ref{tab:F_4pt}.
\item[(4) With Parity, \(S_4^{\text{kin}}\):] Analogous to cases (2) and (3) above, we obtain \(N_{4}^{(h)} = \frac{h^2}{8}(h^2 + 6)\), which reproduces the entry in table~\ref{tab:F_4pt}.
\item[(5) No parity, \(\Sigma_4^{\text{kin,pairs}}\):] \(N_4^{(h)} = \frac{1}{2}h_1h_2(h_1h_2 + 1)\). 
\item[(6) With parity, \(\Sigma_4^{\text{kin,pairs}}\):] \(N_4^{(h)} = \frac{1}{4}h_1h_2(h_1h_2 + 2)\). 
\end{description}

\newpage
\section{EOM for non-linear realizations}
\label{app:eom_nonlin}

In this appendix, we show that the EOM (at the leading order of the EFT) in non-linear representation is
\begin{equation}
D^\mu u_\mu =0 .
\end{equation}
At the leading order, the non-linear Lagrangian is
\begin{equation}
{{\cal L}_K} = f_\pi ^2{\rm{tr}}\left( {{D^\mu }{\xi ^{ - 1}}{D_\mu }\xi } \right) =  - f_\pi ^2{\rm{tr}}\left( {{u^\mu }{u_\mu }} \right) .
\end{equation}
Let us consider a variation in $\xi$ (which is of course induced by variations in $\pi$ fields): $\xi \to \xi + \delta \xi$. This variation will result in a variation in $u_\mu = {\xi ^{ - 1}}{D_\mu }\xi$ as
\begin{align}
\delta {u_\mu } &= \left( {\delta {\xi ^{ - 1}}} \right){D_\mu }\xi  + {\xi ^{ - 1}}{D_\mu }\left( {\delta \xi } \right) =  - {\xi ^{ - 1}}\left( {\delta \xi } \right){\xi ^{ - 1}}{D_\mu }\xi  + {\xi ^{ - 1}}{D_\mu }\left( {\delta \xi } \right) .
\end{align}
It is straightforward to compute the variation of $\mathcal{L}_K$:
\begin{align}
\delta {{\cal L}_K} &=  - 2f_\pi ^2{\rm{tr}}\left( {{u^\mu }\delta {u_\mu }} \right) \nonumber \\
 &=  - 2f_\pi ^2{\rm{tr}}\left\{ {{u^\mu }\left[ { - {\xi ^{ - 1}}\left( {\delta \xi } \right){\xi ^{ - 1}}{D_\mu }\xi  + {\xi ^{ - 1}}{D_\mu }\left( {\delta \xi } \right)} \right]} \right\} \nonumber \\
 &= 2f_\pi ^2{\rm{tr}}\left[ {{u^\mu }{\xi ^{ - 1}}\left( {\delta \xi } \right){u_\mu } + {D_\mu }\left( {{u^\mu }{\xi ^{ - 1}}} \right)\delta \xi } \right] \nonumber \\
 &= 2f_\pi ^2{\rm{tr}}\left[ {{u_\mu }{u^\mu }{\xi ^{ - 1}}\left( {\delta \xi } \right) + {u^\mu }\left( {{D_\mu }{\xi ^{ - 1}}} \right)\delta \xi  + \left( {{D_\mu }{u^\mu }} \right){\xi ^{ - 1}}\delta \xi } \right] \nonumber \\
 &= 2f_\pi ^2{\rm{tr}}\left[ {{u_\mu }{u^\mu }{\xi ^{ - 1}}\left( {\delta \xi } \right) - {u^\mu }{\xi ^{ - 1}}\left( {{D_\mu }\xi } \right){\xi ^{ - 1}}\delta \xi  + \left( {{D_\mu }{u^\mu }} \right){\xi ^{ - 1}}\delta \xi } \right] \nonumber \\
 &= 2f_\pi ^2{\rm{tr}}\left[ {\left( {{D_\mu }{u^\mu }} \right){\xi ^{ - 1}}\delta \xi } \right] .
\end{align}
The EOM is the requirement that for any variation $\delta\xi$, we have $\delta\mathcal{L}_K=0$. Since $\xi^{-1}\delta\xi$ span the space of $\mathfrak{g}/\mathfrak{h}$, we conclude that the EOM is
\begin{equation}
{D_\mu }{u^\mu } = 0 .
\end{equation}

\bibliography{./bibliography}
\bibliographystyle{JHEP}

\end{document}